%% file: thesismain.tex
\documentclass[12pt,beltcrest]{ociamthesis} % use old belt crest logo
\pdfoutput=1

\usepackage{hyperref}
\usepackage{doi}
\usepackage[normalem]{ulem} %added by Harny to remove underline in references
\usepackage{enumerate}

\usepackage{amssymb,nicefrac}
\usepackage{amsfonts}
\usepackage{graphicx}
\usepackage[fleqn]{amsmath}
\usepackage{amsmath}
\usepackage[varg]{txfonts}
\usepackage{stmaryrd}

\usepackage{framed}
\usepackage{color}
\usepackage{ulem}
\usepackage{tikzfig}

\usepackage{turnstile}

\input{tikzstyles.tex}

\input{defs.tex}

\usepackage{placeins}
\usepackage{mathtools}
\newtheorem{Th}{Theorem}[section]
\newtheorem{theorem}[Th]{Theorem}
\newtheorem{proposition}[Th]{Proposition}
\newtheorem{lemma}[Th]{Lemma}
\newtheorem{corollary}[Th]{Corollary}
\newtheorem{definition}[Th]{Definition}

\newtheorem{remark}[Th]{Remark}

\newenvironment{proof}{\textbf{Proof:}}{\hfill$\Box$\newline}

\title{Completeness of the ZX-calculus}   %note \\[1ex] is a line break in the title
\author{Quanlong Wang}             %your name
\college{Wolfson College}  %your college

\degree{Doctor of Philosophy}     %the degree
\degreedate{Hilary 2018}         %the degree date

\begin{document}

%this baselineskip gives sufficient line spacing for an examiner to easily
%markup the thesis with comments
\baselineskip=18pt plus1pt

%set the number of sectioning levels that get number and appear in the contents
\setcounter{secnumdepth}{3}
\setcounter{tocdepth}{3}

\maketitle                  % create a title page from the preamble info
\include{acknowlegements}   % include an acknowledgements.tex file
\include{abstract}          % include the abstract

\begin{romanpages}          % start roman page numbering
\tableofcontents            % generate and include a table of contents
%\listoffigures                    % generate and include a list of figures
\end{romanpages}            % end roman page numbering

%now include the files of latex for each of the chapters etc
\include{chapter1}

\include{chapterbk}

\include{chapter3}

\include{chapter4}

\include{chapter5}

\include{chapter2}

\include{chapter6}

\include{conclusions}

%now enable appendix numbering format and include any appendices
%\appendix
%\include{appendix1}
%\include{appendix2}

%next line adds the Bibliography to the contents page
\addcontentsline{toc}{chapter}{Bibliography}
%uncomment next line to change bibliography name to references
%\renewcommand{\bibname}{References}
\bibliography{thesismain}        %use a bibtex bibliography file refs.bib
\bibliographystyle{plain}  %use the plain bibliography style
 
\end{document}

%% file: tikzstyles.tex
\tikzstyle{env}=[copoint,regular polygon rotate=0,minimum width=0.2cm, fill=black]

\tikzstyle{every picture}=[baseline=-0.25em]
\tikzstyle{dotpic}=[scale=0.5]
\tikzstyle{diredges}=[every to/.style={diredge}]
\tikzstyle{dot graph}=[shorten <=-0.1mm,shorten >=-0.1mm,scale=0.6]
\tikzstyle{plot point}=[circle,fill=black,minimum width=2mm,inner sep=0]

\tikzstyle{braceedge}=[decorate,decoration={brace,amplitude=2mm,raise=-1mm}]
\tikzstyle{small braceedge}=[decorate,decoration={brace,amplitude=1mm,raise=-1mm}]
\tikzstyle{left hook arrow}=[left hook-latex]
\tikzstyle{right hook arrow}=[right hook-latex]

\tikzstyle{dtriangle}=[fill=yellow,draw=black,shape=isosceles triangle,shape border rotate=-90,isosceles triangle stretches=true,inner sep=1pt,minimum width=0.4cm,minimum height=3mm]
\tikzstyle{vtriang}=[fill=yellow,draw=black,shape=isosceles triangle,shape border rotate=180,isosceles triangle stretches=true,inner sep=1pt,minimum width=0.4cm,minimum height=3mm]
\tikzstyle{triangle}=[fill=yellow,draw=black,shape=isosceles triangle,shape border rotate=90,isosceles triangle stretches=true,inner sep=1pt,minimum width=0.4cm,minimum height=3mm]
\tikzstyle{H box}=[rectangle,fill=yellow,draw=black,xscale=1.0,yscale=1.0, inner sep=1.pt]
\tikzstyle{gbox}=[rectangle,fill=green,draw=black,xscale=1.0,yscale=1.0, inner sep=1.pt]
\tikzstyle{rbox}=[rectangle,fill=red,draw=black,xscale=1.0,yscale=1.0, inner sep=1.pt]

\tikzstyle{bn}=[circle,fill=black,draw=black,scale=.4]
\tikzstyle{wn}=[circle,fill=white,draw=black,scale=.6]
\tikzstyle{dn}=[circle,fill=none,draw=gray]

\tikzstyle{black dot}=[inner sep=0.7mm,minimum width=0pt,minimum height=0pt,fill=black,draw=black,shape=circle]

\tikzstyle{dot}=[black dot]
\tikzstyle{smalldot}=[inner sep=0.4mm,minimum width=0pt,minimum height=0pt,fill=black,draw=black,shape=circle]%NEW
\tikzstyle{white dot}=[dot,fill=white]
\tikzstyle{antipode}=[white dot,inner sep=0.3mm,font=\footnotesize]
\tikzstyle{smallwhitedot}=[smalldot,fill=white]%NEW
\tikzstyle{alt white dot}=[white dot,label={[xshift=3.07mm,yshift=-0.05mm,font=\footnotesize]left:$*$}]
\tikzstyle{gray dot}=[dot,fill=gray!40!white]
\tikzstyle{smallgraydot}=[smalldot,fill=gray!40!white]%NEW
\tikzstyle{box vertex}=[draw=black,rectangle]
\tikzstyle{small box}=[box vertex,fill=white]%% added rwd]
\tikzstyle{whitebg}=[fill=white,inner sep=2pt]
\tikzstyle{graph state vertex}=[sg vertex,fill=black]

\tikzstyle{wide copoint}=[fill=white,draw=black,shape=isosceles triangle,shape border rotate=90,isosceles triangle stretches=true,inner sep=1pt,minimum width=1.5cm,minimum height=5mm]
\tikzstyle{wide point}=[fill=white,draw=black,shape=isosceles triangle,shape border rotate=-90,isosceles triangle stretches=true,inner sep=1pt,minimum width=1.5cm,minimum height=4mm]
\tikzstyle{very wide copoint}=[fill=white,draw=black,shape=isosceles triangle,shape border rotate=-90,isosceles triangle stretches=true,inner sep=1pt,minimum width=2.5cm,minimum height=4mm]
\tikzstyle{very wide empty copoint}=[draw=black,shape=isosceles triangle,shape border rotate=-90,isosceles triangle stretches=true,inner sep=1pt,minimum width=2.5cm,minimum height=4mm]
\tikzstyle{symm}=[ultra thick,shorten <=-1mm,shorten >=-1mm]

\tikzstyle{square box}=[rectangle,fill=white,draw=black,minimum height=5mm,minimum width=5mm,font=\small]
\tikzstyle{square gray box}=[rectangle,fill=gray!30,draw=black,minimum height=6mm,minimum width=6mm]
\tikzstyle{copoint}=[regular polygon,regular polygon sides=3,draw=black,scale=0.75,inner sep=-0.5pt,minimum width=7mm,fill=white]
\tikzstyle{point}=[regular polygon,regular polygon sides=3,draw=black,scale=0.75,inner sep=-0.5pt,minimum width=7mm,fill=white,regular polygon rotate=180]
\tikzstyle{gray point}=[point,fill=gray!40!white]
\tikzstyle{gray copoint}=[copoint,fill=gray!40!white]

\newcommand{\edgearrow}{{\arrow[black]{>}}}
\newcommand{\edgetick}{{\arrow[black,scale=0.7,very thick]{|}}}
\tikzstyle{diredge}=[->]
\tikzstyle{rdiredge}=[<-]
\tikzstyle{medium diredge}=[->]

\tikzstyle{short diredge}=[->]
\tikzstyle{halfedge}=[-)]
\tikzstyle{other halfedge}=[(-]
\tikzstyle{freeedge}=[(-)]
\tikzstyle{white edge}=[line width=5pt,white]
\tikzstyle{tick}=[postaction=decorate,decoration={markings, mark=at position 0.5 with \edgetick}]
\tikzstyle{small map edge}=[|-latex, gray!60!blue, shorten <=0.9mm, shorten >=0.5mm]
\tikzstyle{thick dashed edge}=[very thick,dashed,gray!40]
\tikzstyle{map edge}=[|-latex,very thick, gray!40, shorten <=1mm, shorten >=0.5mm]
\tikzstyle{tickedge}=[postaction=decorate,
  decoration={markings, mark=at position 0.5 with \edgetick}]
\tikzstyle{dirtickedge}=[postaction=decorate,
  decoration={markings, mark=at position 0.5 with \edgetick},
  decoration={markings, mark=at position 0.85 with \edgearrow}]
\tikzstyle{dirdoubletickedge}=[postaction=decorate,
  decoration={markings, mark=at position 0.4 with \edgetick},
  decoration={markings, mark=at position 0.6 with \edgetick},
  decoration={markings, mark=at position 0.85 with \edgearrow}]

\makeatletter
\newcommand{\boxshape}[3]{%
\pgfdeclareshape{#1}{
\inheritsavedanchors[from=rectangle] % this is nearly a rectangle
\inheritanchorborder[from=rectangle]
\inheritanchor[from=rectangle]{center}
\inheritanchor[from=rectangle]{north}
\inheritanchor[from=rectangle]{south}
\inheritanchor[from=rectangle]{west}
\inheritanchor[from=rectangle]{east}
\backgroundpath{% this is new
\southwest \pgf@xa=\pgf@x \pgf@ya=\pgf@y
\northeast \pgf@xb=\pgf@x \pgf@yb=\pgf@y

\@tempdima=#2
\@tempdimb=#3

\pgfpathmoveto{\pgfpoint{\pgf@xa - 5pt + \@tempdima}{\pgf@ya}}
\pgfpathlineto{\pgfpoint{\pgf@xa - 5pt - \@tempdima}{\pgf@yb}}
\pgfpathlineto{\pgfpoint{\pgf@xb + 5pt + \@tempdimb}{\pgf@yb}}
\pgfpathlineto{\pgfpoint{\pgf@xb + 5pt - \@tempdimb}{\pgf@ya}}
\pgfpathlineto{\pgfpoint{\pgf@xa - 5pt + \@tempdima}{\pgf@ya}}
\pgfpathclose
}
}}

\boxshape{NEbox}{0pt}{8pt}
\boxshape{SEbox}{0pt}{-8pt}
\boxshape{NWbox}{8pt}{0pt}
\boxshape{SWbox}{-8pt}{0pt}
\makeatother

\tikzstyle{map}=[draw,shape=NEbox,inner sep=7pt]
\tikzstyle{mapdag}=[draw,shape=SEbox,inner sep=7pt]
\tikzstyle{maptrans}=[draw,shape=SWbox,inner sep=7pt]
\tikzstyle{mapconj}=[draw,shape=NWbox,inner sep=7pt]

\tikzstyle{probs}=[shape=semicircle,fill=gray!40!white,draw=black,shape border rotate=180,minimum width=1.2cm]

\tikzstyle{arrs}=[-latex,font=\small,auto]
\tikzstyle{arrow plain}=[arrs]
\tikzstyle{arrow dashed}=[dashed,arrs]
\tikzstyle{arrow bold}=[very thick,arrs]
\tikzstyle{arrow hide}=[draw=white!0,-]
\tikzstyle{arrow reverse}=[latex-]
\tikzstyle{cdnode}=[]

\tikzstyle{gn}=[dot,fill=green,minimum width=0.3cm,inner sep=0pt]
\tikzstyle{rn}=[dot,fill=red,inner sep=0pt,minimum width=0.3cm]
\tikzstyle{rc}=[dot,thick,fill=white,draw = red,minimum width=0.3cm,inner sep=0pt]
\tikzstyle{gc}=[dot,thick,fill=white,draw= green,inner sep=0pt,minimum width=0.3cm]
\tikzstyle{bc}=[dot,thick,fill=white,draw= blue,minimum width=0.3cm]

\tikzstyle{label}=[circle,fill=white,minimum width=0.3cm]

\tikzstyle{clocklabel}=[dot,fill=yellow,draw=black,font=\tiny,inner sep=0.75pt]

\tikzstyle{rsn}=[circle split,draw,fill=red,font=\tiny,inner sep=0.75pt]
\tikzstyle{gsn}=[circle split,draw,fill=green,font=\tiny,inner sep=0.75pt]
\tikzstyle{bsn}=[circle split,draw,fill=blue,font=\tiny,inner sep=0.75pt]

\tikzstyle{rsc}=[circle split,thick,draw= red,draw,fill=white,font=\tiny,inner sep=0.75pt]
\tikzstyle{gsc}=[circle split,thick,draw= green,draw,fill=white,font=\tiny,inner sep=0.75pt]
\tikzstyle{bsc}=[circle split,thick,draw= blue,draw,fill=white,font=\tiny,inner sep=0.75pt]

\tikzstyle{cnot}=[fill=white,shape=circle,inner sep=-1.4pt]
\tikzstyle{wire label}=[font=\tiny, auto]

%% file: defs.tex
\newcommand{\bra}[1]{\ensuremath{\left\langle #1 \right|}}
\newcommand{\ket}[1]{\ensuremath{\left|  #1 \right\rangle}}

\tikzstyle{cdiag}=[matrix of math nodes, row sep=3em, column sep=3em, text height=1.5ex, text depth=0.25ex,inner sep=0.5em]
\tikzstyle{arrow above}=[transform canvas={yshift=0.5ex}]
\tikzstyle{arrow below}=[transform canvas={yshift=-0.5ex}]

%% file: acknowlegements.tex
\begin{acknowledgements}
Firstly, I would like to express my sincere gratitude to my supervisor Bob Coecke for all his huge help, encouragement, discussions and comments. I can not imagine what my life would have been like without his great assistance. 

Great thanks to my colleague, co-auhor and friend Kang Feng Ng, for the valuable cooperation in research and his helpful suggestions in my daily life.

My sincere thanks also goes to Amar Hadzihasanovic, who has kindly shared his idea and agreed to cooperate on writing a paper based one his results. 

Many thanks to Simon Perdrix, from whom I have learned a lot and received much help when I worked with him in Nancy, while still benefitting from this experience in Oxford.

I would also like to thank Miriam Backens for loads of useful discussions,  advertising for my talk in QPL and helping me on latex problems.  

Special thanks to Dan Marsden for his patience and generousness in answering my questions and giving suggestions. 

I would like to thank Xiaoning Bian for always being ready to help me solve problems in using latex and other softwares. 

 I also wish to thank all the people who attended the weekly ZX meeting for many interesting discussions.
 
I am also grateful to my college advisor Jonathan Barrett and department advisor Jamie Vicary, thank you for chatting with me about my research and my life.

I particularly want to thank my examiners, Ross Duncan and Sam Staton, for their very detailed and helpful comments and corrections by which this thesis has been significantly improved.

 I am thankful to the Department for Computer Science and Wolfson College,  for their kind help and support. 

Last but not the least, I would like to thank my family: my parents, my wife and my daughters for supporting me throughout writing this thesis and my life in general.

\end{acknowledgements}

%% file: abstract.tex
\begin{abstract}
The ZX-calculus is an intuitive but also mathematically strict graphical language for quantum computing, which is especially powerful for the framework of quantum circuits. Completeness of the ZX-calculus means any equality of matrices with size powers of $n$ can be derived purely diagrammatically.  

In this thesis, we give the first complete axiomatisation the ZX-calculus for the overall pure qubit quantum mechanics, via a translation from the completeness result of another graphical language for quantum computing-- the ZW-calculus.  This paves the way for automated pictorial quantum computing,  with the aid of some software like Quantomatic. 

Based on this universal completeness, we directly obtain a complete axiomatisation of the ZX-calculus for the Clifford+T quantum mechanics, which is approximatively universal for quantum computing,
by restricting the ring of complex numbers to its subring corresponding to the Clifford+T fragment resting on the completeness theorem of the ZW-calculus for arbitrary commutative ring.

Furthermore, we prove the completeness of the ZX-calculus (with just 9 rules) for 2-qubit Clifford+T circuits by verifying the complete set of 17 circuit relations in diagrammatic rewriting.  This is an important step towards efficient simplification of general n-qubit  Clifford+T circuits, considering that we now have all the necessary rules for diagrammatical quantum reasoning and a very simple construction of Toffoli gate within our axiomatisation framework, which is approximately universal for quantum computation together with the Hadamard gate.

In addition to completeness results within the qubit related formalism, we extend the completeness of the ZX-calculus for qubit stabilizer quantum mechanics  to the qutrit stabilizer system. 

Finally, we show with some examples the application of the ZX-calculus to the proof of generalised supplementarity, the representation of entanglement classification and Toffoli gate, as well as equivalence-checking for the UMA gate.

% the representation of 3-entanglement qubit states, multiple-control Toffoli gates, and equivalence-checking for quantum circuits. 

\end{abstract}

%% file: chapter1.tex
\chapter{Introduction}

There are two paradigms for western metaphysics: substances and processes. The substance metaphysics sees objects as the basic constituents of the universe, while the process metaphysics take processes rather than objects as fundamental \cite{Celso}.
Meanwhile, an influential eastern philosophy--the Madhyamaka philosophy \cite{Nagarjuna}, has the key idea that all things (dharmas) are empty of substance or essence (svabh$\bar{a}$va) because of dependent arising (Prat$\bar{\i}$tyasamutp$\bar{a}$da), thus similar to the opinion of process metaphysics. Following the spirit of  the process philosophy and the Madhyamaka philosophy, we think the theory of processes has scientific and philosophical advantages, especially for its application in quantum physics. In fact, if we treat quantum processes as transformations between different types of quantum systems and highlight the compositions of processes, then we arrive at the theory of categorical quantum mechanics (CQM) proposed by Abramsky and Coecke \cite{Coeckesamson}. 
Therein the theory of processes can be made  strict in the mathematical framework of symmetric monoidal categories  \cite{MacLane}.  It would be of great interest to know how processes could be fundamental while objects being less important in a mathematical formulation of process theory for quantum mechanics. To see this, we need some concepts from category theory. This part will be illustrated in Chapter \ref{backgrd}, standard references for which can be found in    \cite{MacLane}  and   \cite{borceux_1994}.

Now we give a introduction to the main theme of this thesis--the ZX-calculus.
The ZX-calculus introduced by Coecke and Duncan \cite{coeckeduncan2008, CoeckeDuncan}  is an intuitive yet mathematically strict graphical language for quantum computing: it is formulated within the framework of compact closed categories which has a rigorous underpinning for graphical calculus \cite{Joyal}, meanwhile being an important branch of CQM \cite{Coeckesamson}. Notably, it has simple rewriting rules to transform diagrams from one to another. Each diagram in the ZX-calculus has a so-called standard interpretation \cite{PerdrixWang}, in finite-dimensional Hilbert spaces \cite{SELINGER2011113}, thus makes it relevant for quantum computing. For the past ten years, the ZX-calculus has enjoyed success in applying to fields of quantum information and quantum computation (QIC) \cite{Nielsen}, in particular (topological) measurement-based quantum computing  \cite{Duncanpx,  Horsman}  and quantum error correction \cite{DuncanLucas,  ckzh}.  Very recently, the ZX-calculus is also used for reducing the cost of implementing quantum programs \cite{CQC}.

 It is clear that the usefulness of the ZX-calculus is based on the properties of this theory.  There are three main properties of the ZX-calculus: soundness, universality and completeness. Soundness means that all the rules in the  ZX-calculus have a correct standard interpretation in the  Hilbert spaces. Universality is about if there exists a ZX-calculus diagram for every linear map in Hilbert spaces under the standard interpretation.  Completeness refers to whether an equation of diagrams can be derived in the ZX-calculus when their corresponding equation of linear maps under the standard interpretation holds true. 
 
 In the framework of category theory, the ZX-calculus is just a PROP \cite{BONCHI2017144},  and a Hilbert space model of the ZX-calculus is a symmetric monoidal category that is equivalent to a PROP, with objects generated by (tensor powers of) a Hilbert space of dimension $d>1$ (called qubit model for $d=2$ and qudit model for $d>2$). The three main properties of the ZX-calculus  are just about the properties of the interpretation from the  ZX-calculus to its Hilbert space model : soundness means that this interpretation is  a symmetric monoidal functor, while universality and completeness mean this functor is a full and faithful functor respectively. 
 
 The property that the ZX-calculus is sound relative to the qubit and qudit model  has been shown in  \cite{CoeckeDuncan}  and  \cite{Ranchin}  respectively. The universality of the ZX-calculus as to the qubit and qudit model has also been shown in  \cite{CoeckeDuncan}  and  \cite{BianWang2}  respectively.  The ZX-calculus has been proved to be complete for qubit stabilizer quantum mechanics (qubit model) \cite{Miriam1}, and recently further axiomatised to be complete for qubit Clifford +T quantum mechanics \cite {Emmanuel}, an approximatively universal fragment of quantum mechanics which has been widely used in quantum computing \cite{Boykin}. 
 
 The axiomatisation given in \cite {Emmanuel} relies on a complicated translation from the ZX-calculus to another graphical calculus--the ZW-calculus \cite{amar1}.  As a result that the ZX-calculus is not easy to use for exploring properties of multipartite entangled quantum states,  Coecke and Kissinger propose a new graphical calculus called GHZ/W-calculus which is based on the interaction of special commutative Frobenius algebras induced by GHZ-states and  anti-special commutative Frobenius algebras induced by W-states \cite{Coeckealeksmen}. In \cite{amar1}, Hadzihasanovic extends the GHZ/W-calculus into the ZW-calculus with diagrams corresponding to integer matrices, modelling on the ZX-calculus.  Most importantly, the ZW-calculus is proved to be complete for pure qubit states with integer coefficients.  It was just based on this completeness result that Jeandel,  Perdrix and Vilmart were able to give a complete axiomatisation of the ZX-calculus for the Clifford +T quantum mechanics.
 
 However, even with all the rules from the complete axiomatisation for qubit Clifford +T quantum mechanics, the ZX-calculus is still incomplete for the overall qubit  quantum mechanics, as suggested in \cite{Vladimir} and proved in \cite{JPV2}. On the other hand, 
 the ZW-calculus is generalised to a new version $ZW_R$-calculus  where $R$ is an arbitrary commutative ring, and the $ZW_R$-calculus is proved to be complete for $R$-bits ( analogues of qubits with  coefficients in $R$) \cite{amar}.  
  
  In this thesis, further to the result of \cite{Emmanuel}, we give the first complete axiomatisation of the  ZX-calculus for the overall qubit quantum mechanics (which will be called $ZX_{ full}$-calculus in this thesis) \cite{ngwang}, based on the completeness result of $ZW_{ \mathbb{C}}$-calculus, where  $\mathbb{C}$ is the field of complex numbers. In view of our results, there comes the paper \cite{JPV2} afterwards which also give an axiomatisation of the ZX-calculus for the entire qubit quantum mechanics, with different generators of diagrams and rewriting rules.  Chapter 2 of this thesis will show the details of  the  complete axiomatisation of the  $ZX_{ full}$-calculus.
  
  Given the complete axiomatisation of the  $ZX_{ full}$-calculus \cite{ngwang}, we also obtain a complete axiomatisation of the ZX-calculus for the Clifford+T quantum mechanics (which will be called $ZX_{ C+T}$-calculus in this thesis)
by restricting the ring $\mathbb{C}$ to its subring $\mathbb{Z}[i, \frac{1}{\sqrt{2}}]$ which exactly corresponds to the Clifford+T fragment, resting on the completeness theorem of the $ZW_R$-calculus. In contrast to the first complete axiomatisation of the ZX-calculus for the Clifford+T fragment \cite{Emmanuel}, we have two new generators--a triangle and a $\lambda$ box-- as features rather than novelties: the triangle can be employed as an essential component to construct a Toffoli gate in a very simple form, while the  $\lambda$ box can be slightly extended to a generalised phase so that the generalised supplementarity (also called cyclotomic supplementarity) \cite{jpvw} is naturally seen as a special case of the generalised spider rule. In addition, due to the introduction of the new generators, our proof for that the Clifford +T fragment of the ZX-calculus exactly corresponds to matrices over the ring $\mathbb{Z}[i, \frac{1}{\sqrt{2}}]$ is much simpler than the corresponding proof given in \cite{Emmanuel}.
These results are shown in detail in Chapter 3.

Considering that  Clifford+T quantum circuits are most frequently used in quantum computation,  it would be more efficient to use a small set of ZX rules for the purpose of circuit simplification, although the ZX-calculus is complete for both the overall pure qubit quantum mechanics \cite{ngwang} and the Clifford+T pure qubit quantum mechanics \cite{Emmanuel}. In Chapter 4, we prove the completeness of the ZX-calculus (with just 9 rules) for 2-qubit Clifford+T circuits by verifying the complete set of 17 circuit relations  \cite{ptbian} in diagrammatic rewriting. As a consequence, our result can also be seen as a completeness result for single-qubit Clifford+T ZX-calculus \cite{Miriam1ct}.
 In addition, we are able to give an analytic solution for converting from ZXZ to XZX Euler decompositions of single-qubit unitary gates as suggested by Schr\"oder de Witt and Zamdzhiev  \cite{Vladimir}.
  
Since there already exist qutrit  and general qudit versions of the ZX-calculus  \cite{Ranchin, BianWang2}, it is natural to ask if one could generalise the completeness result of the qubit ZX-calculus to the qudit version for arbitrary dimension $d$.   As for the completeness of  the ZX-calculus for the overall and generalised Clifford+T qudit quantum mechanics, there is no result available.  Fortunately, the completeness of the ZX-calculus for qubit stabilizer quantum mechanics can be generalised to the qutrit case, which will be shown explicitly in chapter 5.

Having the completeness results established above, it would be interesting to see how it could be applied. In chapter 6, we show by some examples the application of the ZX-calculus to the proof of generalised supplementarity, the representation of entanglement classification and Toffoli gate, as well as equivalence-checking for the UMA gate.

 %the representation of entanglement classification and Toffoli gate, as well as equivalence-checking for quantum circuits. 

Finally in chapter 7, we conclude this thesis with some open problems.

%% file: chapterbk.tex
\chapter{Background}\label{backgrd}

In this chapter we give the requisite knowledge of this thesis, which includes the concepts related to category theory, ZX-calculus, and ZW-calculus. 

\section{Some concepts from category theory}
In this section, we give some categorical concepts constituting the theoretical framework of this thesis, the standard references for which can be found in  \cite{MacLane}  and   \cite{borceux_1994}.

\subsection*{Category}
A category  $\mathfrak{C}$  consists of:
\begin{itemize}
\item a class of objects $ob(\mathfrak{C})$;
\item for each pair of objects $A, B$, a set  $\mathfrak{C}(A, B)$ of morphisms from $A$ to $B$;
\item for each triple of objects $A, B, C$, a  composition map  
\[
\begin{array}{ccc} 
\mathfrak{C}(B, C) \times \mathfrak{C}(A, B)& \longrightarrow & \mathfrak{C}(A, C)\\
(g, f) & \mapsto & g\circ f;
\end{array}
\]
\item for each object $A$, an identity morphism $1_A \in \mathfrak{C}(A, A)$,
\end{itemize}
satisfying the following axioms:
\begin{itemize}
\item associativity:  for any $f\in  \mathfrak{C}(A, B),  g\in  \mathfrak{C}(B, C), h\in  \mathfrak{C}(C, D)$,  there holds $(h\circ g)\circ f=h\circ (g\circ f)$;
\item identity law: for any $f\in  \mathfrak{C}(A, B),  1_B\circ f=f=f\circ 1_A$.
\end{itemize}

A morphism $f\in  \mathfrak{C}(A, B)$ is an  \textit{  isomorphism} if there exists a  morphism $g\in  \mathfrak{C}(B, A)$ such that $g\circ f=1_A$ and $f\circ g=1_B$.  A \textit{product category} $\mathfrak{A} \times \mathfrak{B}$ can be defined componentwise
by two categories $\mathfrak{A}$ and $\mathfrak{B}$.
%In the spirit of process theory, what is important is the part of processes. A process between two categories will be called as a functor. 
\subsection*{Functor}
Given categories $\mathfrak{C}$ and $\mathfrak{D}$, a functor  $F: \mathfrak{C} \longrightarrow \mathfrak{D}$  consists of:
\begin{itemize}
\item a  mapping  
\[
\begin{array}{ccc} 
\mathfrak{C}& \longrightarrow & \mathfrak{D}\\
A & \mapsto & F(A);
\end{array}
\]

\item for each pair of objects $A, B$ of $\mathfrak{C}$,  a map
\[
\begin{array}{ccc} 
\mathfrak{C}(A, B) & \longrightarrow &\mathfrak{D}(F(A), F(B))\\
f & \mapsto & F(f),
\end{array}
\]

\end{itemize}
satisfying the following axioms:
\begin{itemize}
\item preserving composition:  for any morphisms $f\in  \mathfrak{C}(A, B),  g\in  \mathfrak{C}(B, C)$,  there holds $F(g\circ f)=F(g)\circ F( f))$;
\item preserving identity: for any object $A$ of $\mathfrak{C}$,  $ F(1_A)=1_{F(A)}$.

\end{itemize}

A functor  $F: \mathfrak{C} \longrightarrow \mathfrak{D}$ is  \textit{ faithful (full)} if for each pair  of objects $A, B$ of $\mathfrak{C}$, the map
\[
\begin{array}{ccc} 
\mathfrak{C}(A, B) & \longrightarrow &\mathfrak{D}(F(A), F(B))\\
f & \mapsto & F(f)
\end{array}
\]
is injective (surjective).

\subsection*{Natural transformation}

 Let $F, G: \mathfrak{C}\longrightarrow \mathfrak{D} $ be two functors. A natural transformation $\tau : F \rightarrow G$ is a family $(\tau_A:F(A)\longrightarrow G(A))_{A\in \mathfrak{C}} $ of morphisms in $ \mathfrak{D}$ such that the following square commutes: 

% \tikzfig{diagrams//natural}

\begin{center}

\begin{tikzpicture}

\node (1) at  ( -0.5,4) [] {$F(A)$};
\node  at  ( 0.5,4.2) [] {$\tau_A$};
\node  at  ( -1,3.3) [] {$F(f)$};
\node (2) at  ( 1.5,4) [] {$G(A)$};
\node (3) at ( -0.5,2.5) [] {$F(B)$};
\node  at  ( 0.5,2.72) [] {$\tau_B$};
\node  at  ( 2,3.3) [] {$G(f)$};
\node (4) at ( 1.5,2.5) [] {$G(B)$};

\draw [->] (1) -- (2) {} ;
\draw [->] (3) -- (4) {} ;
\draw [->] (1) -- (3) {} ;
\draw [->] (2) -- (4) {} ;

\end{tikzpicture}

\end{center}

\noindent
for all morphisms $f\in \mathfrak{C}(A,B) $. A natural isomorphism is a natural transformation where each of the $\tau_A$ is an isomorphism.

\subsection*{Strict monoidal category}

A strict monoidal category consists of:

\begin{itemize}

\item a category $\mathfrak{C}$; 

\item a unit object $I\in ob(\mathfrak{C})$;

\item  a bifunctor $- \otimes - : \mathfrak{C} \times \mathfrak{C} \longrightarrow \mathfrak{C} $,

\end{itemize}
satisfying 
\begin{itemize}
\item associativity:  for each triple of objects $A, B, C$ of $\mathfrak{C}$, $A\otimes (B \otimes C)=(A\otimes B) \otimes C$; for each triple of morphisms $f, g, h$ of $\mathfrak{C}$, $f\otimes (g \otimes h)=(f\otimes g) \otimes h$; 

\item unit law: for each object $A$ of $\mathfrak{C}$,  $A\otimes I= A=I\otimes A$; for each morphism $f$ of $\mathfrak{C}$,  $f\otimes 1_I= f=1_I\otimes f$.
\end{itemize}

%\textbf{FinHilb} is a monoidal category. In \textbf{FinHilb} parallel composition is the tensor product of Hilbert spaces. $I$ is the field of complex numbers $\mathbb{C}$, which is a 1-dimensional Hilbert space. 

\subsection*{ Strict symmetric monoidal category}

A strict monoidal category $\mathfrak{C}$ is symmetric if it is equipped with a natural isomorphism

\begin{center}
$\sigma_{A,B} : A \otimes B \rightarrow B \otimes A$

\end{center}

\noindent  for all objects $A, B, C$ of $\mathfrak{C}$  satisfying:
 $$\sigma_{B,A} \circ \sigma_{A,B} =1_{A \otimes B},  ~~\sigma_{A,I}=1_A, ~~ (1_B \otimes \sigma_{A,C}) \circ  (\sigma_{A,B} \otimes 1_C)  =\sigma_{A, B\otimes C}.$$

\iffalse
The swap morphism is graphically represented as the following:

\begin{center}

\begin{tikzpicture}

\node at ( -1.5,3) [] {$\sigma_{A,B}$ $:=$};

\draw [-] ( 0,2.3) .. controls ( 0,3) and ( 1,3.1) .. ( 1,3.8);

\draw [-] ( 1,2.3) .. controls ( 1,3) and ( 0,3.1) .. ( 0,3.8);

\node at ( 1.3,2.5) [] {$B$};
\node at ( -0.3,2.5) [] {$A$};

\node at ( 1.3,3.6) [] {$A$};
\node at ( -0.3,3.6) [] {$B$};

\end{tikzpicture}

\end{center}
\fi

\subsection*{Strict monoidal functor}
Given two strict monoidal categories $\mathfrak{C}$ and $\mathfrak{D}$, a strict monoidal functor  $F: \mathfrak{C} \longrightarrow \mathfrak{D}$  is a functor  $F: \mathfrak{C} \longrightarrow \mathfrak{D}$ such that 
$F(A)\otimes F(B)=F(A\otimes B), F(f)\otimes F(g)=F(f\otimes g), F(I_{\mathfrak{C}})=I_{\mathfrak{D}}$, for any objects $A, B$ of $\mathfrak{C}$, and any morphisms $f\in\mathfrak{C}(A, A_1), g\in\mathfrak{C}(B, B_1)$.

A strict symmetric monoidal functor $F$ is a strict monoidal functor that preserves the symmetry structure, i.e., $ F(\sigma_{A,B})= \sigma_{F(A),F(B)}$.

\subsection*{Self-dual strict compact closed category }
A self-dual strict compact closed category  is a strict symmetric monoidal category $\mathfrak{C}$ such that for each object $A$ of $\mathfrak{C}$, there exists two morphisms
 $$\epsilon_{A} : A \otimes A \rightarrow I,     ~~            \eta_{A} : I \rightarrow  A \otimes A$$
 satisfying:

$$  (\epsilon_{A}   \otimes 1_A ) \circ (1_A \otimes \eta_A) =1_A, ~~         (1_A \otimes  \epsilon_{A}  ) \circ (\eta_A  \otimes 1_A) =1_A.$$

Note that here we use the word ``self-dual" in the same sense as in \cite{Coeckebk}, instead of the sense in Peter Selinger's paper \cite{selinger_selfdual_2010}.

\subsection*{PROP}
`PROP'  is an acronym for `products and permutations', as introduced by Mac Lane \cite{maclane1965}.  A PROP is a strict symmetric monoidal category having the natural numbers as objects, with the tensor product of objects given by addition. A morphism between two PROPs is a strict symmetric monoidal functor that is the identity on objects \cite{BaezCR}. 

As an example, the category  $\mathbf{FinSet}$ is a PROP whose objects are all finite sets and whose morphisms are all functions between them.

Just as any group can be represented by generators and relations, any PROP can be described as a presentation in terms of  generators and relations, which is proved in  \cite{BaezCR}.

\subsection*{Some typical examples of categories in this thesis}
\begin{itemize}
 \item $\mathbf{FdHilb_d}$: the category whose objects are complex Hilbert spaces with dimensions $d^k$, where $d > 1$ is a given integer and $k$ is an arbitrary non-negative integer, and whose morphisms are linear maps between the Hilbert spaces with ordinary composition of linear maps as composition of morphisms. The usual Kronecker tensor product is the monoidal tensor, and the field of complex numbers $\mathbb{C}$ (which is a one-dimensional Hilbert space over itself) is the tensor unit.
  
   For each object of  $\mathbf{FdHilb_d}$, we can choose an orthonormal basis  $\{\ket{i}\}_{0 \leq i \leq d-1}$ denoted in Dirac notation. 
   
   \item $\mathbf{FdHilb_d}/s$: the category which has the same objects as $\mathbf{FdHilb_d}$ but whose morphisms are equivalence classes of $\mathbf{FdHilb_d}$-morphisms, given by the following equivalence relation
   \[
   f \sim g \Leftrightarrow \exists r \in \mathbb{C}\diagdown \{0\} \quad \mbox{such that} \quad  f = r \cdot g  
   \]
   
   \item $\mathbf{Mat}_{R}$: the category whose objects are natural numbers and whose morphisms $M: m \rightarrow n$
 are $n \times m$  matrices taking values in a given commutative ring $R$. The composition is matrix multiplication, the monoidal
product on objects and morphisms are multiplication of natural numbers and the Kronecker product of matrices respectively. 
\end{itemize}
 
 %$\mathbf{FdHilb_d}$ of Hilbert spaces with dimensions $d^k$, and quotient category $\mathbf{FdHilb_d}/s$ modulo non-zero scalars, category of matrices on a given  denoted as ,  relationship between Dirac symbol and matrix.

\section{ZX-calculus}

The ZX-calculus was introduced by Coecke and Duncan in \cite{coeckeduncan2008, CoeckeDuncan}  as a graphical language for describing a pair of complementary quantum observables. The key feature of the ZX is that it has intuitive rewriting rules which allow one to transform diagrams from one to another, instead of performing tedious matrix calculations. The original  ZX-calculus is designed particularly for qubits \cite{Nielsen}, then it is generalised to higher dimensions \cite{BianWang2, Ranchin}. In this section, we will first give a formal definition of ZX-calculus for arbitrary dimension (called qudit ZX-calculus), then present the details of qubit ZX-calculus and qutrit ZX-calculus, including related properties. Finally we display another graphical language--ZW-calculus in terms of generators and rewriting rules.

 \subsection {ZX-calculus in general }\label{zxgeneral}
 We will describe the ZX-calculus in the framework of PROPs in terms of generators and relations following the way presented in \cite{amar1}.  Explicitly, we build the ZX-calculus in the following way: first we give a set $\mathnormal{S}$ consisting of basic diagrams including empty diagram and the straight line as generators, where by diagram we mean a picture composed of a vertex, $n$ incoming wires (inputs) and  $m$ outgoing wires (outputs):
                                \begin{center}
\beginpgfgraphicnamed{diagrams//popmorph}
\InputIfFileExists{diagrams//popmorph.tikz}{}{\input{./figures/diagrams//popmorph.tikz}}
\endpgfgraphicnamed
                                \end{center}

 Note that in this thesis any diagram should be read from top to bottom.
\noindent Let $ZX[\mathnormal{S}]$ be the strict monoidal category freely generated by diagrams of $\mathnormal{S}$ in parallel composition $\otimes$ where any two diagrams $D_1$ and $D_2$ are placed side-by-side with $D_1$ on the left of $D_2$ ($D_1 \otimes D_2$), or in sequential composition $\circ$ where $D_1$ has the same number of outputs as the number of inputs of $D_2$ and  $D_1$ is placed above $D_2$ with the outputs of $D_1$ connected to the inputs of $D_2$ ($D_2 \circ D_1$). It is clear that the empty diagram is a uint of parallel composition and the diagram of a straight line is a unit of the sequential composition. 
 Then we give an equivalence relation $\mathnormal{R}$  of diagrams (morphisms) in $ZX[\mathnormal{S}]$ including the diagrammatical representation (see below) of axioms of a self-dual strict compact closed category.  Let $ZX[\mathnormal{S}]/\mathnormal{R}$ be the PROP obtained from $ZX[\mathnormal{S}]$ modulo the equivalence relation $\mathnormal{R}$. The pairs in $\mathnormal{R}$ will be called the rules of $ZX[\mathnormal{S}]/\mathnormal{R}$. Furthermore, $ZX[\mathnormal{S}]/\mathnormal{R}$ is called a ZX-calculus if 
 \begin{itemize}
\item the generating set  $\mathnormal{S}$ consists of the following basic diagrams:   
    \begin{center}
\beginpgfgraphicnamed{diagrams//quditgspider}
\InputIfFileExists{diagrams//quditgspider.tikz}{}{\input{./figures/diagrams//quditgspider.tikz}}
\endpgfgraphicnamed,\hspace{0.3cm}  %
\beginpgfgraphicnamed{diagrams//quditrspider}
\InputIfFileExists{diagrams//quditrspider.tikz}{}{\input{./figures/diagrams//quditrspider.tikz}}
\endpgfgraphicnamed,\hspace{0.3cm}  %
\beginpgfgraphicnamed{diagrams//HadaDecomSingleslt}
\InputIfFileExists{diagrams//HadaDecomSingleslt.tikz}{}{\input{./figures/diagrams//HadaDecomSingleslt.tikz}}
\endpgfgraphicnamed,\hspace{0.3cm} %
\beginpgfgraphicnamed{diagrams//swap}
\InputIfFileExists{diagrams//swap.tikz}{}{\input{./figures/diagrams//swap.tikz}}
\endpgfgraphicnamed,\hspace{0.3cm} %
\beginpgfgraphicnamed{diagrams//Id}
\InputIfFileExists{diagrams//Id.tikz}{}{\input{./figures/diagrams//Id.tikz}}
\endpgfgraphicnamed,\hspace{0.3cm} %
\beginpgfgraphicnamed{diagrams//emptysquare}
\InputIfFileExists{diagrams//emptysquare.tikz}{}{\input{./figures/diagrams//emptysquare.tikz}}
\endpgfgraphicnamed,\hspace{0.3cm}%
\beginpgfgraphicnamed{diagrams//cap}
\InputIfFileExists{diagrams//cap.tikz}{}{\input{./figures/diagrams//cap.tikz}}
\endpgfgraphicnamed,\hspace{0.3cm}%
\beginpgfgraphicnamed{diagrams//cup}
\InputIfFileExists{diagrams//cup.tikz}{}{\input{./figures/diagrams//cup.tikz}}
\endpgfgraphicnamed
        \end{center}
   where $ m,n\in \mathbb N$, $\overrightarrow{\alpha}=(\alpha_1, \cdots, \alpha_{d-1}), \alpha_i \in [0,  2\pi) $, and the dashed square represents an empty diagram;
  \item  the equivalence relation $\mathnormal{R}$ consists of two types of rules:    
  \begin{enumerate}
   \item  the structure rules for a self-dual compact closed category:
 \begin{equation}\label{compactstructure}
\beginpgfgraphicnamed{diagrams//compactstructure_cap}
\InputIfFileExists{diagrams//compactstructure_cap.tikz}{}{\input{./figures/diagrams//compactstructure_cap.tikz}}
\endpgfgraphicnamed \qquad\quad %
\beginpgfgraphicnamed{diagrams//compactstructure_cup}
\InputIfFileExists{diagrams//compactstructure_cup.tikz}{}{\input{./figures/diagrams//compactstructure_cup.tikz}}
\endpgfgraphicnamed \qquad\quad %
\beginpgfgraphicnamed{diagrams//compactstructure_snake}
\InputIfFileExists{diagrams//compactstructure_snake.tikz}{}{\input{./figures/diagrams//compactstructure_snake.tikz}}
\endpgfgraphicnamed
 \end{equation}
  \begin{equation}\label{compactstructureslide}
\beginpgfgraphicnamed{diagrams//compactstructure_slide}
\InputIfFileExists{diagrams//compactstructure_slide.tikz}{}{\input{./figures/diagrams//compactstructure_slide.tikz}}
\endpgfgraphicnamed
  \end{equation}
  \item non-structural rewriting rules for transforming the generators, typically the spider rule, copy rule, bialgebra rule, Euler decomposition rule, and the colour change rule \cite{CoeckeDuncan}.
\end{enumerate}
  
 \end{itemize}
 
 For convenience, we have the following short notation:
\[
 %\tikzfig{scalars-s//spiderg} := \tikzfig{scalars-s//spidergz}\qquad\qquad\tikzfig{scalars-s//spiderr} := \tikzfig{scalars-s//spiderrz}
 %
\beginpgfgraphicnamed{diagrams//spiderg}
\InputIfFileExists{diagrams//spiderg.tikz}{}{\input{./figures/diagrams//spiderg.tikz}}
\endpgfgraphicnamed := %
\beginpgfgraphicnamed{diagrams//quditgspider0phase}
\InputIfFileExists{diagrams//quditgspider0phase.tikz}{}{\input{./figures/diagrams//quditgspider0phase.tikz}}
\endpgfgraphicnamed\qquad\qquad%
\beginpgfgraphicnamed{diagrams//spiderr}
\InputIfFileExists{diagrams//spiderr.tikz}{}{\input{./figures/diagrams//spiderr.tikz}}
\endpgfgraphicnamed := %
\beginpgfgraphicnamed{diagrams//quditrspider0phase}
\InputIfFileExists{diagrams//quditrspider0phase.tikz}{}{\input{./figures/diagrams//quditrspider0phase.tikz}}
\endpgfgraphicnamed
\]
 
Since we have  the parameter $d$ in the ZX-calculus, we also call it \textit{ qudit ZX-calculus}  which will be meaningful when we give to it a semantics. It is called \textit{ qubit ZX-calculus}  if $d=2$, and called \textit{ qutrit ZX-calculus}  if $d=3$. 
Also we call a diagram with no inputs and no outputs a \textit{ scalar}.  
% It is easy to see that this prop is a self-dual compact closed category. 

 Now we associate to each diagram in the ZX-calculus a standard interpretation $\llbracket \cdot \rrbracket$ in $\mathbf{FdHilb_d}$:
 \[
\left\llbracket %
\beginpgfgraphicnamed{diagrams//quditgspider}
\InputIfFileExists{diagrams//quditgspider.tikz}{}{\input{./figures/diagrams//quditgspider.tikz}}
\endpgfgraphicnamed \right\rrbracket=\sum_{j=0}^{d-1}e^{i\alpha_j}\ket{j}^{\otimes m}\bra{j}^{\otimes n}, \alpha_0=1,
\]

 \[
\left\llbracket %
\beginpgfgraphicnamed{diagrams//quditrspider}
\InputIfFileExists{diagrams//quditrspider.tikz}{}{\input{./figures/diagrams//quditrspider.tikz}}
\endpgfgraphicnamed \right\rrbracket=\sum_{j=0}^{d-1}e^{i\alpha_j}\ket{h_j}^{\otimes m}\bra{h_j}^{\otimes n}, 
\alpha_0=1,
 \ket{h_j}=\frac{1}{\sqrt{d}}\sum_{j=0}^{d-1}\xi^{jk}\ket{k}, \xi=e^{i\frac{2\pi}{d}},
\]

\[
\left\llbracket%
\beginpgfgraphicnamed{diagrams//HadaDecomSingleslt}
\InputIfFileExists{diagrams//HadaDecomSingleslt.tikz}{}{\input{./figures/diagrams//HadaDecomSingleslt.tikz}}
\endpgfgraphicnamed\right\rrbracket=\frac{1}{\sqrt{d}}\sum_{i, j=0}^{d-1}\xi^{ji}\ket{j}\bra{i}, \quad
\quad
   \left\llbracket%
\beginpgfgraphicnamed{diagrams//emptysquare}
\InputIfFileExists{diagrams//emptysquare.tikz}{}{\input{./figures/diagrams//emptysquare.tikz}}
\endpgfgraphicnamed\right\rrbracket=1, \quad
   \left\llbracket%
\beginpgfgraphicnamed{diagrams//Id}
\InputIfFileExists{diagrams//Id.tikz}{}{\input{./figures/diagrams//Id.tikz}}
\endpgfgraphicnamed\right\rrbracket=\sum_{j=0}^{d-1}\ket{j}\bra{j}, 
   \]

\[
 \left\llbracket%
\beginpgfgraphicnamed{diagrams//swap}
\InputIfFileExists{diagrams//swap.tikz}{}{\input{./figures/diagrams//swap.tikz}}
\endpgfgraphicnamed\right\rrbracket=\sum_{i, j=0}^{d-1}\ket{ji}\bra{ij},\quad
  \left\llbracket%
\beginpgfgraphicnamed{diagrams//cap}
\InputIfFileExists{diagrams//cap.tikz}{}{\input{./figures/diagrams//cap.tikz}}
\endpgfgraphicnamed\right\rrbracket=\sum_{j=0}^{d-1}\ket{jj}, \quad
   \left\llbracket%
\beginpgfgraphicnamed{diagrams//cup}
\InputIfFileExists{diagrams//cup.tikz}{}{\input{./figures/diagrams//cup.tikz}}
\endpgfgraphicnamed\right\rrbracket=\sum_{j=0}^{d-1}\bra{jj},
      \]

\[  \llbracket D_1\otimes D_2  \rrbracket =  \llbracket D_1  \rrbracket \otimes  \llbracket  D_2  \rrbracket, \quad 
 \llbracket D_1\circ D_2  \rrbracket =  \llbracket D_1  \rrbracket \circ  \llbracket  D_2  \rrbracket.
  \]
%where 
%$$ \ket{0}= \begin{pmatrix}
%        1  \\
%        0  \\
 %\end{pmatrix}, \quad 
% \bra{0}=\begin{pmatrix}
  %      1 & 0 
    %     \end{pmatrix},
 %\quad  \ket{1}= \begin{pmatrix}
   %     0  \\
 %       1  \\
% \end{pmatrix}, \quad 
%  \bra{1}=\begin{pmatrix}
  %   0 & 1 
  %       \end{pmatrix}.
 %$$
%It can be verified that the interpretation $\llbracket \cdot \rrbracket$ is a monoidal functor. 
 
 Now we are ready to define three important properties of the ZX-calculus: soundness, universality and completeness. Note that if a diagram $D_1$ in the ZX-calculus can be rewritten into another diagram $D_2$ using the ZX rules, then we denote this as $ZX\vdash D_1=D_2$. 
 
 \begin{definition}
 The ZX-calculus is called sound if for any two diagrams $D_1$ and $D_2$, $ZX\vdash D_1=D_2$ must imply that $\llbracket D_1  \rrbracket =  \llbracket  D_2  \rrbracket$.
 
 \end{definition}
 
  \begin{definition}
 The ZX-calculus is called universal if for any linear map $L$ in $\mathbf{FdHilb_d}$, there must exist a diagram  $D$ in the ZX-calculus such that $\llbracket D  \rrbracket = L$.
 
 \end{definition}

  \begin{definition}
 The ZX-calculus is called complete if for  any two diagrams $D_1$ and $D_2$,  $\llbracket D_1  \rrbracket =  \llbracket  D_2  \rrbracket$ must imply that $ZX\vdash D_1=D_2$.
 
 \end{definition}

Among these three properties, soundness means if an equality of diagrams holds in the ZX-calculus then the corresponding linear maps under the standard interpretation must be the same. Since the derivation of true equalities comes from the rewriting rules,
 soundness can be checked on a rule-by-rule basis. Secondly, universality means each linear map can be represented by a diagram in the ZX-calculus. Thirdly, completeness means all true equalities of linear maps can be derived graphically. 

 %The qubit ZX-calculus has been proved to be sound and universal in  \cite{CoeckeDuncan}. More generally, the qudit ZX-calculus has been proved to be sound in \cite{Ranchin}  and to be universal in \cite{BianWang2}.  
 
 \begin{proposition} \cite{Ranchin}
 The qudit ZX-calculus is sound for any $d \geq 2$.
 \end{proposition}
 
  \begin{proposition} \cite{BianWang2}
 The qudit ZX-calculus is universal for any $d \geq 2$.
 \end{proposition}

The proof of the completeness of the ZX-calculus is the main theme of this thesis. 
 
  \subsection*{Scalar-free ZX-calculus}
A scalar $D$ is called non-zero if $\llbracket D  \rrbracket \neq 0$.  The ZX-calculus could has a scalar-free version where all the non-zero scalars can be ignored. 
 \begin{definition}
 A scalar-free ZX-calculus is obtained from the ZX-calculus $ZX[\mathnormal{S}]/\mathnormal{R}$ modulo an equivalence relation $\mathnormal{F}$ where two diagrams $D_1$ and $D_2$ are in the same equivalent class if there exist non-zero scalars $s$ and $t$ such that $s \otimes D_1= t\otimes D_2$. 
  \end{definition}
The standard interpretation $\llbracket \cdot \rrbracket:  ZX[\mathnormal{S}]/\mathnormal{R} \longrightarrow \mathbf{FdHilb_d}$ can be generalised to an interpretation $\llbracket \cdot \rrbracket_{sf}: (ZX[\mathnormal{S}]/\mathnormal{R})/\mathnormal{F} \longrightarrow \mathbf{FdHilb_d}/s$ in a natural way such that the following square commute:

\begin{equation*}
\beginpgfgraphicnamed{diagrams//interpresqure}
\InputIfFileExists{diagrams//interpresqure.tikz}{}{\input{./figures/diagrams//interpresqure.tikz}}
\endpgfgraphicnamed
\end{equation*}

By the construction of the scalar-free ZX-calculus and the soundness of the qudit ZX-calculus, it is easy to see that the scalar-free ZX-calculus is sound  in relative to the interpretation  $\llbracket \cdot \rrbracket_{sf}$.
 \subsection{Qubit ZX-calculus}\label{qubitzxintro}
In this subsection, we describe qubit ZX-calculus in detail and give some of its useful properties.

The qubit ZX-calculus has the following generators:
\begin{table}
\begin{center} % \tikzfig{diagrams//emptysquare-small}
\begin{tabular}{|r@{~}r@{~}c@{~}c|r@{~}r@{~}c@{~}c|}
\hline
%$R_Z^{(n,m)}$&$:$&$n\to m$ & \tikzfig{diagrams//generator_spider2} & $A$&$:$&$ 1\to 1$& \tikzfig{diagrams//alphagate}\\
$R_{Z,\alpha}^{(n,m)}$&$:$&$n\to m$ & %
\beginpgfgraphicnamed{diagrams//generator_spider2}
\InputIfFileExists{diagrams//generator_spider2.tikz}{}{\input{./figures/diagrams//generator_spider2.tikz}}
\endpgfgraphicnamed  & $R_{X,\alpha}^{(n,m)}$&$:$&$n\to m$ & %
\beginpgfgraphicnamed{diagrams//generator_redspider2}
\InputIfFileExists{diagrams//generator_redspider2.tikz}{}{\input{./figures/diagrams//generator_redspider2.tikz}}
\endpgfgraphicnamed\\
%\multicolumn{3}{|c}{ \tikzfig{diagrams//generator_spider2} }&(P1)\\
\hline
$H$&$:$&$1\to 1$ &%
\beginpgfgraphicnamed{diagrams//HadaDecomSingleslt}
\InputIfFileExists{diagrams//HadaDecomSingleslt.tikz}{}{\input{./figures/diagrams//HadaDecomSingleslt.tikz}}
\endpgfgraphicnamed
%& $s$&$:$&$0\to 0$ &\tikzfig{scalars//halfscalar}  &
 &  $\sigma$&$:$&$ 2\to 2$& %
\beginpgfgraphicnamed{diagrams//swap}
\InputIfFileExists{diagrams//swap.tikz}{}{\input{./figures/diagrams//swap.tikz}}
\endpgfgraphicnamed\\\hline
   $\mathbb I$&$:$&$1\to 1$&%
\beginpgfgraphicnamed{diagrams//Id}
\InputIfFileExists{diagrams//Id.tikz}{}{\input{./figures/diagrams//Id.tikz}}
\endpgfgraphicnamed & $e $&$:$&$0 \to 0$& %
\beginpgfgraphicnamed{diagrams//emptysquare}
\InputIfFileExists{diagrams//emptysquare.tikz}{}{\input{./figures/diagrams//emptysquare.tikz}}
\endpgfgraphicnamed\\\hline
   $C_a$&$:$&$ 0\to 2$& %
\beginpgfgraphicnamed{diagrams//cap}
\InputIfFileExists{diagrams//cap.tikz}{}{\input{./figures/diagrams//cap.tikz}}
\endpgfgraphicnamed &$ C_u$&$:$&$ 2\to 0$&%
\beginpgfgraphicnamed{diagrams//cup}
\InputIfFileExists{diagrams//cup.tikz}{}{\input{./figures/diagrams//cup.tikz}}
\endpgfgraphicnamed \\\hline
\end{tabular}\caption{Generators of qubit ZX-calculus} \label{qbzxgenerator}
\end{center}
\end{table}
where $m,n\in \mathbb N$, $\alpha \in [0,  2\pi)$, and $e$ represents an empty diagram. 

The qubit ZX-calculus has non-structural rewriting rules as follows:
\begin{figure}[!h]
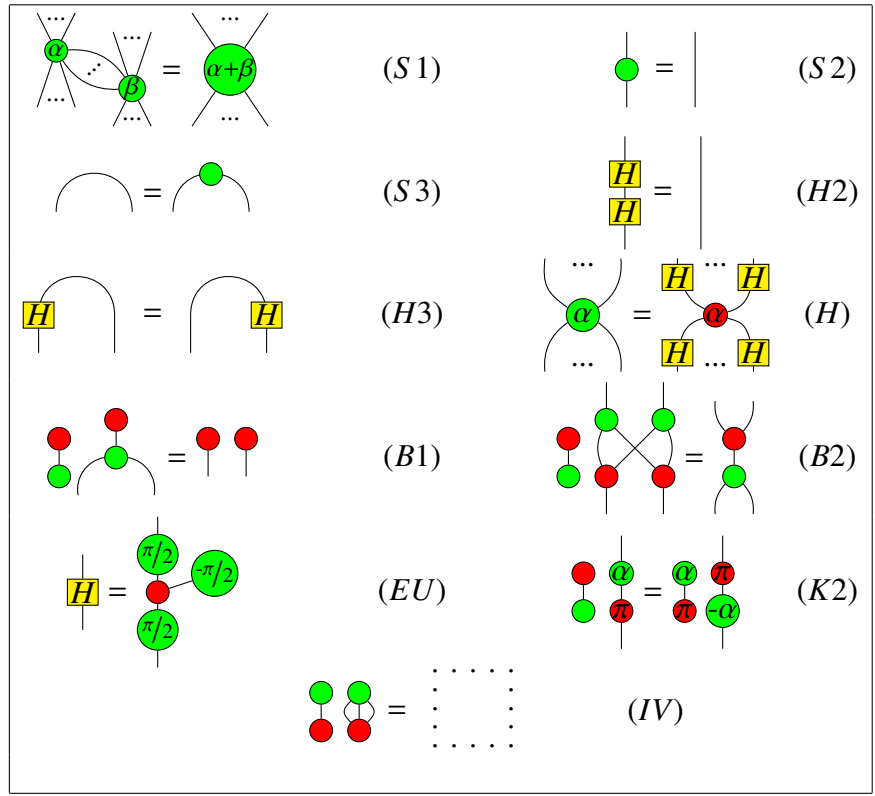

\begin{center}
\[
\quad \qquad\begin{array}{|cccc|}
\hline
\beginpgfgraphicnamed{diagrams//spiderlt}
\InputIfFileExists{diagrams//spiderlt.tikz}{}{\input{./figures/diagrams//spiderlt.tikz}}
\endpgfgraphicnamed=%
\beginpgfgraphicnamed{diagrams//spiderrt}
\InputIfFileExists{diagrams//spiderrt.tikz}{}{\input{./figures/diagrams//spiderrt.tikz}}
\endpgfgraphicnamed &(S1) &%
\beginpgfgraphicnamed{diagrams//s2new}
\InputIfFileExists{diagrams//s2new.tikz}{}{\input{./figures/diagrams//s2new.tikz}}
\endpgfgraphicnamed &(S2)\\
\beginpgfgraphicnamed{diagrams//induced_compact_structure-2wirelt}
\InputIfFileExists{diagrams//induced_compact_structure-2wirelt.tikz}{}{\input{./figures/diagrams//induced_compact_structure-2wirelt.tikz}}
\endpgfgraphicnamed=%
\beginpgfgraphicnamed{diagrams//induced_compact_structure-2wirert}
\InputIfFileExists{diagrams//induced_compact_structure-2wirert.tikz}{}{\input{./figures/diagrams//induced_compact_structure-2wirert.tikz}}
\endpgfgraphicnamed&(S3) & %
\beginpgfgraphicnamed{diagrams//hsquare}
\InputIfFileExists{diagrams//hsquare.tikz}{}{\input{./figures/diagrams//hsquare.tikz}}
\endpgfgraphicnamed &(H2)\\
\beginpgfgraphicnamed{diagrams//hslidecap}
\InputIfFileExists{diagrams//hslidecap.tikz}{}{\input{./figures/diagrams//hslidecap.tikz}}
\endpgfgraphicnamed &(H3) &%
\beginpgfgraphicnamed{diagrams//h2newlt2}
\InputIfFileExists{diagrams//h2newlt2.tikz}{}{\input{./figures/diagrams//h2newlt2.tikz}}
\endpgfgraphicnamed=%
\beginpgfgraphicnamed{diagrams//h2newrt2}
\InputIfFileExists{diagrams//h2newrt2.tikz}{}{\input{./figures/diagrams//h2newrt2.tikz}}
\endpgfgraphicnamed&(H)\\
%&\tikzfig{diagrams//hslidecap} &(H3) &\\
    %
\beginpgfgraphicnamed{diagrams//b1slt}
\InputIfFileExists{diagrams//b1slt.tikz}{}{\input{./figures/diagrams//b1slt.tikz}}
\endpgfgraphicnamed=%
\beginpgfgraphicnamed{diagrams//b1srt}
\InputIfFileExists{diagrams//b1srt.tikz}{}{\input{./figures/diagrams//b1srt.tikz}}
\endpgfgraphicnamed&(B1) & %
\beginpgfgraphicnamed{diagrams//b2slt}
\InputIfFileExists{diagrams//b2slt.tikz}{}{\input{./figures/diagrams//b2slt.tikz}}
\endpgfgraphicnamed=%
\beginpgfgraphicnamed{diagrams//b2srt}
\InputIfFileExists{diagrams//b2srt.tikz}{}{\input{./figures/diagrams//b2srt.tikz}}
\endpgfgraphicnamed&(B2)\\
\beginpgfgraphicnamed{diagrams//HadaDecomSingleslt}
\InputIfFileExists{diagrams//HadaDecomSingleslt.tikz}{}{\input{./figures/diagrams//HadaDecomSingleslt.tikz}}
\endpgfgraphicnamed= %
\beginpgfgraphicnamed{diagrams//HadaDecomSinglesrt2}
\InputIfFileExists{diagrams//HadaDecomSinglesrt2.tikz}{}{\input{./figures/diagrams//HadaDecomSinglesrt2.tikz}}
\endpgfgraphicnamed&(EU)    & %
\beginpgfgraphicnamed{diagrams//k2slt}
\InputIfFileExists{diagrams//k2slt.tikz}{}{\input{./figures/diagrams//k2slt.tikz}}
\endpgfgraphicnamed=%
\beginpgfgraphicnamed{diagrams//k2srt}
\InputIfFileExists{diagrams//k2srt.tikz}{}{\input{./figures/diagrams//k2srt.tikz}}
\endpgfgraphicnamed&(K2)\\
&%
\beginpgfgraphicnamed{diagrams//lemmainv}
\InputIfFileExists{diagrams//lemmainv.tikz}{}{\input{./figures/diagrams//lemmainv.tikz}}
\endpgfgraphicnamed &(IV) &\\
&&&\\ 
\hline
\end{array}\]
\end{center}
  \caption{Non-structural ZX-calculus rules, where $\alpha, \beta\in [0,~2\pi)$.}\label{figurenon}  
  \end{figure}
  \FloatBarrier
Note that all the rules enumerated in Figures \ref{figurenon} still hold when they are flipped upside-down. Due to the rule (H) and (H2), 
 the rules in Figure \ref{figurenon} have a property that they still hold when the colours green and red swapped. 
In this thesis, for simplicity, we won't distinguish a rule with its flipped upside-down version or colour swapped version when it is referred to in a diagrammatic rewriting. The structural rules listed in (\ref{compactstructure}) and (\ref{compactstructureslide}) will also be used without being explicitly stated.  %For this reason,  we won't distinguish any rule in Figure \ref{figurenon} with its colour swapped version when it is referred to in a diagrammatic rewriting.

If we let $d=2$ in the standard interpretation of qudit ZX-clculus, the we have the following interpretation for qubit:
\[
\left\llbracket %
\beginpgfgraphicnamed{diagrams//generator_spider2}
\InputIfFileExists{diagrams//generator_spider2.tikz}{}{\input{./figures/diagrams//generator_spider2.tikz}}
\endpgfgraphicnamed \right\rrbracket=\ket{0}^{\otimes m}\bra{0}^{\otimes n}+e^{i\alpha}\ket{1}^{\otimes m}\bra{1}^{\otimes n},
\left\llbracket %
\beginpgfgraphicnamed{diagrams//generator_redspider2}
\InputIfFileExists{diagrams//generator_redspider2.tikz}{}{\input{./figures/diagrams//generator_redspider2.tikz}}
\endpgfgraphicnamed\right\rrbracket=\ket{+}^{\otimes m}\bra{+}^{\otimes n}+e^{i\alpha}\ket{-}^{\otimes m}\bra{-}^{\otimes n},
\]

\[
\left\llbracket%
\beginpgfgraphicnamed{diagrams//HadaDecomSingleslt}
\InputIfFileExists{diagrams//HadaDecomSingleslt.tikz}{}{\input{./figures/diagrams//HadaDecomSingleslt.tikz}}
\endpgfgraphicnamed\right\rrbracket=\frac{1}{\sqrt{2}}\begin{pmatrix}
        1 & 1 \\
        1 & -1
 \end{pmatrix}, \quad
  \left\llbracket%
\beginpgfgraphicnamed{diagrams//emptysquare}
\InputIfFileExists{diagrams//emptysquare.tikz}{}{\input{./figures/diagrams//emptysquare.tikz}}
\endpgfgraphicnamed\right\rrbracket=1, \quad
\left\llbracket%
\beginpgfgraphicnamed{diagrams//Id}
\InputIfFileExists{diagrams//Id.tikz}{}{\input{./figures/diagrams//Id.tikz}}
\endpgfgraphicnamed\right\rrbracket=\begin{pmatrix}
        1 & 0 \\
        0 & 1
 \end{pmatrix}, 
   \]

\[
 \left\llbracket%
\beginpgfgraphicnamed{diagrams//swap}
\InputIfFileExists{diagrams//swap.tikz}{}{\input{./figures/diagrams//swap.tikz}}
\endpgfgraphicnamed\right\rrbracket=\begin{pmatrix}
        1 & 0 & 0 & 0 \\
        0 & 0 & 1 & 0 \\
        0 & 1 & 0 & 0 \\
        0 & 0 & 0 & 1 
 \end{pmatrix}, \quad
  \left\llbracket%
\beginpgfgraphicnamed{diagrams//cap}
\InputIfFileExists{diagrams//cap.tikz}{}{\input{./figures/diagrams//cap.tikz}}
\endpgfgraphicnamed\right\rrbracket=\begin{pmatrix}
        1  \\
        0  \\
        0  \\
        1  \\
 \end{pmatrix}, \quad
   \left\llbracket%
\beginpgfgraphicnamed{diagrams//cup}
\InputIfFileExists{diagrams//cup.tikz}{}{\input{./figures/diagrams//cup.tikz}}
\endpgfgraphicnamed\right\rrbracket=\begin{pmatrix}
        1 & 0 & 0 & 1 
         \end{pmatrix},
   \]

\[  \llbracket D_1\otimes D_2  \rrbracket =  \llbracket D_1  \rrbracket \otimes  \llbracket  D_2  \rrbracket, \quad 
 \llbracket D_1\circ D_2  \rrbracket =  \llbracket D_1  \rrbracket \circ  \llbracket  D_2  \rrbracket,
  \]
where 
$$ \ket{0}= \begin{pmatrix}
        1  \\
        0  \\
 \end{pmatrix}, \quad 
 \bra{0}=\begin{pmatrix}
        1 & 0 
         \end{pmatrix},
 \quad  \ket{1}= \begin{pmatrix}
        0  \\
        1  \\
 \end{pmatrix}, \quad 
  \bra{1}=\begin{pmatrix}
     0 & 1 
         \end{pmatrix},
 $$
$$ \ket{+}= \frac{1}{\sqrt{2}}\begin{pmatrix}
        1  \\
        1  \\
 \end{pmatrix}, \quad 
 \bra{+}=\frac{1}{\sqrt{2}}\begin{pmatrix}
        1 & 1
         \end{pmatrix},
 \quad  \ket{-}= \frac{1}{\sqrt{2}}\begin{pmatrix}
        1  \\
        -1  \\
 \end{pmatrix}, \quad 
  \bra{-}=\frac{1}{\sqrt{2}}\begin{pmatrix}
     1 & -1 
         \end{pmatrix}.
 $$

Below we give some useful properties of the qubit ZX-calculus.

\begin{lemma}
\begin{equation}\label{lem:hopf}
	\input{diagrams/lemmahopf.tikz}
	\end{equation}
\end{lemma}

\begin{proof}
	Proof in \cite{bpw}. The rules used are $S1$, $S2$, $S3$, $H2$, $H3$, $H$, $B1$, $B2$.
\end{proof}

\begin{lemma}
	\begin{equation}\label{lem:6b}
	\input{diagrams/lemma6b.tikz}
	\end{equation}
\end{lemma}

\begin{proof}
	Proof in \cite{bpw}. The rules and properties used are $S1$, $S2$, $S3$, $H$, $H2$, $H3$, $B1$, $B2$, $EU$, $IV$.
\end{proof}

\begin{lemma}
	\begin{equation}\label{lem:com}
	\input{diagrams/lemmacom.tikz}
	\end{equation}
\end{lemma}

\begin{proof}
	Proof in \cite{bpw}. The rules and properties used are \ref{lem:hopf}, $B2$, $S2$, $H2$, $S1$, $H$, $B1$.
\end{proof}

\begin{lemma}
	\begin{equation}\label{lem:S4}
\beginpgfgraphicnamed{diagrams//alphadeletestate}
\InputIfFileExists{diagrams//alphadeletestate.tikz}{}{\input{./figures/diagrams//alphadeletestate.tikz}}
\endpgfgraphicnamed
	\end{equation}
	\end{lemma}

\begin{proof}
	Proof in \cite{bpw}. The rules and properties used are $S1$, $B1$, $K2$, $\ref{lem:6b}$, $IV$.
\end{proof}

\begin{lemma}
	\begin{equation}\label{lem:1}
	\input{diagrams/lemma1.tikz}
	\end{equation}
\end{lemma}

\begin{proof}~\newline \newline
	\input{diagrams/lemma1_proof.tikz}
\end{proof}

\begin{lemma}
	\begin{equation}\label{cntswap}
	\input{diagrams/cnotswap.tikz}
	\end{equation}
\end{lemma}

\begin{proof}~\newline \newline
	\input{diagrams/cntswap_proof.tikz}
\end{proof}

\begin{lemma}
	\begin{equation}\label{pbtsaa}
\input{diagrams/alphadelete.tikz}	%\input{diagrams/piby2trans.tikz} alphadelete
	\end{equation}
\end{lemma}
\begin{proof}
	Proof in \cite{bpw}. The rules and properties used are $S1$, $B1$, $K2$, \ref{lem:6b}, $IV$.
\end{proof}

\begin{lemma}
	\begin{equation}\label{lem:6}
	\input{diagrams/lemma6.tikz}
	\end{equation}
	%Equivalently,\\
	%$
	%\input{diagrams/lemma6v2.tikz}
	%$
\end{lemma}

\begin{proof}
	\begin{equation*}
	\input{diagrams/lemma6-proof.tikz}
	\end{equation*}
	%Straightforward application of rules $H$ and $EU$.
\end{proof}

\begin{lemma}
	\begin{equation}\label{3crossing}
	\input{diagrams/3crossings.tikz}
	\end{equation}
\end{lemma}

\begin{proof} The proof can be done by simply sliding the middle line (the line connects the second input and the second output) from the left to the right by naturality. 
\end{proof}

\subsection{ Some known completeness results of the qubit ZX-calculus}
 It has been shown in \cite{Vladimir} that the original version of the ZX-calculus  \cite{CoeckeDuncan}  plus the Euler decomposition of Hadamard gate is incomplete for the overall pure qubit quantum mechanics (QM). Since then, plenty of efforts have been devoted to the completion of some fragment of  qubit QM. In fact, the $\pi$-fragment of the ZX-calculus (corresponding to diagrams involving angles multiple of $\pi$) has been proved to be complete for real stabilizer QM in \cite{duncanperdrix}, and 
the $\frac{\pi}{2}$-fragment of the ZX-calculus was shown to be complete for the stabilizer QM in \cite{Miriam1}. Moreover, Backens has given the proof of completeness for  single qubit Clifford+T ZX-calculus in  \cite{Miriam1ct}.

The next chapters of this thesis will fill the gap between the above results and the universal completeness of the ZX-calculus for the whole QM.

 %and Clifford+T QM \cite{Emmanuel}.  stabilzer ZX,  incompleteness results, ...Also  single qubit Clifford+T QM \cite{Miriam1ct}.  

 \section{ZW-calculus}
%{ \bf The ZW-calculus is proved to be sound, universal and complete \cite{amar}. more details here, especially completeness for any ring stated as a theorem.}

The ZW-calculus is another graphical language for quantum computing modelled on the ZX-calculus \cite{amar}.
Let  $R$ be an arbitrary commutative ring. The ZW-calculus with all parameters in $R$ is denoted as $ZW_{R} $-calculus. Like the ZX-calculus, the $ZW_{R} $-calculus is also a self-dual compact closed PROP $\mathfrak{F}$ with a set of rewriting rules. An arbitrary morphism of $\mathfrak{F}$ is a diagram  $D:k\to l$  with source object $k$ and target object $l$, composed of the following basic components:

\begin{center} % \tikzfig{diagrams//emptysquare-small}
\begin{tabular}{|r@{~}r@{~}c@{~}c|r@{~}r@{~}c@{~}c|}
\hline
%$Z^{(n,m)}$&$:$&$n\to m$ & \tikzfig{diagrams//generatorwtspider} & $R$&$:$&$ 1\to 1$& \tikzfig{diagrams//rgatewhite}\\
$Z$&$:$&$1\to 2$ & %
\beginpgfgraphicnamed{diagrams//generatorwtspider2}
\InputIfFileExists{diagrams//generatorwtspider2.tikz}{}{\input{./figures/diagrams//generatorwtspider2.tikz}}
\endpgfgraphicnamed & $R$&$:$&$ 1\to 1$& %
\beginpgfgraphicnamed{diagrams//rgatewhite}
\InputIfFileExists{diagrams//rgatewhite.tikz}{}{\input{./figures/diagrams//rgatewhite.tikz}}
\endpgfgraphicnamed\\
\hline
$\tau$&$:$&$2\to 2$ &%
\beginpgfgraphicnamed{diagrams//corsszw}
\InputIfFileExists{diagrams//corsszw.tikz}{}{\input{./figures/diagrams//corsszw.tikz}}
\endpgfgraphicnamed
%& $s$&$:$&$0\to 0$ &\tikzfig{scalars//halfscalar}  &
 & $P$&$:$&$1\to 1$  &%
\beginpgfgraphicnamed{diagrams//piblack}
\InputIfFileExists{diagrams//piblack.tikz}{}{\input{./figures/diagrams//piblack.tikz}}
\endpgfgraphicnamed \\\hline
  $\sigma$&$:$&$ 2\to 2$& %
\beginpgfgraphicnamed{diagrams//swap}
\InputIfFileExists{diagrams//swap.tikz}{}{\input{./figures/diagrams//swap.tikz}}
\endpgfgraphicnamed &$\mathbb I$&$:$&$1\to 1$&%
\beginpgfgraphicnamed{diagrams//Id}
\InputIfFileExists{diagrams//Id.tikz}{}{\input{./figures/diagrams//Id.tikz}}
\endpgfgraphicnamed \\\hline
   $e $&$:$&$0 \to 0$& %
\beginpgfgraphicnamed{diagrams//emptysquare}
\InputIfFileExists{diagrams//emptysquare.tikz}{}{\input{./figures/diagrams//emptysquare.tikz}}
\endpgfgraphicnamed &$W$&$:$&$1\to 2$&%
\beginpgfgraphicnamed{diagrams//wblack}
\InputIfFileExists{diagrams//wblack.tikz}{}{\input{./figures/diagrams//wblack.tikz}}
\endpgfgraphicnamed 
  \\\hline
   $C_a$&$:$&$ 0\to 2$& %
\beginpgfgraphicnamed{diagrams//cap}
\InputIfFileExists{diagrams//cap.tikz}{}{\input{./figures/diagrams//cap.tikz}}
\endpgfgraphicnamed &$ C_u$&$:$&$ 2\to 0$&%
\beginpgfgraphicnamed{diagrams//cup}
\InputIfFileExists{diagrams//cup.tikz}{}{\input{./figures/diagrams//cup.tikz}}
\endpgfgraphicnamed \\\hline
\end{tabular}
\end{center}
%where $m,n\in \mathbb{N}$,   $r \in  \mathbb{C}$, 
where $r \in  R$,  
and $e$ represents an empty diagram. With these generators, we can define the following diagrams:
\begin{equation}\label{wnodespdef}
%\tikzfig{diagrams//wnodesdef} 
%
\beginpgfgraphicnamed{diagrams//wnodesdefadd}
\InputIfFileExists{diagrams//wnodesdefadd.tikz}{}{\input{./figures/diagrams//wnodesdefadd.tikz}}
\endpgfgraphicnamed 
\end{equation}

The composition of morphisms is  to combine these components in the following two ways: for any two morphisms $D_1:a\to b$ and $D_2: c\to d$, a \textit{ parallel composition} $D_1\otimes D_2 : a+c\to b+d$ is obtained by placing $D_1$ and $D_2$ side-by-side with $D_1$ on the left of $D_2$;
 for any two morphisms $D_1:a\to b$ and $D_2: b\to c$,  a  \textit{ sequential  composition} $D_2\circ D_1 : a\to c$ is obtained by placing $D_1$ above $D_2$, connecting the outputs of $D_1$ to the inputs of $D_2$.

\begin{figure}[!h]
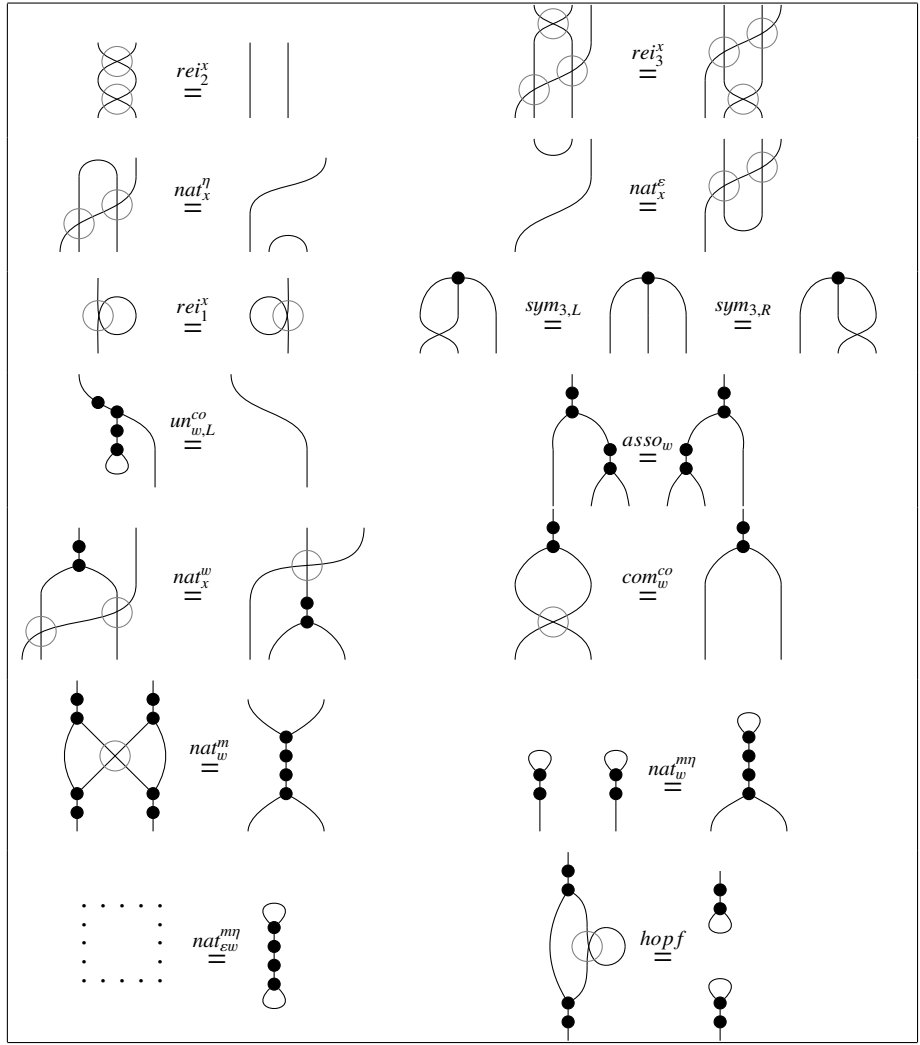

\begin{center}
\[
\quad \qquad\begin{array}{|cccc|}
\hline
\beginpgfgraphicnamed{diagrams//reix2}
\InputIfFileExists{diagrams//reix2.tikz}{}{\input{./figures/diagrams//reix2.tikz}}
\endpgfgraphicnamed& &%
\beginpgfgraphicnamed{diagrams//reix3}
\InputIfFileExists{diagrams//reix3.tikz}{}{\input{./figures/diagrams//reix3.tikz}}
\endpgfgraphicnamed & \\
\beginpgfgraphicnamed{diagrams//natnx}
\InputIfFileExists{diagrams//natnx.tikz}{}{\input{./figures/diagrams//natnx.tikz}}
\endpgfgraphicnamed & & %
\beginpgfgraphicnamed{diagrams//natex}
\InputIfFileExists{diagrams//natex.tikz}{}{\input{./figures/diagrams//natex.tikz}}
\endpgfgraphicnamed &\\
%  \tikzfig{diagrams//reix1} & &\tikzfig{diagrams//uncowR}&\\
 %
\beginpgfgraphicnamed{diagrams//reix1}
\InputIfFileExists{diagrams//reix1.tikz}{}{\input{./figures/diagrams//reix1.tikz}}
\endpgfgraphicnamed & &%
\beginpgfgraphicnamed{diagrams//sym3}
\InputIfFileExists{diagrams//sym3.tikz}{}{\input{./figures/diagrams//sym3.tikz}}
\endpgfgraphicnamed&\\
\beginpgfgraphicnamed{diagrams//uncowL}
\InputIfFileExists{diagrams//uncowL.tikz}{}{\input{./figures/diagrams//uncowL.tikz}}
\endpgfgraphicnamed& & %
\beginpgfgraphicnamed{diagrams//natww-equiv}
\InputIfFileExists{diagrams//natww-equiv.tikz}{}{\input{./figures/diagrams//natww-equiv.tikz}}
\endpgfgraphicnamed&\\
\beginpgfgraphicnamed{diagrams//natwx}
\InputIfFileExists{diagrams//natwx.tikz}{}{\input{./figures/diagrams//natwx.tikz}}
\endpgfgraphicnamed&  & %
\beginpgfgraphicnamed{diagrams//comcow}
\InputIfFileExists{diagrams//comcow.tikz}{}{\input{./figures/diagrams//comcow.tikz}}
\endpgfgraphicnamed&\\
\beginpgfgraphicnamed{diagrams//natmw}
\InputIfFileExists{diagrams//natmw.tikz}{}{\input{./figures/diagrams//natmw.tikz}}
\endpgfgraphicnamed & &%
\beginpgfgraphicnamed{diagrams//natmnw}
\InputIfFileExists{diagrams//natmnw.tikz}{}{\input{./figures/diagrams//natmnw.tikz}}
\endpgfgraphicnamed &\\
\beginpgfgraphicnamed{diagrams//natmnew}
\InputIfFileExists{diagrams//natmnew.tikz}{}{\input{./figures/diagrams//natmnew.tikz}}
\endpgfgraphicnamed& &%
\beginpgfgraphicnamed{diagrams//hopf}
\InputIfFileExists{diagrams//hopf.tikz}{}{\input{./figures/diagrams//hopf.tikz}}
\endpgfgraphicnamed &\\
%\tikzfig{diagrams//sym3} & &  &\\
\hline
\end{array}\]
\end{center}

  \caption{$ZW_{ R}$-calculus rules I}\label{figure3}
\end{figure}

There are two kinds of rules for the morphisms of $\mathfrak{F}$:  the structure rules for $\mathfrak{F}$ as an compact closed category shown in (\ref{compactstructure}) and (\ref{compactstructureslide}), as well as the rewriting rules listed in Figure \ref{figure3}, \ref{figure4}, \ref{figure5}.

Like the ZX-calculus, all the ZW diagrams should be read from top to bottom.

\begin{figure}[!h]
\begin{center}
\[
\quad \qquad\begin{array}{|cccc|}
\hline
%  \tikzfig{diagrams//sym2} &\tikzfig{diagrams//inv}&&\\
  %
\beginpgfgraphicnamed{diagrams//antnx}
\InputIfFileExists{diagrams//antnx.tikz}{}{\input{./figures/diagrams//antnx.tikz}}
\endpgfgraphicnamed  &%
\beginpgfgraphicnamed{diagrams//inv}
\InputIfFileExists{diagrams//inv.tikz}{}{\input{./figures/diagrams//inv.tikz}}
\endpgfgraphicnamed&&\\
%    &\tikzfig{diagrams//symz}&&\\
    \multicolumn{3}{|c}{%
\beginpgfgraphicnamed{diagrams//symz}
\InputIfFileExists{diagrams//symz.tikz}{}{\input{./figures/diagrams//symz.tikz}}
\endpgfgraphicnamed}&\\ \hspace{1.6cm}
\beginpgfgraphicnamed{diagrams//uncozR}
\InputIfFileExists{diagrams//uncozR.tikz}{}{\input{./figures/diagrams//uncozR.tikz}}
\endpgfgraphicnamed  &%
\beginpgfgraphicnamed{diagrams//associaz}
\InputIfFileExists{diagrams//associaz.tikz}{}{\input{./figures/diagrams//associaz.tikz}}
\endpgfgraphicnamed  & & \\
\beginpgfgraphicnamed{diagrams//ph}
\InputIfFileExists{diagrams//ph.tikz}{}{\input{./figures/diagrams//ph.tikz}}
\endpgfgraphicnamed  &%
\beginpgfgraphicnamed{diagrams//natnc}
\InputIfFileExists{diagrams//natnc.tikz}{}{\input{./figures/diagrams//natnc.tikz}}
\endpgfgraphicnamed  & & \\
\beginpgfgraphicnamed{diagrams//natmc}
\InputIfFileExists{diagrams//natmc.tikz}{}{\input{./figures/diagrams//natmc.tikz}}
\endpgfgraphicnamed  &%
\beginpgfgraphicnamed{diagrams//loop}
\InputIfFileExists{diagrams//loop.tikz}{}{\input{./figures/diagrams//loop.tikz}}
\endpgfgraphicnamed  & & \\
\beginpgfgraphicnamed{diagrams//unx}
\InputIfFileExists{diagrams//unx.tikz}{}{\input{./figures/diagrams//unx.tikz}}
\endpgfgraphicnamed  &%
\beginpgfgraphicnamed{diagrams//rng1}
\InputIfFileExists{diagrams//rng1.tikz}{}{\input{./figures/diagrams//rng1.tikz}}
\endpgfgraphicnamed  & & \\
\beginpgfgraphicnamed{diagrams//rng-1}
\InputIfFileExists{diagrams//rng-1.tikz}{}{\input{./figures/diagrams//rng-1.tikz}}
\endpgfgraphicnamed  &%
\beginpgfgraphicnamed{diagrams//rngrsx}
\InputIfFileExists{diagrams//rngrsx.tikz}{}{\input{./figures/diagrams//rngrsx.tikz}}
\endpgfgraphicnamed  & & \\
\hline
\end{array}\]
\end{center}

  \caption{$ZW_{ R}$-calculus rules II}\label{figure4}
\end{figure}

\begin{figure}[!h]
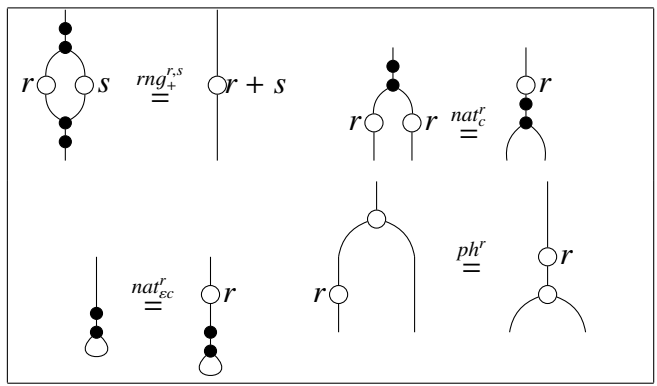

\begin{center}
\[
\quad \qquad\begin{array}{|cccc|}
\hline
\beginpgfgraphicnamed{diagrams//rngrsp}
\InputIfFileExists{diagrams//rngrsp.tikz}{}{\input{./figures/diagrams//rngrsp.tikz}}
\endpgfgraphicnamed  &%
\beginpgfgraphicnamed{diagrams//natrc}
\InputIfFileExists{diagrams//natrc.tikz}{}{\input{./figures/diagrams//natrc.tikz}}
\endpgfgraphicnamed   & & \\
\beginpgfgraphicnamed{diagrams//natrec}
\InputIfFileExists{diagrams//natrec.tikz}{}{\input{./figures/diagrams//natrec.tikz}}
\endpgfgraphicnamed  &%
\beginpgfgraphicnamed{diagrams//phr}
\InputIfFileExists{diagrams//phr.tikz}{}{\input{./figures/diagrams//phr.tikz}}
\endpgfgraphicnamed  & & \\
\hline
\end{array}\]
\end{center}

  \caption{$ZW_{ R}$-calculus rules III}\label{figure5}
\end{figure}

\FloatBarrier

%Note that here we used finite number of generators for the ZW-calculus. However, there is an equivalent set of generators for the ZW-calculus as well. 
Note that here we presented a $ZW_{ R}$-calculus generated by a finite set of diagrams. However, there is an equivalent yet more concise presentation generated by an infinite set of diagrams.
Both of them are proposed in \cite{amar}.  We have the following spider form of the white node due to expressions in (\ref{wnodespdef}) and the rule $rng_1$, which will be used for translations between the ZX-calculus and the ZW-calculus.

\begin{equation}\label{wspiderdef}
 %\tikzfig{diagrams//spiderwhitedef}
 \input{diagrams/spiderwhitedef.tikz}
\end{equation}
By the rules $sym_{z,L}$, $sym_{z,R}$ and $asso_{z}$ listed in Figure \ref{figure4}, this white node spider is commutative.

The diagrams in the $ZW_{ R}$-calculus have a standard interpretation $\llbracket \cdot \rrbracket$ in the category  $\mathbf{Mat}_{R}$. %  of    $\mathbf{FdHilb_2}$:
\[
\left\llbracket %
\beginpgfgraphicnamed{diagrams//generatorwtspider2}
\InputIfFileExists{diagrams//generatorwtspider2.tikz}{}{\input{./figures/diagrams//generatorwtspider2.tikz}}
\endpgfgraphicnamed \right\rrbracket=\begin{pmatrix}
        1 & 0  \\
        0 & 0 \\
        0 & 0  \\
        0 & 1 
 \end{pmatrix},
\quad
\left\llbracket%
\beginpgfgraphicnamed{diagrams//rgatewhite}
\InputIfFileExists{diagrams//rgatewhite.tikz}{}{\input{./figures/diagrams//rgatewhite.tikz}}
\endpgfgraphicnamed\right\rrbracket=\ket{0}\bra{0}+r\ket{1}\bra{1}=\begin{pmatrix}
        1 & 0 \\
        0 & r
 \end{pmatrix}.
\]

\[
\left\llbracket%
\beginpgfgraphicnamed{diagrams//corsszw}
\InputIfFileExists{diagrams//corsszw.tikz}{}{\input{./figures/diagrams//corsszw.tikz}}
\endpgfgraphicnamed\right\rrbracket=\begin{pmatrix}
         1 & 0 & 0 & 0 \\
        0 & 0 & 1 & 0 \\
        0 & 1 & 0 & 0 \\
        0 & 0 & 0 & -1 
 \end{pmatrix}, \quad
  \left\llbracket%
\beginpgfgraphicnamed{diagrams//wblack}
\InputIfFileExists{diagrams//wblack.tikz}{}{\input{./figures/diagrams//wblack.tikz}}
\endpgfgraphicnamed\right\rrbracket=\begin{pmatrix}
        0 & 1  \\
        1 & 0 \\
        1 & 0  \\
        0 & 0 
 \end{pmatrix}, \quad
  \left\llbracket%
\beginpgfgraphicnamed{diagrams//piblack}
\InputIfFileExists{diagrams//piblack.tikz}{}{\input{./figures/diagrams//piblack.tikz}}
\endpgfgraphicnamed\right\rrbracket=\begin{pmatrix}
        0 & 1 \\
        1 & 0
 \end{pmatrix}, \quad
   \left\llbracket%
\beginpgfgraphicnamed{diagrams//emptysquare}
\InputIfFileExists{diagrams//emptysquare.tikz}{}{\input{./figures/diagrams//emptysquare.tikz}}
\endpgfgraphicnamed\right\rrbracket=1.
   \]

\[
\left\llbracket%
\beginpgfgraphicnamed{diagrams//Id}
\InputIfFileExists{diagrams//Id.tikz}{}{\input{./figures/diagrams//Id.tikz}}
\endpgfgraphicnamed\right\rrbracket=\begin{pmatrix}
        1 & 0 \\
        0 & 1
 \end{pmatrix}, \quad
  \left\llbracket%
\beginpgfgraphicnamed{diagrams//swap}
\InputIfFileExists{diagrams//swap.tikz}{}{\input{./figures/diagrams//swap.tikz}}
\endpgfgraphicnamed\right\rrbracket=\begin{pmatrix}
        1 & 0 & 0 & 0 \\
        0 & 0 & 1 & 0 \\
        0 & 1 & 0 & 0 \\
        0 & 0 & 0 & 1 
 \end{pmatrix}, \quad
  \left\llbracket%
\beginpgfgraphicnamed{diagrams//cap}
\InputIfFileExists{diagrams//cap.tikz}{}{\input{./figures/diagrams//cap.tikz}}
\endpgfgraphicnamed\right\rrbracket=\begin{pmatrix}
        1  \\
        0  \\
        0  \\
        1  \\
 \end{pmatrix}, \quad
   \left\llbracket%
\beginpgfgraphicnamed{diagrams//cup}
\InputIfFileExists{diagrams//cup.tikz}{}{\input{./figures/diagrams//cup.tikz}}
\endpgfgraphicnamed\right\rrbracket=\begin{pmatrix}
        1 & 0 & 0 & 1 
         \end{pmatrix}.
   \]

\[  \llbracket D_1\otimes D_2  \rrbracket =  \llbracket D_1  \rrbracket \otimes  \llbracket  D_2  \rrbracket, \quad 
 \llbracket D_1\circ D_2  \rrbracket =  \llbracket D_1  \rrbracket \circ  \llbracket  D_2  \rrbracket.
  \]

Soundness, universality and completeness of the $ZW_{ R}$-calculus can be similarly defined as that for the ZX-calculus. We only recall these properties here.
\begin{theorem}\label{zwsounduniv}\cite{amar}
The  $ZW_{ R}$-calculus is sound and universal.
\end{theorem}

\begin{theorem}\label{zwcomplete}\cite{amar}
The  $ZW_{ R}$-calculus is complete.
\end{theorem}
%It is shown in \cite{amar} that the ZW-calculus is sound, which means the interpretation $\llbracket \cdot \rrbracket$ is a symmetric monoidal functor. 

%% file: diagrams/HadaDecomSingleslt.tikz
\begin{tikzpicture}
	\begin{pgfonlayer}{nodelayer}
		\node [style=H box] (0) at (-0.75, 0) {$H$};
		\node [style=none] (1) at (-0.75, -0.5) {};
		\node [style=none] (2) at (-0.75, 0.5) {};
	\end{pgfonlayer}
	\begin{pgfonlayer}{edgelayer}
		\draw (2.center) to (0);
		\draw (1.center) to (0);
	\end{pgfonlayer}
\end{tikzpicture}

%% file: diagrams/Id.tikz
\begin{tikzpicture}
	\begin{pgfonlayer}{nodelayer}
		\node [style=none] (1) at (0.5, 0.3) {};
		\node [style=none] (2) at (0.5, -0.3) {};
		\node [style=none] (3) at (0.5, -0.5) {};
		\node [style=none] (4) at (0.5, 0.5) {};
	\end{pgfonlayer}
	\begin{pgfonlayer}{edgelayer}
		\draw (1.center) to (2.center);
	\end{pgfonlayer}
\end{tikzpicture}

%% file: diagrams/cap.tikz
\begin{tikzpicture}
	\begin{pgfonlayer}{nodelayer}
		\node [style=none] (0) at (0, -0) {};
		\node [style=none] (1) at (1, -0) {};
	\end{pgfonlayer}
	\begin{pgfonlayer}{edgelayer}
		\draw [bend left=90, looseness=1.50] (0.center) to (1.center);
	\end{pgfonlayer}
\end{tikzpicture}

%% file: diagrams/cup.tikz
\begin{tikzpicture}
	\begin{pgfonlayer}{nodelayer}
		\node [style=none] (0) at (0, 0.5) {};
		\node [style=none] (1) at (1, 0.5) {};
	\end{pgfonlayer}
	\begin{pgfonlayer}{edgelayer}
		\draw [bend right=90, looseness=1.50] (0.center) to (1.center);
	\end{pgfonlayer}
\end{tikzpicture}

%% file: diagrams/alphadelete.tikz
\begin{tikzpicture}
	\begin{pgfonlayer}{nodelayer}
		\node [style=rn] (0) at (0.2, 0.25) {};
		\node [style=gn] (1) at (0.2, -0.25) {$\alpha$};
		\node [style=none] (2) at (0.5, -0) {$=$};
		\node [style=rn] (3) at (0.8, 0.25) {};
		\node [style=gn] (4) at (0.8, -0.25) {};
	\end{pgfonlayer}
	\begin{pgfonlayer}{edgelayer}
		\draw (0) to (1);
		\draw (3) to (4);
	\end{pgfonlayer}
\end{tikzpicture}

%% file: diagrams/rgatewhite.tikz
\begin{tikzpicture}
	\begin{pgfonlayer}{nodelayer}
		\node [style=wn] (0) at (0, 0) {};
		\node [style=none] (1) at (0, -0.5) {};
		\node [style=none] (2) at (0.25, 0) {$r$};
		\node [style=none] (3) at (0, 0.5) {};
	\end{pgfonlayer}
	\begin{pgfonlayer}{edgelayer}
		\draw (3.center) to (1.center);
	\end{pgfonlayer}
\end{tikzpicture}

%% file: diagrams/piblack.tikz
\begin{tikzpicture}
	\begin{pgfonlayer}{nodelayer}
		\node [style=bn] (0) at (0, 0) {};
		\node [style=none] (1) at (0, -0.5) {};
		\node [style=none] (2) at (0, 0.5) {};
	\end{pgfonlayer}
	\begin{pgfonlayer}{edgelayer}
		\draw (2.center) to (1.center);
	\end{pgfonlayer}
\end{tikzpicture}

%% file: diagrams/wblack.tikz
\begin{tikzpicture}
	\begin{pgfonlayer}{nodelayer}
		\node [style=bn] (0) at (0, 0) {};
		\node [style=none] (1) at (-0.25, -0.5) {};
		\node [style=none] (2) at (0.25, -0.5) {};
		\node [style=none] (3) at (0, 0.5) {};
	\end{pgfonlayer}
	\begin{pgfonlayer}{edgelayer}
		\draw (3.center) to (0);
		\draw (0) to (1.center);
		\draw (0) to (2.center);
	\end{pgfonlayer}
\end{tikzpicture}

%% file: chapter3.tex
\chapter{Completeness for full qubit quantum mechanics}\label{chfull}

%\section{Introduction}
%The ZX-calculus introduced by Coecke and Duncan \cite{CoeckeDuncan}  is an graphical language for both quantum foundation and quantum computing, with intuitive rewriting rules transforming diagrams from one to another. As an important branch of categorical quantum mechanics (CQM) pioneered by Abramsky and Coecke \cite{Coeckesamson},  the ZX-calculus has been applied to various fields such as measurement-based quantum computing  \cite{Duncanpx,  Horsman}  and quantum error correction \cite{DuncanLucas,  ckzh}. 

%To realise its greatest advantage, the so-called completeness is of concerned with the ZX-calculus, that is,  any equation of diagrams that holds true when calculated by corresponding matrices can be derived diagrammatically. 

 It has been shown in \cite{Vladimir} that the original version of the ZX-calculus  \cite{CoeckeDuncan}  plus the Euler decomposition of Hadamard gate is incomplete for the overall pure qubit quantum mechanics (QM). Since then, plenty of efforts have been devoted to the completion of some part of QM:  real QM \cite{duncanperdrix}, stabilizer QM \cite{Miriam1}, single qubit Clifford+T QM \cite{Miriam1ct}  and Clifford+T QM \cite{Emmanuel}.  Amongst them, the completeness of ZX-calculus for Clifford+T QM is especially interesting, since it is approximatively universal for QM.  Note that their proof relies on the completeness of ZW-calculus for  ``qubits with integer coefficients" \cite{amar1}.

In this chapter, we give the first complete axiomatisation of the ZX-calculus for the entire qubit QM, i.e., the $ZX_{ full}$-calculus, based on the completeness result of the $ZW_{ \mathbb{C}}$-calculus  \cite{amar}. Firstly, we introduce two new generators: a triangle and a series of  $\lambda$-labeled boxes ($\lambda \geq 0$),  which turns out to be expressible in ZX-calculus  without these symbols.  Then we establish reversible translations from ZX to ZW and vice versa. By checking carefully that all the ZW rewriting rules still hold under translation from ZW to ZX, we finally finished the proof of completeness of $ZX_{ full}$-calculus.  

Throughout this chapter,  the terms ``equation" and ``rewriting rule" will be used interchangeably. 
%The proof of the completeness  results of this chapter are collected from the arXiv paper \cite{ngwang}.
The proof of the completeness of the full qubit ZX-calculus has been published in \cite{amarngwanglics}, with coauthors Amar Hadzihasanovic and Kang Feng Ng.

\section{$ZX_{ full}$-calculus}\label{zxccls}

The $ZX_{ full}$-calculus has generators as listed in Table \ref{qbzxgenerator} plus two new generators given in Table \ref{newgr}.

 % \tikzfig{diagrams//emptysquare-small}
\begin{table}\begin{center}
	\begin{tabular}{|r@{~}r@{~}c@{~}c|r@{~}r@{~}c@{~}c|}
		\hline
		$L$&$:$&$1\to 1$  &%
\beginpgfgraphicnamed{diagrams//lambdabox}
\InputIfFileExists{diagrams//lambdabox.tikz}{}{\input{./figures/diagrams//lambdabox.tikz}}
\endpgfgraphicnamed &$T$&$:$&$1\to 1$&%
\beginpgfgraphicnamed{diagrams//triangle}
\InputIfFileExists{diagrams//triangle.tikz}{}{\input{./figures/diagrams//triangle.tikz}}
\endpgfgraphicnamed \\\hline
	\end{tabular}\caption{New generators with $ \lambda  \geq 0$.} \label{newgr}\end{center}
	\end{table}
\FloatBarrier

%These generators are the same as the normal ones  \cite{CoeckeDuncan, Miriam1}, but it turns out to be very useful to introduce two new generators (although in principle they could be eliminated, see 
%lemma \ref{lem:lamb_tri_decomposition}) as well:
It seems that the $ZX_{ full}$-calculus has more generators than the traditional ZX-calculus. However we will show that they are expressible in red and green nodes in Proposition \ref{lem:lamb_tri_decomposition}.

 Also we define the following notation:
 \begin{equation}\label{downtriangledef}
\beginpgfgraphicnamed{diagrams//downtriangle}
\InputIfFileExists{diagrams//downtriangle.tikz}{}{\input{./figures/diagrams//downtriangle.tikz}}
\endpgfgraphicnamed
\end{equation}
Then it is clear that 
$$ %
\beginpgfgraphicnamed{diagrams//horizontriangle-proof}
\InputIfFileExists{diagrams//horizontriangle-proof.tikz}{}{\input{./figures/diagrams//horizontriangle-proof.tikz}}
\endpgfgraphicnamed$$
Thus it makes sense to draw the following picture:
$$ %
\beginpgfgraphicnamed{diagrams//horizontriangle}
\InputIfFileExists{diagrams//horizontriangle.tikz}{}{\input{./figures/diagrams//horizontriangle.tikz}}
\endpgfgraphicnamed$$

%Moreover, we define the red spider for any $\alpha \in [0,  2\pi)$ as follows:
%\begin{equation}\label{hadamadefine}
%  \tikzfig{diagrams//h2newlt}:=\tikzfig{diagrams//h2newrt}      
%  \end{equation}

%The composition of morphisms is  to combine these components in the following two ways: for any two morphisms $D_1:a\to b$ and $D_2: c\to d$, a \textit{ parallel composition} $D_1\otimes D_2 : a+c\to b+d$ is obtained by placing $D_1$ and $D_2$ side-by-side with $D_1$ on the left of $D_2$; for any two morphisms $D_1:a\to b$ and $D_2: b\to c$,  a  \textit{ sequential  composition} $D_2\circ D_1 : a\to c$ is obtained by placing $D_1$ above $D_2$, connecting the outputs of $D_1$ to the inputs of $D_2$.

%There are two kinds of rules for the morphisms of $\mathfrak{C}$: 
 The $ZX_{ full}$-calculus has the same structural rules as that of the traditional ZX-calculus given in (\ref{compactstructure})  and (\ref{compactstructureslide}). Its non-structural rewriting rules are presented in Figures \ref{figure1} (exactly the same as  Figure  \ref{figurenon}), \ref{figure2} and  \ref{figure0}:

%\begin{itemize}
%\item the structure rules for $\mathfrak{C}$ as an compact closed category which include the naturality rule that any diagram can slide freely along either wire of a swap (the generator $\sigma$) and the following rules for a compact structure:
%\FloatBarrier
%\begin{figure}
 %\centering
% \begin{equation}\label{compactstructure}
 %\tikzfig{diagrams//compactstructure_cap} \qquad\quad \tikzfig{diagrams//compactstructure_cup} \qquad\quad \tikzfig{diagrams//compactstructure_snake}
 %\end{equation}
 %\caption{Some of the equations satisfied by the structural maps in a compact closed category: the caps and cups are symmetrical and they satisfy the \emph{snake equations}.}\label{fig:compact_structure}
%\end{figure}

%\FloatBarrier

%\item  traditional rewriting rules listed in Figure \ref{figure1} and our extended rules listed in Figure \ref{figure2} and Figure \ref{figure0}. 
%\end{itemize}

%Note that in this thesis any diagram should be read from top to bottom, and

\begin{figure}[!h]
\begin{center}
\[
\quad \qquad\begin{array}{|cccc|}
\hline
\beginpgfgraphicnamed{diagrams//spiderlt}
\InputIfFileExists{diagrams//spiderlt.tikz}{}{\input{./figures/diagrams//spiderlt.tikz}}
\endpgfgraphicnamed=%
\beginpgfgraphicnamed{diagrams//spiderrt}
\InputIfFileExists{diagrams//spiderrt.tikz}{}{\input{./figures/diagrams//spiderrt.tikz}}
\endpgfgraphicnamed &(S1) &%
\beginpgfgraphicnamed{diagrams//s2new}
\InputIfFileExists{diagrams//s2new.tikz}{}{\input{./figures/diagrams//s2new.tikz}}
\endpgfgraphicnamed &(S2)\\
\beginpgfgraphicnamed{diagrams//induced_compact_structure-2wirelt}
\InputIfFileExists{diagrams//induced_compact_structure-2wirelt.tikz}{}{\input{./figures/diagrams//induced_compact_structure-2wirelt.tikz}}
\endpgfgraphicnamed=%
\beginpgfgraphicnamed{diagrams//induced_compact_structure-2wirert}
\InputIfFileExists{diagrams//induced_compact_structure-2wirert.tikz}{}{\input{./figures/diagrams//induced_compact_structure-2wirert.tikz}}
\endpgfgraphicnamed&(S3) & %
\beginpgfgraphicnamed{diagrams//hsquare}
\InputIfFileExists{diagrams//hsquare.tikz}{}{\input{./figures/diagrams//hsquare.tikz}}
\endpgfgraphicnamed &(H2)\\
\beginpgfgraphicnamed{diagrams//hslidecap}
\InputIfFileExists{diagrams//hslidecap.tikz}{}{\input{./figures/diagrams//hslidecap.tikz}}
\endpgfgraphicnamed &(H3) &%
\beginpgfgraphicnamed{diagrams//h2newlt2}
\InputIfFileExists{diagrams//h2newlt2.tikz}{}{\input{./figures/diagrams//h2newlt2.tikz}}
\endpgfgraphicnamed=%
\beginpgfgraphicnamed{diagrams//h2newrt2}
\InputIfFileExists{diagrams//h2newrt2.tikz}{}{\input{./figures/diagrams//h2newrt2.tikz}}
\endpgfgraphicnamed&(H)\\
%&\tikzfig{diagrams//hslidecap} &(H3) &\\
    %
\beginpgfgraphicnamed{diagrams//b1slt}
\InputIfFileExists{diagrams//b1slt.tikz}{}{\input{./figures/diagrams//b1slt.tikz}}
\endpgfgraphicnamed=%
\beginpgfgraphicnamed{diagrams//b1srt}
\InputIfFileExists{diagrams//b1srt.tikz}{}{\input{./figures/diagrams//b1srt.tikz}}
\endpgfgraphicnamed&(B1) & %
\beginpgfgraphicnamed{diagrams//b2slt}
\InputIfFileExists{diagrams//b2slt.tikz}{}{\input{./figures/diagrams//b2slt.tikz}}
\endpgfgraphicnamed=%
\beginpgfgraphicnamed{diagrams//b2srt}
\InputIfFileExists{diagrams//b2srt.tikz}{}{\input{./figures/diagrams//b2srt.tikz}}
\endpgfgraphicnamed&(B2)\\
\beginpgfgraphicnamed{diagrams//HadaDecomSingleslt}
\InputIfFileExists{diagrams//HadaDecomSingleslt.tikz}{}{\input{./figures/diagrams//HadaDecomSingleslt.tikz}}
\endpgfgraphicnamed= %
\beginpgfgraphicnamed{diagrams//HadaDecomSinglesrt2}
\InputIfFileExists{diagrams//HadaDecomSinglesrt2.tikz}{}{\input{./figures/diagrams//HadaDecomSinglesrt2.tikz}}
\endpgfgraphicnamed&(EU)    & %
\beginpgfgraphicnamed{diagrams//k2slt}
\InputIfFileExists{diagrams//k2slt.tikz}{}{\input{./figures/diagrams//k2slt.tikz}}
\endpgfgraphicnamed=%
\beginpgfgraphicnamed{diagrams//k2srt}
\InputIfFileExists{diagrams//k2srt.tikz}{}{\input{./figures/diagrams//k2srt.tikz}}
\endpgfgraphicnamed&(K2)\\

%\tikzfig{diagrams//alphadelete}&(S4) & &\\
&&&\\ 
\hline
\end{array}\]
\end{center}
% \caption{ZX-calculus rules I, where $\lambda  \geq 0, \alpha, \beta \in [0,~2\pi);$ in (AD1), $\gamma=\beta+arccos(\frac{\lambda_1+\lambda_2cos(\alpha-\beta)}{\lambda})$,  $\lambda=\sqrt{\lambda_1^2+\lambda_2^2+2\lambda_1\lambda_2cos(\alpha-\beta)}$, $\lambda_1^2+\lambda_2^2 > 0, ~ \alpha-\beta \neq \pi (mod ~2\pi)$.}\label{figure1}
  \caption{Traditional-style ZX-calculus rules, where $\alpha, \beta\in [0,~2\pi)$. The upside-down version and colour swapped version of these rules still hold.}\label{figure1}  
  \end{figure}
  %%%%%%%%%%%%%%%%%%%%%%%%%%%%%%%%%%%%%%%%%%%%
  
  \begin{figure}[!h]
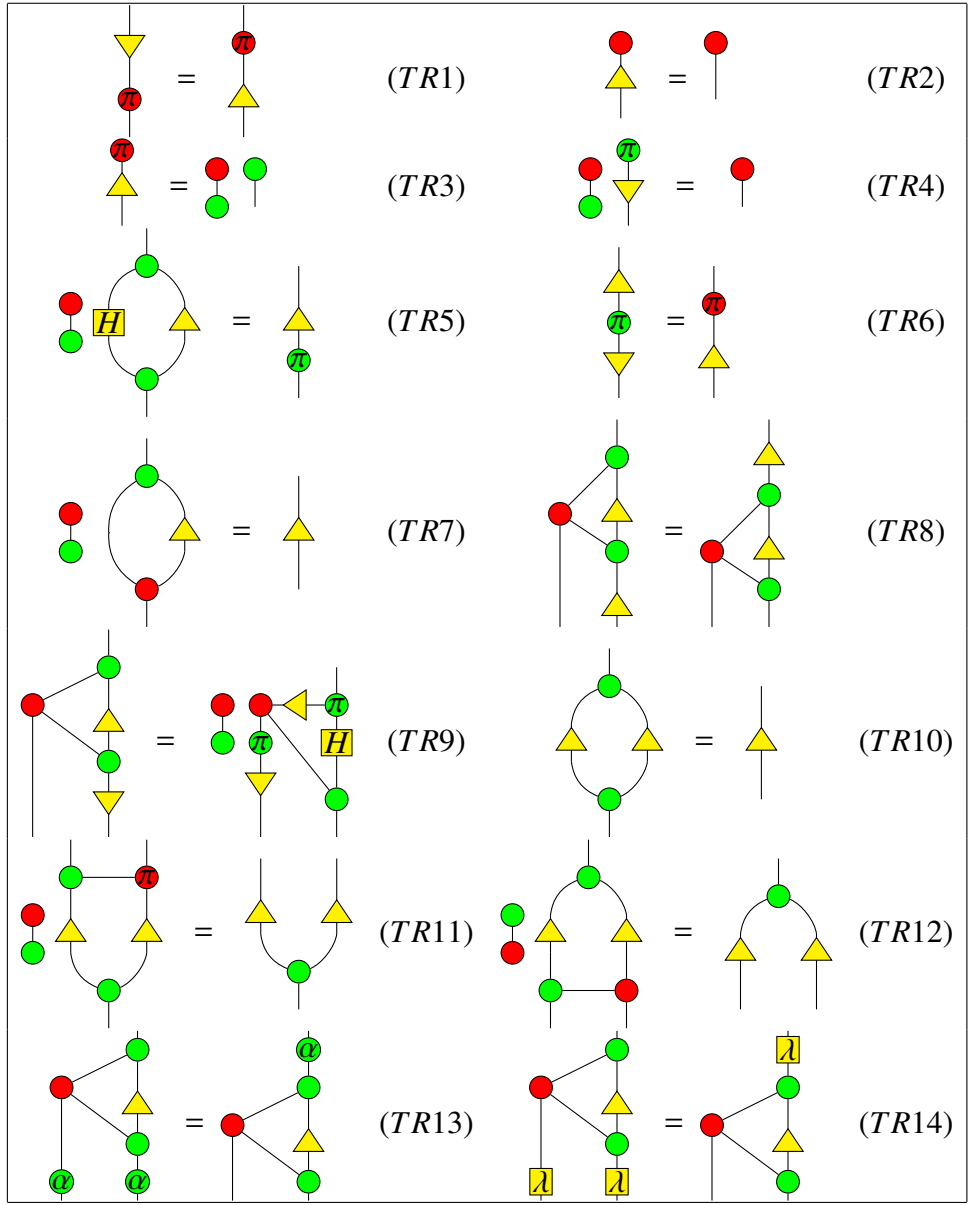

  	\begin{center}
  		\[
  		\quad \qquad\begin{array}{|cccc|}
  		\hline
  		%\tikzfig{diagrams//0plus0}&(AD2) &\tikzfig{diagrams//complemplus}&(AD3)\\
  		%
\beginpgfgraphicnamed{diagrams//trianglepicommute}
\InputIfFileExists{diagrams//trianglepicommute.tikz}{}{\input{./figures/diagrams//trianglepicommute.tikz}}
\endpgfgraphicnamed &(TR1) &%
\beginpgfgraphicnamed{diagrams//triangleocopy}
\InputIfFileExists{diagrams//triangleocopy.tikz}{}{\input{./figures/diagrams//triangleocopy.tikz}}
\endpgfgraphicnamed &(TR2)\\

\beginpgfgraphicnamed{diagrams//trianglepicopy}
\InputIfFileExists{diagrams//trianglepicopy.tikz}{}{\input{./figures/diagrams//trianglepicopy.tikz}}
\endpgfgraphicnamed&(TR3) & %
\beginpgfgraphicnamed{diagrams//trianglegdpicopy}
\InputIfFileExists{diagrams//trianglegdpicopy.tikz}{}{\input{./figures/diagrams//trianglegdpicopy.tikz}}
\endpgfgraphicnamed &(TR4)\\

\beginpgfgraphicnamed{diagrams//trianglehhopf}
\InputIfFileExists{diagrams//trianglehhopf.tikz}{}{\input{./figures/diagrams//trianglehhopf.tikz}}
\endpgfgraphicnamed &(TR5) &%
\beginpgfgraphicnamed{diagrams//gpiintriangles}
\InputIfFileExists{diagrams//gpiintriangles.tikz}{}{\input{./figures/diagrams//gpiintriangles.tikz}}
\endpgfgraphicnamed&(TR6)\\

\beginpgfgraphicnamed{diagrams//trianglehopf}
\InputIfFileExists{diagrams//trianglehopf.tikz}{}{\input{./figures/diagrams//trianglehopf.tikz}}
\endpgfgraphicnamed&(TR7) & %
\beginpgfgraphicnamed{diagrams//2triangleup}
\InputIfFileExists{diagrams//2triangleup.tikz}{}{\input{./figures/diagrams//2triangleup.tikz}}
\endpgfgraphicnamed&(TR8)\\

\beginpgfgraphicnamed{diagrams//2triangledown}
\InputIfFileExists{diagrams//2triangledown.tikz}{}{\input{./figures/diagrams//2triangledown.tikz}}
\endpgfgraphicnamed&(TR9)    & %
\beginpgfgraphicnamed{diagrams//2trianglehopf}
\InputIfFileExists{diagrams//2trianglehopf.tikz}{}{\input{./figures/diagrams//2trianglehopf.tikz}}
\endpgfgraphicnamed&(TR10)\\

\beginpgfgraphicnamed{diagrams//2triangledeloop}
\InputIfFileExists{diagrams//2triangledeloop.tikz}{}{\input{./figures/diagrams//2triangledeloop.tikz}}
\endpgfgraphicnamed &(TR11) &%
\beginpgfgraphicnamed{diagrams//2triangledeloopnopi}
\InputIfFileExists{diagrams//2triangledeloopnopi.tikz}{}{\input{./figures/diagrams//2triangledeloopnopi.tikz}}
\endpgfgraphicnamed &(TR12)\\

\beginpgfgraphicnamed{diagrams//alphacopyw}
\InputIfFileExists{diagrams//alphacopyw.tikz}{}{\input{./figures/diagrams//alphacopyw.tikz}}
\endpgfgraphicnamed&(TR13) &%
\beginpgfgraphicnamed{diagrams//lambdacopyw}
\InputIfFileExists{diagrams//lambdacopyw.tikz}{}{\input{./figures/diagrams//lambdacopyw.tikz}}
\endpgfgraphicnamed &(TR14)\\

  		\hline
  		\end{array}\]
  	\end{center}
  	
  	\caption{Extended ZX-calculus rules for triangle, where $\lambda  \geq 0, \alpha \in [0,~2\pi).$ The upside-down version of these rules still hold.}\label{figure2}
  \end{figure}

%%%%%%%%%%%%%%%%%%%%%%%%%%%%%%%%%%%%%%%%%%%%%%%%%  
  
  \begin{figure}[!h]
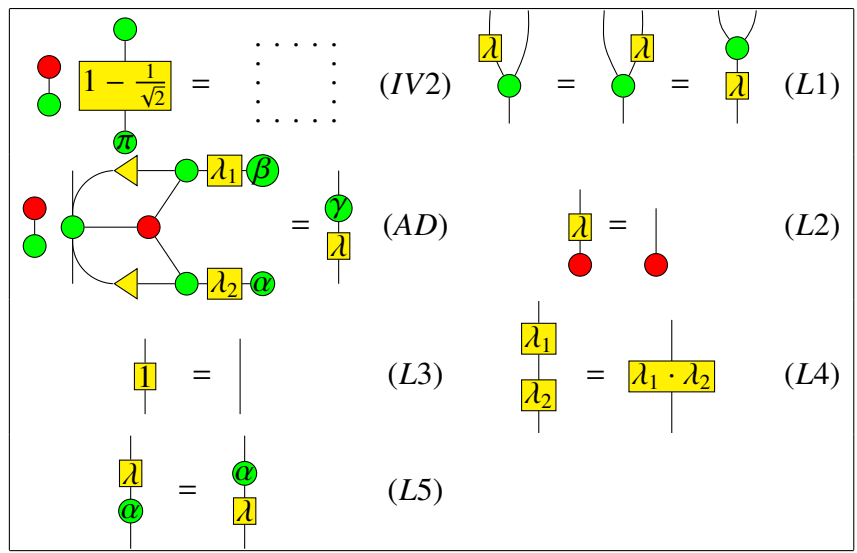

\begin{center}
\[
\quad \qquad\begin{array}{|cccc|}
\hline
\beginpgfgraphicnamed{diagrams//emptyrule}
\InputIfFileExists{diagrams//emptyrule.tikz}{}{\input{./figures/diagrams//emptyrule.tikz}}
\endpgfgraphicnamed &(IV2) &%
\beginpgfgraphicnamed{diagrams//lambbranch}
\InputIfFileExists{diagrams//lambbranch.tikz}{}{\input{./figures/diagrams//lambbranch.tikz}}
\endpgfgraphicnamed &(L1)\\

\beginpgfgraphicnamed{diagrams//plus}
\InputIfFileExists{diagrams//plus.tikz}{}{\input{./figures/diagrams//plus.tikz}}
\endpgfgraphicnamed&(AD) &%
\beginpgfgraphicnamed{diagrams//lambdadelete}
\InputIfFileExists{diagrams//lambdadelete.tikz}{}{\input{./figures/diagrams//lambdadelete.tikz}}
\endpgfgraphicnamed &(L2)\\

\beginpgfgraphicnamed{diagrams//sqr1is1}
\InputIfFileExists{diagrams//sqr1is1.tikz}{}{\input{./figures/diagrams//sqr1is1.tikz}}
\endpgfgraphicnamed&(L3) &%
\beginpgfgraphicnamed{diagrams//lambdatimes}
\InputIfFileExists{diagrams//lambdatimes.tikz}{}{\input{./figures/diagrams//lambdatimes.tikz}}
\endpgfgraphicnamed&(L4)\\

\beginpgfgraphicnamed{diagrams//lambdaalpha}
\InputIfFileExists{diagrams//lambdaalpha.tikz}{}{\input{./figures/diagrams//lambdaalpha.tikz}}
\endpgfgraphicnamed&(L5) &&\\
\hline
\end{array}\]
\end{center}
  \caption{Extended ZX-calculus rules for $\lambda$ and addition, where $\lambda, \lambda_1,  \lambda_2 \geq 0, \alpha, \beta, \gamma \in [0,~2\pi);$ in (AD), $\lambda e^{i\gamma} 
  =\lambda_1 e^{i\beta}+ \lambda_2 e^{i\alpha}$. The upside-down version of these rules still hold.}\label{figure0}  
  \end{figure}

%%%%%%%%%%%%%%%%%%%%%%%%%%%%%%%%%%%%%%%%%%%%%%%%%%%%

\FloatBarrier

%all the rules enumerated in Figures \ref{figure1}, \ref{figure2} and  \ref{figure0} still hold when they are flipped upside-down. For simplicity, we won't distinguish a rule with its flipped upside-down version when it is referred to in a diagrammatic rewriting. The structural rules will also be used without being explicitly stated in this thesis.

%By the definition of red spider in (\ref{hadamadefine}) and the rule (H2), we have the following colour-change rule that will be frequently used in this thesis:

%\begin{equation*}\label{Hrule2}
%\tikzfig{diagrams//h2newlt2}=\tikzfig{diagrams//h2newrt2}  \quad \quad  (H)
%\end{equation*}
%For convenience, we also refer to the definition  (\ref{hadamadefine}) as rule (H). Due to the  rule (H) and (H2), 
 %the rules in Figure \ref{figure1} have a property that they still hold when the colours green and red swapped. For this reason,  we won't distinguish any rule in Figure \ref{figure1} with its colour swapped version when it is referred to in a diagrammatic rewriting.

The diagrams in the $ZX_{ full}$-calculus have a standard interpretation composed of two parts: the standard interpretation of the traditional qubit ZX-calculus described in Section \ref{qubitzxintro} as well as the interpretation of the new generators triangle and $\lambda$ box to be given below. For simplicity, we still use the notation $\llbracket \cdot \rrbracket$ to denote the standard interpretation for the $ZX_{ full}$-calculus.
 
%in the category $\mathbf{FdHilb_2}$ of Hilbert spaces with dimensions $2^k$ :
%\[
%\left\llbracket \tikzfig{diagrams//generator_spider2} \right\rrbracket=\ket{0}^{\otimes m}\bra{0}^{\otimes n}+e^{i\alpha}\ket{1}^{\otimes m}\bra{1}^{\otimes n},
%\quad
%\left\llbracket\tikzfig{diagrams//alphagate}\right\rrbracket=\ket{0}\bra{0}+e^{i\alpha}\ket{1}\bra{1}=\begin{pmatrix}
   %     1 & 0 \\
      %  0 & e^{i\alpha}
 %\end{pmatrix},
%\]

\begin{equation}\label{interpretl}
 \left\llbracket%
\beginpgfgraphicnamed{diagrams//triangle}
\InputIfFileExists{diagrams//triangle.tikz}{}{\input{./figures/diagrams//triangle.tikz}}
\endpgfgraphicnamed\right\rrbracket=\begin{pmatrix}
        1 & 1 \\
        0 & 1
 \end{pmatrix}, \quad
  \left\llbracket%
\beginpgfgraphicnamed{diagrams//lambdabox}
\InputIfFileExists{diagrams//lambdabox.tikz}{}{\input{./figures/diagrams//lambdabox.tikz}}
\endpgfgraphicnamed\right\rrbracket=\begin{pmatrix}
        1 & 0 \\
        0 & \lambda
 \end{pmatrix}.
 \end{equation}

%It can be verified that the interpretation $\llbracket \cdot \rrbracket$ is a monoidal functor. 
\subsection*{Useful derivable results}

Now we derive some identities that will be useful in this thesis. 

%comparing to those used in \cite{Emmanuel}, we will make it clear the differences and the coincidence in the following. For convenience, we denote the ZX-calculus described in \cite{Emmanuel} as $ZX_{frqb}$-calculus.

%First, we list the differences:
%\begin{itemize}
%\item $ZX_{ full}$-calculus  has two more generators than $ZX_{frqb}$-calculus.
%\item s
%\end{itemize}

\begin{lemma}
	\begin{equation}\label{lem:inv}
	\input{diagrams/lemmainv.tikz}
	\end{equation}
\end{lemma}
\begin{proof}~\newline
First we have
\begin{equation}\label{lem:inv2}
	\input{diagrams/lemmainv_proof1.tikz}
	\end{equation}
Then ~\newline
\input{diagrams/lemmainv_proof2.tikz}
\end{proof}

\begin{lemma}
	\begin{equation}\label{lem:4}
	\input{diagrams/lemma4.tikz}
	\end{equation}
\end{lemma}

\begin{proof}
\begin{equation*}
	\input{diagrams/lemma4proof.tikz}
	\end{equation*}
	%Here we make a reference to the proof in \cite{Emmanuel} lemma $26$. %The rules used are $B2, S1, TR7$.
\end{proof}

\begin{lemma}
	\begin{equation}\label{lem:5}
	\input{diagrams/lemma5.tikz}
	\end{equation}
\end{lemma}

\begin{proof}
	\begin{equation*}
	\input{diagrams/lemma5-proof.tikz}
	\end{equation*}
	%Straightforward application of rules and properties $\ref{lem:com}$ and $TR8$.
\end{proof}

\begin{lemma}
	\begin{equation}\label{lem:7}
	\input{diagrams/lemma7.tikz}
	\end{equation}
\end{lemma}

\begin{proof}~\newline
	$
	\input{diagrams/lemma7_proof.tikz}
	$
\end{proof}

\section{Simplification of the rules of from the $ZX_{ full}$-calculus}\label{simplyrules}
%Commutativity:

%\tikzfig{diagrams//sumcommutativity}
The rules for the $ZX_{ full}$-calculus as listed in  Figure \ref{figure2}, and Figure \ref{figure0}
 can be further simplified, and some rules can be derived from others. We did not introduce the simplified version of rules at the beginning because we want to show the developing process of the 
 theory of the ZX-calculus. In this section, we will exhibit how rules could be simplified or derived.

First we show that the addition rule (AD) in Figure \ref{figure0} can be simplified:
%By the symmetry of \tikzfig{diagrams//wstate} and the rule (TR10), we have
\begin{equation}
\beginpgfgraphicnamed{diagrams//sumsimplify}
\InputIfFileExists{diagrams//sumsimplify.tikz}{}{\input{./figures/diagrams//sumsimplify.tikz}}
\endpgfgraphicnamed
\end{equation}
This means
\begin{equation}\label{addsimplify}
\beginpgfgraphicnamed{diagrams//sumsimplify2}
\InputIfFileExists{diagrams//sumsimplify2.tikz}{}{\input{./figures/diagrams//sumsimplify2.tikz}}
\endpgfgraphicnamed
\end{equation}

From now on, we will call the simplified addition rule (AD$^{\prime}$):
\begin{equation*}
\beginpgfgraphicnamed{diagrams//plusnew}
\InputIfFileExists{diagrams//plusnew.tikz}{}{\input{./figures/diagrams//plusnew.tikz}}
\endpgfgraphicnamed
\end{equation*}

As a consequence, we have the following commutativity of addition:
\begin{equation}\label{sumcommutativity}
\beginpgfgraphicnamed{diagrams//sumcommutativity2}
\InputIfFileExists{diagrams//sumcommutativity2.tikz}{}{\input{./figures/diagrams//sumcommutativity2.tikz}}
\endpgfgraphicnamed
\end{equation}

%\section{Properties of ZX-rules}
Next we prove that some rules in Figure \ref{figure2} are derivable. 
\begin{lemma}
\begin{equation}\label{lem:3}   %\label{tr3equiv}
\beginpgfgraphicnamed{diagrams//tr3equivl}
\InputIfFileExists{diagrams//tr3equivl.tikz}{}{\input{./figures/diagrams//tr3equivl.tikz}}
\endpgfgraphicnamed
\end{equation}
\end{lemma}

\begin{proof}
\begin{equation*}
\beginpgfgraphicnamed{diagrams//tr3equivlproof}
\InputIfFileExists{diagrams//tr3equivlproof.tikz}{}{\input{./figures/diagrams//tr3equivlproof.tikz}}
\endpgfgraphicnamed
\end{equation*}
\end{proof}

\begin{lemma}\label{redundantrules}
The rules (TR4), (TR10), and (TR11) can be derived from other rules.
\end{lemma}

\begin{proof}
For the derivation of (TR4), we have

\begin{equation}\label{TR4}
%\tikzfig{diagrams//TR4derive}
\input{diagrams/TR4derive.tikz}
\end{equation}

%where we used (TR3) for the second equality, (TR6) for the third equality, and (TR2) for the last equality.

For the derivation of (TR10), we have
\begin{equation}\label{TR10}
\beginpgfgraphicnamed{diagrams//TR10derive}
\InputIfFileExists{diagrams//TR10derive.tikz}{}{\input{./figures/diagrams//TR10derive.tikz}}
\endpgfgraphicnamed
\end{equation}
%where we used (TR12) for the second equality, (TR3) for the third equality.

For the derivation of (TR11), we have
\begin{equation}\label{TR11}
\beginpgfgraphicnamed{diagrams//TR11derive}
\InputIfFileExists{diagrams//TR11derive.tikz}{}{\input{./figures/diagrams//TR11derive.tikz}}
\endpgfgraphicnamed
\end{equation}
%where we used (TR12) for the fourth equality and (TR1) several times.
\end{proof}

%As we will see in Figure \ref{figure2t},  we need  the following new rule called (TR$10^{\prime}$) for the completeness of the ZX-calculus for Clifford+T quantum mechanics:
The following property will be very useful for deriving  (TR5) here and in later sections:
\begin{lemma}
\begin{equation*}
\beginpgfgraphicnamed{diagrams//tr10prime}
\InputIfFileExists{diagrams//tr10prime.tikz}{}{\input{./figures/diagrams//tr10prime.tikz}}
\endpgfgraphicnamed ~~~~  (TR10^{\prime})
\end{equation*}
\end{lemma}

\begin{proof}
\begin{equation}\label{TR10primederive2}
%\tikzfig{diagrams//TR10primederive}
%
\beginpgfgraphicnamed{diagrams//TR10primederive22}
\InputIfFileExists{diagrams//TR10primederive22.tikz}{}{\input{./figures/diagrams//TR10primederive22.tikz}}
\endpgfgraphicnamed 
\end{equation}
Then
\begin{equation*}
\beginpgfgraphicnamed{diagrams//TR10primederive23}
\InputIfFileExists{diagrams//TR10primederive23.tikz}{}{\input{./figures/diagrams//TR10primederive23.tikz}}
\endpgfgraphicnamed 
\end{equation*}
\end{proof}

%With this new rule (TR$10^{\prime}$),  we can show that the rule TR (5) is also derivable. 

\begin{lemma}
\begin{equation}\label{TR5derived2}
\beginpgfgraphicnamed{diagrams//TR5derivepre}
\InputIfFileExists{diagrams//TR5derivepre.tikz}{}{\input{./figures/diagrams//TR5derivepre.tikz}}
\endpgfgraphicnamed
\end{equation}
\end{lemma}

\begin{proof}
\begin{equation*}
\beginpgfgraphicnamed{diagrams//TR5derive2}
\InputIfFileExists{diagrams//TR5derive2.tikz}{}{\input{./figures/diagrams//TR5derive2.tikz}}
\endpgfgraphicnamed
\end{equation*}
\end{proof}

\begin{lemma}
\begin{equation}\label{TR5derived3}
\beginpgfgraphicnamed{diagrams//TR5derivepre2}
\InputIfFileExists{diagrams//TR5derivepre2.tikz}{}{\input{./figures/diagrams//TR5derivepre2.tikz}}
\endpgfgraphicnamed
\end{equation}
\end{lemma}
\begin{proof}
\begin{equation*}
\beginpgfgraphicnamed{diagrams//TR5derivepre2proof}
\InputIfFileExists{diagrams//TR5derivepre2proof.tikz}{}{\input{./figures/diagrams//TR5derivepre2proof.tikz}}
\endpgfgraphicnamed
\end{equation*}
\end{proof}

\begin{lemma} \label{tr5derived}
The rule (TR5) can be derived.
\end{lemma}

\begin{proof}

\beginpgfgraphicnamed{diagrams//TR5derive}
\InputIfFileExists{diagrams//TR5derive.tikz}{}{\input{./figures/diagrams//TR5derive.tikz}}
\endpgfgraphicnamed
\end{proof}
%where for the third equality we used the following diagrammatic reasoning via rules (TR1), (K2), (TR9) and (TR$10^{\prime}$) :

%\tikzfig{diagrams//TR5derive2}
\begin{lemma} \label{TR7}
The rule (TR7) can be derived.
\end{lemma}

\begin{proof}

\beginpgfgraphicnamed{diagrams//TR7derive}
\InputIfFileExists{diagrams//TR7derive.tikz}{}{\input{./figures/diagrams//TR7derive.tikz}}
\endpgfgraphicnamed
\end{proof}

\begin{lemma}
The rules  (TR13) and (TR14) can be combined.
\end{lemma}
\begin{proof}
Obviously, (TR13) and (TR14) can be combined into a single rule called (TR$13^{\prime}$)  as follows:
\begin{equation}\label{TR1314com}
\beginpgfgraphicnamed{diagrams//TR1314combine}
\InputIfFileExists{diagrams//TR1314combine.tikz}{}{\input{./figures/diagrams//TR1314combine.tikz}}
\endpgfgraphicnamed
\end{equation}
\end{proof}

\begin{lemma}\label{L1}
The rule  (L1) can be derived.
%\tikzfig{diagrams//lambbranch}  (L1)
%can be simplified as follows:
%\begin{equation}\label{l1simplify}
%\tikzfig{diagrams//l1simplify}
%\end{equation}
\end{lemma}
\begin{proof}
By the addition rule, we have 
\begin{equation*}
% \tikzfig{diagrams//l1simplifyproof} 
 %
\beginpgfgraphicnamed{diagrams//l1deriveproof}
\InputIfFileExists{diagrams//l1deriveproof.tikz}{}{\input{./figures/diagrams//l1deriveproof.tikz}}
\endpgfgraphicnamed 
\end{equation*}
Then  (L1)  directly follows from the spider rule (S1).
\end{proof}

\begin{lemma}\label{L2}
The rule  (L2) can be derived.

\end{lemma}

\begin{proof}

First we can write the non-negative number $\lambda$ as a sum of its integer part and remainder part: $\lambda= [\lambda] +\{\lambda\}$, where $ [\lambda]$ is a non-negative integer and $0\leq\{\lambda\}<1$.
Let $n= [\lambda],  \alpha=arccos\frac{\{\lambda\}}{2}$.  It follows that $\{\lambda\}=2\cos\alpha=e^{i\alpha}+e^{-i\alpha}$. If $n=0$, then  we have
\begin{equation*}
\beginpgfgraphicnamed{diagrams//lambdaderive1}
\InputIfFileExists{diagrams//lambdaderive1.tikz}{}{\input{./figures/diagrams//lambdaderive1.tikz}}
\endpgfgraphicnamed 
\end{equation*}
If $n>0$, then we have
%by induction on $n$ and the simplified addition rule (AD$^{\prime}$), we have
\begin{equation}\label{nrepresent}
\beginpgfgraphicnamed{diagrams//lambdaderive2}
\InputIfFileExists{diagrams//lambdaderive2.tikz}{}{\input{./figures/diagrams//lambdaderive2.tikz}}
\endpgfgraphicnamed 
\end{equation}
We show this by induction on $n$. When $n=1$, we have
\begin{equation*}
\beginpgfgraphicnamed{diagrams//lambd1tr}
\InputIfFileExists{diagrams//lambd1tr.tikz}{}{\input{./figures/diagrams//lambd1tr.tikz}}
\endpgfgraphicnamed 
\end{equation*}
 Suppose (\ref{nrepresent}) holds for $n$. Then for $n+1$ we have
 \begin{equation*}
\beginpgfgraphicnamed{diagrams//lambdn1tr}
\InputIfFileExists{diagrams//lambdn1tr.tikz}{}{\input{./figures/diagrams//lambdn1tr.tikz}}
\endpgfgraphicnamed 
\end{equation*}
 This completes the induction.
Therefore,
\begin{equation*}
\beginpgfgraphicnamed{diagrams//lambdaderive3}
\InputIfFileExists{diagrams//lambdaderive3.tikz}{}{\input{./figures/diagrams//lambdaderive3.tikz}}
\endpgfgraphicnamed 
\end{equation*}
%where we used axioms (AD), (TR10$^{\prime}$), (TR2), (B1), (S1).
\end{proof}

\begin{lemma}
The rule (L5) can be derived.
\end{lemma}
\begin{proof}
\begin{equation}\label{L5}
\beginpgfgraphicnamed{diagrams//l5proof}
\InputIfFileExists{diagrams//l5proof.tikz}{}{\input{./figures/diagrams//l5proof.tikz}}
\endpgfgraphicnamed 
\end{equation}
\end{proof}

\subsection{Simplified rules for the $ZX_{ full}$-calculus}

Now can summarise the results obtained in Section \ref{simplyrules} and give the simplified version of all the non-structural rewriting rules for the $ZX_{ full}$-calculus in the following Figures \ref{figure1sfy} and \ref{figure2sfy} .

\begin{figure}[!h]
\begin{center}
\[
\quad \qquad\begin{array}{|cccc|}
\hline
\beginpgfgraphicnamed{diagrams//spiderlt}
\InputIfFileExists{diagrams//spiderlt.tikz}{}{\input{./figures/diagrams//spiderlt.tikz}}
\endpgfgraphicnamed=%
\beginpgfgraphicnamed{diagrams//spiderrt}
\InputIfFileExists{diagrams//spiderrt.tikz}{}{\input{./figures/diagrams//spiderrt.tikz}}
\endpgfgraphicnamed &(S1) &%
\beginpgfgraphicnamed{diagrams//s2new}
\InputIfFileExists{diagrams//s2new.tikz}{}{\input{./figures/diagrams//s2new.tikz}}
\endpgfgraphicnamed &(S2)\\
\beginpgfgraphicnamed{diagrams//induced_compact_structure-2wirelt}
\InputIfFileExists{diagrams//induced_compact_structure-2wirelt.tikz}{}{\input{./figures/diagrams//induced_compact_structure-2wirelt.tikz}}
\endpgfgraphicnamed=%
\beginpgfgraphicnamed{diagrams//induced_compact_structure-2wirert}
\InputIfFileExists{diagrams//induced_compact_structure-2wirert.tikz}{}{\input{./figures/diagrams//induced_compact_structure-2wirert.tikz}}
\endpgfgraphicnamed&(S3) & %
\beginpgfgraphicnamed{diagrams//hsquare}
\InputIfFileExists{diagrams//hsquare.tikz}{}{\input{./figures/diagrams//hsquare.tikz}}
\endpgfgraphicnamed &(H2)\\
\beginpgfgraphicnamed{diagrams//hslidecap}
\InputIfFileExists{diagrams//hslidecap.tikz}{}{\input{./figures/diagrams//hslidecap.tikz}}
\endpgfgraphicnamed &(H3) &%
\beginpgfgraphicnamed{diagrams//h2newlt}
\InputIfFileExists{diagrams//h2newlt.tikz}{}{\input{./figures/diagrams//h2newlt.tikz}}
\endpgfgraphicnamed=%
\beginpgfgraphicnamed{diagrams//h2newrt}
\InputIfFileExists{diagrams//h2newrt.tikz}{}{\input{./figures/diagrams//h2newrt.tikz}}
\endpgfgraphicnamed&(H)\\
% & \tikzfig{diagrams//hslidecap} &(H3) &\\
    %
\beginpgfgraphicnamed{diagrams//b1slt}
\InputIfFileExists{diagrams//b1slt.tikz}{}{\input{./figures/diagrams//b1slt.tikz}}
\endpgfgraphicnamed=%
\beginpgfgraphicnamed{diagrams//b1srt}
\InputIfFileExists{diagrams//b1srt.tikz}{}{\input{./figures/diagrams//b1srt.tikz}}
\endpgfgraphicnamed&(B1) & %
\beginpgfgraphicnamed{diagrams//b2slt}
\InputIfFileExists{diagrams//b2slt.tikz}{}{\input{./figures/diagrams//b2slt.tikz}}
\endpgfgraphicnamed=%
\beginpgfgraphicnamed{diagrams//b2srt}
\InputIfFileExists{diagrams//b2srt.tikz}{}{\input{./figures/diagrams//b2srt.tikz}}
\endpgfgraphicnamed&(B2)\\
\beginpgfgraphicnamed{diagrams//HadaDecomSingleslt}
\InputIfFileExists{diagrams//HadaDecomSingleslt.tikz}{}{\input{./figures/diagrams//HadaDecomSingleslt.tikz}}
\endpgfgraphicnamed= %
\beginpgfgraphicnamed{diagrams//HadaDecomSinglesrt2}
\InputIfFileExists{diagrams//HadaDecomSinglesrt2.tikz}{}{\input{./figures/diagrams//HadaDecomSinglesrt2.tikz}}
\endpgfgraphicnamed&(EU)    & %
\beginpgfgraphicnamed{diagrams//k2slt}
\InputIfFileExists{diagrams//k2slt.tikz}{}{\input{./figures/diagrams//k2slt.tikz}}
\endpgfgraphicnamed=%
\beginpgfgraphicnamed{diagrams//k2srt}
\InputIfFileExists{diagrams//k2srt.tikz}{}{\input{./figures/diagrams//k2srt.tikz}}
\endpgfgraphicnamed&(K2)\\
%& \tikzfig{diagrams//newemptyrl} &(IV')  &\\
%& \tikzfig{diagrams//emptyrule} &(IV) &\\
%\tikzfig{diagrams//alphadelete}&(S4) & &\\
&&&\\ 
\hline
\end{array}\]
\end{center}
% \caption{ZX-calculus rules I, where $\lambda  \geq 0, \alpha, \beta \in [0,~2\pi);$ in (AD1), $\gamma=\beta+arccos(\frac{\lambda_1+\lambda_2cos(\alpha-\beta)}{\lambda})$,  $\lambda=\sqrt{\lambda_1^2+\lambda_2^2+2\lambda_1\lambda_2cos(\alpha-\beta)}$, $\lambda_1^2+\lambda_2^2 > 0, ~ \alpha-\beta \neq \pi (mod ~2\pi)$.}\label{figure1}
   \caption{Traditional ZX-calculus rules, where $\alpha, \beta\in [0,~2\pi)$. The upside-down version and colour swapped version of these rules still hold.}\label{figure1sfy}  
  \end{figure}
  %%%%%%%%%%%%%%%%%%%%%%%%%%%%%%%%%%%%%%%%%%%%

 \begin{figure}[!h]
  	\begin{center}
  		\[
  		\quad \qquad\begin{array}{|cccc|}
  		\hline
  		%\tikzfig{diagrams//0plus0}&(AD2) &\tikzfig{diagrams//complemplus}&(AD3)\\
  		%
\beginpgfgraphicnamed{diagrams//trianglepicommute}
\InputIfFileExists{diagrams//trianglepicommute.tikz}{}{\input{./figures/diagrams//trianglepicommute.tikz}}
\endpgfgraphicnamed &(TR1) &%
\beginpgfgraphicnamed{diagrams//triangleocopy}
\InputIfFileExists{diagrams//triangleocopy.tikz}{}{\input{./figures/diagrams//triangleocopy.tikz}}
\endpgfgraphicnamed &(TR2)\\
  		
  		%\tikzfig{diagrams//trianglepicopy}&(TR3) & \tikzfig{diagrams//tr4prime} &(TR4')\\
  		%&\tikzfig{diagrams//trianglepicopy}&(TR3)  &\\
		%
\beginpgfgraphicnamed{diagrams//trianglepicopy}
\InputIfFileExists{diagrams//trianglepicopy.tikz}{}{\input{./figures/diagrams//trianglepicopy.tikz}}
\endpgfgraphicnamed&(TR3) & %
\beginpgfgraphicnamed{diagrams//emptyrule}
\InputIfFileExists{diagrams//emptyrule.tikz}{}{\input{./figures/diagrams//emptyrule.tikz}}
\endpgfgraphicnamed &(IV2)\\
		
  		%\tikzfig{diagrams//tr5prime} &(TR5') &\tikzfig{diagrams//gpiintriangles}&(TR6)\\
		%\tikzfig{diagrams//gpiintriangles}&(TR6) & \tikzfig{diagrams//trianglehopf}&(TR7)\\
		%
\beginpgfgraphicnamed{diagrams//gpiintriangles}
\InputIfFileExists{diagrams//gpiintriangles.tikz}{}{\input{./figures/diagrams//gpiintriangles.tikz}}
\endpgfgraphicnamed&(TR6) & %
\beginpgfgraphicnamed{diagrams//2triangleup}
\InputIfFileExists{diagrams//2triangleup.tikz}{}{\input{./figures/diagrams//2triangleup.tikz}}
\endpgfgraphicnamed&(TR8)\\

\beginpgfgraphicnamed{diagrams//2triangledown}
\InputIfFileExists{diagrams//2triangledown.tikz}{}{\input{./figures/diagrams//2triangledown.tikz}}
\endpgfgraphicnamed&(TR9)&%
\beginpgfgraphicnamed{diagrams//2triangledeloopnopi}
\InputIfFileExists{diagrams//2triangledeloopnopi.tikz}{}{\input{./figures/diagrams//2triangledeloopnopi.tikz}}
\endpgfgraphicnamed &(TR12) \\
  		
  		%\tikzfig{diagrams//2triangledown}&(TR9)    & \tikzfig{diagrams//tr10prime}&(TR10')\\
  		
  		%\tikzfig{diagrams//2triangledeloopnopi} &(TR12) &\tikzfig{diagrams//alphacopyw} &(TR13)\\
  		%
\beginpgfgraphicnamed{diagrams//TR1314combine}
\InputIfFileExists{diagrams//TR1314combine.tikz}{}{\input{./figures/diagrams//TR1314combine.tikz}}
\endpgfgraphicnamed &(TR13')&%
\beginpgfgraphicnamed{diagrams//plusnew}
\InputIfFileExists{diagrams//plusnew.tikz}{}{\input{./figures/diagrams//plusnew.tikz}}
\endpgfgraphicnamed&(AD')\\
  		%\tikzfig{diagrams//alphacopyw}&(TR13) &\tikzfig{diagrams//lambdacopyw} &(TR14)\\ 
  		%\tikzfig{diagrams//alphacopyw}&(TR13) &&\\ 
  		%\tikzfig{diagrams//plusnew}&(AD')&\tikzfig{diagrams//l1simplify} &(L1)\\
		
		%
\beginpgfgraphicnamed{diagrams//sqr1is1}
\InputIfFileExists{diagrams//sqr1is1.tikz}{}{\input{./figures/diagrams//sqr1is1.tikz}}
\endpgfgraphicnamed&(L3) &%
\beginpgfgraphicnamed{diagrams//lambdatimes}
\InputIfFileExists{diagrams//lambdatimes.tikz}{}{\input{./figures/diagrams//lambdatimes.tikz}}
\endpgfgraphicnamed&(L4)\\
  		\hline
  		\end{array}\]
  	\end{center}
  	
  	\caption{Extended ZX-calculus rules, where  $\lambda, \lambda_1,  \lambda_2 \geq 0, \alpha, \beta, \gamma \in [0,~2\pi);$ in (AD$^{\prime}$), $\lambda e^{i\gamma} 
  =\lambda_1 e^{i\beta}+ \lambda_2 e^{i\alpha}$.The upside-down version of these rules still hold.}\label{figure2sfy}
  \end{figure}

\FloatBarrier
With these simplified rules, we have 
\begin{proposition}
The $ZX_{ full}$-calculus is sound.
\end{proposition}
\begin{proof}
By the construction of general ZX-calculus described in Section \ref{zxgeneral}, any two diagrams $D_1$ and $D_2$ of the $ZX_{ full}$-calculus are equal modulo the equivalence relations given in (\ref{compactstructure}) and (\ref{compactstructureslide}) as structural rules and in Figures \ref{figure1sfy} and \ref{figure2sfy}  as non-structural rewriting rules. That is, $D_1=D_2$ if and only if $D_1$ can be rewritten into $D_2$ using finitely many of these rules. In addition, 
$  \llbracket D_1\otimes D_2  \rrbracket =  \llbracket D_1  \rrbracket \otimes  \llbracket  D_2  \rrbracket, \quad 
 \llbracket D_1\circ D_2  \rrbracket =  \llbracket D_1  \rrbracket \circ  \llbracket  D_2  \rrbracket.
  $
% the sense that 
%Since all the diagrams in the $ZX_{ full}$-calculus are generated parallelly or sequentially  from generators enumerated in Table \ref{qbzxgenerator} and Table \ref{newgr}, 
Therefore, to prove the soundness of the $ZX_{ full}$-calculus, it suffices to verify that all the rules listed in (\ref{compactstructure}) and (\ref{compactstructureslide}) and Figures \ref{figure1sfy} and \ref{figure2sfy} still hold under the standard interpretation  $\llbracket \cdot \rrbracket$ including (\ref{interpretl}).  The structural rules have been proved to be sound in \cite{CoeckeDuncan}. It is a routine check that the rules in Figures \ref{figure1sfy} and \ref{figure2sfy} are sound.

\end{proof}

\section{Interpretations between the $ZX_{ full}$-calculus and the $ZW_{ \mathbb{C}}$-calculus}\label{transzxzw}
The $ZX_{ full}$-calculus and the $ZW_{ \mathbb{C}}$-calculus are not irrelevant to each other. In fact, there exists an invertible translation between them. 
Note that a spider can be decomposed into a  form consisting of a phase-free spider and a pure phase gate, which will be used for translation between the ZX-calculus and the ZW-calculus:
%\begin{equation}\label{gspiderdecom}
\begin{align}\label{gspiderdecom}
\beginpgfgraphicnamed{diagrams//gspiderdecomp}
\InputIfFileExists{diagrams//gspiderdecomp.tikz}{}{\input{./figures/diagrams//gspiderdecomp.tikz}}
\endpgfgraphicnamed\quad\quad %
\beginpgfgraphicnamed{diagrams//rspiderdecomp}
\InputIfFileExists{diagrams//rspiderdecomp.tikz}{}{\input{./figures/diagrams//rspiderdecomp.tikz}}
\endpgfgraphicnamed
%\end{equation}
\end{align}

First we define the interpretation $\llbracket \cdot \rrbracket_{XW}$   from $ZX_{ full}$-calculus to  $ZW_{ \mathbb{C}}$-calculus  as follows:
\[
 \left\llbracket%
\beginpgfgraphicnamed{diagrams//emptysquare}
\InputIfFileExists{diagrams//emptysquare.tikz}{}{\input{./figures/diagrams//emptysquare.tikz}}
\endpgfgraphicnamed\right\rrbracket_{XW}=  %
\beginpgfgraphicnamed{diagrams//emptysquare}
\InputIfFileExists{diagrams//emptysquare.tikz}{}{\input{./figures/diagrams//emptysquare.tikz}}
\endpgfgraphicnamed,  \quad
  \left\llbracket%
\beginpgfgraphicnamed{diagrams//Id}
\InputIfFileExists{diagrams//Id.tikz}{}{\input{./figures/diagrams//Id.tikz}}
\endpgfgraphicnamed\right\rrbracket_{XW}=  %
\beginpgfgraphicnamed{diagrams//Id}
\InputIfFileExists{diagrams//Id.tikz}{}{\input{./figures/diagrams//Id.tikz}}
\endpgfgraphicnamed,   \quad
 \left\llbracket%
\beginpgfgraphicnamed{diagrams//cap}
\InputIfFileExists{diagrams//cap.tikz}{}{\input{./figures/diagrams//cap.tikz}}
\endpgfgraphicnamed\right\rrbracket_{XW}=  %
\beginpgfgraphicnamed{diagrams//cap}
\InputIfFileExists{diagrams//cap.tikz}{}{\input{./figures/diagrams//cap.tikz}}
\endpgfgraphicnamed,   \quad
  \left\llbracket%
\beginpgfgraphicnamed{diagrams//cup}
\InputIfFileExists{diagrams//cup.tikz}{}{\input{./figures/diagrams//cup.tikz}}
\endpgfgraphicnamed\right\rrbracket_{XW}=  %
\beginpgfgraphicnamed{diagrams//cup}
\InputIfFileExists{diagrams//cup.tikz}{}{\input{./figures/diagrams//cup.tikz}}
\endpgfgraphicnamed,  
  \]
  
  \[
   \left\llbracket%
\beginpgfgraphicnamed{diagrams//swap}
\InputIfFileExists{diagrams//swap.tikz}{}{\input{./figures/diagrams//swap.tikz}}
\endpgfgraphicnamed\right\rrbracket_{XW}=  %
\beginpgfgraphicnamed{diagrams//swap}
\InputIfFileExists{diagrams//swap.tikz}{}{\input{./figures/diagrams//swap.tikz}}
\endpgfgraphicnamed,   \quad
   \left\llbracket%
\beginpgfgraphicnamed{diagrams//generator_spider-nonum}
\InputIfFileExists{diagrams//generator_spider-nonum.tikz}{}{\input{./figures/diagrams//generator_spider-nonum.tikz}}
\endpgfgraphicnamed\right\rrbracket_{XW}=  %
\beginpgfgraphicnamed{diagrams//spiderwhite}
\InputIfFileExists{diagrams//spiderwhite.tikz}{}{\input{./figures/diagrams//spiderwhite.tikz}}
\endpgfgraphicnamed,   \quad 
     \left\llbracket%
\beginpgfgraphicnamed{diagrams//alphagate}
\InputIfFileExists{diagrams//alphagate.tikz}{}{\input{./figures/diagrams//alphagate.tikz}}
\endpgfgraphicnamed\right\rrbracket_{XW}=  %
\beginpgfgraphicnamed{diagrams//alphagatewhite}
\InputIfFileExists{diagrams//alphagatewhite.tikz}{}{\input{./figures/diagrams//alphagatewhite.tikz}}
\endpgfgraphicnamed,   \quad 
       \left\llbracket%
\beginpgfgraphicnamed{diagrams//lambdabox}
\InputIfFileExists{diagrams//lambdabox.tikz}{}{\input{./figures/diagrams//lambdabox.tikz}}
\endpgfgraphicnamed\right\rrbracket_{XW}=  %
\beginpgfgraphicnamed{diagrams//lambdagatewhiteld}
\InputIfFileExists{diagrams//lambdagatewhiteld.tikz}{}{\input{./figures/diagrams//lambdagatewhiteld.tikz}}
\endpgfgraphicnamed,   
         \]

 \[
  \left\llbracket%
\beginpgfgraphicnamed{diagrams//HadaDecomSingleslt}
\InputIfFileExists{diagrams//HadaDecomSingleslt.tikz}{}{\input{./figures/diagrams//HadaDecomSingleslt.tikz}}
\endpgfgraphicnamed\right\rrbracket_{XW}=  %
\beginpgfgraphicnamed{diagrams//Hadamardwhite}
\InputIfFileExists{diagrams//Hadamardwhite.tikz}{}{\input{./figures/diagrams//Hadamardwhite.tikz}}
\endpgfgraphicnamed,   \quad
    \left\llbracket%
\beginpgfgraphicnamed{diagrams//triangle}
\InputIfFileExists{diagrams//triangle.tikz}{}{\input{./figures/diagrams//triangle.tikz}}
\endpgfgraphicnamed\right\rrbracket_{XW}=  %
\beginpgfgraphicnamed{diagrams//trianglewhite}
\InputIfFileExists{diagrams//trianglewhite.tikz}{}{\input{./figures/diagrams//trianglewhite.tikz}}
\endpgfgraphicnamed,
       \left\llbracket%
\beginpgfgraphicnamed{diagrams//alphagatered}
\InputIfFileExists{diagrams//alphagatered.tikz}{}{\input{./figures/diagrams//alphagatered.tikz}}
\endpgfgraphicnamed\right\rrbracket_{XW}= \left\llbracket%
\beginpgfgraphicnamed{diagrams//HadaDecomSingleslt}
\InputIfFileExists{diagrams//HadaDecomSingleslt.tikz}{}{\input{./figures/diagrams//HadaDecomSingleslt.tikz}}
\endpgfgraphicnamed\right\rrbracket_{XW}\circ \left(%
\beginpgfgraphicnamed{diagrams//alphagatewhite}
\InputIfFileExists{diagrams//alphagatewhite.tikz}{}{\input{./figures/diagrams//alphagatewhite.tikz}}
\endpgfgraphicnamed\right) \circ \left\llbracket%
\beginpgfgraphicnamed{diagrams//HadaDecomSingleslt}
\InputIfFileExists{diagrams//HadaDecomSingleslt.tikz}{}{\input{./figures/diagrams//HadaDecomSingleslt.tikz}}
\endpgfgraphicnamed\right\rrbracket_{XW},   \quad 
     \]
    
     \[
       \left\llbracket%
\beginpgfgraphicnamed{diagrams//generator_redspider2cged}
\InputIfFileExists{diagrams//generator_redspider2cged.tikz}{}{\input{./figures/diagrams//generator_redspider2cged.tikz}}
\endpgfgraphicnamed\right\rrbracket_{XW}= \left\llbracket \left( %
\beginpgfgraphicnamed{diagrams//HadaDecomSingleslt}
\InputIfFileExists{diagrams//HadaDecomSingleslt.tikz}{}{\input{./figures/diagrams//HadaDecomSingleslt.tikz}}
\endpgfgraphicnamed\right)^{\otimes m} \right\rrbracket_{XW}\circ \left(%
\beginpgfgraphicnamed{diagrams//spiderwhiteindex}
\InputIfFileExists{diagrams//spiderwhiteindex.tikz}{}{\input{./figures/diagrams//spiderwhiteindex.tikz}}
\endpgfgraphicnamed\right) \circ \left\llbracket \left( %
\beginpgfgraphicnamed{diagrams//HadaDecomSingleslt}
\InputIfFileExists{diagrams//HadaDecomSingleslt.tikz}{}{\input{./figures/diagrams//HadaDecomSingleslt.tikz}}
\endpgfgraphicnamed\right)^{\otimes n} \right\rrbracket_{XW}, 
     \]
    
    \[ \llbracket D_1\otimes D_2  \rrbracket_{XW} =  \llbracket D_1  \rrbracket_{XW} \otimes  \llbracket  D_2  \rrbracket_{XW}, \quad 
 \llbracket D_1\circ D_2  \rrbracket_{XW} =  \llbracket D_1  \rrbracket_{XW} \circ  \llbracket  D_2  \rrbracket_{XW},
 \]
where $ \alpha \in [0,~2\pi), ~ \lambda  \geq 0$.

The interpretation $\llbracket \cdot \rrbracket_{XW}$ preserves the standard interpretation:

\begin{lemma}\label{xtowpreservesemantics}
Suppose $D$ is an arbitrary diagram in the $ZX_{ full}$-calculus. Then  $\llbracket \llbracket D \rrbracket_{XW}\rrbracket = \llbracket D \rrbracket$.
\end{lemma}

\begin{proof}
Since each diagram in the  $ZX_{ full}$-calculus is generated in parallel composition $\otimes$ and sequential composition $\circ$ by generators in Table \ref{qbzxgenerator} and Table \ref{newgr}, and the interpretation $\llbracket \cdot \rrbracket_{XW}$ respects
these two compositions, it suffices to prove $\llbracket \llbracket D \rrbracket_{XW}\rrbracket = \llbracket D \rrbracket$ whend $D$ is a generator diagram. This is a routine check, we omit the verification details here. 
\end{proof}
%%%%%%%%%%%%%%%%%%%%%%%%%%%%%%%%%%%%%%%%%%%%%%%%%%%%%%%%

Next we define the interpretation $\llbracket \cdot \rrbracket_{WX}$   from $ZW_{ \mathbb{C}}$-calculus to  $ZX_{ full}$-calculus  as follows:

\[
 \left\llbracket%
\beginpgfgraphicnamed{diagrams//emptysquare}
\InputIfFileExists{diagrams//emptysquare.tikz}{}{\input{./figures/diagrams//emptysquare.tikz}}
\endpgfgraphicnamed\right\rrbracket_{WX}=  %
\beginpgfgraphicnamed{diagrams//emptysquare}
\InputIfFileExists{diagrams//emptysquare.tikz}{}{\input{./figures/diagrams//emptysquare.tikz}}
\endpgfgraphicnamed,  \quad
  \left\llbracket%
\beginpgfgraphicnamed{diagrams//Id}
\InputIfFileExists{diagrams//Id.tikz}{}{\input{./figures/diagrams//Id.tikz}}
\endpgfgraphicnamed\right\rrbracket_{WX}=  %
\beginpgfgraphicnamed{diagrams//Id}
\InputIfFileExists{diagrams//Id.tikz}{}{\input{./figures/diagrams//Id.tikz}}
\endpgfgraphicnamed,   \quad
 \left\llbracket%
\beginpgfgraphicnamed{diagrams//cap}
\InputIfFileExists{diagrams//cap.tikz}{}{\input{./figures/diagrams//cap.tikz}}
\endpgfgraphicnamed\right\rrbracket_{WX}=  %
\beginpgfgraphicnamed{diagrams//cap}
\InputIfFileExists{diagrams//cap.tikz}{}{\input{./figures/diagrams//cap.tikz}}
\endpgfgraphicnamed,   \quad
  \left\llbracket%
\beginpgfgraphicnamed{diagrams//cup}
\InputIfFileExists{diagrams//cup.tikz}{}{\input{./figures/diagrams//cup.tikz}}
\endpgfgraphicnamed\right\rrbracket_{WX}=  %
\beginpgfgraphicnamed{diagrams//cup}
\InputIfFileExists{diagrams//cup.tikz}{}{\input{./figures/diagrams//cup.tikz}}
\endpgfgraphicnamed,  
  \]
  
  \[
   \left\llbracket%
\beginpgfgraphicnamed{diagrams//swap}
\InputIfFileExists{diagrams//swap.tikz}{}{\input{./figures/diagrams//swap.tikz}}
\endpgfgraphicnamed\right\rrbracket_{WX}=  %
\beginpgfgraphicnamed{diagrams//swap}
\InputIfFileExists{diagrams//swap.tikz}{}{\input{./figures/diagrams//swap.tikz}}
\endpgfgraphicnamed,   \quad
   \left\llbracket%
\beginpgfgraphicnamed{diagrams//spiderwhite}
\InputIfFileExists{diagrams//spiderwhite.tikz}{}{\input{./figures/diagrams//spiderwhite.tikz}}
\endpgfgraphicnamed\right\rrbracket_{WX}=  %
\beginpgfgraphicnamed{diagrams//generator_spider-nonum}
\InputIfFileExists{diagrams//generator_spider-nonum.tikz}{}{\input{./figures/diagrams//generator_spider-nonum.tikz}}
\endpgfgraphicnamed,   \quad 
     \left\llbracket%
\beginpgfgraphicnamed{diagrams//rgatewhite}
\InputIfFileExists{diagrams//rgatewhite.tikz}{}{\input{./figures/diagrams//rgatewhite.tikz}}
\endpgfgraphicnamed\right\rrbracket_{WX}=  %
\beginpgfgraphicnamed{diagrams//alphalambdagate}
\InputIfFileExists{diagrams//alphalambdagate.tikz}{}{\input{./figures/diagrams//alphalambdagate.tikz}}
\endpgfgraphicnamed,   \quad 
       \left\llbracket%
\beginpgfgraphicnamed{diagrams//piblack}
\InputIfFileExists{diagrams//piblack.tikz}{}{\input{./figures/diagrams//piblack.tikz}}
\endpgfgraphicnamed\right\rrbracket_{WX}=  %
\beginpgfgraphicnamed{diagrams//pired}
\InputIfFileExists{diagrams//pired.tikz}{}{\input{./figures/diagrams//pired.tikz}}
\endpgfgraphicnamed,   
         \]

 \[
  \left\llbracket%
\beginpgfgraphicnamed{diagrams//corsszw}
\InputIfFileExists{diagrams//corsszw.tikz}{}{\input{./figures/diagrams//corsszw.tikz}}
\endpgfgraphicnamed\right\rrbracket_{WX}=  %
\beginpgfgraphicnamed{diagrams//crossxz}
\InputIfFileExists{diagrams//crossxz.tikz}{}{\input{./figures/diagrams//crossxz.tikz}}
\endpgfgraphicnamed,   \quad \quad
    \left\llbracket%
\beginpgfgraphicnamed{diagrams//wblack}
\InputIfFileExists{diagrams//wblack.tikz}{}{\input{./figures/diagrams//wblack.tikz}}
\endpgfgraphicnamed\right\rrbracket_{WX}=  %
\beginpgfgraphicnamed{diagrams//winzx}
\InputIfFileExists{diagrams//winzx.tikz}{}{\input{./figures/diagrams//winzx.tikz}}
\endpgfgraphicnamed, \]

    \[ \llbracket D_1\otimes D_2  \rrbracket_{WX} =  \llbracket D_1  \rrbracket_{WX} \otimes  \llbracket  D_2  \rrbracket_{WX}, \quad 
 \llbracket D_1\circ D_2  \rrbracket_{WX} =  \llbracket D_1  \rrbracket_{WX} \circ  \llbracket  D_2  \rrbracket_{WX}.
 \]
where $r=\lambda e^{i\alpha},~   \alpha \in [0,~2\pi), ~ \lambda  \geq 0$.

The interpretation $\llbracket \cdot \rrbracket_{WX}$ preserves the standard interpretation as well:

\begin{lemma}\label{wtoxpreservesemantics}
Suppose $D$ is an arbitrary diagram in $ZW_{ \mathbb{C}}$-calculus. Then  $\llbracket \llbracket D \rrbracket_{WX}\rrbracket = \llbracket D \rrbracket$.
\end{lemma}
\begin{proof}
The proof is similar to that of Lemma \ref{xtowpreservesemantics}.
\end{proof}

Thus  both $\llbracket \cdot \rrbracket_{WX}$ and $\llbracket \cdot \rrbracket_{XW}$  preserve the standard interpretation. Combining with the completeness of the $ZW_{ \mathbb{C}}$-calculus, we have
\begin{lemma}\label{wtoxisfunctor}
If  $ZX_{ full}\vdash D_1=D_2$, then  $ZW_{ \mathbb{C}}\vdash \llbracket D_1 \rrbracket_{XW} =\llbracket D_2 \rrbracket_{XW}$.
\end{lemma}

\begin{proof}
Suppose $ZX_{ full}\vdash D_1=D_2$, Then by the soundness of the $ZX_{ full}$, we have $\llbracket D_1 \rrbracket =\llbracket D_2 \rrbracket$. 
Therefore  $\llbracket \llbracket D_1 \rrbracket_{XW}\rrbracket = \llbracket D_1 \rrbracket  =\llbracket D_2 \rrbracket =\llbracket \llbracket D_2 \rrbracket_{XW}\rrbracket $ by Lemma \ref{xtowpreservesemantics}. Then it follows that
$ZW_{ \mathbb{C}}\vdash \llbracket D_1 \rrbracket_{XW} =\llbracket D_2 \rrbracket_{XW}$ by the completeness of the $ZW_{ \mathbb{C}}$-calculus.
\end{proof}

This means the interpretation $\llbracket \cdot \rrbracket_{XW}$ is a well-defined functor from the PROP $ZX_{ full}$ to the PROP $ZW_{ \mathbb{C}}$. 

%For the same reason, we get 
Moreover,  we have
\begin{lemma}\label{zwinverse}
For any diagram $G\in ZW_{ \mathbb{C}}$,
\begin{equation}\label{zwinvert}
ZW_{ \mathbb{C}}\vdash \llbracket \llbracket G \rrbracket_{WX}\rrbracket_{XW} =G
\end{equation}
\end{lemma}

\begin{proof}
By Lemma \ref{xtowpreservesemantics} and Lemma \ref{wtoxpreservesemantics}, the interpretations $\llbracket \cdot \rrbracket_{XW}$ and $\llbracket \cdot \rrbracket_{WX}$ preserve the standard interpretation $\llbracket \cdot \rrbracket$. Thus 
$\llbracket G \rrbracket = \llbracket \llbracket G \rrbracket_{WX} \rrbracket = \llbracket  \llbracket \llbracket G \rrbracket_{WX}\rrbracket_{XW} \rrbracket $. By the completeness of the $ZW_{ \mathbb{C}}$-calculus, it must be that $ZW_{ \mathbb{C}}\vdash \llbracket \llbracket G \rrbracket_{WX}\rrbracket_{XW} =G$.
\end{proof}

On the other hand, we have

\begin{lemma}\label{interpretationreversible}
Suppose $D$ is an arbitrary diagram in the $ZX_{ full}$-calculus. Then  $ZX_{ full}\vdash \llbracket \llbracket D \rrbracket_{XW}\rrbracket_{WX} =D$.
\end{lemma}

\begin{proof}
By the construction of  $\llbracket \cdot  \rrbracket_{XW}$  and $\llbracket \cdot  \rrbracket_{WX}$, we only need to prove for the generators of the $ZX_{ full}$-calculus. Here we consider  all the generators translated at the beginning of this section.
  The first six generators  are the same as the  first six generators in  the $ZW_{ \mathbb{C}}$-calculus, so we just need to check for the last six generators. Since the red phase gate and the red spider  are translated in terms of the translation of Hadamard gate, green phase gate and green spider, we only need to care for the four generators: Hadamard gate, green phase gate, $\lambda$ box and the triangle. 

Firstly,
\[
 \left\llbracket%
\beginpgfgraphicnamed{diagrams//alphagate}
\InputIfFileExists{diagrams//alphagate.tikz}{}{\input{./figures/diagrams//alphagate.tikz}}
\endpgfgraphicnamed\right\rrbracket_{XW}=  %
\beginpgfgraphicnamed{diagrams//alphagatewhite}
\InputIfFileExists{diagrams//alphagatewhite.tikz}{}{\input{./figures/diagrams//alphagatewhite.tikz}}
\endpgfgraphicnamed,
\]
so we have 
\[
 \left\llbracket \left\llbracket%
\beginpgfgraphicnamed{diagrams//alphagate}
\InputIfFileExists{diagrams//alphagate.tikz}{}{\input{./figures/diagrams//alphagate.tikz}}
\endpgfgraphicnamed\right\rrbracket_{XW}\right\rrbracket_{WX}= \left\llbracket%
\beginpgfgraphicnamed{diagrams//alphagatewhite}
\InputIfFileExists{diagrams//alphagatewhite.tikz}{}{\input{./figures/diagrams//alphagatewhite.tikz}}
\endpgfgraphicnamed\right\rrbracket_{WX}=%
\beginpgfgraphicnamed{diagrams//alphagate}
\InputIfFileExists{diagrams//alphagate.tikz}{}{\input{./figures/diagrams//alphagate.tikz}}
\endpgfgraphicnamed,
\]
by the definition of  $\llbracket \cdot  \rrbracket_{WX}$ and the ZX rule (L3).
Similarly, we can easily check that 
\[
 \left\llbracket \left\llbracket%
\beginpgfgraphicnamed{diagrams//lambdabox}
\InputIfFileExists{diagrams//lambdabox.tikz}{}{\input{./figures/diagrams//lambdabox.tikz}}
\endpgfgraphicnamed\right\rrbracket_{XW}\right\rrbracket_{WX}=%
\beginpgfgraphicnamed{diagrams//lambdabox}
\InputIfFileExists{diagrams//lambdabox.tikz}{}{\input{./figures/diagrams//lambdabox.tikz}}
\endpgfgraphicnamed. 
 \]
 
 Finally,
 \[
  \left\llbracket \left\llbracket%
\beginpgfgraphicnamed{diagrams//HadaDecomSingleslt}
\InputIfFileExists{diagrams//HadaDecomSingleslt.tikz}{}{\input{./figures/diagrams//HadaDecomSingleslt.tikz}}
\endpgfgraphicnamed\right\rrbracket_{XW}\right\rrbracket_{WX}=   \left\llbracket  %
\beginpgfgraphicnamed{diagrams//Hadamardwhite}
\InputIfFileExists{diagrams//Hadamardwhite.tikz}{}{\input{./figures/diagrams//Hadamardwhite.tikz}}
\endpgfgraphicnamed \right\rrbracket_{WX} 
  =  %
\beginpgfgraphicnamed{diagrams//Hadamardwhiteint}
\InputIfFileExists{diagrams//Hadamardwhiteint.tikz}{}{\input{./figures/diagrams//Hadamardwhiteint.tikz}}
\endpgfgraphicnamed    \overset{IV}{=}   %
\beginpgfgraphicnamed{diagrams//Hadamardwhiteint2}
\InputIfFileExists{diagrams//Hadamardwhiteint2.tikz}{}{\input{./figures/diagrams//Hadamardwhiteint2.tikz}}
\endpgfgraphicnamed \overset{S1,S2}{=}  %
\beginpgfgraphicnamed{diagrams//HadaDecomSingleslt}
\InputIfFileExists{diagrams//HadaDecomSingleslt.tikz}{}{\input{./figures/diagrams//HadaDecomSingleslt.tikz}}
\endpgfgraphicnamed,
  \]
  
  \[
   \left\llbracket \left\llbracket%
\beginpgfgraphicnamed{diagrams//triangle}
\InputIfFileExists{diagrams//triangle.tikz}{}{\input{./figures/diagrams//triangle.tikz}}
\endpgfgraphicnamed\right\rrbracket_{XW}\right\rrbracket_{WX}= \left\llbracket   %
\beginpgfgraphicnamed{diagrams//trianglewhite}
\InputIfFileExists{diagrams//trianglewhite.tikz}{}{\input{./figures/diagrams//trianglewhite.tikz}}
\endpgfgraphicnamed \right\rrbracket_{WX} =  %
\beginpgfgraphicnamed{diagrams//trianinzx}
\InputIfFileExists{diagrams//trianinzx.tikz}{}{\input{./figures/diagrams//trianinzx.tikz}}
\endpgfgraphicnamed  \overset{B1,S1, S2}{=} %
\beginpgfgraphicnamed{diagrams//triangle}
\InputIfFileExists{diagrams//triangle.tikz}{}{\input{./figures/diagrams//triangle.tikz}}
\endpgfgraphicnamed.
\]

\end{proof}

Therefore, by Lemma \ref{zwinverse} and Lemma \ref{interpretationreversible}, the interpretations between the $ZX_{ full}$-calculus and the  $ZW_{ \mathbb{C}}$-calculus are mutually invertible to each other.

\section{Completeness}
\begin{proposition}\label{zwrulesholdinzx}
If  $ZW_{ \mathbb{C}}\vdash D_1=D_2$, then  $ZX_{ full}\vdash \llbracket D_1 \rrbracket_{WX} =\llbracket D_2 \rrbracket_{WX}$.

\end{proposition}

\begin{proof}
By the construction of the ZW-calculus, any two ZW diagrams are equal if and only if one of them can be rewritten into another.  So here we need only to prove that $ZX_{ full} \vdash \left\llbracket D_1\right\rrbracket_{WX} = \left\llbracket D_2\right\rrbracket_{WX}$ where  $D_1=D_2$ is a rewriting rule of $ZW_{ \mathbb{C}}$-calculus. This proof is quite lengthy, we put it at the end of this chapter for the convenience of reading.
\end{proof}
  
This proposition means the interpretation $\llbracket \cdot \rrbracket_{WX}$ is a well-defined functor from the PROP $ZW_{ \mathbb{C}}$ to the PROP $ZX_{ full}$. Together with  lemma \ref{zwinverse} and lemma \ref{interpretationreversible}, now we can say that there are invertible functors between  the PROP $ZW_{ \mathbb{C}}$ and the PROP $ZX_{ full}$, which means the two calculi $ZW_{ \mathbb{C}}$ and $ZX_{ full}$ are isomorphic. 

At last, we prove the main theorem of this chapter. 
\begin{theorem}\label{maintheorem}
The $ZX_{ full}$-calculus is complete for the entire pure qubit quantum mechanics:
If $\llbracket D_1 \rrbracket =\llbracket D_2 \rrbracket$, then $ZX_{ full}\vdash D_1=D_2$,

\end{theorem}

\begin{proof}
Suppose $D_1,  D_2 \in ZX_{ full}$ and  $\llbracket D_1 \rrbracket =\llbracket D_2 \rrbracket$. Then by lemma \ref{xtowpreservesemantics},  $\llbracket \llbracket D_1 \rrbracket_{XW}\rrbracket = \llbracket D_1 \rrbracket= \llbracket D_2 \rrbracket=\llbracket \llbracket D_2 \rrbracket_{XW}\rrbracket $.  Thus by the completeness of the $ZW_{ \mathbb{C}}$-calculus \cite{amar},  $ZW_{ \mathbb{C}}\vdash \llbracket D_2 \rrbracket_{XW}=  \llbracket D_2 \rrbracket_{XW}$.  Now by proposition \ref{zwrulesholdinzx},  $ZX_{ full}\vdash \llbracket \llbracket D_1 \rrbracket_{XW}\rrbracket_{WX} =\llbracket \llbracket D_2 \rrbracket_{XW}\rrbracket_{WX}$.
Finally, by lemma \ref{interpretationreversible},  $ZX_{ full}\vdash D_1=D_2$.
\end{proof}

Now we can express the new generators in terms of green and red nodes.
\begin{proposition}\label{lem:lamb_tri_decomposition}
The triangle %
\beginpgfgraphicnamed{diagrams//triangle}
\InputIfFileExists{diagrams//triangle.tikz}{}{\input{./figures/diagrams//triangle.tikz}}
\endpgfgraphicnamed and the lambda box %
\beginpgfgraphicnamed{diagrams//lambdabox}
\InputIfFileExists{diagrams//lambdabox.tikz}{}{\input{./figures/diagrams//lambdabox.tikz}}
\endpgfgraphicnamed  are expressible in Z and X phases.
\end{proposition}

\begin{proof}
The semantic representation of the triangle %
\beginpgfgraphicnamed{diagrams//triangle}
\InputIfFileExists{diagrams//triangle.tikz}{}{\input{./figures/diagrams//triangle.tikz}}
\endpgfgraphicnamed in terms of  Z and X phases in the ZX-calculus has been clearly described in \cite{Emmanuel}. Interestingly, another representation of the triangle was implicitly given in \cite{Coeckebk} by a slash-labeled box in the following form:
\begin{equation}\label{triangleslash}
\beginpgfgraphicnamed{diagrams//triangledecompose}
\InputIfFileExists{diagrams//triangledecompose.tikz}{}{\input{./figures/diagrams//triangledecompose.tikz}}
\endpgfgraphicnamed 
\end{equation}
It can be directly verified that the two diagrams on both sides of (\ref{triangleslash}) have the same standard interpretation. Thus by Theorem \ref{maintheorem} the identity (\ref{triangleslash}) can be derived in the $ZX_{ full}$-calculus.

 Now we consider the representation of the lambda box. 
Since $\lambda$  is a non-negative real number, we can write $\lambda$ as a sum of its integer part and remainder part: $\lambda= [\lambda] +\{\lambda\}$, where $ [\lambda]$ is a non-negative integer and $0\leq\{\lambda\}<1$.
Let $n= [\lambda]$. 
% If $n=1$, then by rule (L3), 
%$$\tikzfig{diagrams//sqr1is1} $$
If  $n=0$, then 
\begin{equation}\label{lemma0decom}
 %\tikzfig{diagrams//lemma0} 
 %
\beginpgfgraphicnamed{diagrams//lemma02v}
\InputIfFileExists{diagrams//lemma02v.tikz}{}{\input{./figures/diagrams//lemma02v.tikz}}
\endpgfgraphicnamed 
\end{equation}
%$$\tikzfig{diagrams//lemma0} $$
 
  If $n\geq 1$,  then by (\ref{nrepresent}),  we have
  \begin{equation*}
\beginpgfgraphicnamed{diagrams//lambdaderive2}
\InputIfFileExists{diagrams//lambdaderive2.tikz}{}{\input{./figures/diagrams//lambdaderive2.tikz}}
\endpgfgraphicnamed 
\end{equation*}
% If $n\geq 2$, then by the rule (AD$^{\prime}$) and induction on $n$, we have 
% $$\tikzfig{diagrams//lexpress1}, \quad \cdots, \quad \tikzfig{diagrams//lexpress2}.$$
 %Again by rule (AD),   we have
 Since $0\leq\{\lambda\}<1$,  we could let $\alpha=arccos\frac{\{\lambda\}}{2}$. Then $\{\lambda\}=2\cos\alpha=e^{i\alpha}+e^{-i\alpha}$, and
 % $$\tikzfig{diagrams//lexpress3}.$$
  $$%
\beginpgfgraphicnamed{diagrams//lexpress3nd}
\InputIfFileExists{diagrams//lexpress3nd.tikz}{}{\input{./figures/diagrams//lexpress3nd.tikz}}
\endpgfgraphicnamed.$$
  Therefore, we have
    %$$\tikzfig{diagrams//lexpress4}.$$
  $$%
\beginpgfgraphicnamed{diagrams//lexpress4nd}
\InputIfFileExists{diagrams//lexpress4nd.tikz}{}{\input{./figures/diagrams//lexpress4nd.tikz}}
\endpgfgraphicnamed.$$

\end{proof}

%%%%%%%%%%%%%%%%%%%%%%%%%%%%%%%%%%%%%%%%%%%%%%%%%%

\section{Proof of proposition \ref{zwrulesholdinzx}}

\begin{lemma}
	\begin{equation}\label{lem:2}
	\left\llbracket~
	\input{diagrams/lemma2left.tikz}
	~\right\rrbracket_{WX}=
	\input{diagrams/lemma2right.tikz}
	\end{equation}
\end{lemma}

\begin{proof}
	%Proof in \cite{Emmanuel} lemma $33$. The rules and properties used are \ref{lem:hopf}, \ref{lem:inv}, $B1$, $TR2$.
	\begin{equation*}%\label{lem:2}
	\left\llbracket~
	\input{diagrams/lemma2left.tikz}
	~\right\rrbracket_{WX}=
	\input{diagrams/lemma2right2nd.tikz}
	\end{equation*}
\end{proof}

\begin{proposition}(ZW rule $rei^x_2$)~\newline \\
	$
	ZX\vdash
	\left\llbracket~
	\input{diagrams/reix2_LHS.tikz}
	~\right\rrbracket_{WX}
	=
	\left\llbracket~
	\input{diagrams/reix2_RHS.tikz}
	~\right\rrbracket_{WX}
	$
\end{proposition}

\begin{proof}~\newline
	\input{diagrams/reix2_proof.tikz}
\end{proof}

%\begin{lemma}
%	\begin{equation}\label{3crossing}
%	\input{diagrams/3crossings.tikz}
%	\end{equation}
%\end{lemma}

%\begin{proof} The proof can be done by simply sliding the middle line from the left to the right by naturality. 
%\end{proof}

%\input{Def_5.3/reix2/prop_reix2}
\begin{proposition}(ZW rule $rei^x_3$)~\newline \\
	$
	ZX\vdash
	\left\llbracket~
	\input{diagrams/reix3_LHS.tikz}
	~\right\rrbracket_{WX}
	=
	\left\llbracket~
	\input{diagrams/reix3_RHS.tikz}
	~\right\rrbracket_{WX}
	$
\end{proposition}

\begin{proof}~\newline
	\input{diagrams/reix3_proof.tikz}
\end{proof}

\begin{proposition}\label{prop:natnx}(ZW rule $nat^\eta_x$)~\newline \\
	$
	ZX\vdash
	\left\llbracket~
	\input{diagrams/natnx_LHS.tikz}
	~\right\rrbracket_{WX}
	=
	\left\llbracket~
	\input{diagrams/natnx_RHS.tikz}
	~\right\rrbracket_{WX}
	$
\end{proposition}

\begin{proof}~\newline
	\input{diagrams/natnx_proof.tikz}
\end{proof}

\begin{proposition}(ZW rule $nat^\varepsilon_x$)~\newline \\
	$
	ZX\vdash
	\left\llbracket~
	\input{diagrams/natex_LHS.tikz}
	~\right\rrbracket_{WX}
	=
	\left\llbracket~
	\input{diagrams/natex_RHS.tikz}
	~\right\rrbracket_{WX}
	$
\end{proposition}

\begin{proof}
	This is the upside-down version of  Proposition \ref{prop:natnx}, thus the proof is similar. We omit the proof but note that the rules and properties used are $S1$, $S2$, $S3$, $H$, \ref{lem:hopf}.
\end{proof}

\begin{proposition}(ZW rule $rei^x_1$)~\newline \\
	$
	ZX\vdash
	\left\llbracket~
	\input{diagrams/reix1_LHS.tikz}
	~\right\rrbracket_{WX}
	=
	\left\llbracket~
	\input{diagrams/reix1_RHS.tikz}
	~\right\rrbracket_{WX}
	$
\end{proposition}

\begin{proof}
\begin{equation}\label{prop:reix1}
	\input{diagrams/reix1_proof.tikz}
\end{equation}	
\end{proof}

\begin{proposition}(ZW rule $un^{co}_{w,L}$)~\newline \\
	$
	ZX\vdash
	\left\llbracket~
	\input{diagrams/uncowL_LHS.tikz}
	~\right\rrbracket_{WX}
	=
	\left\llbracket~
	\input{diagrams/uncowL_RHS.tikz}
	~\right\rrbracket_{WX}
	$
\end{proposition}

\begin{proof}
	% Proof in \cite{Emmanuel}, proposition 7 part $1b$. The rules used are $\ref{lem:inv}$, $S1$, $H2$, $H$, $S2$, $TR7$, $\ref{lem:hopf}$, $TR2$, $\ref{lem:3}$.
\begin{equation*}%\label{prop:reix1}
\beginpgfgraphicnamed{diagrams//uncowLproof}
\InputIfFileExists{diagrams//uncowLproof.tikz}{}{\input{./figures/diagrams//uncowLproof.tikz}}
\endpgfgraphicnamed
\end{equation*}	

\end{proof}

%\input{Def_5.4/uncowL/prop_uncowL}
%\begin{proposition}(ZW rule $un^{co}_{w,R}$)~\newline \\
%	$
%	ZX\vdash
%	\left\llbracket~
%	\input{diagrams/uncowR_LHS.tikz}
%	~\right\rrbracket_{WX}
%	=
%	\left\llbracket~
%	\input{diagrams/uncowR_RHS.tikz}
%	~\right\rrbracket_{WX}
%	$
%\end{proposition}

%\begin{proof}
%	Similar proof in \cite{Emmanuel}, proposition 7 part $1b$. The rules used are $\ref{lem:inv}$, $S1$, $H2$, $H$, $S2$, $TR7$, $\ref{lem:hopf}$, $TR2$, $\ref{lem:3}$.
%\end{proof}

%\input{Def_5.4/uncowR/prop_uncowR}
\begin{proposition}(ZW rule $asso_w$)~\newline\\
	$
     ZX\vdash    \left\llbracket~
	\input{diagrams/natww_simon_RHS.tikz}
	~\right\rrbracket_{WX}
	=
	\left\llbracket~
	\input{diagrams/natww_simon_LHS.tikz}
	~\right\rrbracket_{WX}
	$
\end{proposition}

\begin{proof}
%Proof in \cite{Emmanuel}, proposition 7 part $1a$. The rules used are $S1$, $B2$, $\ref{lem:4}$, $\ref{lem:5}$, $TR8$.
\begin{equation*}
\beginpgfgraphicnamed{diagrams//assowproof}
\InputIfFileExists{diagrams//assowproof.tikz}{}{\input{./figures/diagrams//assowproof.tikz}}
\endpgfgraphicnamed
\end{equation*}	

\end{proof}

\begin{proposition}(ZW rule $nat^w_x$)~\newline\\
	$
	ZX\vdash
	\left\llbracket~
	\input{diagrams/natwx_LHS.tikz}
	~\right\rrbracket_{WX}
	=
	\left\llbracket~
	\input{diagrams/natwx_RHS.tikz}
	~\right\rrbracket_{WX}
	$
\end{proposition}

\begin{proof}
	%Proof in \cite{Emmanuel}, proposition 7 part $7a$. The rules used are $H$, $\ref{lem:1}$, $S1$, $B2$.
	\begin{equation*}
\beginpgfgraphicnamed{diagrams//natwxproof}
\InputIfFileExists{diagrams//natwxproof.tikz}{}{\input{./figures/diagrams//natwxproof.tikz}}
\endpgfgraphicnamed
\end{equation*}	
\end{proof}

\begin{proposition}(ZW rule $com^{co}_w$)~\newline \\
	$
	ZX\vdash
	\left\llbracket~
	\input{diagrams/comcow_LHS.tikz}
	~\right\rrbracket_{WX}
	=
	\left\llbracket~
	\input{diagrams/comcow_RHS.tikz}
	~\right\rrbracket_{WX}
	$
\end{proposition}

\begin{proof}~\newline
	\input{diagrams/comcow_proof.tikz}\\
\end{proof}

\begin{proposition}(ZW rule $nat^m_w$)~\newline \\
	$
	ZX\vdash
	\left\llbracket~
	\input{diagrams/natmw_LHS.tikz}
	~\right\rrbracket_{WX}
	=
	\left\llbracket~
	\input{diagrams/natmw_RHS.tikz}
	~\right\rrbracket_{WX}
	$
\end{proposition}

\begin{proof}
	This has been proved in \cite{Emmanuel}, proposition 7, part $5a$. The proof is quite complicated, lots of rules and lemmas are employed. Fortunately, all the rules used for this proof in \cite{Emmanuel} are either a part of the rules in the $ZX_{ full}$-calculus or derivatives from the rules of  the $ZX_{ full}$-calculus. Also, all the lemmas applied to this proof are either a part of the rules in the $ZX_{ full}$-calculus or derived properties from the rules of the $ZX_{ full}$-calculus. Therefore, we need not to repeat the proof here, but just indicate part by part the rules and lemmas used in \cite{Emmanuel}, proposition 7, part $5a$,  with the correspondence between lemmas in \cite{Emmanuel} and rules and properties in the $ZX_{ full}$-calculus shown in brackets.
	
	   There are eight parts in the proof: (\textit{i}), (\textit{ii}), (\textit{iii}), (\textit{iv}), (\textit{v}), (\textit{vi}), (\textit{vii}), (\textit{viii}), as well as a final derivation at the end.
	
	The rules used in part (\textit{i}) are $B2$, $S1$, $H$,  lemma 23 ($\ref{tr5derived}$ in this thesis),  lemma 3 ($\ref{lem:6b}$ in this thesis).
	
	The rules used in part (\textit{ii}) are lemma 26 ($\ref{lem:4}$ in this thesis), $S1$, $B2$, $S1$, lemma 32 ($\ref{TR11}$ in this thesis), lemma 3 ($\ref{lem:6b}$ in this thesis), lemma 16 ($TR1$ in this thesis).
	
	The rules used in part (\textit{iii}) are lemma 24 ($TR6$ in this thesis),  lemma 26 ($\ref{lem:4}$ in this thesis), $S1$,  lemma 27 ($TR8$ in this thesis),  lemma 3 ($\ref{lem:6b}$ in this thesis).
	
	The rules used in part (\textit{iv}) are lemma 26 ($\ref{lem:4}$ in this thesis), $S1$, $B2$, lemma 3 ($\ref{lem:6b}$ in this thesis), lemma 32 ($\ref{TR11}$ in this thesis).
	
	The rules used in part (\textit{v}) are lemma 3 ($\ref{lem:6b}$ in this thesis), part (\textit{v}).
	
	The rules used in part (\textit{vi}) are part (\textit{i}), lemma 26 ($\ref{lem:4}$ in this thesis), lemma 28 ($TR9$ in this thesis), $S1$, lemma 3 ($\ref{lem:6b}$ in this thesis), lemma 2 ($\ref{lem:hopf}$ in this thesis), lemma 16 ($TR1$ in this thesis), $K2$, part (\textit{iii}), part (\textit{v}), part (\textit{iv}), part (\textit{ii}).
	
	The rules used in part (\textit{vii}) are lemma 3 ($\ref{lem:6b}$ in this thesis), $S1$, lemma 16 ($TR1$ in this thesis), $B2$, lemma 25 ($TR7$ in this thesis), lemma 26 ($\ref{lem:4}$ in this thesis), lemma 32 ($\ref{TR11}$ in this thesis).
	
	The rules used in part (\textit{viii}) are lemma 3 ($\ref{lem:6b}$ in this thesis), $H$, $B2$, $S1$, lemma 2 ($\ref{lem:hopf}$ in this thesis), lemma 8 ($\ref{lem:6}$ in this thesis).
	
	The final derivation of the proof of this proposition uses  lemma 26 ($\ref{lem:4}$ in this thesis), $S1$, part (\textit{viii}), lemma 2 ($\ref{lem:hopf}$ in this thesis), part (\textit{vii}), part (\textit{vi}), $B2$, $\ref{lem:com}$ and $\ref{lem:7}$ in this thesis, lemma 3 ($\ref{lem:6b}$ in this thesis), lemma 16 ($TR1$ in this thesis).

\end{proof}

\begin{proposition}\label{blakbialg}(ZW rule $nat^{m \eta}_{w}$)~\newline\\
	$
	ZX\vdash
	\left\llbracket~
	\input{diagrams/natmnw_RHS.tikz}
	~\right\rrbracket_{WX}
	=
	\left\llbracket~
	\input{diagrams/natmnw_LHS.tikz}
	~\right\rrbracket_{WX}
	$
\end{proposition}

\begin{proof}
	%Proof in \cite{Emmanuel}, proposition 7 part $5c$. The rules used are $\ref{lem:2}$, $B1$, $TR2$.
	\begin{equation*}
\beginpgfgraphicnamed{diagrams//natmnwproof}
\InputIfFileExists{diagrams//natmnwproof.tikz}{}{\input{./figures/diagrams//natmnwproof.tikz}}
\endpgfgraphicnamed
\end{equation*}	
\end{proof}

\begin{proposition}(ZW rule $nat^{m \eta}_{\varepsilon ,w}$)~\newline\\
	$
	ZX\vdash
	\left\llbracket~
	\input{diagrams/natmnew_RHS.tikz}
	~\right\rrbracket_{WX}
	=
	\left\llbracket~
	\input{diagrams/natmnew_LHS.tikz}
	~\right\rrbracket_{WX}
	$
\end{proposition}

\begin{proof}
	%Proof in \cite{Emmanuel}, proposition 7 part 5c. The rules used are $\ref{lem:2}$, $S1$, $B1$, $\ref{lem:inv}$.
\begin{equation*}
\beginpgfgraphicnamed{diagrams//natmnewpproof}
\InputIfFileExists{diagrams//natmnewpproof.tikz}{}{\input{./figures/diagrams//natmnewpproof.tikz}}
\endpgfgraphicnamed
\end{equation*}	
\end{proof}

\begin{proposition}(ZW rule $hopf$)~\newline \\
	$
	ZX\vdash
	\left\llbracket~
	\input{diagrams/hopf_LHS.tikz}
	~\right\rrbracket_{WX}
	=
	\left\llbracket~
	\input{diagrams/hopf_RHS.tikz}
	~\right\rrbracket_{WX}
	$
\end{proposition}

\begin{proof}
	%Proof in \cite{Emmanuel}, proposition 7 part $5d$. The rules used are $\ref{prop:rng-1}$, $\ref{lem:hopf}$, $S1$, $TR10$, $TR4$, $B1$.
	\begin{equation*}
\beginpgfgraphicnamed{diagrams//zwhopfproof}
\InputIfFileExists{diagrams//zwhopfproof.tikz}{}{\input{./figures/diagrams//zwhopfproof.tikz}}
\endpgfgraphicnamed
\end{equation*}	

\end{proof}
\begin{proposition}(ZW rules $sym_{3,L}$ and $sym_{3,R}$)~\newline \\
	$
	ZX\vdash
	\left\llbracket~
	\input{diagrams/sym3_LHS.tikz}
	~\right\rrbracket_{WX}
	=
	\left\llbracket~
	\input{diagrams/sym3_middle.tikz}
	~\right\rrbracket_{WX}
	=
	\left\llbracket~
	\input{diagrams/sym3_RHS.tikz}
	~\right\rrbracket_{WX}
	$
\end{proposition}

\begin{proof}
	%Proof in \cite{Emmanuel}, proposition 7 part $0b$ and $0b'$ respectively. The rules used are $\ref{lem:4}$, $\ref{lem:com}$, $\ref{lem:6b}$, $S1$, $TR1$.
	%\input{Def_5.7/sym3/sym3_proof.tikz}
	\begin{equation*}
\beginpgfgraphicnamed{diagrams//sym3pfproof}
\InputIfFileExists{diagrams//sym3pfproof.tikz}{}{\input{./figures/diagrams//sym3pfproof.tikz}}
\endpgfgraphicnamed
\end{equation*}	
\end{proof}

\begin{proposition}(ZW rule $inv$)~\newline \\
	$
	ZX\vdash
	\left\llbracket~
	\input{diagrams/inv_LHS.tikz}
	~\right\rrbracket_{WX}
	=
	\left\llbracket~
	\input{diagrams/inv_RHS.tikz}
	~\right\rrbracket_{WX}
	$
\end{proposition}

\begin{proof}
	%Proof in \cite{Emmanuel}, proposition 7 part $2a$. The rule used is $S1$.
	\begin{equation*}
\beginpgfgraphicnamed{diagrams//zwinvpfproof}
\InputIfFileExists{diagrams//zwinvpfproof.tikz}{}{\input{./figures/diagrams//zwinvpfproof.tikz}}
\endpgfgraphicnamed
\end{equation*}	

\end{proof}

\begin{proposition}(ZW rule $ant^\eta_x$)~\newline \\
	$
	ZX\vdash
	\left\llbracket~
	\input{diagrams/antnx_LHS.tikz}
	~\right\rrbracket_{WX}
	=
	\left\llbracket~
	\input{diagrams/antnx_RHS.tikz}
	~\right\rrbracket_{WX}
	$
\end{proposition}

\begin{proof}
	%Proof in \cite{Emmanuel}, proposition 7 part $7b$. The rules used are $\ref{lem:6b}$, $H$, $S1$.
\begin{equation*}
\beginpgfgraphicnamed{diagrams//antnxproof}
\InputIfFileExists{diagrams//antnxproof.tikz}{}{\input{./figures/diagrams//antnxproof.tikz}}
\endpgfgraphicnamed
\end{equation*}	
\end{proof}

\begin{proposition}(ZW rules $sym_{z,L}$ and $sym_{z,R}$)~\newline \\
	$
	ZX\vdash
	\left\llbracket~
	\input{diagrams/symz_LHS.tikz}
	~\right\rrbracket_{WX}
	=
	\left\llbracket~
	\input{diagrams/symz_middle.tikz}
	~\right\rrbracket_{WX}
	=
	\left\llbracket~
	\input{diagrams/symz_RHS.tikz}
	~\right\rrbracket_{WX}
	$
\end{proposition}

\begin{proof}
	%Proof follows directly from rules $S1$ and $\ref{lem:com}$.
\begin{equation*}
\beginpgfgraphicnamed{diagrams//symzdproof}
\InputIfFileExists{diagrams//symzdproof.tikz}{}{\input{./figures/diagrams//symzdproof.tikz}}
\endpgfgraphicnamed
\end{equation*}	
\end{proof}
\begin{proposition}(ZW rule $un^{co}_{z,R}$)~\newline \\
	$
	ZX\vdash
	\left\llbracket~
	\input{diagrams/uncozR_LHS.tikz}
	~\right\rrbracket_{WX}
	%=\tikzfig{diagrams//uncozR_LHS_zx}
	=
	\left\llbracket~
	\input{diagrams/uncozR_RHS.tikz}
	~\right\rrbracket_{WX}
	%=\input{diagrams/uncozR_RHS.tikz}
	$
\end{proposition}

\begin{proof}
	%Proof follows directly from rule $S1$.
\begin{equation*}
\beginpgfgraphicnamed{diagrams//uncozRdproof}
\InputIfFileExists{diagrams//uncozRdproof.tikz}{}{\input{./figures/diagrams//uncozRdproof.tikz}}
\endpgfgraphicnamed
\end{equation*}	
\end{proof}
\begin{proposition}(ZW rule $asso_z$)~\newline \\
	$
	ZX\vdash
	\left\llbracket~
	\input{diagrams/assoz_LHS.tikz}
	~\right\rrbracket_{WX}
	=
	\left\llbracket~
	\input{diagrams/assoz_RHS.tikz}
	~\right\rrbracket_{WX}
	$
\end{proposition}

\begin{proof}
	%Proof follows directly from rules $S1$.
	\begin{equation*}
\beginpgfgraphicnamed{diagrams//assozdproof}
\InputIfFileExists{diagrams//assozdproof.tikz}{}{\input{./figures/diagrams//assozdproof.tikz}}
\endpgfgraphicnamed
\end{equation*}	
\end{proof}
\begin{proposition}\label{prop:ph}(ZW rule $ph$)~\newline \\
	$
	ZX\vdash
	\left\llbracket~
	\input{diagrams/ph_LHS.tikz}
	~\right\rrbracket_{WX}
	=
	\left\llbracket~
	\input{diagrams/ph_RHS.tikz}
	~\right\rrbracket_{WX}
	$
\end{proposition}

\begin{proof}~\newline
	\begin{equation*}
\beginpgfgraphicnamed{diagrams//ph_proof2}
\InputIfFileExists{diagrams//ph_proof2.tikz}{}{\input{./figures/diagrams//ph_proof2.tikz}}
\endpgfgraphicnamed
\end{equation*}	
\end{proof}
\begin{proposition}(ZW rule $nat^n_c$)~\newline \\
	$
	ZX\vdash
	\left\llbracket~
	\input{diagrams/natnc_LHS.tikz}
	~\right\rrbracket_{WX}
	=
	\left\llbracket~
	\input{diagrams/natnc_RHS.tikz}
	~\right\rrbracket_{WX}
	$
\end{proposition}

\begin{proof}
%	Proof in \cite{Emmanuel}, proposition 7 part $3a$. The rule used is $\ref{lem:6b}$.
	\begin{equation*}
\beginpgfgraphicnamed{diagrams//natncproof2}
\InputIfFileExists{diagrams//natncproof2.tikz}{}{\input{./figures/diagrams//natncproof2.tikz}}
\endpgfgraphicnamed
\end{equation*}	
\end{proof}

\begin{proposition}\label{zxnatmc}(ZW rule $nat^m_c$)~\newline \\
	$
	ZX\vdash
	\left\llbracket~
	\input{diagrams/natmc_LHS.tikz}
	~\right\rrbracket_{WX}
	=
	\left\llbracket~
	\input{diagrams/natmc_RHS.tikz}
	~\right\rrbracket_{WX}
	$
\end{proposition}

\begin{proof}
	%Proof in \cite{Emmanuel}, proposition 7 part $6a$. The rules used are $S1$, $B2$, $TR10$.
\begin{equation*}
\beginpgfgraphicnamed{diagrams//natmcproof}
\InputIfFileExists{diagrams//natmcproof.tikz}{}{\input{./figures/diagrams//natmcproof.tikz}}
\endpgfgraphicnamed
\end{equation*}	

\end{proof}

\begin{proposition}(ZW rule $loop$)~\newline \\
	$
	ZX\vdash
	\left\llbracket~
	\input{diagrams/loop_LHS.tikz}
	~\right\rrbracket_{WX}
	=
	\left\llbracket~
	\input{diagrams/loop_RHS.tikz}
	~\right\rrbracket_{WX}
	$
\end{proposition}

\begin{proof}
	%Proof in \cite{Emmanuel}, proposition 7 part $6c$. The rules used are $\ref{lem:hopf}$, $B1$, $TR2$.
\begin{equation*}
\beginpgfgraphicnamed{diagrams//loopproof}
\InputIfFileExists{diagrams//loopproof.tikz}{}{\input{./figures/diagrams//loopproof.tikz}}
\endpgfgraphicnamed
\end{equation*}		
\end{proof}

\begin{proposition}(ZW rule $unx$)~\newline \\
	$
	ZX\vdash
	\left\llbracket~
	\input{diagrams/unx_LHS.tikz}
	~\right\rrbracket_{WX}
	=
	\left\llbracket~
	\input{diagrams/unx_RHS.tikz}
	~\right\rrbracket_{WX}
	$
\end{proposition}

\begin{proof}
	%Proof in \cite{Emmanuel}, proposition 7 part $X$. The rules used are $\ref{lem:1}$, $S1$, $B2$, $\ref{lem:6}$, $TR5$.
\begin{equation*}
\beginpgfgraphicnamed{diagrams//unxproof}
\InputIfFileExists{diagrams//unxproof.tikz}{}{\input{./figures/diagrams//unxproof.tikz}}
\endpgfgraphicnamed
\end{equation*}	
\end{proof}

\begin{proposition}(ZW rule $rng_1$)~\newline \\
	$
	ZX\vdash
	\left\llbracket~
	\input{diagrams/rng1_LHS.tikz}
	~\right\rrbracket_{WX}
	=
	\left\llbracket~
	\input{diagrams/rng1_RHS.tikz}
	~\right\rrbracket_{WX}
	$
\end{proposition}

\begin{proof}
	%Proof follows directly from rule $L3$.
\begin{equation*}
\beginpgfgraphicnamed{diagrams//rng1dproof}
\InputIfFileExists{diagrams//rng1dproof.tikz}{}{\input{./figures/diagrams//rng1dproof.tikz}}
\endpgfgraphicnamed
\end{equation*}	
\end{proof}
\begin{proposition}\label{prop:rng-1}(ZW rule $rng_{-1}$)~\newline \\
	$
	ZX\vdash
	\left\llbracket~
	\input{diagrams/rng-1_RHS.tikz}
	~\right\rrbracket_{WX}
	=
	\left\llbracket~
	\input{diagrams/rng-1_LHS.tikz}
	~\right\rrbracket_{WX}
	$
\end{proposition}

\begin{proof}
	%Similar to \ref{prop:ph}.
	\begin{equation*}
\beginpgfgraphicnamed{diagrams//rng-1proof}
\InputIfFileExists{diagrams//rng-1proof.tikz}{}{\input{./figures/diagrams//rng-1proof.tikz}}
\endpgfgraphicnamed
\end{equation*}	
\end{proof}

\begin{proposition}(ZW rule $rng^{r,s}_{\times}$)~\newline \\
	$
	ZX\vdash
	\left\llbracket~
	\input{diagrams/rngrsx_LHS.tikz}
	~\right\rrbracket_{WX}
	=
	\left\llbracket~
	\input{diagrams/rngrsx_RHS.tikz}
	~\right\rrbracket_{WX}
	$
\end{proposition}

\begin{proof}
%	Proof follows directly from rules $S1$, $L4$, $L5$.
\begin{equation*}
\beginpgfgraphicnamed{diagrams//rngrsxdproof}
\InputIfFileExists{diagrams//rngrsxdproof.tikz}{}{\input{./figures/diagrams//rngrsxdproof.tikz}}
\endpgfgraphicnamed
\end{equation*}	
where $s=\lambda_s e^{i\alpha}, r=\lambda_r e^{i\beta}$.
\end{proof}
\begin{proposition}(ZW rule $rng^{r,s}_{+}$)~\newline \\
	$
	ZX\vdash
	\left\llbracket~
	\input{diagrams/rngrsp_LHS.tikz}
	~\right\rrbracket_{WX}
	=
	\left\llbracket~
	\input{diagrams/rngrsp_RHS.tikz}
	~\right\rrbracket_{WX}
	$
\end{proposition}

\begin{proof}
	%Proof follows directly from rule $AD$.
	\begin{equation*}
\beginpgfgraphicnamed{diagrams//rngrspdproof}
\InputIfFileExists{diagrams//rngrspdproof.tikz}{}{\input{./figures/diagrams//rngrspdproof.tikz}}
\endpgfgraphicnamed
\end{equation*}	
where $s=\lambda_s e^{i\alpha}, r=\lambda_r e^{i\beta},  r+s=\lambda e^{i\gamma}$.
\end{proof}
\begin{proposition}(ZW rule $nat^{r}_{c}$)~\newline \\
	$
	ZX\vdash
	\left\llbracket~
	\input{diagrams/natrc_LHS.tikz}
	~\right\rrbracket_{WX}
	=
	\left\llbracket~
	\input{diagrams/natrc_RHS.tikz}
	~\right\rrbracket_{WX}
	$
\end{proposition}

\begin{proof}
	%Proof follows directly from rules $TR13$, $TR14$.
	\begin{equation*}
\beginpgfgraphicnamed{diagrams//natrcdproof}
\InputIfFileExists{diagrams//natrcdproof.tikz}{}{\input{./figures/diagrams//natrcdproof.tikz}}
\endpgfgraphicnamed
\end{equation*}	
where $ r=\lambda_r e^{i\beta}$.

\end{proof}
\begin{proposition}(ZW rule $nat^{r}_{\varepsilon c}$)~\newline \\
	$
	ZX\vdash
	\left\llbracket~
	\input{diagrams/natrec_LHS.tikz}
	~\right\rrbracket_{WX}
	=
	\left\llbracket~
	\input{diagrams/natrec_RHS.tikz}
	~\right\rrbracket_{WX}
	$
\end{proposition}

\begin{proof}
	%Proof follows directly from rules $\ref{lem:S4}$, $L2$.
	\begin{equation*}
\beginpgfgraphicnamed{diagrams//natrecdproof}
\InputIfFileExists{diagrams//natrecdproof.tikz}{}{\input{./figures/diagrams//natrecdproof.tikz}}
\endpgfgraphicnamed
\end{equation*}	
where $ r=\lambda_r e^{i\beta}$.
\end{proof}
\begin{proposition}(ZW rule $ph^{r}$)~\newline \\
	$
	ZX\vdash
	\left\llbracket~
	\input{diagrams/phr_RHS.tikz}
	~\right\rrbracket_{WX}
	=
	\left\llbracket~
	\input{diagrams/phr_LHS.tikz}
	~\right\rrbracket_{WX}
	$
\end{proposition}

\begin{proof}
	%Proof follows directly from rules $S1$, $L1$.
	\begin{equation*}
\beginpgfgraphicnamed{diagrams//phrdproof}
\InputIfFileExists{diagrams//phrdproof.tikz}{}{\input{./figures/diagrams//phrdproof.tikz}}
\endpgfgraphicnamed
\end{equation*}	
where $ r=\lambda_r e^{i\beta}$.	
\end{proof}

%% file: diagrams/triangle.tikz
\begin{tikzpicture}
	\begin{pgfonlayer}{nodelayer}
		\node [style=none] (0) at (0, 0.5) {};
		\node [style=triangle] (1) at (0, 0) {};
		\node [style=none] (2) at (0, -0.5) {};
	\end{pgfonlayer}
	\begin{pgfonlayer}{edgelayer}
		\draw (0.center) to (2.center);
	\end{pgfonlayer}
\end{tikzpicture}

%% file: diagrams/lambdabox.tikz
\begin{tikzpicture}
	\begin{pgfonlayer}{nodelayer}
		\node [style=H box] (0) at (0, 0) {$\lambda$};
		\node [style=none] (1) at (0, -0.5) {};
		\node [style=none] (2) at (0, 0.5) {};
	\end{pgfonlayer}
	\begin{pgfonlayer}{edgelayer}
		\draw (2.center) to (0);
		\draw (1.center) to (0);
	\end{pgfonlayer}
\end{tikzpicture}

%% file: diagrams/alphagate.tikz
\begin{tikzpicture}
	\begin{pgfonlayer}{nodelayer}
		\node [style=none] (0) at (0, -0.5) {};
		\node [style=none] (1) at (0, 0.5) {};
		\node [style=gn] (2) at (0, 0) {$\alpha$};
	\end{pgfonlayer}
	\begin{pgfonlayer}{edgelayer}
		\draw (1.center) to (0.center);
	\end{pgfonlayer}
\end{tikzpicture}

%% file: diagrams/alphagatewhite.tikz
\begin{tikzpicture}
	\begin{pgfonlayer}{nodelayer}
		\node [style=wn] (0) at (0.25, 0) {};
		\node [style=none] (1) at (0.25, -0.5) {};
		\node [style=none] (2) at (0.25, 0.5) {};
		\node [style=none] (3) at (0.75, 0) {$e^{i\alpha}$};
	\end{pgfonlayer}
	\begin{pgfonlayer}{edgelayer}
		\draw (2.center) to (1.center);
	\end{pgfonlayer}
\end{tikzpicture}

%% file: diagrams/lambdagatewhiteld.tikz
\begin{tikzpicture}
	\begin{pgfonlayer}{nodelayer}
		\node [style=none] (0) at (0, -0.5) {};
		\node [style=none] (1) at (0.25, 0) {$\lambda$};
		\node [style=wn] (2) at (0, 0) {};
		\node [style=none] (3) at (0, 0.5) {};
	\end{pgfonlayer}
	\begin{pgfonlayer}{edgelayer}
		\draw (3.center) to (0.center);
	\end{pgfonlayer}
\end{tikzpicture}

%% file: diagrams/alphagatered.tikz
\begin{tikzpicture}
	\begin{pgfonlayer}{nodelayer}
		\node [style=none] (0) at (0, 0.5) {};
		\node [style=rn] (1) at (0, 0) {$\alpha$};
		\node [style=none] (2) at (0, -0.5) {};
	\end{pgfonlayer}
	\begin{pgfonlayer}{edgelayer}
		\draw [style=none] (2.center) to (1);
		\draw [style=none] (1) to (0.center);
	\end{pgfonlayer}
\end{tikzpicture}

%% file: diagrams/alphalambdagate.tikz
\begin{tikzpicture}
	\begin{pgfonlayer}{nodelayer}
		\node [style=H box] (0) at (0, -0.25) {$\lambda$};
		\node [style=none] (1) at (0, 0.75) {};
		\node [style=gn] (2) at (0, 0.25) {$\alpha$};
		\node [style=none] (3) at (0, -0.75) {};
	\end{pgfonlayer}
	\begin{pgfonlayer}{edgelayer}
		\draw (1.center) to (3.center);
	\end{pgfonlayer}
\end{tikzpicture}

%% file: diagrams/pired.tikz
\begin{tikzpicture}
	\begin{pgfonlayer}{nodelayer}
		\node [style=none] (0) at (0, -0.5) {};
		\node [style=none] (1) at (0, 0.5) {};
		\node [style=rn] (2) at (0, 0) {$\pi$};
	\end{pgfonlayer}
	\begin{pgfonlayer}{edgelayer}
		\draw (1.center) to (0.center);
	\end{pgfonlayer}
\end{tikzpicture}

%% file: diagrams/lemma2left.tikz
\begin{tikzpicture}
	\begin{pgfonlayer}{nodelayer}
		\node [style=bn] (0) at (0, 0) {};
		\node [style=none] (1) at (0, 0.5) {};
	\end{pgfonlayer}
	\begin{pgfonlayer}{edgelayer}
		\draw [in=-135, out=-45, loop] (0) to ();
		\draw (0) to (1.center);
	\end{pgfonlayer}
\end{tikzpicture}

%% file: diagrams/reix2_RHS.tikz
\begin{tikzpicture}
	\begin{pgfonlayer}{nodelayer}
		\node [style=none] (0) at (-0.25, -0.5000001) {};
		\node [style=none] (1) at (0.25, -0.5000001) {};
		\node [style=none] (2) at (-0.25, 0.5) {};
		\node [style=none] (3) at (0.25, 0.5) {};
	\end{pgfonlayer}
	\begin{pgfonlayer}{edgelayer}
		\draw (2.center) to (0.center);
		\draw (3.center) to (1.center);
	\end{pgfonlayer}
\end{tikzpicture}

%% file: diagrams/uncowL_RHS.tikz
\begin{tikzpicture}
	\begin{pgfonlayer}{nodelayer}
		\node [style=none] (0) at (0.5000001, -0.7500001) {};
		\node [style=none] (1) at (-0.5000001, 0.75) {};
		\node [style=none] (2) at (0.5000001, -0.2499999) {};
	\end{pgfonlayer}
	\begin{pgfonlayer}{edgelayer}
		\draw [in=90, out=-90, looseness=1.00] (1.center) to (2.center);
		\draw (2.center) to (0.center);
	\end{pgfonlayer}
\end{tikzpicture}

%% file: diagrams/inv_LHS.tikz
\begin{tikzpicture}
	\begin{pgfonlayer}{nodelayer}
		\node [style=bn] (0) at (0, -0.2500001) {};
		\node [style=none] (1) at (0, 1) {};
		\node [style=none] (2) at (0, -1) {};
		\node [style=bn] (3) at (0, 0.2500001) {};
	\end{pgfonlayer}
	\begin{pgfonlayer}{edgelayer}
		\draw (3) to (1.center);
		\draw (3) to (0);
		\draw (0) to (2.center);
	\end{pgfonlayer}
\end{tikzpicture}

%% file: diagrams/inv_RHS.tikz
\begin{tikzpicture}
	\begin{pgfonlayer}{nodelayer}
		\node [style=none] (0) at (0, 1) {};
		\node [style=none] (1) at (0, -1) {};
	\end{pgfonlayer}
	\begin{pgfonlayer}{edgelayer}
		\draw (0.center) to (1.center);
	\end{pgfonlayer}
\end{tikzpicture}

%% file: diagrams/uncozR_RHS.tikz
\begin{tikzpicture}
	\begin{pgfonlayer}{nodelayer}
		\node [style=none] (0) at (0.5, 0.75) {};
		\node [style=none] (1) at (-0.5, -0.25) {};
		\node [style=none] (2) at (-0.5, -0.75) {};
	\end{pgfonlayer}
	\begin{pgfonlayer}{edgelayer}
		\draw [in=90, out=-90, looseness=1.00] (0.center) to (1.center);
		\draw (1.center) to (2.center);
	\end{pgfonlayer}
\end{tikzpicture}

%% file: diagrams/rng1_LHS.tikz
\begin{tikzpicture}
	\begin{pgfonlayer}{nodelayer}
		\node [style=wn] (0) at (0, -0) {};
		\node [style=none] (1) at (0.25, -0) {$1$};
		\node [style=none] (2) at (0, -0.5) {};
		\node [style=none] (3) at (0, 0.5) {};
	\end{pgfonlayer}
	\begin{pgfonlayer}{edgelayer}
		\draw (3.center) to (0);
		\draw (0) to (2.center);
	\end{pgfonlayer}
\end{tikzpicture}

%% file: diagrams/rng1_RHS.tikz
\begin{tikzpicture}
	\begin{pgfonlayer}{nodelayer}
		\node [style=none] (0) at (0, 0.5) {};
		\node [style=none] (1) at (0, -0.5) {};
	\end{pgfonlayer}
	\begin{pgfonlayer}{edgelayer}
		\draw (0.center) to (1.center);
	\end{pgfonlayer}
\end{tikzpicture}

%% file: diagrams/rng-1_LHS.tikz
\begin{tikzpicture}
	\begin{pgfonlayer}{nodelayer}
		\node [style=none] (0) at (0, 0.5) {};
		\node [style=wn] (1) at (0, -0) {};
		\node [style=none] (2) at (0, -0.5) {};
		\node [style=none] (3) at (0.25, -0) {$-1$};
	\end{pgfonlayer}
	\begin{pgfonlayer}{edgelayer}
		\draw (0.center) to (1);
		\draw (1) to (2.center);
	\end{pgfonlayer}
\end{tikzpicture}

%% file: diagrams/rngrsx_RHS.tikz
\begin{tikzpicture}
	\begin{pgfonlayer}{nodelayer}
		\node [style=none] (0) at (0, 0.75) {};
		\node [style=none] (1) at (0, -0.7499999) {};
		\node [style=none] (2) at (0.5, -0) {$rs$};
		\node [style=wn] (3) at (0, -0) {};
	\end{pgfonlayer}
	\begin{pgfonlayer}{edgelayer}
		\draw (3) to (0.center);
		\draw (3) to (1.center);
	\end{pgfonlayer}
\end{tikzpicture}

%% file: diagrams/rngrsp_RHS.tikz
\begin{tikzpicture}
	\begin{pgfonlayer}{nodelayer}
		\node [style=none] (0) at (0, -1) {};
		\node [style=wn] (1) at (0, -0) {};
		\node [style=none] (2) at (0.5, -0) {$r+s$};
		\node [style=none] (3) at (0, 1) {};
	\end{pgfonlayer}
	\begin{pgfonlayer}{edgelayer}
		\draw (1) to (3.center);
		\draw (1) to (0.center);
	\end{pgfonlayer}
\end{tikzpicture}

%% file: diagrams/natrec_LHS.tikz
\begin{tikzpicture}
	\begin{pgfonlayer}{nodelayer}
		\node [style=bn] (0) at (0, -0) {};
		\node [style=none] (1) at (0, 0.75) {};
		\node [style=bn] (2) at (0, -0.25) {};
	\end{pgfonlayer}
	\begin{pgfonlayer}{edgelayer}
		\draw [in=-135, out=-45, loop] (2) to ();
		\draw (2) to (0);
		\draw (0) to (1.center);
	\end{pgfonlayer}
\end{tikzpicture}

%% file: chapter4.tex
\chapter{Completeness for Clifford+T qubit quantum mechanics}\label{cplustch}

%\section{Introduction}
Clifford+T qubit quantum mechanics (QM) is an approximately universal fragment of QM, which has been widely used in quantum computing. 
%In contrast to the traditional way of matrix calculation,  the ZX-calculus introduced by Coecke and Duncan \cite{CoeckeDuncan}  is a diagrammatical axiomatisation of quantum computing.  
One of the main open problems of the ZX-calculus  is to give a complete axiomatisation for the Clifford+T QM \cite{cqmwiki}.
After the first completeness result on this fragment
\textemdash the completeness of the ZX-calculus for single qubit Clifford+T QM \cite{Miriam1ct}, there finally comes an completion of the ZX-calculus for the whole Clifford+T QM  \cite{Emmanuel}, 
which contributes a solution to the above mentioned open problem. However, this complete axiomatisation for the Clifford+T QM relies on a very complicated translation from the ZX-calculus to the ZW-calculus.

Further to this complete axiomatisation for the Clifford+T fragment QM \cite{Emmanuel}, we have given a complete axiomatisation of the ZX-calculus for the overall pure qubit QM (i.e., the $ZX_{ full}$-calculus) in the previous chapter. Then a natural question arises: can we just make a restriction on the generators and rules of the $ZX_{ full}$-calculus obtained in Chapter \ref{chfull} to trivially get a complete axiomatisation of the ZX-calculus for the Clifford+T QM (called $ZX_{ C+T}$-calculus)? The answer is negative: we will show in this chapter that we can have a complete $ZX_{ C+T}$-calculus by restricting on the generators and rules of the $ZX_{ full}$-calculus, but some modifications like changing or adding rules have to be made.   We will illustrate this by a counterexample. To do this, we need to determine the range of the value of $\lambda$ in a $\lambda$ box which will be used as a generator in the $ZX_{ C+T}$-calculus. Let $\mathbb{T}:=\mathbb{Z}[\frac{1}{2}, e^{i\frac{\pi}{4}}]$ be the ring extension of
 $\mathbb{Z}$ in $\mathbb{C}$ generated by $\frac{1}{2}, e^{i\frac{\pi}{4}}$. Similarly we can define the ring $\mathbb{Z}[i, \frac{1}{\sqrt{2}}]$. It is not difficult to see that $\mathbb{T}$ is a commutative ring and $\mathbb{Z}[\frac{1}{2}, e^{i\frac{\pi}{4}}]=\mathbb{Z}[i, \frac{1}{\sqrt{2}}]$.
 
  Denote by $ZX_{\frac{\pi}{4}}$ the ZX-calculus which has only traditional generators as given in Table \ref{qbzxgenerator}  and angles multiple of $\frac{\pi}{4}$ in any green or red spider. 
 Now we recall a result on the relation between  $ZX_{ \frac{\pi}{4}}$ diagrams and their corresponding matrices (we will give a simpler proof later in this Chapter):
%we discuss the range of the value in a $\labmda$ box.
% resting on the paper \cite{ngwang}.

%\section*{Correspondence between  $ZX_{ C+T}$ diagrams and matrices}
%Now we can explain why we choose the ring $\mathbb{T}=\mathbb{Z}[i, \frac{1}{\sqrt{2}}]$ for the proof of the completeness of the $ZX_{ C+T}$-calculus.

\begin{proposition}\label{cliftmatix} \cite{Emmanuel}
The diagrams of the $ZX_{\frac{\pi}{4}}$-calculus exactly corresponds to the matrices over the ring $\mathbb{T}$.
\end{proposition}
This means if we want to introduce the $\lambda$ box as a generator in the the same way ($\lambda$ being the magnitude of a complex number) as in Chapter \ref{chfull} to make a $ZX_{ C+T}$-calculus, then 
 it must be that each $\lambda$ is a non-negative real number and  $\lambda \in \mathbb{T}$. Furthermore, as pointed out in \cite{Emmanuel}, each element $r$ of $\mathbb{T}$ can be uniquely written as the form $r=a_0+a_1e^{i\frac{\pi}{4}}+a_2(e^{i\frac{\pi}{4}})^2+a_3(e^{i\frac{\pi}{4}})^3,~    a_j \in \mathbb Z[\frac{1}{2}] $.  Equivalently,  $r=a_0e^{i\alpha_0}+a_1e^{i\alpha_1}+a_2e^{i\alpha_2}+a_3e^{i\alpha_3},~   0\leq a_j \in \mathbb Z[\frac{1}{2}], ~\alpha_j=j\frac{\pi}{4} ~or~ j\frac{\pi}{4}+\pi,~~ j=0, 1, 2, 3$. 
 Therefore, if an arbitrary complex number in $\mathbb{T}$  has phase restricted to multiple of $\frac{\pi}{4}$, then its magnitude $\lambda$  must satisfy that $ 0\leq \lambda \in \mathbb Z[\frac{1}{2}]$. On the other hand, we have the following identity in the $ZX_{ full}$-calculus by the rule (AD$^{\prime}$):
 \begin{equation}\label{conuterexam}
\beginpgfgraphicnamed{diagrams//countexam}
\InputIfFileExists{diagrams//countexam.tikz}{}{\input{./figures/diagrams//countexam.tikz}}
\endpgfgraphicnamed
 \end{equation}
The left side of (\ref{conuterexam}) is already in the range of   $ZX_{ C+T}$-calculus, but the right side of  (\ref{conuterexam}) is beyond the   $ZX_{ C+T}$-calculus ($\lambda=\sqrt{2} \notin \mathbb Z[\frac{1}{2}]$). This means we can not have a $ZX_{ C+T}$-calculus that is complete for the Clifford+T QM  just by restricting on the generators and rules of the $ZX_{ full}$-calculus.

   In this chapter, we  propose a complete axiomatisation of the ZX-calculus for the Clifford+T quantum mechanics (the  $ZX_{ C+T}$-calculus), not only  making a restriction on the generators and rules of the $ZX_{ full}$-calculus, but also modifying and adding some rules. As before, we still need the completeness result of the ZW-calculus, but will restrict the parameter ring $\mathbb{C}$ to its subring $\mathbb{T}$ in the ZW-calculus (will be called $ZW_{\mathbb{T}}$-calculus afterwards in this thesis) \cite{amar}.  
   %In comparison to the completeness proof in the previous chapter,  a modification of the interpretation from the ZW-calculus to the ZX-calculus is made here. 

The results of this chapter are collected from the arXiv paper \cite{ngwang2} coauthored with Kang Feng Ng.
The proof of the completeness of the $ZX_{ C+T}$-calculus has been published in \cite{amarngwanglics}, with coauthors Amar Hadzihasanovic and Kang Feng Ng.

\section{$ZX_{ C+T}$-calculus}
In this section, we give for the $ZX_{ C+T}$-calculus all the rewriting rules which will be shown to be sufficient for the completeness with regard to the Clifford+T QM.

 The $ZX_{ C+T}$-calculus is still a kind of ZX-calculus as described in the previous chapter. Its generators are listed in  Table \ref{qbzxgenerator} and Table \ref{newgr}, with the restriction that $\alpha \in \{\frac{k\pi}{4}| k=0, 1, \cdots, 7\}, 0\leqslant \lambda,  \in \mathbb Z[\frac{1}{2}]$.

There are two kinds of rules for the $ZX_{ C+T}$-calculus:  the categorical structure rules  as listed in (\ref{compactstructure}) and (\ref{compactstructureslide}), and the non-structural rewriting rules including
% Taking into account of the simplification of rules we made in the previous section, we give
  the traditional rules in Figure \ref{figure1t} and the extended rules in Figure \ref{figure2t}.% and Figure \ref{figure0t}. 

%Note that all the diagrams should be read from top to bottom.

\begin{figure}[!h]
\begin{center}
\[
\quad \qquad\begin{array}{|cccc|}
\hline
\beginpgfgraphicnamed{diagrams//spiderlt}
\InputIfFileExists{diagrams//spiderlt.tikz}{}{\input{./figures/diagrams//spiderlt.tikz}}
\endpgfgraphicnamed=%
\beginpgfgraphicnamed{diagrams//spiderrt}
\InputIfFileExists{diagrams//spiderrt.tikz}{}{\input{./figures/diagrams//spiderrt.tikz}}
\endpgfgraphicnamed &(S1) &%
\beginpgfgraphicnamed{diagrams//s2new}
\InputIfFileExists{diagrams//s2new.tikz}{}{\input{./figures/diagrams//s2new.tikz}}
\endpgfgraphicnamed &(S2)\\
\beginpgfgraphicnamed{diagrams//induced_compact_structure-2wirelt}
\InputIfFileExists{diagrams//induced_compact_structure-2wirelt.tikz}{}{\input{./figures/diagrams//induced_compact_structure-2wirelt.tikz}}
\endpgfgraphicnamed=%
\beginpgfgraphicnamed{diagrams//induced_compact_structure-2wirert}
\InputIfFileExists{diagrams//induced_compact_structure-2wirert.tikz}{}{\input{./figures/diagrams//induced_compact_structure-2wirert.tikz}}
\endpgfgraphicnamed&(S3) & %
\beginpgfgraphicnamed{diagrams//hsquare}
\InputIfFileExists{diagrams//hsquare.tikz}{}{\input{./figures/diagrams//hsquare.tikz}}
\endpgfgraphicnamed &(H2)\\
  &&&\\ 
\beginpgfgraphicnamed{diagrams//hslidecap}
\InputIfFileExists{diagrams//hslidecap.tikz}{}{\input{./figures/diagrams//hslidecap.tikz}}
\endpgfgraphicnamed &(H3) &%
\beginpgfgraphicnamed{diagrams//h2newlt}
\InputIfFileExists{diagrams//h2newlt.tikz}{}{\input{./figures/diagrams//h2newlt.tikz}}
\endpgfgraphicnamed=%
\beginpgfgraphicnamed{diagrams//h2newrt}
\InputIfFileExists{diagrams//h2newrt.tikz}{}{\input{./figures/diagrams//h2newrt.tikz}}
\endpgfgraphicnamed&(H)\\
%  \tikzfig{diagrams//hslidecap} &(H3) & \tikzfig{diagrams//newemptyrl} &(IV')\\
  &&&\\ 
\beginpgfgraphicnamed{diagrams//b1slt}
\InputIfFileExists{diagrams//b1slt.tikz}{}{\input{./figures/diagrams//b1slt.tikz}}
\endpgfgraphicnamed=%
\beginpgfgraphicnamed{diagrams//b1srt}
\InputIfFileExists{diagrams//b1srt.tikz}{}{\input{./figures/diagrams//b1srt.tikz}}
\endpgfgraphicnamed&(B1) & %
\beginpgfgraphicnamed{diagrams//b2slt}
\InputIfFileExists{diagrams//b2slt.tikz}{}{\input{./figures/diagrams//b2slt.tikz}}
\endpgfgraphicnamed=%
\beginpgfgraphicnamed{diagrams//b2srt}
\InputIfFileExists{diagrams//b2srt.tikz}{}{\input{./figures/diagrams//b2srt.tikz}}
\endpgfgraphicnamed&(B2)\\
\beginpgfgraphicnamed{diagrams//HadaDecomSingleslt}
\InputIfFileExists{diagrams//HadaDecomSingleslt.tikz}{}{\input{./figures/diagrams//HadaDecomSingleslt.tikz}}
\endpgfgraphicnamed= %
\beginpgfgraphicnamed{diagrams//HadaDecomSinglesrt2}
\InputIfFileExists{diagrams//HadaDecomSinglesrt2.tikz}{}{\input{./figures/diagrams//HadaDecomSinglesrt2.tikz}}
\endpgfgraphicnamed&(EU)    & %
\beginpgfgraphicnamed{diagrams//k2slt}
\InputIfFileExists{diagrams//k2slt.tikz}{}{\input{./figures/diagrams//k2slt.tikz}}
\endpgfgraphicnamed=%
\beginpgfgraphicnamed{diagrams//k2srt}
\InputIfFileExists{diagrams//k2srt.tikz}{}{\input{./figures/diagrams//k2srt.tikz}}
\endpgfgraphicnamed&(K2)\\
& %
\beginpgfgraphicnamed{diagrams//newemptyrl}
\InputIfFileExists{diagrams//newemptyrl.tikz}{}{\input{./figures/diagrams//newemptyrl.tikz}}
\endpgfgraphicnamed &(IV')  &\\
%\tikzfig{diagrams//alphadelete}&(S4) & &\\
&&&\\ 
\hline
\end{array}\]
\end{center}
% \caption{ZX-calculus rules I, where $\lambda  \geq 0, \alpha, \beta \in [0,~2\pi);$ in (AD1), $\gamma=\beta+arccos(\frac{\lambda_1+\lambda_2cos(\alpha-\beta)}{\lambda})$,  $\lambda=\sqrt{\lambda_1^2+\lambda_2^2+2\lambda_1\lambda_2cos(\alpha-\beta)}$, $\lambda_1^2+\lambda_2^2 > 0, ~ \alpha-\beta \neq \pi (mod ~2\pi)$.}\label{figure1}
  \caption{ Traditional-style $ZX_{ C+T}$-calculus rules, where $\alpha, \beta\in  \{\frac{k\pi}{4}| k=0, 1, \cdots, 7\}$. The upside-down version and colour swapped version of these rules still hold.}\label{figure1t}  
  \end{figure}
  %%%%%%%%%%%%%%%%%%%%%%%%%%%%%%%%%%%%%%%%%%%%

 \begin{figure}[!h]
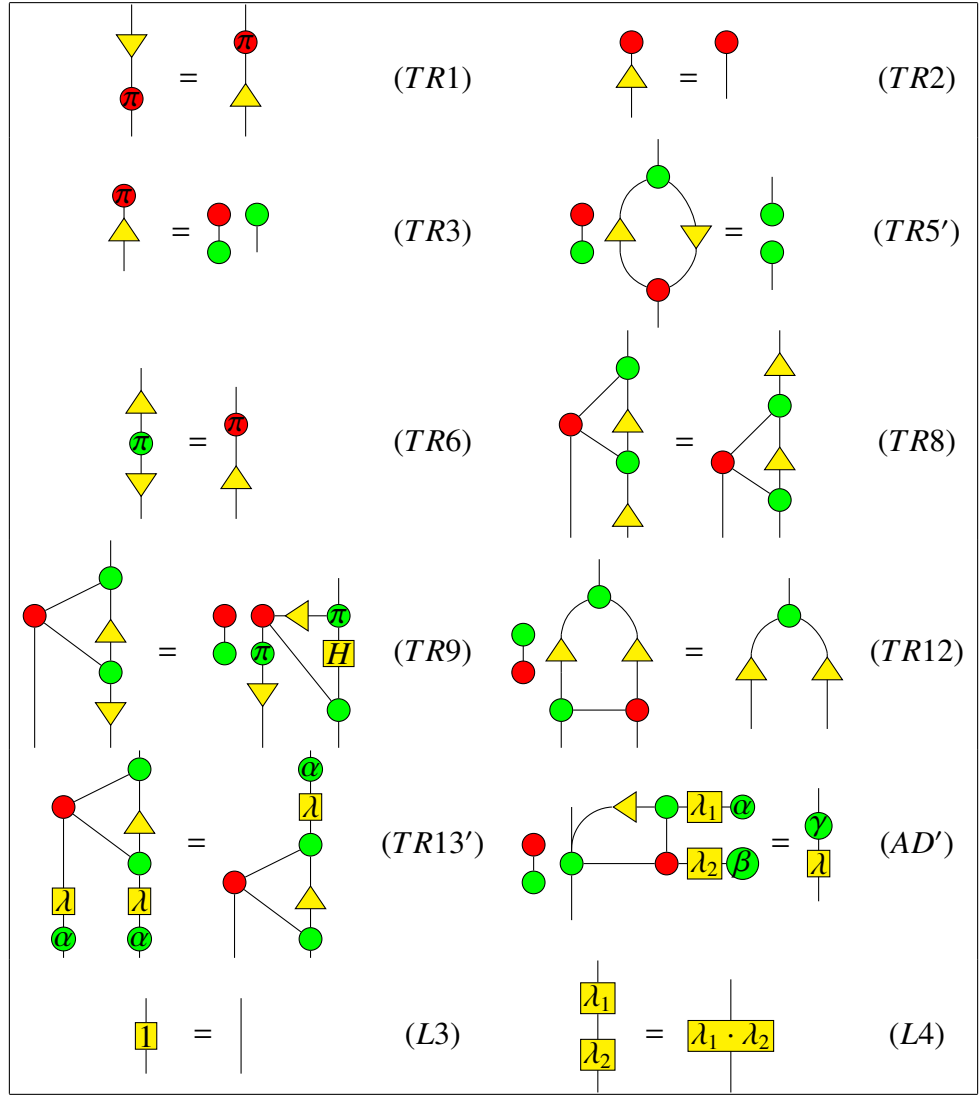

  	\begin{center}
  		\[
  		\quad \qquad\begin{array}{|cccc|}
  		\hline
  		%\tikzfig{diagrams//0plus0}&(AD2) &\tikzfig{diagrams//complemplus}&(AD3)\\
  		%
\beginpgfgraphicnamed{diagrams//trianglepicommute}
\InputIfFileExists{diagrams//trianglepicommute.tikz}{}{\input{./figures/diagrams//trianglepicommute.tikz}}
\endpgfgraphicnamed &(TR1) &%
\beginpgfgraphicnamed{diagrams//triangleocopy}
\InputIfFileExists{diagrams//triangleocopy.tikz}{}{\input{./figures/diagrams//triangleocopy.tikz}}
\endpgfgraphicnamed &(TR2)\\
  		
  		%\tikzfig{diagrams//trianglepicopy}&(TR3) & \tikzfig{diagrams//tr4prime} &(TR4')\\
  		%
\beginpgfgraphicnamed{diagrams//trianglepicopy}
\InputIfFileExists{diagrams//trianglepicopy.tikz}{}{\input{./figures/diagrams//trianglepicopy.tikz}}
\endpgfgraphicnamed&(TR3)  &%
\beginpgfgraphicnamed{diagrams//tr5prime}
\InputIfFileExists{diagrams//tr5prime.tikz}{}{\input{./figures/diagrams//tr5prime.tikz}}
\endpgfgraphicnamed &(TR5') \\
\beginpgfgraphicnamed{diagrams//gpiintriangles}
\InputIfFileExists{diagrams//gpiintriangles.tikz}{}{\input{./figures/diagrams//gpiintriangles.tikz}}
\endpgfgraphicnamed&(TR6) & %
\beginpgfgraphicnamed{diagrams//2triangleup}
\InputIfFileExists{diagrams//2triangleup.tikz}{}{\input{./figures/diagrams//2triangleup.tikz}}
\endpgfgraphicnamed&(TR8)\\

\beginpgfgraphicnamed{diagrams//2triangledown}
\InputIfFileExists{diagrams//2triangledown.tikz}{}{\input{./figures/diagrams//2triangledown.tikz}}
\endpgfgraphicnamed&(TR9)&%
\beginpgfgraphicnamed{diagrams//2triangledeloopnopi}
\InputIfFileExists{diagrams//2triangledeloopnopi.tikz}{}{\input{./figures/diagrams//2triangledeloopnopi.tikz}}
\endpgfgraphicnamed &(TR12) \\
  		
  		%\tikzfig{diagrams//2triangledown}&(TR9)    & \tikzfig{diagrams//tr10prime}&(TR10')\\
  		
  		%\tikzfig{diagrams//2triangledeloopnopi} &(TR12) &\tikzfig{diagrams//alphacopyw} &(TR13)\\
  		%
\beginpgfgraphicnamed{diagrams//TR1314combine}
\InputIfFileExists{diagrams//TR1314combine.tikz}{}{\input{./figures/diagrams//TR1314combine.tikz}}
\endpgfgraphicnamed &(TR13')&%
\beginpgfgraphicnamed{diagrams//plusnew}
\InputIfFileExists{diagrams//plusnew.tikz}{}{\input{./figures/diagrams//plusnew.tikz}}
\endpgfgraphicnamed&(AD')\\
  		%\tikzfig{diagrams//alphacopyw}&(TR13) &\tikzfig{diagrams//lambdacopyw} &(TR14)\\ 
  		%\tikzfig{diagrams//alphacopyw}&(TR13) &&\\ 
  		%\tikzfig{diagrams//plusnew}&(AD')&\tikzfig{diagrams//l1simplify} &(L1)\\
		
		%
\beginpgfgraphicnamed{diagrams//sqr1is1}
\InputIfFileExists{diagrams//sqr1is1.tikz}{}{\input{./figures/diagrams//sqr1is1.tikz}}
\endpgfgraphicnamed&(L3) &%
\beginpgfgraphicnamed{diagrams//lambdatimes}
\InputIfFileExists{diagrams//lambdatimes.tikz}{}{\input{./figures/diagrams//lambdatimes.tikz}}
\endpgfgraphicnamed&(L4)\\

  		%\tikzfig{diagrams//gpiintriangles}&(TR6)&\tikzfig{diagrams//trianglehopf}&(TR7) \\
  		
  	%	 \tikzfig{diagrams//2triangleup}&(TR8) &\tikzfig{diagrams//2triangledown}&(TR9)  \\
  		
  		  %& \tikzfig{diagrams//tr10prime}&(TR10')\\
  		%\tikzfig{diagrams//2triangledeloopnopi} &(TR12) &\tikzfig{diagrams//alphacopyw} &(TR13)\\
  	%	\tikzfig{diagrams//2triangledeloopnopi} &(TR12) &\tikzfig{diagrams//TR1314combine} &(TR13')\\
  		%\tikzfig{diagrams//alphacopyw}&(TR13) &\tikzfig{diagrams//lambdacopyw} &(TR14)\\ 
  		%\tikzfig{diagrams//alphacopyw}&(TR13) &&\\ 
  	%	\tikzfig{diagrams//plusnew2}&(AD'')&\tikzfig{diagrams//l1simplify} &(L1)\\
		
	%	\tikzfig{diagrams//sqr1is1}&(L3) &\tikzfig{diagrams//lambdatimes}&(L4)\\
  		\hline
  		\end{array}\]
  	\end{center}
  	
  	\caption{ $ZX_{ C+T}$-calculus rules with triangle and $\lambda$ box, where $0\leqslant \lambda, \lambda_1,  \lambda_2 \in \mathbb Z[\frac{1}{2}], \alpha \in \{\frac{k\pi}{4}| k=0, 1, \cdots, 7\}, \alpha\equiv \beta \equiv \gamma ~(mod~ \pi)$ in (AD$^{\prime}$). The upside-down version of these rules still hold.}\label{figure2t}
  \end{figure}

%%%%%%%%%%%%%%%%%%%%%%%%%%%%%%%%%%%%%%%%%%%%%%%%%  
  \iffalse
  \begin{figure}[!h]
\begin{center}
\[
\quad \qquad\begin{array}{|cccc|}
\hline
%\tikzfig{diagrams//emptyrulecplust} &(IV) &\tikzfig{diagrams//lambbranch} &(L1)\\
%\tikzfig{diagrams//newemptyrl} &(IV') &\tikzfig{diagrams//lambbranch} &(L1)\\
%
\beginpgfgraphicnamed{diagrams//plusnew2}
\InputIfFileExists{diagrams//plusnew2.tikz}{}{\input{./figures/diagrams//plusnew2.tikz}}
\endpgfgraphicnamed&(AD'')&%
\beginpgfgraphicnamed{diagrams//l1simplify}
\InputIfFileExists{diagrams//l1simplify.tikz}{}{\input{./figures/diagrams//l1simplify.tikz}}
\endpgfgraphicnamed &(L1)\\
%\tikzfig{diagrams//plus}&(AD) &\tikzfig{diagrams//lambdadelete} &(L2)\\ 
%\tikzfig{diagrams//plusnew}&(AD) &\tikzfig{diagrams//lambdadelete} &(L2)\\ 

%
\beginpgfgraphicnamed{diagrams//sqr1is1}
\InputIfFileExists{diagrams//sqr1is1.tikz}{}{\input{./figures/diagrams//sqr1is1.tikz}}
\endpgfgraphicnamed&(L3) &%
\beginpgfgraphicnamed{diagrams//lambdatimes}
\InputIfFileExists{diagrams//lambdatimes.tikz}{}{\input{./figures/diagrams//lambdatimes.tikz}}
\endpgfgraphicnamed&(L4)\\

%\tikzfig{diagrams//lambdaalpha}&(L5) &\tikzfig{diagrams//lambdacopyw} &(TR14)\\
%\tikzfig{diagrams//lambdaalpha}&(L5) &&\\
\hline
\end{array}\]
\end{center}
  \caption{Extended $\frac{\pi}{4}$-fragment ZX-calculus rules for $\lambda$ and addition, where $0\leqslant \lambda, \lambda_1,  \lambda_2 \in \mathbb Z[\frac{1}{2}], \alpha, \beta, \gamma \in \{\frac{k\pi}{4}| k=0, 1, \cdots, 7\}$.}\label{figure0t}  
  \end{figure}
\fi

%%%%%%%%%%%%%%%%%%%%%%%%%%%%%%%%%%%%%%%%%%%%%%%%%%%%

\FloatBarrier

%Note that the upside-down versions of all the above listed rules still hold. 

%\begin{lemma}
%We have the following property: the addition rule can be simplified and the rule (TR4') can be derived.
%\begin{equation}
%\tikzfig{diagrams//TR4primederive}
%\end{equation}
%\end{lemma}

In comparison to the rules of the $ZX_{ full}$-calculus, except for the restrictions described above, we made the following modifications in the
 $ZX_{ C+T}$-calculus: 
 \begin{itemize}
\item The empty rule (IV2) for the $ZX_{ full}$-calculus shown in Figure \ref{figure2sfy} is changed to the form of rule (IV$^{\prime}$) in Figure \ref{figure1t}. 
\item The rule (TR5$^{\prime}$) is added in Figure \ref{figure1t}. 
\item The condition $\lambda e^{i\gamma} =\lambda_1 e^{i\beta}+ \lambda_2 e^{i\alpha}$ for the rule (AD$^{\prime}$) of the $ZX_{ full}$-calculus shown in Figure \ref{figure2sfy} has been changed to an equivalence condition 
$\alpha\equiv \beta \equiv \gamma ~(mod~ \pi)$  for the rule (AD$^{\prime}$) of the  $ZX_{ C+T}$-calculus shown in Figure \ref{figure2t}.
 \end{itemize} 

Next we explain theses modifications are needed. Firstly, the complex number $1- \frac{1}{\sqrt{2}}$ in the $\lambda$ box of the rule (IV2) does not exist in the $ZX_{ C+T}$-calculus, so a modified empty rule is required
 for the $ZX_{ C+T}$-calculus satisfying that the translation of the Hadamard gate to a ZW diagram is reversible. Therefore we have the rule (IV$^{\prime}$)  and the following useful property. 

%\tikzfig{diagrams//newemptyrl}

\begin{lemma}
The frequently used empty rule can be derived from the $ZX_{\frac{\pi}{4}}$-calculus:
\begin{equation}\label{emptypiby4sim}
\beginpgfgraphicnamed{diagrams//emptyoften}
\InputIfFileExists{diagrams//emptyoften.tikz}{}{\input{./figures/diagrams//emptyoften.tikz}}
\endpgfgraphicnamed
\end{equation}
\end{lemma}

\begin{proof}
\begin{equation*}
\beginpgfgraphicnamed{diagrams//newemptytoold}
\InputIfFileExists{diagrams//newemptytoold.tikz}{}{\input{./figures/diagrams//newemptytoold.tikz}}
\endpgfgraphicnamed
\end{equation*}
\end{proof}

Secondly,  the rule (TR5$^{\prime}$) is introduced to make the $\frac{1}{2}$-box expressible in the $ZX_{ C+T}$-calculus:
\begin{lemma}\label{halfboxlm}
\begin{equation}\label{halfbox}
\beginpgfgraphicnamed{diagrams//lambda1by2}
\InputIfFileExists{diagrams//lambda1by2.tikz}{}{\input{./figures/diagrams//lambda1by2.tikz}}
\endpgfgraphicnamed
 \end{equation}
 \end{lemma}
\begin{proof}
First, it is clear that 
\begin{equation*}
\beginpgfgraphicnamed{diagrams//boxof2}
\InputIfFileExists{diagrams//boxof2.tikz}{}{\input{./figures/diagrams//boxof2.tikz}}
\endpgfgraphicnamed 
\end{equation*}
Then
\begin{equation*}
\beginpgfgraphicnamed{diagrams//boxof2timehalf}
\InputIfFileExists{diagrams//boxof2timehalf.tikz}{}{\input{./figures/diagrams//boxof2timehalf.tikz}}
\endpgfgraphicnamed 
\end{equation*}
\end{proof}

Finally, it is easy to check that the condition $\lambda e^{i\gamma} =\lambda_1 e^{i\beta}+ \lambda_2 e^{i\alpha}$ is equivalent to the condition $\alpha\equiv \beta \equiv \gamma ~(mod~ \pi)$, when 
$0\leqslant \lambda, \lambda_1,  \lambda_2 \in \mathbb Z[\frac{1}{2}], \alpha, \beta, \gamma \in \{\frac{k\pi}{4}| k=0, 1, \cdots, 7\}$.

Following the soundness of the $ZX_{ full}$-calculus,  it suffice to check that the rules (IV$^{\prime}$) and (TR5$^{\prime}$) are sound in order to have the soundness of $ZX_{ C+T}$-calculus. This is just a routine verification,  we omit the details here. 

Also we mention that the rules (TR10$^{\prime}$), (TR5) and (L5) proved in Chapter \ref{chfull} for the $ZX_{ full}$-calculus still hold for the  $ZX_{ C+T}$-calculus, since each rule applied in those proofs  resides in the  $ZX_{ C+T}$-calculus as well. 

%As the case of the $ZX_{ full}$-calculus, the new generators can be represented by traditional generators of the ZX-calculus.
In the previous chapter, the lambda box has been described in terms of triangle, green nods and red nodes with angles in  $[0,~2\pi)$. While for the $ZX_{ C+T}$-calculus, we have  

\begin{lemma}\label{lem:lamb_tri_decomposition2}
%The triangle \tikzfig{diagrams//triangle} and the
The lambda box %
\beginpgfgraphicnamed{diagrams//lambdabox}
\InputIfFileExists{diagrams//lambdabox.tikz}{}{\input{./figures/diagrams//lambdabox.tikz}}
\endpgfgraphicnamed  is expressible  in terms of triangle, green nods and red nodes with angles in $\{\frac{k\pi}{4}| k=0, 1, \cdots, 7\}$.
% in the $ZX_{\frac{\pi}{4}}$-calculus.
\end{lemma}

\begin{proof}
%The triangle \tikzfig{diagrams//triangle} has been represented in the $ZX_{\frac{\pi}{4}}$-calculus in \cite{Coeckebk,Emmanuel}, we give these two decompositions as follows:
%\begin{equation}\label{triangleslash}
 %\tikzfig{diagrams//triangledecompose}, \quad \quad \quad \quad \tikzfig{diagrams//triangledecompose2nd} 
%\end{equation}

 %Now we deal with the lambda box. 
First we can write $\lambda$ as a sum of its integer part and remainder part: $\lambda= [\lambda] +\{\lambda\}$, where $ [\lambda]$ is a non-negative integer and $0\leq\{\lambda\}<1$. Since  $\lambda \in \mathbb Z[\frac{1}{2}]$,
$\{\lambda\}$ can be uniquely written as a binary expansion of the form $a_1\frac{1}{2}+\cdots+a_s\frac{1}{2^s}$, where $a_i\in \{ 0, 1\} , i=1, \cdots, s.$  For the integer part $[\lambda]$, the corresponding $\lambda$ box 
has been represented in terms of triangle, green nods and red nodes with angles multiples of $\frac{k\pi}{4}$ as shown in Lemma \ref{lem:lamb_tri_decomposition}. For the remainder part $\{\lambda\}$,   it is sufficient to express the $\lambda$ box for
$\lambda=\frac{1}{2^k}$ in terms of triangle and $Z, X$ phases for any positive integer $k$, since one can apply the addition rule (AD$^{\prime}$) recursively.  In fact, we have

\begin{equation}\label{halfnbox}
\beginpgfgraphicnamed{diagrams//lambda1by2k}
\InputIfFileExists{diagrams//lambda1by2k.tikz}{}{\input{./figures/diagrams//lambda1by2k.tikz}}
\endpgfgraphicnamed 
\end{equation}

 We prove this by induction on $k$. When $k=1$, it is just the identity (\ref{halfbox}) proved in Lemma \ref{halfboxlm}. Suppose (\ref{halfnbox}) holds for $k$. Then for $k+1$ we have 
 \begin{equation*}
\beginpgfgraphicnamed{diagrams//halfnboxproof}
\InputIfFileExists{diagrams//halfnboxproof.tikz}{}{\input{./figures/diagrams//halfnboxproof.tikz}}
\endpgfgraphicnamed 
\end{equation*}
  This completes the induction.
  Therefore, 
    $$%
\beginpgfgraphicnamed{diagrams//lexpress4new}
\InputIfFileExists{diagrams//lexpress4new.tikz}{}{\input{./figures/diagrams//lexpress4new.tikz}}
\endpgfgraphicnamed.$$

\end{proof}

\section{Translations between the $ZX_{ C+T}$-calculus and the  $ZW_{\mathbb{T}}$-calculus}\label{zwtozx2}
%The $ZW_T$-calculus is a restriction of the $ZW_{ \mathbb{C}}$-calculus,  thus remains the same as that given in Section \ref{zwcalcul} of Chapter \ref{chfull}, except that now $r$ in the generator \tikzfig{diagrams//rgatewhite}  lies in  the ring $\mathbb{Z}[i, \frac{1}{\sqrt{2}}]$.

 The interpretation $\llbracket \cdot \rrbracket_{XW}$   from the $ZX_{ C+T}$-calculus to  the  $ZW_{\mathbb{T}}$-calculus is just a restriction of the interpretation $\llbracket \cdot \rrbracket_{XW}$  from the $ZX_{ full}$-calculus to  the  $ZW_{\mathbb{C}}$-calculus:
 %  for Clifford +T QM remains the same as that for the entire qubit QM:
\[
 \left\llbracket%
\beginpgfgraphicnamed{diagrams//emptysquare}
\InputIfFileExists{diagrams//emptysquare.tikz}{}{\input{./figures/diagrams//emptysquare.tikz}}
\endpgfgraphicnamed\right\rrbracket_{XW}=  %
\beginpgfgraphicnamed{diagrams//emptysquare}
\InputIfFileExists{diagrams//emptysquare.tikz}{}{\input{./figures/diagrams//emptysquare.tikz}}
\endpgfgraphicnamed,  \quad
  \left\llbracket%
\beginpgfgraphicnamed{diagrams//Id}
\InputIfFileExists{diagrams//Id.tikz}{}{\input{./figures/diagrams//Id.tikz}}
\endpgfgraphicnamed\right\rrbracket_{XW}=  %
\beginpgfgraphicnamed{diagrams//Id}
\InputIfFileExists{diagrams//Id.tikz}{}{\input{./figures/diagrams//Id.tikz}}
\endpgfgraphicnamed,   \quad
 \left\llbracket%
\beginpgfgraphicnamed{diagrams//cap}
\InputIfFileExists{diagrams//cap.tikz}{}{\input{./figures/diagrams//cap.tikz}}
\endpgfgraphicnamed\right\rrbracket_{XW}=  %
\beginpgfgraphicnamed{diagrams//cap}
\InputIfFileExists{diagrams//cap.tikz}{}{\input{./figures/diagrams//cap.tikz}}
\endpgfgraphicnamed,   \quad
  \left\llbracket%
\beginpgfgraphicnamed{diagrams//cup}
\InputIfFileExists{diagrams//cup.tikz}{}{\input{./figures/diagrams//cup.tikz}}
\endpgfgraphicnamed\right\rrbracket_{XW}=  %
\beginpgfgraphicnamed{diagrams//cup}
\InputIfFileExists{diagrams//cup.tikz}{}{\input{./figures/diagrams//cup.tikz}}
\endpgfgraphicnamed,  
  \]
  
  \[
   \left\llbracket%
\beginpgfgraphicnamed{diagrams//swap}
\InputIfFileExists{diagrams//swap.tikz}{}{\input{./figures/diagrams//swap.tikz}}
\endpgfgraphicnamed\right\rrbracket_{XW}=  %
\beginpgfgraphicnamed{diagrams//swap}
\InputIfFileExists{diagrams//swap.tikz}{}{\input{./figures/diagrams//swap.tikz}}
\endpgfgraphicnamed,   \quad
   \left\llbracket%
\beginpgfgraphicnamed{diagrams//generator_spider-nonum}
\InputIfFileExists{diagrams//generator_spider-nonum.tikz}{}{\input{./figures/diagrams//generator_spider-nonum.tikz}}
\endpgfgraphicnamed\right\rrbracket_{XW}=  %
\beginpgfgraphicnamed{diagrams//spiderwhite}
\InputIfFileExists{diagrams//spiderwhite.tikz}{}{\input{./figures/diagrams//spiderwhite.tikz}}
\endpgfgraphicnamed,   \quad 
     \left\llbracket%
\beginpgfgraphicnamed{diagrams//alphagate}
\InputIfFileExists{diagrams//alphagate.tikz}{}{\input{./figures/diagrams//alphagate.tikz}}
\endpgfgraphicnamed\right\rrbracket_{XW}=  %
\beginpgfgraphicnamed{diagrams//alphagatewhite}
\InputIfFileExists{diagrams//alphagatewhite.tikz}{}{\input{./figures/diagrams//alphagatewhite.tikz}}
\endpgfgraphicnamed,   \quad 
       \left\llbracket%
\beginpgfgraphicnamed{diagrams//lambdabox}
\InputIfFileExists{diagrams//lambdabox.tikz}{}{\input{./figures/diagrams//lambdabox.tikz}}
\endpgfgraphicnamed\right\rrbracket_{XW}=  %
\beginpgfgraphicnamed{diagrams//lambdagatewhiteld}
\InputIfFileExists{diagrams//lambdagatewhiteld.tikz}{}{\input{./figures/diagrams//lambdagatewhiteld.tikz}}
\endpgfgraphicnamed,   
         \]

 \[
  \left\llbracket%
\beginpgfgraphicnamed{diagrams//HadaDecomSingleslt}
\InputIfFileExists{diagrams//HadaDecomSingleslt.tikz}{}{\input{./figures/diagrams//HadaDecomSingleslt.tikz}}
\endpgfgraphicnamed\right\rrbracket_{XW}=  %
\beginpgfgraphicnamed{diagrams//Hadamardwhite}
\InputIfFileExists{diagrams//Hadamardwhite.tikz}{}{\input{./figures/diagrams//Hadamardwhite.tikz}}
\endpgfgraphicnamed,   \quad
    \left\llbracket%
\beginpgfgraphicnamed{diagrams//triangle}
\InputIfFileExists{diagrams//triangle.tikz}{}{\input{./figures/diagrams//triangle.tikz}}
\endpgfgraphicnamed\right\rrbracket_{XW}=  %
\beginpgfgraphicnamed{diagrams//trianglewhite}
\InputIfFileExists{diagrams//trianglewhite.tikz}{}{\input{./figures/diagrams//trianglewhite.tikz}}
\endpgfgraphicnamed, 
     \left\llbracket%
\beginpgfgraphicnamed{diagrams//alphagatered}
\InputIfFileExists{diagrams//alphagatered.tikz}{}{\input{./figures/diagrams//alphagatered.tikz}}
\endpgfgraphicnamed\right\rrbracket_{XW}= \left\llbracket%
\beginpgfgraphicnamed{diagrams//HadaDecomSingleslt}
\InputIfFileExists{diagrams//HadaDecomSingleslt.tikz}{}{\input{./figures/diagrams//HadaDecomSingleslt.tikz}}
\endpgfgraphicnamed\right\rrbracket_{XW}\circ \left(%
\beginpgfgraphicnamed{diagrams//alphagatewhite}
\InputIfFileExists{diagrams//alphagatewhite.tikz}{}{\input{./figures/diagrams//alphagatewhite.tikz}}
\endpgfgraphicnamed\right) \circ \left\llbracket%
\beginpgfgraphicnamed{diagrams//HadaDecomSingleslt}
\InputIfFileExists{diagrams//HadaDecomSingleslt.tikz}{}{\input{./figures/diagrams//HadaDecomSingleslt.tikz}}
\endpgfgraphicnamed\right\rrbracket_{XW},   \]
    \[
       \left\llbracket%
\beginpgfgraphicnamed{diagrams//generator_redspider2cged}
\InputIfFileExists{diagrams//generator_redspider2cged.tikz}{}{\input{./figures/diagrams//generator_redspider2cged.tikz}}
\endpgfgraphicnamed\right\rrbracket_{XW}= \left\llbracket \left( %
\beginpgfgraphicnamed{diagrams//HadaDecomSingleslt}
\InputIfFileExists{diagrams//HadaDecomSingleslt.tikz}{}{\input{./figures/diagrams//HadaDecomSingleslt.tikz}}
\endpgfgraphicnamed\right)^{\otimes m} \right\rrbracket_{XW}\circ \left(%
\beginpgfgraphicnamed{diagrams//spiderwhiteindex}
\InputIfFileExists{diagrams//spiderwhiteindex.tikz}{}{\input{./figures/diagrams//spiderwhiteindex.tikz}}
\endpgfgraphicnamed\right) \circ \left\llbracket \left( %
\beginpgfgraphicnamed{diagrams//HadaDecomSingleslt}
\InputIfFileExists{diagrams//HadaDecomSingleslt.tikz}{}{\input{./figures/diagrams//HadaDecomSingleslt.tikz}}
\endpgfgraphicnamed\right)^{\otimes n} \right\rrbracket_{XW},  \]
       \[
      \llbracket D_1\otimes D_2  \rrbracket_{XW} =  \llbracket D_1  \rrbracket_{XW} \otimes  \llbracket  D_2  \rrbracket_{XW}, \quad 
 \llbracket D_1\circ D_2  \rrbracket_{XW} =  \llbracket D_1  \rrbracket_{XW} \circ  \llbracket  D_2  \rrbracket_{XW},
 \]
where $0\leqslant \lambda \in \mathbb Z[\frac{1}{2}], \alpha \in \{\frac{k\pi}{4}| k=0, 1, \cdots, 7\}$.

Since  $\mathbb{T} =  \mathbb{Z}[\frac{1}{2}, e^{i\frac{\pi}{4}}]=\mathbb{Z}[i, \frac{1}{\sqrt{2}}] $, we have  $\frac{\sqrt{2}-2}{2} \in \mathbb{T} $.  Then it is clear that this restricted translation will always result in a well-defined $ZW_{\mathbb{T}}$ diagram.

By Lemma \ref{xtowpreservesemantics} the interpretation $\llbracket \cdot \rrbracket_{XW}$ preserves the standard interpretation.
%\begin{lemma}\label{xtowpreservesemantics2}
%Suppose $D$ is an arbitrary diagram in the $ZX_{ C+T}$-calculus. Then  $\llbracket \llbracket D \rrbracket_{XW}\rrbracket = \llbracket D \rrbracket$.
%\end{lemma}
%The proof is easy.

%%%%%%%%%%%%%%%%%%%%%%%%%%%%%%%%%%%%%%%%%%%%%%%%%%%%%%%%

On the other hand, the interpretation $\llbracket \cdot \rrbracket_{WX}$  from the  $ZW_{\mathbb{T}}$-calculus to the $ZX_{ C+T}$-calculus  is the same as the interpretation $\llbracket \cdot \rrbracket_{WX}$  from the $ZX_{ full}$-calculus to the  $ZW_{\mathbb{C}}$-calculus except for the $r$-phase part:

\[
 \left\llbracket%
\beginpgfgraphicnamed{diagrams//emptysquare}
\InputIfFileExists{diagrams//emptysquare.tikz}{}{\input{./figures/diagrams//emptysquare.tikz}}
\endpgfgraphicnamed\right\rrbracket_{WX}=  %
\beginpgfgraphicnamed{diagrams//emptysquare}
\InputIfFileExists{diagrams//emptysquare.tikz}{}{\input{./figures/diagrams//emptysquare.tikz}}
\endpgfgraphicnamed,  \quad
  \left\llbracket%
\beginpgfgraphicnamed{diagrams//Id}
\InputIfFileExists{diagrams//Id.tikz}{}{\input{./figures/diagrams//Id.tikz}}
\endpgfgraphicnamed\right\rrbracket_{WX}=  %
\beginpgfgraphicnamed{diagrams//Id}
\InputIfFileExists{diagrams//Id.tikz}{}{\input{./figures/diagrams//Id.tikz}}
\endpgfgraphicnamed,   \quad
 \left\llbracket%
\beginpgfgraphicnamed{diagrams//cap}
\InputIfFileExists{diagrams//cap.tikz}{}{\input{./figures/diagrams//cap.tikz}}
\endpgfgraphicnamed\right\rrbracket_{WX}=  %
\beginpgfgraphicnamed{diagrams//cap}
\InputIfFileExists{diagrams//cap.tikz}{}{\input{./figures/diagrams//cap.tikz}}
\endpgfgraphicnamed,   \quad
  \left\llbracket%
\beginpgfgraphicnamed{diagrams//cup}
\InputIfFileExists{diagrams//cup.tikz}{}{\input{./figures/diagrams//cup.tikz}}
\endpgfgraphicnamed\right\rrbracket_{WX}=  %
\beginpgfgraphicnamed{diagrams//cup}
\InputIfFileExists{diagrams//cup.tikz}{}{\input{./figures/diagrams//cup.tikz}}
\endpgfgraphicnamed,  
  \]
  
  \[
   \left\llbracket%
\beginpgfgraphicnamed{diagrams//swap}
\InputIfFileExists{diagrams//swap.tikz}{}{\input{./figures/diagrams//swap.tikz}}
\endpgfgraphicnamed\right\rrbracket_{WX}=  %
\beginpgfgraphicnamed{diagrams//swap}
\InputIfFileExists{diagrams//swap.tikz}{}{\input{./figures/diagrams//swap.tikz}}
\endpgfgraphicnamed,   \quad
   \left\llbracket%
\beginpgfgraphicnamed{diagrams//spiderwhite}
\InputIfFileExists{diagrams//spiderwhite.tikz}{}{\input{./figures/diagrams//spiderwhite.tikz}}
\endpgfgraphicnamed\right\rrbracket_{WX}=  %
\beginpgfgraphicnamed{diagrams//generator_spider-nonum}
\InputIfFileExists{diagrams//generator_spider-nonum.tikz}{}{\input{./figures/diagrams//generator_spider-nonum.tikz}}
\endpgfgraphicnamed,   \quad 
  %   \left\llbracket\tikzfig{diagrams//rgatewhite}\right\rrbracket_{WX}=  \tikzfig{diagrams//alphalambdagate},   \quad 
       \left\llbracket%
\beginpgfgraphicnamed{diagrams//piblack}
\InputIfFileExists{diagrams//piblack.tikz}{}{\input{./figures/diagrams//piblack.tikz}}
\endpgfgraphicnamed\right\rrbracket_{WX}=  %
\beginpgfgraphicnamed{diagrams//pired}
\InputIfFileExists{diagrams//pired.tikz}{}{\input{./figures/diagrams//pired.tikz}}
\endpgfgraphicnamed,   
         \]

 \[
  \left\llbracket%
\beginpgfgraphicnamed{diagrams//corsszw}
\InputIfFileExists{diagrams//corsszw.tikz}{}{\input{./figures/diagrams//corsszw.tikz}}
\endpgfgraphicnamed\right\rrbracket_{WX}=  %
\beginpgfgraphicnamed{diagrams//crossxz}
\InputIfFileExists{diagrams//crossxz.tikz}{}{\input{./figures/diagrams//crossxz.tikz}}
\endpgfgraphicnamed,   \quad \quad
    \left\llbracket%
\beginpgfgraphicnamed{diagrams//wblack}
\InputIfFileExists{diagrams//wblack.tikz}{}{\input{./figures/diagrams//wblack.tikz}}
\endpgfgraphicnamed\right\rrbracket_{WX}=  %
\beginpgfgraphicnamed{diagrams//winzx}
\InputIfFileExists{diagrams//winzx.tikz}{}{\input{./figures/diagrams//winzx.tikz}}
\endpgfgraphicnamed, \]
    
     \[
 \left\llbracket%
\beginpgfgraphicnamed{diagrams//rgatewhite}
\InputIfFileExists{diagrams//rgatewhite.tikz}{}{\input{./figures/diagrams//rgatewhite.tikz}}
\endpgfgraphicnamed\right\rrbracket_{WX}=%
\beginpgfgraphicnamed{diagrams//rwphasenew2}
\InputIfFileExists{diagrams//rwphasenew2.tikz}{}{\input{./figures/diagrams//rwphasenew2.tikz}}
\endpgfgraphicnamed, 
  % \tikzfig{diagrams//rgateclt},   
 \]

    \[ \llbracket D_1\otimes D_2  \rrbracket_{WX} =  \llbracket D_1  \rrbracket_{WX} \otimes  \llbracket  D_2  \rrbracket_{WX}, \quad 
 \llbracket D_1\circ D_2  \rrbracket_{WX} =  \llbracket D_1  \rrbracket_{WX} \circ  \llbracket  D_2  \rrbracket_{WX},
 \]
%where $r=a_0+a_1e^{i\frac{\pi}{4}}+a_2e^{i\frac{2\pi}{4}}+a_3e^{i\frac{3\pi}{4}},~   a_j \in \mathbb Z[\frac{1}{2}], ~~ j=0, 1, 2, 3$. 
where $r=a_0e^{i\alpha_0}+a_1e^{i\alpha_1}+a_2e^{i\alpha_2}+a_3e^{i\alpha_3},~   0\leq a_j \in \mathbb Z[\frac{1}{2}], ~\alpha_j=j\frac{\pi}{4} ~or~ j\frac{\pi}{4}+\pi,~~ j=0, 1, 2, 3$. 
Note that the representation of $a_j$ box is described in Lemma \ref{lem:lamb_tri_decomposition2}.

 It seems that the interpretation of the  $r$-phase is somehow complicated. However, the corresponding interpreted ZX diagram is actually the result of three applications of the addition rule (AD$^{\prime}$) according to the identity $r=a_0e^{i\alpha_0}+a_1e^{i\alpha_1}+a_2e^{i\alpha_2}+a_3e^{i\alpha_3}$, which clearly has the same standard interpretation as that of the $r$-phase if we take this as happened in the $ZX_{ full}$-calculus. Then by Lemma \ref{wtoxpreservesemantics}, the interpretation
   $\llbracket \cdot \rrbracket_{WX}$ from the  $ZW_{\mathbb{T}}$-calculus to the  $ZX_{ C+T}$-calculus preserves the standard interpretation.
%\begin{lemma}\label{wtoxpreservesemantics2}
%Suppose $D$ is an arbitrary diagram in the $ZW_{\mathbb{T}}$-calculus. Then  $\llbracket \llbracket D \rrbracket_{WX}\rrbracket = \llbracket D \rrbracket$.
%\end{lemma}
%The proof is easy.

Next we will prove that the effect of successive translations from ZX to ZW and then back to ZW is the same as no translations. 

\begin{lemma}
 \begin{equation}\label{greenpitriangl}
\beginpgfgraphicnamed{diagrams//greenpitriangle}
\InputIfFileExists{diagrams//greenpitriangle.tikz}{}{\input{./figures/diagrams//greenpitriangle.tikz}}
\endpgfgraphicnamed
\end{equation}
\end{lemma}

\begin{proof}
 \begin{equation*}
\beginpgfgraphicnamed{diagrams//greenpitriangleproof}
\InputIfFileExists{diagrams//greenpitriangleproof.tikz}{}{\input{./figures/diagrams//greenpitriangleproof.tikz}}
\endpgfgraphicnamed
 \end{equation*}
\end{proof}

\begin{lemma}
 \begin{equation}\label{redpilefttriangl}
\beginpgfgraphicnamed{diagrams//redpilefttriangle}
\InputIfFileExists{diagrams//redpilefttriangle.tikz}{}{\input{./figures/diagrams//redpilefttriangle.tikz}}
\endpgfgraphicnamed
\end{equation}
\end{lemma}
\begin{proof}
 \begin{equation*}
\beginpgfgraphicnamed{diagrams//redpilefttriangleprf}
\InputIfFileExists{diagrams//redpilefttriangleprf.tikz}{}{\input{./figures/diagrams//redpilefttriangleprf.tikz}}
\endpgfgraphicnamed
 \end{equation*}
\end{proof}

\begin{lemma}
 \begin{equation}\label{emptyvariant}
\beginpgfgraphicnamed{diagrams//emptyvariants}
\InputIfFileExists{diagrams//emptyvariants.tikz}{}{\input{./figures/diagrams//emptyvariants.tikz}}
\endpgfgraphicnamed
\end{equation}
\end{lemma}
\begin{proof}
 \begin{equation*}
\beginpgfgraphicnamed{diagrams//emptyvariantsprf}
\InputIfFileExists{diagrams//emptyvariantsprf.tikz}{}{\input{./figures/diagrams//emptyvariantsprf.tikz}}
\endpgfgraphicnamed
 \end{equation*}
\end{proof}

\begin{lemma}\label{interpretationreversible2}
Suppose $D$ is an arbitrary diagram in the $ZX_{ C+T}$-calculus. Then  $ZX_{ C+T}\vdash \llbracket \llbracket D \rrbracket_{XW}\rrbracket_{WX} =D$.
\end{lemma}

\begin{proof}
By the construction of  $\llbracket \cdot  \rrbracket_{XW}$  and $\llbracket \cdot  \rrbracket_{WX}$, we only need to prove for the generators of the $ZX_{ C+T}$-calculus.  Here we consider  all the generators translated at the beginning of this section.
  The first six generators  are the same as the  first six generators in  the $ZW_{ \mathbb{T}}$-calculus, so we just need to check for the last six generators. Since the red phase gate and the red spider  are translated in terms of the translation of Hadamard gate, green phase gate and green spider, we only need to care for the four generators: Hadamard gate, green phase gate, $\lambda$ box and the triangle. 

%The first six generators in the $ZX_{ C+T}$-calculus are the same as the first six generators in the $ZW_{\mathbb{T}}$-calculus, so we just need to check for the last four generators in the $ZX_{ C+T}$-calculus, i.e., the green phase gate, the Hadamard gate, the $\lambda$ box and the triangle.

For the green phase gate,  if we write $e^{i\alpha}$ in the form $r=a_0+a_1e^{i\frac{\pi}{4}}+a_2e^{i\frac{2\pi}{4}}+a_3e^{i\frac{3\pi}{4}}$, there is exactly one $a_i$ which is non-zero. By commutative property of addition \ref{sumcommutativity},  we  have 
\[
 \left\llbracket \left\llbracket%
\beginpgfgraphicnamed{diagrams//alphagate}
\InputIfFileExists{diagrams//alphagate.tikz}{}{\input{./figures/diagrams//alphagate.tikz}}
\endpgfgraphicnamed\right\rrbracket_{XW}\right\rrbracket_{WX}= \left\llbracket%
\beginpgfgraphicnamed{diagrams//alphagatewhite}
\InputIfFileExists{diagrams//alphagatewhite.tikz}{}{\input{./figures/diagrams//alphagatewhite.tikz}}
\endpgfgraphicnamed\right\rrbracket_{WX}\]
 \[
\beginpgfgraphicnamed{diagrams//alphadrvctnew}
\InputIfFileExists{diagrams//alphadrvctnew.tikz}{}{\input{./figures/diagrams//alphadrvctnew.tikz}}
\endpgfgraphicnamed
% \tikzfig{diagrams//alphagate},
\]
For the Hadamard gate,  we write  $\frac{\sqrt{2}-2}{2}=-1+\frac{1}{2}e^{i\frac{\pi}{4}}+0e^{i\frac{2\pi}{4}}-\frac{1}{2}e^{i\frac{3\pi}{4}}=-1+\frac{1}{2}e^{i\frac{\pi}{4}}+0e^{i\frac{2\pi}{4}}+\frac{1}{2}e^{i\frac{-\pi}{4}}$, then

\[
  \left\llbracket \left\llbracket%
\beginpgfgraphicnamed{diagrams//HadaDecomSingleslt}
\InputIfFileExists{diagrams//HadaDecomSingleslt.tikz}{}{\input{./figures/diagrams//HadaDecomSingleslt.tikz}}
\endpgfgraphicnamed\right\rrbracket_{XW}\right\rrbracket_{WX}=  
  \left\llbracket  %
\beginpgfgraphicnamed{diagrams//Hadamardwhite}
\InputIfFileExists{diagrams//Hadamardwhite.tikz}{}{\input{./figures/diagrams//Hadamardwhite.tikz}}
\endpgfgraphicnamed  \right\rrbracket_{WX} =    %
\beginpgfgraphicnamed{diagrams//Hadascalar}
\InputIfFileExists{diagrams//Hadascalar.tikz}{}{\input{./figures/diagrams//Hadascalar.tikz}}
\endpgfgraphicnamed \left\llbracket  %
\beginpgfgraphicnamed{diagrams//Hadamardwhitescalar}
\InputIfFileExists{diagrams//Hadamardwhitescalar.tikz}{}{\input{./figures/diagrams//Hadamardwhitescalar.tikz}}
\endpgfgraphicnamed  \right\rrbracket_{WX}\]
  \[
\beginpgfgraphicnamed{diagrams//Hadamardxwx-new}
\InputIfFileExists{diagrams//Hadamardxwx-new.tikz}{}{\input{./figures/diagrams//Hadamardxwx-new.tikz}}
\endpgfgraphicnamed 
\]

%Here we used  $\frac{\sqrt{2}-2}{2}=-1+\frac{1}{2}e^{i\frac{\pi}{4}}+0e^{i\frac{2\pi}{4}}-\frac{1}{2}e^{i\frac{3\pi}{4}}=-1+\frac{1}{2}e^{i\frac{\pi}{4}}+0e^{i\frac{2\pi}{4}}+\frac{1}{2}e^{i\frac{-\pi}{4}}.$

%\iffalse
Finally, it is easy to check that 
\[
 \left\llbracket \left\llbracket%
\beginpgfgraphicnamed{diagrams//lambdabox}
\InputIfFileExists{diagrams//lambdabox.tikz}{}{\input{./figures/diagrams//lambdabox.tikz}}
\endpgfgraphicnamed\right\rrbracket_{XW}\right\rrbracket_{WX}=%
\beginpgfgraphicnamed{diagrams//lambdabox}
\InputIfFileExists{diagrams//lambdabox.tikz}{}{\input{./figures/diagrams//lambdabox.tikz}}
\endpgfgraphicnamed,\quad 
 % \left\llbracket \left\llbracket\tikzfig{diagrams//HadaDecomSingleslt}\right\rrbracket_{XW}\right\rrbracket_{WX}=\tikzfig{diagrams//HadaDecomSingleslt},\quad 
   \left\llbracket \left\llbracket%
\beginpgfgraphicnamed{diagrams//triangle}
\InputIfFileExists{diagrams//triangle.tikz}{}{\input{./figures/diagrams//triangle.tikz}}
\endpgfgraphicnamed\right\rrbracket_{XW}\right\rrbracket_{WX}=%
\beginpgfgraphicnamed{diagrams//triangle}
\InputIfFileExists{diagrams//triangle.tikz}{}{\input{./figures/diagrams//triangle.tikz}}
\endpgfgraphicnamed.
\]
%\fi

\end{proof}

For the same reason as we have shown in Section \ref{transzxzw} of Chapter  \ref{chfull}, the translations $\llbracket \cdot  \rrbracket_{XW}$  and $\llbracket \cdot  \rrbracket_{WX}$ are mutually invertible to each other.

\section{Completeness}
\begin{proposition}\label{zwrulesholdinzx2}
If  $ZW_{\mathbb{T}}\vdash D_1=D_2$, then  $ZX_{ C+T}\vdash \llbracket D_1 \rrbracket_{WX} =\llbracket D_2 \rrbracket_{WX}$.

\end{proposition}

\begin{proof}
Since the derivation of equalities in ZW and ZX is made by rewriting rules,  we need only to prove that $ZX_{ C+T} \vdash \left\llbracket D_1\right\rrbracket_{WX} = \left\llbracket D_2\right\rrbracket_{WX}$ where  $D_1=D_2$ is a rewriting rule of the $ZW_{\mathbb{T}}$-calculus. Most proofs of this proposition have been done in the proof for completeness of the  $ZX_{ full}$-calculus in the previous chapter, we only need to check  for the last 5 rules $rng^{r,s}_{\times}$,  $rng^{r,s}_{+}$, $nat^{r}_{c}$, $nat^{r}_{\varepsilon c}$, $ph^{r}$,  which involve white phases in the $ZW_{\mathbb{T}}$-calculus. The rules $nat^{r}_{\varepsilon c}$ and $ph^{r}$ are easy to check, we just deal with the rules  $rng^{r,s}_{\times}$,  $rng^{r,s}_{+}$ and $nat^{r}_{c}$ at the end of this section, for the sake of easy reading.
\end{proof}

\begin{theorem}\label{maintheorem2}
The $ZX_{ C+T}$-calculus is complete for Clifford+T QM:
if $\llbracket D_1 \rrbracket =\llbracket D_2 \rrbracket$, then $ZX_{ C+T}\vdash D_1=D_2$.

\end{theorem}

\begin{proof}
Suppose $D_1,  D_2 \in ZX$ and  $\llbracket D_1 \rrbracket =\llbracket D_2 \rrbracket$. Then by lemma \ref{xtowpreservesemantics2},  $\llbracket \llbracket D_1 \rrbracket_{XW}\rrbracket = \llbracket D_1 \rrbracket= \llbracket D_2 \rrbracket=\llbracket \llbracket D_2 \rrbracket_{XW}\rrbracket $.  Thus by the completeness of the $ZW_{ \mathbb{T}}$-calculus \cite{amar},  $ZW_{ \mathbb{T}}\vdash \llbracket D_2 \rrbracket_{XW}=  \llbracket D_2 \rrbracket_{XW}$.  Now by proposition \ref{zwrulesholdinzx2},  $ZX_{ C+T}\vdash \llbracket \llbracket D_1 \rrbracket_{XW}\rrbracket_{WX} =\llbracket \llbracket D_2 \rrbracket_{XW}\rrbracket_{WX}$.
Finally, by lemma \ref{interpretationreversible2},  $ZX_{ C+T}\vdash D_1=D_2$.
\end{proof}

Now we can give a new proof for Proposition \ref{cliftmatix}.
\begin{proposition}\label{rings}
The  diagrams of the $ZX_{ C+T}$-calculus exactly corresponds to the matrices over the ring $ \mathbb{Z}[i, \frac{1}{\sqrt{2}}]$.
\end{proposition}
\begin{proof}
The triangle %
\beginpgfgraphicnamed{diagrams//triangle}
\InputIfFileExists{diagrams//triangle.tikz}{}{\input{./figures/diagrams//triangle.tikz}}
\endpgfgraphicnamed has been represented in the $ZX_{\frac{\pi}{4}}$-calculus in \cite{Coeckebk} as follows:
\begin{equation}\label{triangleslash2}
\beginpgfgraphicnamed{diagrams//triangledecompose}
\InputIfFileExists{diagrams//triangledecompose.tikz}{}{\input{./figures/diagrams//triangledecompose.tikz}}
\endpgfgraphicnamed%, \quad \quad \quad \quad \tikzfig{diagrams//triangledecompose2nd} 
\end{equation}
It can be directly verified that the two diagrams on both sides of (\ref{triangleslash2}) have the same standard interpretation. Thus by Theorem \ref{maintheorem2} the identity (\ref{triangleslash2}) can be derived in the $ZX_{ C+T}$-calculus.

Obviuosly each traditional generator of the  $ZX_{ C+T}$-calculus corresponds to a matrix over the ring $\mathbb{Z}[i, \frac{1}{\sqrt{2}}]$ via the standard interpretation. By lemma \ref{lem:lamb_tri_decomposition2}, the new generators can also be represented by traditional generators and the triangle (described in terms of green and red noeds in (\ref{triangleslash2})), thus correspond to matrices over $\mathbb{Z}[i, \frac{1}{\sqrt{2}}]$ as well. Therefore, each diagram of the $ZX_{ C+T}$-calculus which is composed of those generators must correspond to a matrix over the ring $\mathbb{Z}[i, \frac{1}{\sqrt{2}}]$.

% It is also clear that each generator of the $\frac{\pi}{4}$-fragment ZX-calculus corresponds to a matrix over the ring $\mathbb{Z}[i, \frac{1}{\sqrt{2}}]$, 
%thus each diagram of the $\frac{\pi}{4}$-fragment ZX-calculus must correspond to a matrix over the ring $\mathbb{Z}[i, \frac{1}{\sqrt{2}}]$. 
Conversely, since $\mathbb{Z}[i, \frac{1}{\sqrt{2}}]=\mathbb{Z}[\frac{1}{2}, e^{i\frac{\pi}{4}}]$, each matrix over the ring $\mathbb{T}$ can be first turned into a non-normalized quantum state by map-state duality, then represented by a normal form 
in the $ZW_{\mathbb{T}}$-calculus with phases belong to the ring $\mathbb{Z}[\frac{1}{2}, e^{i\frac{\pi}{4}}]$ \cite{amar}. Furthermore, by the interpretation $\llbracket \cdot \rrbracket_{WX}$ given in section \ref{zwtozx2}, this ZW normal form can be translated
to a diagram in the $ZX_{ C+T}$-calculus.
% hence can be represented by a diagram of the $\frac{\pi}{4}$-fragment ZX-calculus via the translation from ZW to ZX as described in  \cite{ngwang}.
\end{proof}

In \cite{GilesSelinger}, Giles and Selinger have established a correspondence between Clifford +T quantum circuits with some finite number of ancillas  and the ring $\mathbb{Z}[i, \frac{1}{\sqrt{2}}]$. Here we generalise this result to the case of the $ZX_{ C+T}$-calculus which is not restrained by unitarity. Our proof  here is more direct thus simpler than the one  presented in \cite{Emmanuel}.
%Also there is another proof for this proposition as presented in \cite{Emmanuel}, but our proof is more direct, thus simpler. 

\section*{Proof of proposition \ref{zwrulesholdinzx2}}
First we note that all the ZX diagrams translated by $\llbracket \cdot \rrbracket_{WX}$ won't be given in this proof for the sake of readability.  For simplicity, we will use ZW rules in the diagrammatic reasoning instead of listing all the ZX rules which are used to derive the ZW rules under the interpretation  $\llbracket \cdot \rrbracket_{WX}$. All the ZW rules used in this proof are $cut_w, cut_z, ba_{zw}, sym^x_w, ba_w$, where  $cut_w, cut_z, ba_{zw},  ba_w$ are just the spider forms of the rules  $asso_w, asso_z, , nat^m_c, nat^m_w$ respectively as presented in Figure \ref{figure3} and Figure \ref{figure4}.  We use these  spider forms to avoid plenty of repetitive application of their corresponding non-spider forms \cite{amar}. 

% Also, we will use the rules  spider form of the rules   for great simplicity.
\begin{proposition}(ZW rule $nat^{r}_{c}$)~\newline \\
	\begin{equation}\label{wpscopy}
	ZX\vdash
	\left\llbracket~
	\input{diagrams/natrc_LHS.tikz}
	~\right\rrbracket_{WX}
	=
	\left\llbracket~
	\input{diagrams/natrc_RHS.tikz}
	~\right\rrbracket_{WX},
	\end{equation}
	% where $r=a_0+a_1e^{i\frac{\pi}{4}}+a_2e^{i\frac{2\pi}{4}}+a_3e^{i\frac{3\pi}{4}},   a_j \in \mathbb Z[\frac{1}{2}], ~ j=0, 1, 2, 3$.
	where $r=a_0e^{i\alpha_0}+a_1e^{i\alpha_1}+a_2e^{i\alpha_2}+a_3e^{i\alpha_3},~   0\leq a_j \in \mathbb Z[\frac{1}{2}], ~\alpha_j=j\frac{\pi}{4} ~or~ j\frac{\pi}{4}+\pi,~~ j=0, 1, 2, 3$. 

\end{proposition}

\begin{proof}
	Let $c_k=a_ke^{i\alpha_k},  k=0, 1, 2, 3$. Then
	\begin{equation}\label{wphaseint}
\begin{array}{ll}
\left\llbracket~
	\input{diagrams/sinzwphase.tikz}
	~\right\rrbracket_{WX}
	=%
\beginpgfgraphicnamed{diagrams//rwphasenew2}
\InputIfFileExists{diagrams//rwphasenew2.tikz}{}{\input{./figures/diagrams//rwphasenew2.tikz}}
\endpgfgraphicnamed
	%\stackrel{nat_w^w}{=}
	\stackrel{cut_w}{=}
\left\llbracket~
\input{diagrams/zwphasefuse.tikz}
~\right\rrbracket_{WX}&\vspace{0.5cm}\\
\stackrel{cut_z}{=}
\left\llbracket~
\input{diagrams/zwphasecopy22.tikz}
~\right\rrbracket_{WX}
\stackrel{ba_{zw}}{=}
\left\llbracket~
\input{diagrams/zwphasecopy33.tikz}
~\right\rrbracket_{WX}
\stackrel{cut_z}{=}
\left\llbracket~
\input{diagrams/zwphasecopy44.tikz}
~\right\rrbracket_{WX}
\end{array}
	\end{equation}

%\left\llbracket~
%\input{diagrams/zwphasecopy.tikz}
%~\right\rrbracket_{WX}
%\stackrel{cut_z}{=}
%\left\llbracket~
%\input{diagrams/zwphasecopy2.tikz}
%~\right\rrbracket_{WX}
%\stackrel{ba_{zw}}{=}
%\left\llbracket~
%\input{diagrams/zwphasecopy3.tikz}
%~\right\rrbracket_{WX}

Thus
\begin{equation}
\begin{array}{ll}
\left\llbracket~
	\input{diagrams/natrc_RHS.tikz}
	~\right\rrbracket_{WX}
\stackrel{(\ref{wphaseint})}{=}
\left\llbracket~
\input{diagrams/zwphasecopy4.tikz}
~\right\rrbracket_{WX}
\stackrel{ba_{w}}{=}
\left\llbracket~
\input{diagrams/zwphasecopy5.tikz}
~\right\rrbracket_{WX}
&\vspace{0.5cm}\\
\stackrel{cut_w,sym_w^x, TR13, TR14}{=}
\left\llbracket~
\input{diagrams/zwphasecopy6.tikz}
~\right\rrbracket_{WX}
\stackrel{(\ref{wphaseint})}{=}
\left\llbracket~
	\input{diagrams/natrc_LHS.tikz}
	~\right\rrbracket_{WX}&
\end{array}
\end{equation}   
%Note that all the ZW rules we applied here have been proved to be true as well under the interpretation $\llbracket \cdot \rrbracket_{WX}$.
\iffalse
On the other hand,
\begin{equation}
\begin{array}{ll}
\left\llbracket~
	\input{diagrams/natrc_LHS.tikz}
	~\right\rrbracket_{WX}
	=\left\llbracket~
	\input{diagrams/zwphasecopy2l.tikz}
	~\right\rrbracket_{WX}
	=\left\llbracket~
	\input{diagrams/zwphasecopy3l.tikz}
	~\right\rrbracket_{WX}&\vspace{0.5cm}\\
	=\left\llbracket~
	\input{diagrams/zwphasecopy4l.tikz}
	~\right\rrbracket_{WX}
	=\left\llbracket~
	\input{diagrams/zwphasecopy5l.tikz}
	~\right\rrbracket_{WX}&
	\end{array}
\end{equation}   	
Therefore (\ref{wpscopy}) is proved.
\fi
\end{proof}

\begin{proposition}(ZW rule $rng^{r,s}_{+}$)~\newline \\
	\begin{equation}\label{zwplus}
	ZX\vdash
	\left\llbracket~
	\input{diagrams/rngrsp_LHS.tikz}
	~\right\rrbracket_{WX}
	=
	\left\llbracket~
	\input{diagrams/rngrsp_RHS.tikz}
	~\right\rrbracket_{WX},
	\end{equation}
% where $r=a_0+a_1e^{i\frac{\pi}{4}}+a_2e^{i\frac{2\pi}{4}}+a_3e^{i\frac{3\pi}{4}},~s=b_0+b_1e^{i\frac{\pi}{4}}+b_2e^{i\frac{2\pi}{4}}+b_3e^{i\frac{3\pi}{4}},   a_j, b_j \in \mathbb Z[\frac{1}{2}], ~ j=0, 1, 2, 3$.
 where $r=a_0e^{i\alpha_0}+a_1e^{i\alpha_1}+a_2e^{i\alpha_2}+a_3e^{i\alpha_3},~s=b_0e^{i\beta_0}+b_1e^{i\beta_1}+b_2e^{i\beta_2}+b_3e^{i\beta_3},   0\leq a_j, b_j \in \mathbb Z[\frac{1}{2}], ~\alpha_j, \beta_j=j\frac{\pi}{4} ~or~ j\frac{\pi}{4}+\pi,~ j=0, 1, 2, 3$.
\end{proposition}

\begin{proof}
Let $c_k=a_ke^{i\alpha_k}, d_k=b_ke^{i\alpha_k}, k=0, 1, 2, 3$. Then we have 
\begin{equation}
\begin{array}{ll}
\left\llbracket~
	\input{diagrams/rngrsp_LHS.tikz}
	~\right\rrbracket_{WX}
\stackrel{(\ref{wphaseint})}{=}
\left\llbracket~
	\input{diagrams/zwadd.tikz}
	~\right\rrbracket_{WX}
\stackrel{cut_w}{=}
\left\llbracket~
	\input{diagrams/zwadd1.tikz}
	~\right\rrbracket_{WX}&\vspace{0.5cm}\\
\stackrel{cut_w}{=}
\left\llbracket~
	\input{diagrams/zwadd2.tikz}
	~\right\rrbracket_{WX}
\stackrel{(AD^{\prime})}{=}
\left\llbracket~
	\input{diagrams/zwadd3.tikz}
	~\right\rrbracket_{WX}
	\stackrel{(\ref{wphaseint})}{=} 
	\left\llbracket~
	\input{diagrams/rngrsp_RHS.tikz}
	~\right\rrbracket_{WX}&	
\end{array}
\end{equation}
%Note that all the ZW rules we applied here have been proved to be true as well under the interpretation $\llbracket \cdot \rrbracket_{WX}$.

\end{proof}

\begin{proposition}\label{zwphasefusionclt}(ZW rule $rng^{r,s}_{\times}$)~\newline \\
	\begin{equation}
	ZX\vdash
	\left\llbracket~
	\input{diagrams/rngrsx_LHS.tikz}
	~\right\rrbracket_{WX}
	=
	\left\llbracket~
	\input{diagrams/rngrsx_RHS.tikz}
	~\right\rrbracket_{WX},
	\end{equation}
	 %where $r=a_0+a_1e^{i\frac{\pi}{4}}+a_2e^{i\frac{2\pi}{4}}+a_3e^{i\frac{3\pi}{4}},~s=b_0+b_1e^{i\frac{\pi}{4}}+b_2e^{i\frac{2\pi}{4}}+b_3e^{i\frac{3\pi}{4}},   a_j, b_j \in \mathbb Z[\frac{1}{2}], ~ j=0, 1, 2, 3$.
	where $r=a_0e^{i\alpha_0}+a_1e^{i\alpha_1}+a_2e^{i\alpha_2}+a_3e^{i\alpha_3},~s=b_0e^{i\beta_0}+b_1e^{i\beta_1}+b_2e^{i\beta_2}+b_3e^{i\beta_3},   0\leq a_j, b_j \in \mathbb Z[\frac{1}{2}], ~\alpha_j, \beta_j=j\frac{\pi}{4} ~or~ j\frac{\pi}{4}+\pi,~ j=0, 1, 2, 3$. 
\end{proposition}

\begin{proof}
Let $c_k=a_ke^{i\alpha_k}, d_k=b_ke^{i\alpha_k}, k=0, 1, 2, 3$. Then by (\ref{wphaseint}) we have 
\begin{equation}
\left\llbracket~
	\input{diagrams/sinzwphase.tikz}
	~\right\rrbracket_{WX}
	%=\tikzfig{diagrams//rwphasenew}
	=
\left\llbracket~
\input{diagrams/zwphasecopy44.tikz}
~\right\rrbracket_{WX},~~
\left\llbracket~
	\input{diagrams/sinzwphase2.tikz}
	~\right\rrbracket_{WX}
	=
\left\llbracket~
\input{diagrams/zwphasefuse22.tikz}
~\right\rrbracket_{WX}
	\end{equation}   
Therefore,
\begin{equation}
\begin{array}{ll}
	\left\llbracket~
	\input{diagrams/rngrsx_LHS.tikz}
	~\right\rrbracket_{WX}
	\stackrel{(\ref{wphaseint})}{=} 
	\left\llbracket~
	\input{diagrams/zwphasefuse33.tikz}
	~\right\rrbracket_{WX}
	%\stackrel{cut_z}{=}
	%\left\llbracket~
	%\input{diagrams/zwphasefuse4.tikz}
	%~\right\rrbracket_{WX}
	%\stackrel{ba_zw}{=}
	%\left\llbracket~
	%\input{diagrams/zwphasefuse5.tikz}
	%~\right\rrbracket_{WX}&\vspace{0.5cm}\\
	%\stackrel{cut_z}{=}\left\llbracket~
	%\input{diagrams/zwphasefuse6.tikz}
	%~\right\rrbracket_{WX}
	\stackrel{ba_w}{=}
	\left\llbracket~
	\input{diagrams/zwphasefuse44.tikz}
	~\right\rrbracket_{WX}&\vspace{0.5cm}\\
	\stackrel{TR13,TR14,cut_w}{=}
	\left\llbracket~
	\input{diagrams/zwphasefuse55.tikz}
	~\right\rrbracket_{WX}&\vspace{0.5cm}\\
	\stackrel{TR13,TR14,cut_w}{=}
	\left\llbracket~
	\input{diagrams/zwphasefuse66.tikz}
	~\right\rrbracket_{WX}&\vspace{0.5cm}\\
	\stackrel{(\ref{zwplus})}{=}
	\left\llbracket~
	\input{diagrams/rngrsx_RHS.tikz}
	~\right\rrbracket_{WX}&
	\end{array}
	\end{equation}
%Note that all the ZW rules we applied here have been proved to be true as well under the interpretation $\llbracket \cdot \rrbracket_{WX}$.

\end{proof}

%% file: diagrams/Hadamardwhitescalar.tikz
\begin{tikzpicture}
	\begin{pgfonlayer}{nodelayer}
		\node [style=wn] (0) at (0, 0) {};
		\node [style=wn] (1) at (0, 0.5) {};
		\node [style=wn] (2) at (0, -0.5) {};
		\node [style=none] (3) at (-0.5, 0) {$\frac{\sqrt{2}-2}{2}$};
	\end{pgfonlayer}
	\begin{pgfonlayer}{edgelayer}
		\draw (1) to (0);
		\draw (0) to (2);
	\end{pgfonlayer}
\end{tikzpicture}

%% file: diagrams/sinzwphase.tikz
\begin{tikzpicture}
	\begin{pgfonlayer}{nodelayer}
		\node [style=none] (0) at (0, -0.5) {};
		\node [style=wn] (1) at (0, 0) {};
		\node [style=none] (2) at (0.25, 0) {$r$};
		\node [style=none] (3) at (0, 0.5) {};
	\end{pgfonlayer}
	\begin{pgfonlayer}{edgelayer}
		\draw (1) to (3.center);
		\draw (1) to (0.center);
	\end{pgfonlayer}
\end{tikzpicture}

%% file: diagrams/sinzwphase2.tikz
\begin{tikzpicture}
	\begin{pgfonlayer}{nodelayer}
		\node [style=wn] (0) at (0, 0) {};
		\node [style=none] (1) at (0.25, 0) {$s$};
		\node [style=none] (2) at (0, 0.5) {};
		\node [style=none] (3) at (0, -0.5) {};
	\end{pgfonlayer}
	\begin{pgfonlayer}{edgelayer}
		\draw (0) to (2.center);
		\draw (0) to (3.center);
	\end{pgfonlayer}
\end{tikzpicture}

%% file: chapter5.tex
\chapter{Completeness for 2-qubit Clifford+T circuits}

%\section{Introduction}

Quantum computing is powerful in outperforming classical computing at solving important  problems like factoring large numbers by quantum algorithms. But before implementing a quantum algorithm, one needs to turn an algorithm into elementary gates. In principle, one could use any approximately universal set of elementary gates for gate-synthesis purpose. However, the most widely used approximately universal set of elementary gates is the Clifford+T gate set, which means any unitary
transformation can be approximated with an arbitrary precision by a Clifford+T gate. And a useful cost measure for realising 
Clifford+T circuits is  the T count ---- the number of T gates in the circuit. 

With the motivation to minimise the T count, Selinger and Bian set up a set of relations of circuits which is complete for Clifford+T 2-qubit circuits \cite{ptbian}.  Here by complete it means any two 2-qubit Clifford+T circuits that are equal in their corresponding matrices can be proved to be equal circuits only using the set of relations. Unlike the single-qubit case, there is no known algorithm that is both optimal and efficient for multi-qubit unitary synthesis. 

As evident from the main theorem in  \cite{ptbian}, it is not a simple task to use the circuit relations to simplify the quantum circuits. In contrast, the ZX-calculus has intuitive and simple rewriting rules to transform diagrams from one to another.
%As one can see from the main theorem in  \cite{ptbian}, a shortage of the notation of circuit lies in that it is inconvenient to use circuit relations to simplify quantum circuits. While the ZX-calculus, a graphical language for quantum computing \cite{CoeckeDuncan}, has intuitional and simple rewriting rules to transform diagrams from one to another. 
It is possible to use the ZX-calculus to efficiently simplify Clifford+T circuits. Although the ZX-calculus is complete for both the overall pure qubit quantum mechanics(QM) and the Clifford+T pure qubit QM as we have shown in the last two chapters,  it would be more efficient to use a small set of ZX rules for the purpose of circuit simplification.

The first step towards this goal is from Backens' work on completeness of  the ZX-calculus  for the single-qubit Clifford+T group \cite{Miriam1ct}.  However, the strategy for the single-qubit Clifford+T group does not apply to multi-qubit circuits. 

In this chapter, we verify all the circuit relations (17 equations) in \cite{ptbian} by a small set of simple ZX-calculus rules (9 rules).  Since the 17 relations are complete for 2-qubit Clifford+T circuits, so is the ZX-calculus with the 9 rules. Obviously, any single-qubit Clifford+T group can be seen as a 2-qubit Clifford+T circuit with one line empty of quantum gates. Therefore, our result can also be seen as a completeness result for single-qubit Clifford+T ZX-calculus. %Backen's result is generalised by our result.
 In comparing to the normal ZX rules \cite{CoeckeDuncan, bpw}, we just added the (P) rule, which is a property of the general Euler decomposition in ZXZ and XZX forms.  The problem of giving an analytic solution for converting from ZXZ to XZX Euler decompositions of single-qubit unitary gates has been proposed by Schr\"oder de Witt and Zamdzhiev in \cite{Vladimir}. Here we first give an explicit formula for the relation of  ZXZ and XZX Euler decompositions of generalised Z and X phases, then obtain as a corollary the formula for normal Z and X phases. Note that we do not have to know the precise values of the angles in the (P) rule to verify those circuit relations,  which makes it much simpler for simplifying circuits. Also, in the ZX computation at intermediate stages, we have gone beyond the range of Clifford+T and broken thorough the constraint of unitarity.  These can be seen as the advantages of the ZX-calculus in comparison to quantum circuits. 

Given that the ZX-calculus is universally complete, our result here is an important step towards efficient simplification of arbitrary n-qubit  Clifford+T circuits.

For convenience, in this chapter we use the scalar-free version of the ZX-calculus as described in Section \ref{zxgeneral}.

The results of this chapter has been published in the paper \cite{coeckewang},  with the coauthor Bob Coecke.

\section{Complete relations for 2-qubit Clifford+T quantum circuits}
In 2015,  a set of relations that is complete for 2-qubit Clifford+T quantum circuits was given by Selinger and Bian \cite{ptbian}. We list these relations in Theorem \ref{ptbianthm} in the language of ZX-calculus ignoring the non-zero scalars. Note that given a 2-qubit Clifford+T quantum circuit, it is easy to translate it into a ZX diagram. On the contrary, given an arbitrary ZX diagram with two inputs and two outputs, it is usually hardy to decide whether it is a 2-qubit circuit. 

\begin{theorem}[\cite{ptbian}]\label{ptbianthm}
The following set of relations is complete for 2-qubit Clifford+T circuits:
\begin{equation}\label{cmrns-2}
\beginpgfgraphicnamed{diagrams//completerelationlist-2}
\InputIfFileExists{diagrams//completerelationlist-2.tikz}{}{\input{./figures/diagrams//completerelationlist-2.tikz}}
\endpgfgraphicnamed
\end{equation}
\begin{equation}\label{cmrns-1}
\beginpgfgraphicnamed{diagrams//completerelationlist-1}
\InputIfFileExists{diagrams//completerelationlist-1.tikz}{}{\input{./figures/diagrams//completerelationlist-1.tikz}}
\endpgfgraphicnamed
\end{equation}
\begin{equation}\label{cmrns0}
\beginpgfgraphicnamed{diagrams//completerelationlist0}
\InputIfFileExists{diagrams//completerelationlist0.tikz}{}{\input{./figures/diagrams//completerelationlist0.tikz}}
\endpgfgraphicnamed
\end{equation}
\begin{equation}\label{cmrns1}
\beginpgfgraphicnamed{diagrams//completerelationlist1}
\InputIfFileExists{diagrams//completerelationlist1.tikz}{}{\input{./figures/diagrams//completerelationlist1.tikz}}
\endpgfgraphicnamed
\end{equation}
\begin{equation}\label{cmrns2}
\beginpgfgraphicnamed{diagrams//completerelationlist2}
\InputIfFileExists{diagrams//completerelationlist2.tikz}{}{\input{./figures/diagrams//completerelationlist2.tikz}}
\endpgfgraphicnamed
\end{equation}
\begin{equation}\label{cmrns3}
\beginpgfgraphicnamed{diagrams//completerelationlist3}
\InputIfFileExists{diagrams//completerelationlist3.tikz}{}{\input{./figures/diagrams//completerelationlist3.tikz}}
\endpgfgraphicnamed
\end{equation}
\begin{equation}\label{cmrns4}
\beginpgfgraphicnamed{diagrams//completerelationlist4}
\InputIfFileExists{diagrams//completerelationlist4.tikz}{}{\input{./figures/diagrams//completerelationlist4.tikz}}
\endpgfgraphicnamed
\end{equation}
\begin{equation}\label{cmrns5}
\beginpgfgraphicnamed{diagrams//completerelationlist5}
\InputIfFileExists{diagrams//completerelationlist5.tikz}{}{\input{./figures/diagrams//completerelationlist5.tikz}}
\endpgfgraphicnamed
\end{equation}
\begin{equation}\label{cmrns6}
\beginpgfgraphicnamed{diagrams//completerelationlist6}
\InputIfFileExists{diagrams//completerelationlist6.tikz}{}{\input{./figures/diagrams//completerelationlist6.tikz}}
\endpgfgraphicnamed
\end{equation}
\begin{equation}\label{cmrns7}
\beginpgfgraphicnamed{diagrams//completerelationlist7}
\InputIfFileExists{diagrams//completerelationlist7.tikz}{}{\input{./figures/diagrams//completerelationlist7.tikz}}
\endpgfgraphicnamed
\end{equation}
\begin{equation}\label{cmrns8}
\beginpgfgraphicnamed{diagrams//completerelationlist8}
\InputIfFileExists{diagrams//completerelationlist8.tikz}{}{\input{./figures/diagrams//completerelationlist8.tikz}}
\endpgfgraphicnamed
\end{equation}
\begin{equation}\label{cmrns9}
\beginpgfgraphicnamed{diagrams//completerelationlist9}
\InputIfFileExists{diagrams//completerelationlist9.tikz}{}{\input{./figures/diagrams//completerelationlist9.tikz}}
\endpgfgraphicnamed
\end{equation}
\begin{equation}\label{cmrns10}
\beginpgfgraphicnamed{diagrams//completerelationlist10}
\InputIfFileExists{diagrams//completerelationlist10.tikz}{}{\input{./figures/diagrams//completerelationlist10.tikz}}
\endpgfgraphicnamed
\end{equation}
\begin{equation}\label{cmrns11}
\beginpgfgraphicnamed{diagrams//completerelationlist11}
\InputIfFileExists{diagrams//completerelationlist11.tikz}{}{\input{./figures/diagrams//completerelationlist11.tikz}}
\endpgfgraphicnamed
\end{equation}
\begin{equation}\label{cmrns12}
\left(%
\beginpgfgraphicnamed{diagrams//pbct1}
\InputIfFileExists{diagrams//pbct1.tikz}{}{\input{./figures/diagrams//pbct1.tikz}}
\endpgfgraphicnamed\right)^2=%
\beginpgfgraphicnamed{diagrams//completerelationlist12}
\InputIfFileExists{diagrams//completerelationlist12.tikz}{}{\input{./figures/diagrams//completerelationlist12.tikz}}
\endpgfgraphicnamed
\end{equation}
\begin{equation}\label{cmrns13}
\left(%
\beginpgfgraphicnamed{diagrams//pbctb}
\InputIfFileExists{diagrams//pbctb.tikz}{}{\input{./figures/diagrams//pbctb.tikz}}
\endpgfgraphicnamed\right)^2=%
\beginpgfgraphicnamed{diagrams//completerelationlist12}
\InputIfFileExists{diagrams//completerelationlist12.tikz}{}{\input{./figures/diagrams//completerelationlist12.tikz}}
\endpgfgraphicnamed
\end{equation}
\begin{equation}\label{cmrns14}
\begin{array}{ll}
\beginpgfgraphicnamed{diagrams//completerelationlist141}
\InputIfFileExists{diagrams//completerelationlist141.tikz}{}{\input{./figures/diagrams//completerelationlist141.tikz}}
\endpgfgraphicnamed&\\
\beginpgfgraphicnamed{diagrams//completerelationlist142}
\InputIfFileExists{diagrams//completerelationlist142.tikz}{}{\input{./figures/diagrams//completerelationlist142.tikz}}
\endpgfgraphicnamed&\\
=%
\beginpgfgraphicnamed{diagrams//completerelationlist12}
\InputIfFileExists{diagrams//completerelationlist12.tikz}{}{\input{./figures/diagrams//completerelationlist12.tikz}}
\endpgfgraphicnamed
\end{array}
\end{equation}
\end{theorem}

\section{The ZX-calculus for 2-qubit Clifford+T quantum circuits}
Although we have  several complete axiomatisations of the  ZX-calculus for both entire qubit quantum mechanics and the Clifford +T fragment, it is more practical to have a small set of rewriting rules just for 2-qubit Clifford+T quantum circuits.
In this section, we use the scalar-free version of the ZX-calculus which was described in Section \ref{zxgeneral}. The generators of the  ZX-calculus for 2-qubit Clifford+T quantum circuits is the traditional generators of the qubit ZX-calculus as given in Table \ref{qbzxgenerator}. The structural rules are just those given in (\ref{compactstructure}) and  (\ref{compactstructureslide}). The non-structural rewriting rules of the ZX-calculus for 2-qubit Clifford+T quantum circuits are listed in the following:
\begin{figure}[h]
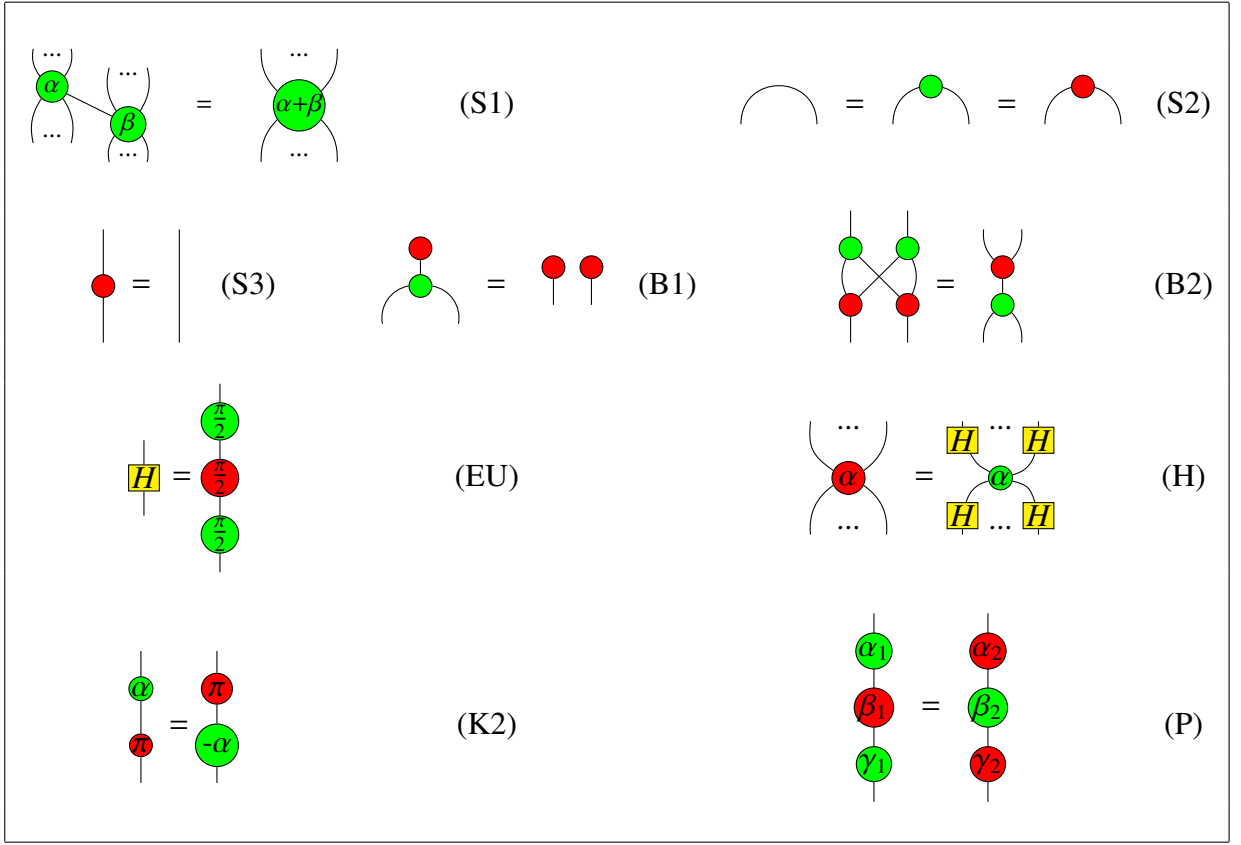

 \centering
 \begin{tabular}{|ccccc|}
  \hline
  &&&& \\
\beginpgfgraphicnamed{diagrams/spider-bis}
\InputIfFileExists{diagrams/spider-bis.tikz}{}{\input{./figures/diagrams/spider-bis.tikz}}
\endpgfgraphicnamed&(S1)&$\qquad$&%
\beginpgfgraphicnamed{diagrams/induced_compact_structure}
\InputIfFileExists{diagrams/induced_compact_structure.tikz}{}{\input{./figures/diagrams/induced_compact_structure.tikz}}
\endpgfgraphicnamed&(S2)\\
  &&&& \\
 % \tikzfig{diagrams/b1}&(B1)&&\tikzfig{diagrams/b2s}&(B2)\\
 %
\beginpgfgraphicnamed{diagrams/redidentity}
\InputIfFileExists{diagrams/redidentity.tikz}{}{\input{./figures/diagrams/redidentity.tikz}}
\endpgfgraphicnamed\quad(S3)& %
\beginpgfgraphicnamed{diagrams/b1}
\InputIfFileExists{diagrams/b1.tikz}{}{\input{./figures/diagrams/b1.tikz}}
\endpgfgraphicnamed&(B1)&%
\beginpgfgraphicnamed{diagrams/b2s}
\InputIfFileExists{diagrams/b2s.tikz}{}{\input{./figures/diagrams/b2s.tikz}}
\endpgfgraphicnamed&(B2)\\
  &&&& \\
\beginpgfgraphicnamed{diagrams/hadamdecom}
\InputIfFileExists{diagrams/hadamdecom.tikz}{}{\input{./figures/diagrams/hadamdecom.tikz}}
\endpgfgraphicnamed&(EU)&&%
\beginpgfgraphicnamed{diagrams/clchge}
\InputIfFileExists{diagrams/clchge.tikz}{}{\input{./figures/diagrams/clchge.tikz}}
\endpgfgraphicnamed&(H)\\
  &&&& \\
  % \tikzfig{diagrams/k2}&(K2)&&\tikzfig{diagrams/clswapaa}&(P)\\
  %
\beginpgfgraphicnamed{diagrams/k2}
\InputIfFileExists{diagrams/k2.tikz}{}{\input{./figures/diagrams/k2.tikz}}
\endpgfgraphicnamed&(K2)&&%
\beginpgfgraphicnamed{diagrams/zxztoxzx}
\InputIfFileExists{diagrams/zxztoxzx.tikz}{}{\input{./figures/diagrams/zxztoxzx.tikz}}
\endpgfgraphicnamed&(P)\\
  &&&& \\
  \hline
 \end{tabular}
 \caption{Rules of  ZX-calculus for 2-qubit Clifford+T quantum circuits, where $\alpha, \beta \in[0, 2\pi)$. The exact formula for the rule (P) is given in (\ref{zxztoxzxcreq}). The red-green colour swapped version and  upside-down flipped version of these rules still hold.}\label{figure2qbt}
\end{figure}

\FloatBarrier
%In addition to the above explicit rules, we also have a meta-rule: `only the topology matters' \cite{CoeckeDuncan, bpw} as a consequence from \cite{bpw22}, which means if one diagram is transformed into another by moving components around without changing their connections, then they are still equal within the ZX-calculus. It then follows that the frequently used rules (H2) and the $\pi$ copy rule (\ref{lem:6b}) are derivable  \cite{bpw}, so are the red-green colour swapped version and  upside-down flipped version of the rules in Figure \ref{figure2qbt}. We will use these derivable rules freely in this chapter.
The main difference between the rules in Figure \ref{figure2qbt} for 2-qubit Clifford+T quantum circuits and the traditional rules in Figure \ref{figure1sfy} for the ZX$_{full}$-calculus is that the former has a new identity called (P) rule. The soundness of the (P) rule
is proved in Corollary \ref{zxztoxzxcr}, however in this thesis we only need to know that the (P) rule hold and to use the property given in Corollary \ref{zxztoxzxcr} that  if $\alpha_1=\gamma_1$, then $\alpha_2=\gamma_2$, and if $\alpha_1=-\gamma_1$, then $\alpha_2=\pi+\gamma_2$.  The exact values of the angles in the (P) rule need not to be known for the proof of the last three equations in the previous section.  We add this (P) rule of the ZX-calculus for 2-qubit Clifford+T quantum circuits because we have no efficient way to prove equations (\ref{cmrns12}), (\ref{cmrns13}), (\ref{cmrns14}) in the ZX-calculus.

\section{ Verification of the complete relations in the ZX-calculus}
\subsection{ Derivation of the (P) rule}
Firstly, we explain how the ZX rule (P) is obtained. 

For arbitrary complex numbers $a, b$, define %
\beginpgfgraphicnamed{diagrams//greenbxa}
\InputIfFileExists{diagrams//greenbxa.tikz}{}{\input{./figures/diagrams//greenbxa.tikz}}
\endpgfgraphicnamed  as a general green phase which has the matrix form
\[
 \begin{pmatrix}
        1 & 0 \\
        0 & a \end{pmatrix},
 \] 
define %
\beginpgfgraphicnamed{diagrams//redbxb}
\InputIfFileExists{diagrams//redbxb.tikz}{}{\input{./figures/diagrams//redbxb.tikz}}
\endpgfgraphicnamed as a general red phase which  has the matrix form
\[
 \begin{pmatrix}
        1+b & 1-b \\
        1-b & 1+b \end{pmatrix}.
        \]

\begin{lemma}\label{phaseclcg}(General phases colour-swap  law)

Let
\begin{equation}\label{taouvst}
\begin{array}{ll}
\tau=  (1-\lambda_2)(\lambda_1+\lambda_3)+(1+\lambda_2)(1+\lambda_1\lambda_3),&\\
U=(1+\lambda_2)(\lambda_1\lambda_3-1), &V=( 1-\lambda_2)(\lambda_1-\lambda_3),\\
S=( 1-\lambda_2)(\lambda_1+\lambda_3)-(1+\lambda_2)(1+\lambda_1\lambda_3), & T=\tau(U^2-V^2).
\end{array}
\end{equation}

% $$S=\tau-2U -4(1+\lambda_2) , ~~T=\tau(U^2-V^2).$$
 Then
\begin{equation}\label{colorspps}
\beginpgfgraphicnamed{diagrams//gpcswp}
\InputIfFileExists{diagrams//gpcswp.tikz}{}{\input{./figures/diagrams//gpcswp.tikz}}
\endpgfgraphicnamed
\end{equation}
where

$$
\sigma_1=-i(U+V)\sqrt{\frac{S}{T}},~~
\sigma_2=\frac{\tau+i\sqrt{\frac{T}{S}}}{\tau-i\sqrt{\frac{T}{S}}}
,~~ \sigma_3=-i(U-V)\sqrt{\frac{S}{T}}.
 $$
 
Especially, if $\lambda_1=\lambda_3,$ then  $\sigma_1=\sigma_3$;  if $\lambda_1\lambda_3=1$, then  $\sigma_1=-\sigma_3$;  if $\lambda_1\lambda_3=-1$, then  $\sigma_1\sigma_3=-1$;  if $\lambda_1=-\lambda_3,$ then  $\sigma_1\sigma_3=1$.

\end{lemma}
\begin{proof}
The matrix of the left-hand-side of (\ref{colorspps}) is 
\[
\begin{pmatrix}
        1 & 0 \\
        0 & \lambda_3
 \end{pmatrix}
 \begin{pmatrix}
       1+\lambda_2 &  1-\lambda_2 \\
        1-\lambda_2 &  1+\lambda_2
 \end{pmatrix}
 \begin{pmatrix}
        1 & 0 \\
        0 & \lambda_1
 \end{pmatrix}
 =\begin{pmatrix}
         1+\lambda_2 &   \lambda_1( 1-\lambda_2)\\
        \lambda_3( 1-\lambda_2) &  \lambda_1 \lambda_3( 1+\lambda_2)
 \end{pmatrix}
 \]
 
 The matrix of the right-hand-hand-side of (\ref{colorspps}) is 
$$
 \begin{pmatrix}
       1+\sigma_3 &  1-\sigma_3 \\
        1-\sigma_3 &  1+\sigma_3
 \end{pmatrix}
 \begin{pmatrix}
        1 & 0 \\
        0 & \sigma_2
 \end{pmatrix}
 \begin{pmatrix}
       1+\sigma_1 &  1-\sigma_1 \\
        1-\sigma_1 &  1+\sigma_1
 \end{pmatrix}
 $$
\[
=\begin{pmatrix}
      (1+\sigma_3) (1+\sigma_1) +(1-\sigma_3) \sigma_2(1-\sigma_1)&  
       (1+\sigma_3) (1-\sigma_1) +(1-\sigma_3) \sigma_2(1+\sigma_1) \\
         (1-\sigma_3) (1+\sigma_1) +(1+\sigma_3) \sigma_2(1-\sigma_1) & 
           (1-\sigma_3) (1-\sigma_1) +(1+\sigma_3) \sigma_2(1+\sigma_1)
 \end{pmatrix}
 :=\begin{pmatrix}
      X &  Y \\
        Z &  W
 \end{pmatrix}
\]
To let the equality (\ref{colorspps})  hold, there must exist a non-zero complex number $k$ such that
\begin{equation}\label{mtrixk}
\begin{pmatrix}
      X &  Y \\
        Z &  W
 \end{pmatrix}=
 k\begin{pmatrix}
         1+\lambda_2 &   \lambda_1( 1-\lambda_2)\\
        \lambda_3( 1-\lambda_2) &  \lambda_1 \lambda_3( 1+\lambda_2)
 \end{pmatrix}
\end{equation}

Then 
$$X+Y=2(1+\sigma_2+\sigma_3-\sigma_2\sigma_3)=k[1+\lambda_2 +  \lambda_1( 1-\lambda_2)]$$
\[
Z+W=[(1-\sigma_3)2+(1+\sigma_3) \sigma_22]=2(1+\sigma_2-\sigma_3+\sigma_2\sigma_3)=k[ \lambda_3( 1-\lambda_2)+\lambda_1\lambda_3(1+\lambda_2) ]
\]
Thus
\[
X+Y+Z+W=4(1+\sigma_2)=k[(1+\lambda_2)(1+\lambda_1\lambda_3)+( 1-\lambda_2)(\lambda_1+\lambda_3)], 
\]
i.e.,
\[
\sigma_2=\frac{k}{4}[(1+\lambda_2)(1+\lambda_1\lambda_3)+( 1-\lambda_2)(\lambda_1+\lambda_3)]-1=\frac{k}{4}\tau -1, 
\]
and

\[
Z+W-(X+Y)=4\sigma_3(\sigma_2-1)=k[(1+\lambda_2)(\lambda_1\lambda_3-1)+( 1-\lambda_2)(\lambda_3-\lambda_1)], 
\]
i.e.,
\[
\sigma_3=\frac{\frac{k}{4}[(1+\lambda_2)(\lambda_1\lambda_3-1)+( 1-\lambda_2)(\lambda_3-\lambda_1)]}{\frac{k}{4}\tau -2}=\frac{k[(1+\lambda_2)(\lambda_1\lambda_3-1)+( 1-\lambda_2)(\lambda_3-\lambda_1)]}{k\tau -8}.
\]
Similarly,
$$X+Z=2(1+\sigma_1+\sigma_2-\sigma_1\sigma_2)=k[1+\lambda_2 +  \lambda_3( 1-\lambda_2)]$$
\[
Y+W=2(1+\sigma_2-\sigma_1+\sigma_2\sigma_1)=k[ \lambda_1( 1-\lambda_2)+\lambda_1\lambda_3(1+\lambda_2) ]
\]

\[
Y+W-(X+Z)=4\sigma_1(\sigma_2-1)=k[(1+\lambda_2)(\lambda_1\lambda_3-1)+( 1-\lambda_2)(\lambda_1-\lambda_3)], 
\]

i.e.,
\[
\sigma_1=\frac{\frac{k}{4}[(1+\lambda_2)(\lambda_1\lambda_3-1)+( 1-\lambda_2)(\lambda_1-\lambda_3)]}{\frac{k}{4}\tau -2}=\frac{k[(1+\lambda_2)(\lambda_1\lambda_3-1)+( 1-\lambda_2)(\lambda_1-\lambda_3)]}{k\tau -8},
\]
Now we decide the value of $k$.
Let $U=(1+\lambda_2)(\lambda_1\lambda_3-1), ~~V=( 1-\lambda_2)(\lambda_1-\lambda_3).$ Then
\[
\sigma_1+\sigma_3=\frac{2kU}{k\tau -8}, ~~\sigma_1\sigma_3=\frac{k^2(U^2-V^2)}{(k\tau -8)^2}, 
\]
Furthermore,
\[
X=(1+\sigma_3) (1+\sigma_1) +(1-\sigma_3) \sigma_2(1-\sigma_1)=k(1+\lambda_2),
\]
i.e.,
\[
1+\sigma_1+\sigma_2 +\sigma_3 +\sigma_1\sigma_3-\sigma_1\sigma_2-\sigma_2\sigma_3+\sigma_1\sigma_2\sigma_3=k(1+\lambda_2),
\]
by rearrangement, we have
\[
(\sigma_1 +\sigma_3)(1 -\sigma_2)+(1+\sigma_2)(1+\sigma_1\sigma_3)=k(1+\lambda_2).
\]

Therefore,
\[
\frac{2kU}{k\tau -8}(2 -\frac{k}{4}\tau)+\frac{k}{4}\tau(1+\frac{k^2(U^2-V^2)}{(k\tau -8)^2})=k(1+\lambda_2).
\]
Divide by $k$ on both sides, then multiply by $(k\tau -8)^2$  on both sides,  we obtain a quadratic equation of $k$:
\[
2U(k\tau -8)(2 -\frac{k}{4}\tau)+\frac{1}{4}\tau[(k\tau -8)^2+k^2(U^2-V^2)]=(k\tau -8)^2(1+\lambda_2).
\]
By rearrangement, we have
\[
(k\tau -8)^2[\tau-2U -4(1+\lambda_2)]+k^2\tau(U^2-V^2)=0.
\]
Let 
\[
\begin{array}{l}
S=\tau-2U -4(1+\lambda_2)\\
=(1+\lambda_2)(1+\lambda_1\lambda_3)+( 1-\lambda_2)(\lambda_1+\lambda_3)-2[(1+\lambda_2)(\lambda_1\lambda_3-1)+2(1+\lambda_2)]\\
=
( 1-\lambda_2)(\lambda_1+\lambda_3)-(1+\lambda_2)(1+\lambda_1\lambda_3),\\
T=\tau(U^2-V^2).
\end{array}
\]
Then the equation can be rewritten as
\[
(S\tau^2+T)k^2-16S\tau k+64S=0.
\]
Solving this equation, we have
\[
k=\frac{8S\tau \pm 8\sqrt{-ST}}{S\tau^2+T}.
\]
When we calculate the square root,   we do not have to consider its sign, hence we can write $ k$ as%we will consider its sign, so here we  can just write $k$ as 
\[
k=\frac{8S\tau + 8\sqrt{-ST}}{S\tau^2+T}.
\]
Now 
\[
\frac{8}{k}=\frac{8}{\frac{8S\tau + 8\sqrt{-ST}}{S\tau^2+T}}=\frac{S\tau^2+T}{S\tau + \sqrt{-ST}}=\frac{(\sqrt{S}\tau+i\sqrt{T})(\sqrt{S}\tau-i\sqrt{T})}{\sqrt{S}(\sqrt{S}\tau+i\sqrt{T})}=\tau-i\sqrt{\frac{T}{S}},
\]
i.e.,
$$
k=\frac{8}{\tau-i\sqrt{\frac{T}{S}}}.
$$

Then
$$
\sigma_1=\frac{k(U+V)}{k\tau-8}=\frac{U+V}{\tau-\frac{8}{k}}=\frac{U+V}{i\sqrt{\frac{T}{S}}}
=-i(U+V)\sqrt{\frac{S}{T}}.$$
$$
\sigma_3=-i(U-V)\sqrt{\frac{S}{T}}, ~~\sigma_2=\frac{\tau+i\sqrt{\frac{T}{S}}}{\tau-i\sqrt{\frac{T}{S}}}.
$$
 If $\lambda_1\lambda_3=-1$, then clearly $\tau=S, T=S(U^2-V^2)$. Thus  $\frac{S}{T}=\frac{1}{U^2-V^2}$. Therefore, 
  $\sigma_1\sigma_3=[-i(U+V)\sqrt{\frac{S}{T}}][-i(U-V)\sqrt{\frac{S}{T}}]=-(U^2-V^2)\frac{S}{T}=-1$. Similarly,   if $\lambda_1=-\lambda_3,$ then  $\sigma_1\sigma_3=1$.

\end{proof}

%\iffalse
%\begin{corollary}\label{zxztoxzxcr}
%For normal $Z$ and $X$ phases,  we have

%\begin{equation}
%\tikzfig{diagrams//zxztoxzx}
%\tikzfig{diagrams//zxztoxzx2nd}
%\end{equation}
%with $\alpha, \beta, \gamma \in (0, ~2\pi), \alpha_1=\arg z+\arg z_1, \gamma_1=\arg z-\arg z_1,  \beta_1=2\arg (|\frac{z}{z_1}|+i)$,  where
% $z=\cos\frac{\beta}{2}\cos\frac{\alpha+\gamma}{2}+i\sin\frac{\beta}{2}\cos\frac{\alpha-\gamma}{2}$, $z_1=\cos\frac{\beta}{2}\sin\frac{\alpha+\gamma}{2}-i\sin\frac{\beta}{2}\sin\frac{\alpha-\gamma}{2}$, $z_2=|\frac{z}{z_1}|+i $. If $\alpha=\gamma$, then $\alpha_1=\gamma_1$;  If $\alpha=-\gamma$, then $\alpha_1=\pi+\gamma_1(Mod ~2\pi)$; If $\alpha=\pi-\gamma$, then $\alpha_1=\pi-\gamma_1(Mod ~2\pi)$

%\end{corollary}
%\fi

\begin{corollary}\label{zxztoxzxcr}
For $\alpha_1, \beta_1, \gamma_1 \in (0, ~2\pi)$ we have:
\begin{equation}\label{zxztoxzxcreq}
\beginpgfgraphicnamed{diagrams//zxztoxzx}
\InputIfFileExists{diagrams//zxztoxzx.tikz}{}{\input{./figures/diagrams//zxztoxzx.tikz}}
\endpgfgraphicnamed\qquad\mbox{with}\quad
\left\{
\begin{array}{l}
\alpha_2=\arg z+\arg z_1\\
\beta_2=2\arg (|\frac{z}{z_1}|+i)\\
\gamma_2=\arg z-\arg z_1
\end{array}
\right.
\end{equation}
where:
\[
\begin{array}{l}
z=\cos\frac{\beta_1}{2}\cos\frac{\alpha_1+\gamma_1}{2}+i\sin\frac{\beta_1}{2}\cos\frac{\alpha_1-\gamma_1}{2}
\qquad
z_1=\cos\frac{\beta_1}{2}\sin\frac{\alpha_1+\gamma_1}{2}-i\sin\frac{\beta_1}{2}\sin\frac{\alpha_1-\gamma_1}{2}
\end{array}
\]
So if $\alpha_1=\gamma_1$, then $\alpha_2=\gamma_2$, and if $\alpha_1=-\gamma_1$, 
 then $\alpha_2=\pi+\gamma_2$. 
\end{corollary}

\begin{proof}
In (\ref{colorspps}), let $\lambda_1=e^{i\alpha_1}, \lambda_2=e^{i\beta_1}, \lambda_3=e^{i\gamma_1}.$  Then for the values of $ U, V, S, \tau$ in (\ref{taouvst}) we have
$$
\begin{array}{ll}
U=4ie^{i\frac{\alpha_1+\beta_1+\gamma_1}{2}}\cos\frac{\beta_1}{2}\sin\frac{\alpha_1+\gamma_1}{2}, &
V=4e^{i\frac{\alpha_1+\beta_1+\gamma_1}{2}}\sin\frac{\beta_1}{2}\sin\frac{\alpha_1-\gamma_1}{2},\\
S=4e^{i\frac{\alpha_1+\beta_1+\gamma_1}{2}}z, & \tau=4e^{i\frac{\alpha_1+\beta_1+\gamma_1}{2}}\overline{z},
\end{array}
$$
where $z=\cos\frac{\beta_1}{2}\cos\frac{\alpha_1+\gamma_1}{2}+i\sin\frac{\beta_1}{2}\cos\frac{\alpha_1-\gamma_1}{2}$, $\overline{z}$ is the complex conjugate of $z$. 
Also, if we let $z_1=\cos\frac{\beta_1}{2}\sin\frac{\alpha_1+\gamma_1}{2}-i\sin\frac{\beta_1}{2}\sin\frac{\alpha_1-\gamma_1}{2}$,  then 
$$
U+V=4ie^{i\frac{\alpha_1+\beta_1+\gamma_1}{2}}z_1, ~~U-V=4ie^{i\frac{\alpha_1+\beta_1+\gamma_1}{2}}\overline{z}_1.
$$
Thus 
$$
\frac{U+V}{U-V}=\frac{z_1}{\overline{z}_1}=\frac{z_1^2}{|z_1|^2},  ~~\sqrt{\frac{U+V}{U-V}}=\frac{z_1}{|z_1|}=e^{i\theta}.
$$
where $|z_1|$ is the magnitude of the complex number $z_1$, and $\theta=\arg z_1\in [0, 2\pi)$ is the phase of $z_1$.
Similarly, we have 
$$
\sqrt{\frac{z}{\overline{z}}}=\frac{z}{|z|}=e^{i\phi},
$$ 
where $\phi=\arg z\in [0, 2\pi)$ is the phase of $z$.
Therefore,
$$
\begin{array}{ll}
\sigma_1=-i(U+V)\sqrt{\frac{S}{T}}=-i(U+V)\sqrt{\frac{S}{\tau(U+V)(U-V)}}&\\
=-i\sqrt{\frac{S}{\tau}\frac{(U+V)}{(U-V)}}=-i\sqrt{-\frac{z}{\overline{z}}\frac{z_1}{\overline{z_1}}}=-iie^{i\phi}e^{i\theta}=e^{i(\phi+\theta)},
\end{array}
$$
$$
\begin{array}{ll}
\sigma_3=-i(U-V)\sqrt{\frac{S}{T}}=-i(U-V)\sqrt{\frac{S}{\tau(U+V)(U-V)}}&\\
=-i\sqrt{\frac{S}{\tau}\frac{(U-V)}{(U+V)}}=-i\sqrt{-\frac{z}{\overline{z}}\frac{\overline{z_1}}{z_1}}=-iie^{i\phi}e^{-i\theta}=e^{i(\phi-\theta)},
\end{array}
$$
$$
\begin{array}{ll}
\sigma_2=\frac{\tau+i\sqrt{\frac{T}{S}}}{\tau-i\sqrt{\frac{T}{S}}}=\frac{\tau\sqrt{\frac{S}{T}}+i}{\tau\sqrt{\frac{S}{T}}-i}=\frac{\sqrt{\frac{S\tau^2}{\tau(U^2-V^2)}}+i}{\sqrt{\frac{S\tau^2}{\tau(U^2-V^2)}}-i}\\
=\frac{\sqrt{\frac{S\tau}{(U+V)(U-V)}}+i}{\sqrt{\frac{S\tau}{(U+V)(U-V)}}-i}=\frac{\sqrt{\frac{z\overline{z}}{z_1\overline{z_1}}}+i}{\sqrt{\frac{z\overline{z}}{z_1\overline{z_1}}}-i}=\frac{|\frac{z}{z_1}|+i}{|\frac{z}{z_1}|-i}=\frac{z_2}{\overline{z_2}}=(\frac{z_2}{|z_2|})^2=e^{i2\varphi},
\end{array}
$$
where $z_2=|\frac{z}{z_1}|+i, ~\varphi=\arg z_2$ is the phase of $z_2$.  Let $\alpha_2=\phi+\theta,  \beta_2=2\varphi, \gamma_2=\phi-\theta$.
Apparently, if $\alpha_1=\gamma_1$, then $V=0,$ i.e., $e^{i\theta}=1,$ thus $\theta=0$. It follows that 
$\alpha_2=\gamma_2$. If $\alpha_1=-\gamma_1$, then $U=0,$ thus
 $e^{i(\phi+\theta)}=-e^{i(\phi-\theta)}=e^{i(\pi+\phi-\theta)}$, i.e.,  $e^{i\alpha_2 }=e^{i(\pi+\gamma_2)}$. 
 %Thus $\alpha_2\equiv \pi+\gamma_2(Mod ~2\pi)$.
 Thus $\alpha_2= \pi+\gamma_2$.

\end{proof}

\subsection{Proof of completeness for 2-qubit Clifford+T quantum circuits}
%\fi
In this subsection we show that the ZX-calculus with rules in Figure \ref{figure2qbt} is complete for  2-qubit Clifford+T quantum circuits. 
First we prove as lemmas the correctness of the three complicated circuit relations (\ref{cmrns12}), (\ref{cmrns13}) and (\ref {cmrns14}) within the ZX-calculus.

\begin{lemma}
Let $A=$
\[
\beginpgfgraphicnamed{diagrams//pbct1}
\InputIfFileExists{diagrams//pbct1.tikz}{}{\input{./figures/diagrams//pbct1.tikz}}
\endpgfgraphicnamed
\]
then $A^2=I.$
\end{lemma}

\begin{proof}
First we have $A=$
\[
%\tikzfig{diagrams//pbct2}
%\tikzfig{diagrams//pbct22}
%
\beginpgfgraphicnamed{diagrams//pbct22new}
\InputIfFileExists{diagrams//pbct22new.tikz}{}{\input{./figures/diagrams//pbct22new.tikz}}
\endpgfgraphicnamed
\]
By the rule (P), we can assume that

\begin{equation}\label{pbct32}
\beginpgfgraphicnamed{diagrams//pbct32new}
\InputIfFileExists{diagrams//pbct32new.tikz}{}{\input{./figures/diagrams//pbct32new.tikz}}
\endpgfgraphicnamed
\end{equation}

Since $e^{i\frac{-\pi}{4}} e^{i\frac{\pi}{4}}=1 $, we could let $\gamma=\alpha+\pi$.
Also note that 
$$
\beginpgfgraphicnamed{diagrams//pbct421new}
\InputIfFileExists{diagrams//pbct421new.tikz}{}{\input{./figures/diagrams//pbct421new.tikz}}
\endpgfgraphicnamed=\left(%
\beginpgfgraphicnamed{diagrams//pbct422new}
\InputIfFileExists{diagrams//pbct422new.tikz}{}{\input{./figures/diagrams//pbct422new.tikz}}
\endpgfgraphicnamed\right)^{-1}
$$
Thus
\begin{equation}\label{pbct52}
\beginpgfgraphicnamed{diagrams//pbct52new}
\InputIfFileExists{diagrams//pbct52new.tikz}{}{\input{./figures/diagrams//pbct52new.tikz}}
\endpgfgraphicnamed
\end{equation}

Therefore, $A=$
$$
\beginpgfgraphicnamed{diagrams//pbct62}
\InputIfFileExists{diagrams//pbct62.tikz}{}{\input{./figures/diagrams//pbct62.tikz}}
\endpgfgraphicnamed
$$
Finally,
$A^2=$
$$
\beginpgfgraphicnamed{diagrams//pbct722}
\InputIfFileExists{diagrams//pbct722.tikz}{}{\input{./figures/diagrams//pbct722.tikz}}
\endpgfgraphicnamed
$$
\end{proof}

\begin{lemma}
Let $B=$
\[
\beginpgfgraphicnamed{diagrams//pbctb}
\InputIfFileExists{diagrams//pbctb.tikz}{}{\input{./figures/diagrams//pbctb.tikz}}
\endpgfgraphicnamed
\]
then $B^2=I$.
\end{lemma}

\begin{proof}
Firstly we have
\[
%\tikzfig{diagrams//pbctb2}
%
\beginpgfgraphicnamed{diagrams//pbctb22new}
\InputIfFileExists{diagrams//pbctb22new.tikz}{}{\input{./figures/diagrams//pbctb22new.tikz}}
\endpgfgraphicnamed
\]

By the rule (P), we can assume that

\begin{equation}\label{pbctb32}
\beginpgfgraphicnamed{diagrams//pbctb32new}
\InputIfFileExists{diagrams//pbctb32new.tikz}{}{\input{./figures/diagrams//pbctb32new.tikz}}
\endpgfgraphicnamed
\end{equation}

Since $e^{i\frac{-\pi}{4}} e^{i\frac{\pi}{4}}=1 $, we could let $\gamma=\alpha+\pi$.

Also note that 
$$
\beginpgfgraphicnamed{diagrams//pbctb421new}
\InputIfFileExists{diagrams//pbctb421new.tikz}{}{\input{./figures/diagrams//pbctb421new.tikz}}
\endpgfgraphicnamed=\left(%
\beginpgfgraphicnamed{diagrams//pbctb422new}
\InputIfFileExists{diagrams//pbctb422new.tikz}{}{\input{./figures/diagrams//pbctb422new.tikz}}
\endpgfgraphicnamed\right)^{-1}
$$
Thus
\begin{equation}\label{pbctb52}
\beginpgfgraphicnamed{diagrams//pbctb52new}
\InputIfFileExists{diagrams//pbctb52new.tikz}{}{\input{./figures/diagrams//pbctb52new.tikz}}
\endpgfgraphicnamed
\end{equation}

Therefore, $B=$
$$
\beginpgfgraphicnamed{diagrams//pbctb62}
\InputIfFileExists{diagrams//pbctb62.tikz}{}{\input{./figures/diagrams//pbctb62.tikz}}
\endpgfgraphicnamed
$$

Finally, $B^2=$
$$
\beginpgfgraphicnamed{diagrams//pbctb72new}
\InputIfFileExists{diagrams//pbctb72new.tikz}{}{\input{./figures/diagrams//pbctb72new.tikz}}
\endpgfgraphicnamed
$$
\end{proof}

\begin{lemma}
Let $C=$
\[
\beginpgfgraphicnamed{diagrams//pbctc1}
\InputIfFileExists{diagrams//pbctc1.tikz}{}{\input{./figures/diagrams//pbctc1.tikz}}
\endpgfgraphicnamed
\]
$D=$
\[
\beginpgfgraphicnamed{diagrams//pbctc2}
\InputIfFileExists{diagrams//pbctc2.tikz}{}{\input{./figures/diagrams//pbctc2.tikz}}
\endpgfgraphicnamed
\]
then $D\circ C=I$.
\end{lemma}

\begin{proof}
Firstly we simplify the circuit $C$ as follows:
\[
\beginpgfgraphicnamed{diagrams//pbctc3new}
\InputIfFileExists{diagrams//pbctc3new.tikz}{}{\input{./figures/diagrams//pbctc3new.tikz}}
\endpgfgraphicnamed
\]
By the rule (P), we can assume that 
\begin{equation}\label{circuitceq}
\beginpgfgraphicnamed{diagrams//pbctc32new}
\InputIfFileExists{diagrams//pbctc32new.tikz}{}{\input{./figures/diagrams//pbctc32new.tikz}}
\endpgfgraphicnamed
\end{equation}
Then we have % $C=$
\begin{equation}\label{circuitceq4}
\beginpgfgraphicnamed{diagrams//pbctc34new}
\InputIfFileExists{diagrams//pbctc34new.tikz}{}{\input{./figures/diagrams//pbctc34new.tikz}}
\endpgfgraphicnamed
\end{equation}

Secondly, we simplify the circuit $D$ as follows:
\begin{equation*}
\beginpgfgraphicnamed{diagrams//pbctc4new}
\InputIfFileExists{diagrams//pbctc4new.tikz}{}{\input{./figures/diagrams//pbctc4new.tikz}}
\endpgfgraphicnamed
\end{equation*}

By the rule (P), we have
\begin{equation}\label{circuitceq2}
\beginpgfgraphicnamed{diagrams//pbctc33new}
\InputIfFileExists{diagrams//pbctc33new.tikz}{}{\input{./figures/diagrams//pbctc33new.tikz}}
\endpgfgraphicnamed
\end{equation}
Therefore, 
\begin{equation}\label{circuitceq5}
\beginpgfgraphicnamed{diagrams//pbctc35new}
\InputIfFileExists{diagrams//pbctc35new.tikz}{}{\input{./figures/diagrams//pbctc35new.tikz}}
\endpgfgraphicnamed
\end{equation}
Then we obtain the composition %$D\circ C=$
\begin{equation}\label{circuitceq6}
\beginpgfgraphicnamed{diagrams//pbctc36new}
\InputIfFileExists{diagrams//pbctc36new.tikz}{}{\input{./figures/diagrams//pbctc36new.tikz}}
\endpgfgraphicnamed
\end{equation}

By Corollary \ref{zxztoxzxcr}, we can assume that
\begin{equation}\label{circuitceq7}
\beginpgfgraphicnamed{diagrams//pbctc37new}
\InputIfFileExists{diagrams//pbctc37new.tikz}{}{\input{./figures/diagrams//pbctc37new.tikz}}
\endpgfgraphicnamed
\end{equation}

Then for its inverse, we have 
\begin{equation}\label{circuitceq8}
\beginpgfgraphicnamed{diagrams//pbctc38new}
\InputIfFileExists{diagrams//pbctc38new.tikz}{}{\input{./figures/diagrams//pbctc38new.tikz}}
\endpgfgraphicnamed
\end{equation}

Also we can obtian that
\begin{equation}\label{circuitceq9}
\beginpgfgraphicnamed{diagrams//pbctc39new}
\InputIfFileExists{diagrams//pbctc39new.tikz}{}{\input{./figures/diagrams//pbctc39new.tikz}}
\endpgfgraphicnamed
\end{equation}

As a consequence, we have the inverse for both sides of (\ref{circuitceq9}):

\begin{equation}\label{circuitceq10}
\beginpgfgraphicnamed{diagrams//pbctc310new}
\InputIfFileExists{diagrams//pbctc310new.tikz}{}{\input{./figures/diagrams//pbctc310new.tikz}}
\endpgfgraphicnamed
\end{equation}

Now we can rewrite  $D\circ C$ as 

\begin{equation}\label{circuitceq11}
\beginpgfgraphicnamed{diagrams//pbctc311new}
\InputIfFileExists{diagrams//pbctc311new.tikz}{}{\input{./figures/diagrams//pbctc311new.tikz}}
\endpgfgraphicnamed
\end{equation}

We can depict the dashed part of (\ref{circuitceq11}) in a form of connected octagons:

%\[
%\tikzfig{diagrams//octagon}%\tikzfig{diagrams//octagonsimpl}
%\]
\begin{equation}\label{octagon1eq}
\beginpgfgraphicnamed{diagrams//octagon1}
\InputIfFileExists{diagrams//octagon1.tikz}{}{\input{./figures/diagrams//octagon1.tikz}}
\endpgfgraphicnamed
\end{equation}

%\[
%\tikzfig{diagrams//octagon1}
%\]

To deal with these octagons, we need to cope with hexagons. The following property of  hexagons was first given in \cite{duncan_graph_2009}:

\begin{equation}\label{hexagon1eq}
\beginpgfgraphicnamed{diagrams//hexagon1new}
\InputIfFileExists{diagrams//hexagon1new.tikz}{}{\input{./figures/diagrams//hexagon1new.tikz}}
\endpgfgraphicnamed
\end{equation}

%\[
%\tikzfig{diagrams//hexagon1}
%\]
By colour change rule (H), 
we have 
\begin{equation}\label{hexagon2eq}
\beginpgfgraphicnamed{diagrams//hexagon2}
\InputIfFileExists{diagrams//hexagon2.tikz}{}{\input{./figures/diagrams//hexagon2.tikz}}
\endpgfgraphicnamed
\end{equation}

Then we can rewrite   (\ref{octagon1eq}) as follows:
\begin{equation}\label{octagon2checkeq}
\beginpgfgraphicnamed{diagrams//octagon2checknew}
\InputIfFileExists{diagrams//octagon2checknew.tikz}{}{\input{./figures/diagrams//octagon2checknew.tikz}}
\endpgfgraphicnamed
\end{equation}
%\tikzfig{diagrams//octagon2}

%where we used property of hexagons in the third, fourth and fifth equalities.

%\\[tikzfig{diagrams//octagonsimplcorrect}\]

By the (P) rule, we have 
\begin{equation}\label{octasim2eq}
\beginpgfgraphicnamed{diagrams//octasim2new}
\InputIfFileExists{diagrams//octasim2new.tikz}{}{\input{./figures/diagrams//octasim2new.tikz}}
\endpgfgraphicnamed
\end{equation}
where $z=x+\pi$.  Then we take inverse for each side of (\ref{octasim2eq}) and obtain that

\begin{equation}\label{octasim3eq}
\beginpgfgraphicnamed{diagrams//octasim3new}
\InputIfFileExists{diagrams//octasim3new.tikz}{}{\input{./figures/diagrams//octasim3new.tikz}}
\endpgfgraphicnamed
\end{equation}

By rearranging the phases on both sides of (\ref{octasim2eq}), we have
\begin{equation}\label{octasim4eq}
\beginpgfgraphicnamed{diagrams//octasim4new}
\InputIfFileExists{diagrams//octasim4new.tikz}{}{\input{./figures/diagrams//octasim4new.tikz}}
\endpgfgraphicnamed
\end{equation}

Thus
\begin{equation}\label{octasim5eq}
\beginpgfgraphicnamed{diagrams//octasim5new}
\InputIfFileExists{diagrams//octasim5new.tikz}{}{\input{./figures/diagrams//octasim5new.tikz}}
\endpgfgraphicnamed
\end{equation}

Therefore,
\begin{equation}\label{octasim6eq}
\beginpgfgraphicnamed{diagrams//octasim6new}
\InputIfFileExists{diagrams//octasim6new.tikz}{}{\input{./figures/diagrams//octasim6new.tikz}}
\endpgfgraphicnamed
\end{equation}

%By  (\ref{octasim6eq}), we have 
It then follows that
\begin{equation}\label{octasim7eq}
\beginpgfgraphicnamed{diagrams//octasim7new}
\InputIfFileExists{diagrams//octasim7new.tikz}{}{\input{./figures/diagrams//octasim7new.tikz}}
\endpgfgraphicnamed
\end{equation}

If we take the inverse of the left-hand-side of (\ref{octasim7eq}), then we have

\begin{equation}\label{octasim8eq}
\beginpgfgraphicnamed{diagrams//octasim8new}
\InputIfFileExists{diagrams//octasim8new.tikz}{}{\input{./figures/diagrams//octasim8new.tikz}}
\endpgfgraphicnamed
\end{equation}

Now we can further simplify the final diagram in  (\ref{octagon2checkeq}) as follows:

\begin{equation}\label{octasim9eq}
\beginpgfgraphicnamed{diagrams//octasim9new}
\InputIfFileExists{diagrams//octasim9new.tikz}{}{\input{./figures/diagrams//octasim9new.tikz}}
\endpgfgraphicnamed
\end{equation}

Finally, the composite circuit $D\circ C$ as  can be simplified as follows: 

\begin{equation}\label{octasim10eq}
\beginpgfgraphicnamed{diagrams//octasim10new}
\InputIfFileExists{diagrams//octasim10new.tikz}{}{\input{./figures/diagrams//octasim10new.tikz}}
\endpgfgraphicnamed
\end{equation}
%where we used (\ref{circuitceq9}) to get 
where we used the following property:
\begin{equation}\label{octasim11eq}
\beginpgfgraphicnamed{diagrams//octasim11new}
\InputIfFileExists{diagrams//octasim11new.tikz}{}{\input{./figures/diagrams//octasim11new.tikz}}
\endpgfgraphicnamed
\end{equation}
\end{proof}

%\fi

The other relations are easy to be verified by the ZX-calculus, we omit those verification here. 
Since the circuit relations are proved to be complete for 2-qubit Clifford+T quantum circuits \cite{ptbian}, we have the following 
\begin{theorem}
The ZX-calculus is complete for  2-qubit Clifford+T quantum circuits.
\end{theorem}

In this chapter,  the essential technique we employed is that we go beyond the range of Clifford+T and broke thorough the constraint of unitarity at intermediate stages by using the new (P) rule. Although we do not have a general general strategy for simplifying arbitrary quantum circuits by the ZX-calculus, we would expect the  (P) rule to have more utilities in quantum computing and information. 

%Apparently, this theorem generalised Backen's result of completeness of the ZX-calculus for single-qubit Clifford+T group \cite{Miriam1ct}, because  any single-qubit Clifford+T group can be seen as a 2-qubit Clifford+T circuit with one line empty of quantum gates. 

%% file: diagrams/completerelationlist12.tikz
\begin{tikzpicture}
	\begin{pgfonlayer}{nodelayer}
		\node [style=none] (0) at (-0.25, 0.5) {};
		\node [style=none] (1) at (0.5, 0.5) {};
		\node [style=none] (2) at (-0.25, -0.5) {};
		\node [style=none] (3) at (0.5, -0.5) {};
	\end{pgfonlayer}
	\begin{pgfonlayer}{edgelayer}
		\draw (0.center) to (1.center);
		\draw (2.center) to (3.center);
	\end{pgfonlayer}
\end{tikzpicture}

%% file: diagrams/greenbxa.tikz
\begin{tikzpicture}
	\begin{pgfonlayer}{nodelayer}
		\node [style=none] (0) at (0, 0.25) {};
		\node [style=gbox] (1) at (0, 0) {$a$};
		\node [style=none] (2) at (0, -0.25) {};
	\end{pgfonlayer}
	\begin{pgfonlayer}{edgelayer}
		\draw (0.center) to (2.center);
	\end{pgfonlayer}
\end{tikzpicture}

%% file: chapter2.tex
\chapter{Completeness for qutrit stabilizer quantum mechanics}

%\section{Introduction}
The theory of quantum information and quantum computation (QIC) is traditionally based on binary logic (qubits). However,  multi-valued logic has been recently proposed for quantum computing, using linear ion traps \cite{Muthukrishnan}, cold atoms \cite{Smith}, and entangled photons \cite{Malik}. In particular, metaplectic non-Abelian anyons were shown to be naturally connected to ternary (qutrit) logic in contrast to binary logic in topological quantum computing, based on which there comes the Metaplectic Topological Quantum Computer platform \cite{Shawn}. %Furthermore,  qutrit-based computers are in certain sense more space-optimal in comparison with quantum computers based on other dimensions of systems \cite{amk}.

%The current theoretical tools for qutrit-based QIC are dominated by quantum circuits. However, the quantum circuit notation has  a major defect: the transformation from one circuit diagram into another under the rules of a set of circuit equations. In contrast, the ZX-calculus introduced by Coecke and Duncan \cite{CoeckeDuncan}  is a more intuitive graphical language. Despite being easy to manipulate and seemingly un-mathematical, it is well formulated within the framework of compacted closed categories, which is an important branch of categorical quantum mechanics (CQM) pioneered in 2004 by Abramsky and Coecke \cite{Coeckesamson}. It is intuitive due to its simple rewriting rules which are represented by equations of diagrams for quantum computing. Most importantly, 
%the ZX-calculus has been shown to be complete for Clifford+T quantum mechanics (QM) \cite{Emmanuel,ngwang2} and the entire qubit QM \cite{JPV2,ngwang}.
% has been successfully applied to QIC in the fields of   (topological) measurement-based quantum computing  \cite{Duncanpx,  Horsman}  and quantum error correction \cite{DuncanLucas,  ckzh}. 
%In particular, the ZX-calculus is pretty useful for qubit stabilizer quantum mechanics because of its completeness for this sub-theory, i.e. any equality that can be derived using matrices can also be derived diagrammatically. 

Taking into consideration the practicality of qutrits and the fruitful results on completeness of the ZX-calculus that have been obtained in the previous chapters, it is natural to give a qutrit version of the ZX-calculus. However, the generalisation from qubits to qutrits is not trivial, since the qutrit-based structures are usually much more complicated then the qubit-based structures. For instance,  the local Clifford group for qubits has only 24 elements, while the local Clifford group for qutrits has 216 elements. Thus it is no surprise that, as presented in   \cite{GongWang},  the rules of qutrit ZX-calculus are significantly different from that of the qubit case: each phase gate has a pair of phase angles, the operator commutation rule is more complicated, the Hadamard gate is not self-adjoint, the colour-change rule is doubled, and the dualiser has a much richer structure than being just an identity.  
Despite being already well established in \cite{BianWang1, BianWang2} and independently introduced as a typical special case of qudit ZX-calculus in \cite{Ranchin}, to the best of our knowledge, there are no completeness results available for qutrit ZX-calculus. Without this kind of results, how can we even know that the rules of  a so-called  ZX-calculus are useful enough for quantum computing?

In this chapter, based on the rules and results in \cite{GongWang}, we show that  the qutrit ZX-calculus is complete for pure qutrit stabilizer quantum mechanics (QM). The strategy we used here mirrors that of the qubit case in \cite{Miriam1}, although it is technically more complicated, especially for the  completeness of the single qutrit Clifford group $\mathcal{C}_1$. Firstly, we show that  any stabilizer state diagram is equal to some GS-LC diagram within the ZX-calculus, 
where a GS-LC diagram consists of a graph state diagram with arbitrary single-qutrit Clifford operators applied to each output. We then show that any stabilizer state diagram can be further brought into a reduced form of the GS-LC diagram. Finally, for any two  stabilizer state diagrams on the same number of qutrits, we make them into a simplified pair of reduced GS-LC diagram such that they are equal under the standard interpretation in Hilbert spaces  if and only if they are identical in the sense that they are constructed by the same element constituents in the same way. By the map-state duality,% (or more general, the Choi-Jamiolkowski isomorphism \cite{paulsen2002completely}),
 the case for operators represented by diagrams are also covered, thus we have shown the completeness of the ZX-calculus for  all pure qutrit stabilizer QM.

%A natural question will arise at the end of this chapter: is there a general proof of completeness of the ZX-calculus for arbitrary dimensional (qudit) stabilizer QM?  This is the problem we would like to address next, but we should also mention some challenges we may face. With exception of the increase of the order of local Clifford groups, the main difficulty comes from the fact that that it is not known whether any stabilizer state is equivalent to a graph state under local Clifford group for the dimension $d$ having multiple prime factors \cite{Hostens}. As far as we know, it is true in the case that $d$ has only single prime factors  \cite{ShiangGr}.  Furthermore, the sufficient and necessary condition for two qudit graph states to be  equivalent under local Clifford group in terms of operations on graphs is unknown in the case that $d$ is non-prime. Finally, the generalised Euler decomposition rule for the generalised Hadamard gate can not be trivially derived, and other uncommon rules might be needed for general $d$.

The results of this chapter are collected from the paper \cite{qlwang} without any coauthor. In addition, all of the proofs are included here.

\section{Qutrit Stabilizer quantum mechanics}

\subsection{ The generalized Pauli group and Clifford group }

 The notions of  Pauli group and Clifford group for qubits can be generalised to 
 qutrits in a natural way:
In the 3-dimensional Hilbert space $H_3$, we define the \textit{ generalised Pauli operators}  $X$ and $Z$ as follows
\begin{equation}
X\ket{j}=\ket{j+1}, ~~Z\ket{j}=\omega^j\ket{j},
\end{equation}
where $ j \in \mathbb{Z}_3 $ (the ring of integers modulo 3), $\omega=e^{i\frac{2}{3}\pi}$, and the
 addition is a modulo 3 operation. We will use the same denotation for
 tensor products of these operators as is presented in \cite{Hostens}: for
$v, w \in \mathbb{Z}_3^n$ ($n$-dimensional vector space over  $\mathbb{Z}_3 $), let
 $$
 v=\left(
\begin{array}{c}
v_1 \\
\vdots \\
v_n
\end{array}
\right),    \quad
w=\left(
\begin{array}{c}
w_1 \\
\vdots \\
w_n
\end{array}
\right),   \quad
a=\left(
\begin{array}{c}
v \\
w
\end{array}
\right)\in \mathbb{Z}_3^{2n},
$$

\begin{equation}\label{xza}
XZ(a)= X^{v_1}Z^{w_1}\otimes \cdots \otimes X^{v_n}Z^{w_n}.
\end{equation}

We define the  \textit{generalized Pauli group} $\mathcal{P}_n$ on $n$ qutrits as
$$
\mathcal{P}_n=\{ \omega^{\delta}XZ(a)| a\in \mathbb{Z}^{2n}_3,  \delta\in \mathbb{Z}_3 \}.
$$
%to contain all $3^{2n}$ tensor products (\ref{xza}) with an additional complex phase factor $\omega^{\delta}$, where $\delta\in \mathbb{Z}_3$. The order of an arbitrary element of this Pauli group is never equal to 6.

\begin{definition}
The  generalized  Clifford group $\mathcal{C}_n$ on n qutrits is defined as the normalizer of $\mathcal{P}_n$ in the group of unitaries:
$
\mathcal{C}_n=\{ Q| Q^{\dagger}=Q^{-1}, Q\mathcal{P}_n Q^{\dagger}= \mathcal{P}_n \}.
$
Especially, for $n=1$, $\mathcal{C}_1$ is called the generalized local Clifford group.

\end{definition}
%operation $Q$ is a unitary operation that maps the Pauligroup on $n$ qutrits to itself under conjugation, or $Q\mathcal{P}_n Q^{\dagger}= \mathcal{P}_n$.
Similar to the qubit case, it can be shown that the generalized Clifford group is generated by  the gate $\mathcal{S}=\ket{0}\bra{0}+\ket{1}\bra{1}+ \omega\ket{2}\bra{2}$,
the generalized Hadamard gate $H=\frac{1}{\sqrt{3}}\sum_{k, j=0}^{2}\omega^{kj}\ket{k}\bra{j}$, and the SUM gate $\Lambda=\sum_{i, j=0}^{2}\ket{i, i+j(mod 3)}\bra{ij}$  \cite{Hostens, Shawn}. In particular, the local Clifford group $\mathcal{C}_1$ is generated by the gate $\mathcal{S}$ and the generalized Hadamard gate $H$ \cite{Hostens}, with the group order being $3^3(3^2-1)=216$, up to global phases \cite{Mark}. 

%\subsection{ Stabilizer states }
We define the \textit{stabilizer code} as  the non-zero joint eigenspace to the eigenvalue $1$ of a subgroup of the generalized Pauli group $\mathcal{P}_n$ \cite{Mohsen}. A  \textit{stabilizer state} $\ket{\psi}$  is a stabilizer code of dimension 1, which   
 is therefore stabilized by an abelian subgroup of order $3^{n}$ of the Pauli group excluding multiples of the identity other than the identity itself \cite{Hostens}. We
call this subgroup the \textit{stabilizer}  $\mathfrak{S}$ of $\ket{\psi}$.

Finally, the \textit{qutrit stabilizer quantum mechanics} can be described as the fragment of the pure qutrit quantum theory where only states represented in computational basis, Clifford unitaries and measurements in computational basis are considered.

\subsection{ Graph states }
Graph states are special stabilizer states which are constructed based on undirected graphs without loops. However, it turns out that they are not far from stabilizer states. 

\begin{definition}\cite{Dan}
A  $\mathbb{Z}_3 $-weighted graph is a pair $G = (V,E)$ where $V$ is a set of $n$ vertices and $E$ is a  collection of weighted edges specified by the adjacency matrix $\Gamma$, which is a symmetric $n$ by $n$ matrix with zero diagonal entries, each  matrix element ${\Gamma_{lm}}\in \mathbb{Z}_3$ representing the weight of the edge connecting vertex $l$ with vertex $m$.
\end{definition}

\begin{definition}\label{Graph State}\cite{Griffiths}
Given a   $\mathbb{Z}_3 $-weighted graph $G = (V,E)$ with $n$ vertices and  adjacency matrix $\Gamma$, 
 the corresponding qutrit graph state can be defined as 
$$\ket{G}=\mathcal{U}\ket{+}^{\otimes n},$$ 
where $\ket{+}=\frac{1}{\sqrt{3}}(\ket{0}+\ket{1}+\ket{2})$,
$\mathcal{U}=\prod_{\{l,m\}\in E}(C_{lm})^{\Gamma_{lm}}$,  $C_{lm}=\Sigma_{j=0}^{2}\Sigma_{k=0}^{2}\omega^{jk}\ket{j}\bra{j}_l\otimes \ket{k}\bra{k}_m$, subscripts indicate to which qutrit the operator is applied.
\end{definition}

\begin{lemma}\cite{Dan}
The qutrit graph state $\ket{G}$ is the unique (up to a global phase) joint $+1$ eigenstate of the group generated by the operators
$$
X_v\prod_{u\in V}(Z_{u})^{\Gamma_{uv}} ~~\mbox{for all}~ v \in V.
$$
\end{lemma}

Therefore,  graph states must be stabilizer states.  On the contrary, stabilizer states are equivalent to graph states in the following sense.
\begin{definition}\cite{Miriam1}
 Two $n$-qutrit stabilizer states $\ket{\psi}$ and $\ket{\phi}$ are equivalent with respect to the local Clifford group if there exists $U \in \mathcal{C}_1^{\otimes n}$ such that $\ket{\psi}=U\ket{\phi}$.
 
 \end{definition}
 
\begin{lemma}\cite{Mohsen} \label{stabilizerequivgraphstate}
 Every qutrit stabilizer state is equivalent to a graph state with respect to the local Clifford group.
 
\end{lemma}

 Below we describe some operations on graphs corresponding to graph states. Theses operations will play a central role in the proof of the completeness of ZX-calculus for qutrit stabilizer quantum mechanics.   

\begin{definition}\cite{Mohsen}
 Let $G = (V,E)$ be a $\mathbb{Z}_3 $-weighted graph with $n$ vertices and  adjacency matrix $\Gamma$. For every vertex $v$, and $0\neq b \in \mathbb{Z}_3$, define the operator $\circ_b v$ on the graph as follows: $G\circ_b v$ is the graph on the same vertex set, with adjacency  matrix $I(v, b)\Gamma I(v, b)$, where $I(v, b) = diag(1,1,...,b,...,1)$, $b$ being on the $v$-th entry.  For every vertex $v$ and $a  \in \mathbb{Z}_3$, define the operator $\ast_a  v $ on the graph as follows: $G\ast_a  v $ is the graph on the same vertex set, with adjacency matrix $\Gamma^{'}$, where
  $\Gamma^{'}_{jk}=\Gamma_{jk}+a\Gamma_{vj}\Gamma_{vk}$ for $j\neq k$, and $\Gamma^{'}_{jj}=0$ for all $j$. The operator $\ast_a  v $ is also called the $a$-local complementation at the vertex $v$ \cite{mmp} .
 
 \end{definition}
 
 Now the equivalence of graph states can be described in terms of these operations on graphs.

 \begin{theorem}\cite{Mohsen} \label{graphstateeuivalence}
 Two graph states  $\ket{G}$  and  $\ket{H}$  with adjacency matrices $M$ and $N$ over $\mathbb{Z}_3 $, are equivalent under local Clifford group if and only if there exists a sequence of $\ast$ and $\circ$ operators acting on one of them to obtain the other.
 \end{theorem}

\section{Qutrit ZX-calculus}
\subsection{The ZX-calculus for general pure state qutrit quantum mechanics}
%In this paper, we will work in a symmetric monoidal category and ignore non-zero scalars. (give the details of the category)
As described in Section \ref{zxgeneral},  the qutrit ZX-calculus is a special case of the qudit ZX-caculus where $d=3$. Especially, we use the scalar-free version of the the qutrit ZX-calculus throughout this chapter.  The generators of the qutrit ZX-calculus is explicitly given as follows:
%The ZX-calculus is founded on a symmetric monoidal category (SMC) $\mathfrak{C}$. The objects of $\mathfrak{C}$ are natural numbers: $0, 1, 2,  \cdots$; the tensor of 
%objects is just addition of numbers: $m \otimes n = m+n$. The morphisms of $\mathfrak{C}$ are diagrams of the qutrit ZX-calculus. A general diagram  $D:k\to l$   with $k$ inputs and $l$ outputs is generated by:

%\begin{center}
%\begin{tabular}{|r@{~}r@{~}c@{~}c|r@{~}r@{~}c@{~}c|}
%\hline
%$R_Z^{(n,m)}(\frac{\alpha}{\beta})$&$:$&$n\to m$ & \tikzfig{Qutrits//generator_spider} & $R_X^{(n,m)}(\frac{\alpha}{\beta})$&$:$&$ n\to m$& \tikzfig{Qutrits//generator_spider_gray}\\
%\hline
%$H$&$:$&$1\to 1$ &\tikzfig{Qutrits//RGg_Hada}
%& $s$&$:$&$0\to 0$ &\tikzfig{scalars//halfscalar}  &
 %& $H^{\dagger}$&$:$&$1\to 1$  &\tikzfig{Qutrits//RGg_Hadad} \\\hline
  %$\sigma$&$:$&$ 2\to 2$& \tikzfig{Qutrits//swap} &$\mathbb I$&$:$&$1\to 1$&\tikzfig{Qutrits//Id} \\\hline
   %$e $&$:$&$0 \to 0$& \tikzfig{Qutrits//emptysquare-small} &&& &  
  %\\\hline
%\end{tabular}
%\end{center}

\begin{table}
\begin{center} % \tikzfig{diagrams//emptysquare-small}
\begin{tabular}{|r@{~}r@{~}c@{~}c|r@{~}r@{~}c@{~}c|}
\hline
%$R_Z^{(n,m)}$&$:$&$n\to m$ & \tikzfig{diagrams//generator_spider2} & $A$&$:$&$ 1\to 1$& \tikzfig{diagrams//alphagate}\\
$R_Z^{(n,m)}(\frac{\alpha}{\beta})$&$:$&$n\to m$ & %
\beginpgfgraphicnamed{Qutrits//generator_spider}
\InputIfFileExists{Qutrits//generator_spider.tikz}{}{\input{./figures/Qutrits//generator_spider.tikz}}
\endpgfgraphicnamed  & $R_X^{(n,m)}(\frac{\alpha}{\beta})$&$:$&$n\to m$ & %
\beginpgfgraphicnamed{Qutrits//generator_spider_gray}
\InputIfFileExists{Qutrits//generator_spider_gray.tikz}{}{\input{./figures/Qutrits//generator_spider_gray.tikz}}
\endpgfgraphicnamed\\
%\multicolumn{3}{|c}{ \tikzfig{diagrams//generator_spider2} }&(P1)\\
\hline
$H$&$:$&$1\to 1$ &%
\beginpgfgraphicnamed{diagrams//HadaDecomSingleslt}
\InputIfFileExists{diagrams//HadaDecomSingleslt.tikz}{}{\input{./figures/diagrams//HadaDecomSingleslt.tikz}}
\endpgfgraphicnamed
%& $s$&$:$&$0\to 0$ &\tikzfig{scalars//halfscalar}  &
 &  $\sigma$&$:$&$ 2\to 2$& %
\beginpgfgraphicnamed{diagrams//swap}
\InputIfFileExists{diagrams//swap.tikz}{}{\input{./figures/diagrams//swap.tikz}}
\endpgfgraphicnamed\\\hline
   $\mathbb I$&$:$&$1\to 1$&%
\beginpgfgraphicnamed{diagrams//Id}
\InputIfFileExists{diagrams//Id.tikz}{}{\input{./figures/diagrams//Id.tikz}}
\endpgfgraphicnamed & $e $&$:$&$0 \to 0$& %
\beginpgfgraphicnamed{diagrams//emptysquare}
\InputIfFileExists{diagrams//emptysquare.tikz}{}{\input{./figures/diagrams//emptysquare.tikz}}
\endpgfgraphicnamed\\\hline
   $C_a$&$:$&$ 0\to 2$& %
\beginpgfgraphicnamed{diagrams//cap}
\InputIfFileExists{diagrams//cap.tikz}{}{\input{./figures/diagrams//cap.tikz}}
\endpgfgraphicnamed &$ C_u$&$:$&$ 2\to 0$&%
\beginpgfgraphicnamed{diagrams//cup}
\InputIfFileExists{diagrams//cup.tikz}{}{\input{./figures/diagrams//cup.tikz}}
\endpgfgraphicnamed \\\hline
\end{tabular}\caption{Generators of qubit ZX-calculus} \label{qbzxgenerator}
\end{center}
\end{table}
\FloatBarrier
where $m,n\in \mathbb N$, $\alpha, \beta \in [0,  2\pi)$, and $e$ is denoted by an empty diagram.
  
%The composition of morphisms is  to combine these components in the following two ways: for any two morphisms $D_1:a\to b$ and $D_2: c\to d$, a \textit{ parallel composition} $D_1\otimes D_2 : a+c\to b+d$ is obtained by placing $D_1$ and $D_2$ side-by-side with $D_1$ on the left of $D_2$;
 %for any two morphisms $D_1:a\to b$ and $D_2: b\to c$,  a  \textit{ sequential  composition} $D_2\circ D_1 : a\to c$ is obtained by placing $D_1$ above $D_2$, connecting the outputs of $D_1$ to the inputs of $D_2$.

%Spiders with all zero phase angles are simply denoted as follows:
For simplicity, we have the following notations:
$$%
\beginpgfgraphicnamed{Qutrits//spiderg}
\InputIfFileExists{Qutrits//spiderg.tikz}{}{\input{./figures/Qutrits//spiderg.tikz}}
\endpgfgraphicnamed := %
\beginpgfgraphicnamed{Qutrits//spidergz}
\InputIfFileExists{Qutrits//spidergz.tikz}{}{\input{./figures/Qutrits//spidergz.tikz}}
\endpgfgraphicnamed \qquad%
\beginpgfgraphicnamed{Qutrits//spiderr}
\InputIfFileExists{Qutrits//spiderr.tikz}{}{\input{./figures/Qutrits//spiderr.tikz}}
\endpgfgraphicnamed := %
\beginpgfgraphicnamed{Qutrits//spiderrz}
\InputIfFileExists{Qutrits//spiderrz.tikz}{}{\input{./figures/Qutrits//spiderrz.tikz}}
\endpgfgraphicnamed \qquad%
\beginpgfgraphicnamed{Qutrits//RGg_Hadad}
\InputIfFileExists{Qutrits//RGg_Hadad.tikz}{}{\input{./figures/Qutrits//RGg_Hadad.tikz}}
\endpgfgraphicnamed :=%
\beginpgfgraphicnamed{diagrams//hcubic}
\InputIfFileExists{diagrams//hcubic.tikz}{}{\input{./figures/diagrams//hcubic.tikz}}
\endpgfgraphicnamed $$

%There are two kinds of rules for the morphisms of $\mathfrak{C}$:  the structure rules for $\mathfrak{C}$ as an SMC, and 
%the rewriting rules listed in Figure \ref{figure1}.

\begin{figure}[!h]
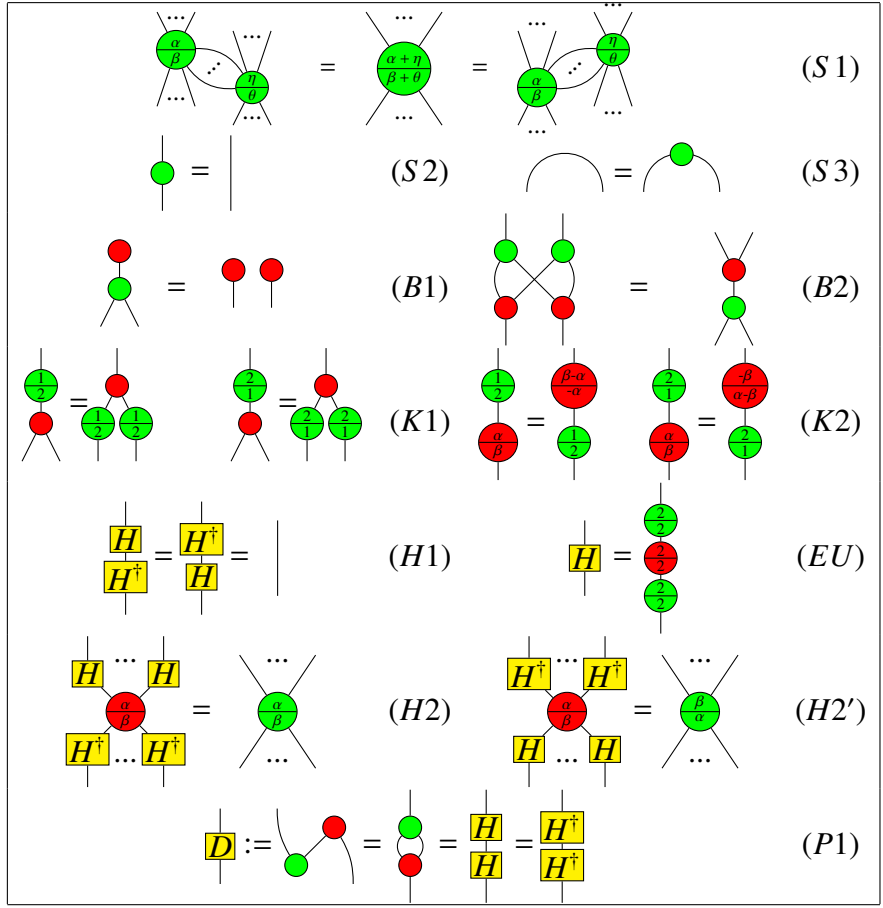

\begin{center}
\[
\quad \qquad\begin{array}{|cccc|}
\hline
\multicolumn{3}{|c}{%
\beginpgfgraphicnamed{Qutrits//spidernew}
\InputIfFileExists{Qutrits//spidernew.tikz}{}{\input{./figures/Qutrits//spidernew.tikz}}
\endpgfgraphicnamed}&(S1)\\
%\tikzfig{Qutrits//s2}&(S2)&\tikzfig{Qutrits//cupswap}&(S3)\\
%
\beginpgfgraphicnamed{diagrams//s2new}
\InputIfFileExists{diagrams//s2new.tikz}{}{\input{./figures/diagrams//s2new.tikz}}
\endpgfgraphicnamed  &(S2)& %
\beginpgfgraphicnamed{diagrams//induced_compact_structure-2wirelt}
\InputIfFileExists{diagrams//induced_compact_structure-2wirelt.tikz}{}{\input{./figures/diagrams//induced_compact_structure-2wirelt.tikz}}
\endpgfgraphicnamed=%
\beginpgfgraphicnamed{diagrams//induced_compact_structure-2wirert}
\InputIfFileExists{diagrams//induced_compact_structure-2wirert.tikz}{}{\input{./figures/diagrams//induced_compact_structure-2wirert.tikz}}
\endpgfgraphicnamed&(S3)\\
\beginpgfgraphicnamed{Qutrits//b1}
\InputIfFileExists{Qutrits//b1.tikz}{}{\input{./figures/Qutrits//b1.tikz}}
\endpgfgraphicnamed&(B1)&%
\beginpgfgraphicnamed{Qutrits//b2}
\InputIfFileExists{Qutrits//b2.tikz}{}{\input{./figures/Qutrits//b2.tikz}}
\endpgfgraphicnamed&(B2)\\
\beginpgfgraphicnamed{Qutrits//k1n}
\InputIfFileExists{Qutrits//k1n.tikz}{}{\input{./figures/Qutrits//k1n.tikz}}
\endpgfgraphicnamed&(K1)&%
\beginpgfgraphicnamed{Qutrits//k2n}
\InputIfFileExists{Qutrits//k2n.tikz}{}{\input{./figures/Qutrits//k2n.tikz}}
\endpgfgraphicnamed&(K2)\\
%\multicolumn{3}{|c}{\tikzfig{Qutrits//k1}}&(K1)\\
%\multicolumn{3}{|c}{\tikzfig{Qutrits//k2}}&(K2)\\
%
\beginpgfgraphicnamed{Qutrits//h1}
\InputIfFileExists{Qutrits//h1.tikz}{}{\input{./figures/Qutrits//h1.tikz}}
\endpgfgraphicnamed&(H1)&%
\beginpgfgraphicnamed{Qutrits//HadaDecom}
\InputIfFileExists{Qutrits//HadaDecom.tikz}{}{\input{./figures/Qutrits//HadaDecom.tikz}}
\endpgfgraphicnamed&(EU)\\
%\multicolumn{3}{|c}{\tikzfig{Qutrits//p1s}}&(P1)\\
%\tikzfig{Qutrits//deri2_rotate}&(R)&\tikzfig{Qutrits//h2prime}&(H2^\prime)\\\tikzfig{Qutrits//greencommute}&(S3)
%
\beginpgfgraphicnamed{Qutrits//h2}
\InputIfFileExists{Qutrits//h2.tikz}{}{\input{./figures/Qutrits//h2.tikz}}
\endpgfgraphicnamed&(H2)&%
\beginpgfgraphicnamed{Qutrits//h2prime}
\InputIfFileExists{Qutrits//h2prime.tikz}{}{\input{./figures/Qutrits//h2prime.tikz}}
\endpgfgraphicnamed&(H2^\prime)\\
\multicolumn{3}{|c}{%
\beginpgfgraphicnamed{Qutrits//p1s}
\InputIfFileExists{Qutrits//p1s.tikz}{}{\input{./figures/Qutrits//p1s.tikz}}
\endpgfgraphicnamed}&(P1)\\
\hline
\end{array}\]
\end{center}

  \caption{Qutrit ZX-calculus rewriting rules}\label{figuretrit}
\end{figure}

 The qutrit ZX-calculus has the structural rules as given in (\ref{compactstructure})  and (\ref{compactstructureslide}). Its non-structural rewriting rules are presented in Figure \ref{figuretrit}.

Note that all the diagrams should be read from top to bottom.

\FloatBarrier

For the easiness of reading, we will denote the
frequently used angles $\frac{2 \pi}{3}$ and   $\frac{4 \pi}{3}$ by $1$ and $2$ respectively. Note that the red spider rule still holds, for simplicity, we also refer to it as rule (S1).

The diagrams in Qutrit ZX-calculus have a standard interpretation (up to a non-zero scalar) $\llbracket \cdot \rrbracket$ in the Hilbert spaces:

\begin{center}
\[
\left\llbracket%
\beginpgfgraphicnamed{Qutrits//generator_spider}
\InputIfFileExists{Qutrits//generator_spider.tikz}{}{\input{./figures/Qutrits//generator_spider.tikz}}
\endpgfgraphicnamed\right\rrbracket=\ket{0}^{\otimes m}\bra{0}^{\otimes n}+e^{i\alpha}\ket{1}^{\otimes m}\bra{1}^{\otimes n}+e^{i\beta}\ket{2}^{\otimes m}\bra{2}^{\otimes n}
\]

\[
\left\llbracket%
\beginpgfgraphicnamed{Qutrits//generator_spider_gray}
\InputIfFileExists{Qutrits//generator_spider_gray.tikz}{}{\input{./figures/Qutrits//generator_spider_gray.tikz}}
\endpgfgraphicnamed\right\rrbracket=\ket{+}^{\otimes m}\bra{+}^{\otimes n}+e^{i\alpha}\ket{\omega}^{\otimes m}\bra{\omega}^{\otimes n}+e^{i\beta}\ket{\bar{\omega}}^{\otimes m}\bra{\bar{\omega}}^{\otimes n}
\]

\[
\left\llbracket%
\beginpgfgraphicnamed{Qutrits//RGg_Hada}
\InputIfFileExists{Qutrits//RGg_Hada.tikz}{}{\input{./figures/Qutrits//RGg_Hada.tikz}}
\endpgfgraphicnamed\right\rrbracket=\ket{+}\bra{0}+ \ket{\omega}\bra{1}+\ket{\bar{\omega}}\bra{2}=\ket{0}\bra{+}+ \ket{1}\bra{\bar{\omega}}+\ket{2}\bra{\omega}\]
\[
\left\llbracket%
\beginpgfgraphicnamed{Qutrits//RGg_Hadad}
\InputIfFileExists{Qutrits//RGg_Hadad.tikz}{}{\input{./figures/Qutrits//RGg_Hadad.tikz}}
\endpgfgraphicnamed\right\rrbracket=\ket{0}\bra{+}+ \ket{1}\bra{\omega}+\ket{2}\bra{\bar{\omega}}=\ket{+}\bra{0}+ \ket{\omega}\bra{2}+\ket{\bar{\omega}}\bra{1}
\]
\end{center}

where $\bar{\omega}=e^{\frac{4}{3}\pi i}=\omega^2$, and
 \begin{center}
 $
   \left\{\begin{array}{rcl}
      \ket{+} & = & \ket{0}+\ket{1}+\ket{2}\\
      \ket{\omega} & = &  \ket{0}+\omega \ket{1}+\bar{\omega}\ket{2}\\
      \ket{\bar{\omega}} & = & \ket{0}+\bar{\omega}\ket{1}+\omega\ket{2}
    \end{array}\right.
   $
 \end{center}

with  $\omega$ satisfying $\omega^3=1,~~1+\omega+\bar{\omega}=0$. 

For convenience, we also use the following matrix form:
  \begin{equation}\label{phasematrix}
  \left\llbracket%
\beginpgfgraphicnamed{Qutrits//RGg_zph_ab}
\InputIfFileExists{Qutrits//RGg_zph_ab.tikz}{}{\input{./figures/Qutrits//RGg_zph_ab.tikz}}
\endpgfgraphicnamed\right\rrbracket=
  \left(
\begin{array}{ccc}
 1 & 0 & 0 \\
 0 & e^{i\alpha} & 0\\
 0 & 0 & e^{i\beta}
\end{array}
\right)
\end{equation}
  %$$
%$$
  \begin{equation}\label{phasematrix2}
\left\llbracket%
\beginpgfgraphicnamed{Qutrits//RGg_xph_ab}
\InputIfFileExists{Qutrits//RGg_xph_ab.tikz}{}{\input{./figures/Qutrits//RGg_xph_ab.tikz}}
\endpgfgraphicnamed\right\rrbracket=
  \left(
\begin{array}{ccc}
1+e^{i\alpha}+e^{i\beta} & 1+\bar{\omega} e^{i\alpha}+\omega e^{i\beta}& 1+\omega e^{i\alpha}+\bar{\omega}e^{i\beta} \\
1+\omega e^{i\alpha}+\bar{\omega}e^{i\beta}& 1+e^{i\alpha}+e^{i\beta} & 1+\bar{\omega}e^{i\alpha}+\omega e^{i\beta}\\
 1+\bar{\omega} e^{i\alpha}+\omega e^{i\beta}&1+\omega e^{i\alpha}+\bar{\omega}e^{i\beta} &1+e^{i\alpha}+e^{i\beta}
\end{array}
\right)
  \end{equation}
  
\begin{equation}\label{phasematrix3}
\left\llbracket%
\beginpgfgraphicnamed{Qutrits//RGg_Hada}
\InputIfFileExists{Qutrits//RGg_Hada.tikz}{}{\input{./figures/Qutrits//RGg_Hada.tikz}}
\endpgfgraphicnamed\right\rrbracket=\begin{pmatrix}
        1 & 1 & 1 \\
        1 & \omega & \bar{\omega}\\
        1 & \bar{\omega} & \omega
 \end{pmatrix}
\end{equation}
  
Like the qubit case, there are three important properties about the qutrit ZX-calculus, i.e. universality,  soundness, and completeness: Universality is about if there exists a ZX-calculus diagram for every corresponding linear map in Hilbert spaces under the standard interpretation. Soundness means that all the rules in the qutrit ZX-calculus have a correct standard interpretation in the  Hilbert spaces. Completeness is concerned with whether  an equation of diagrams can be derived in the ZX-calculus when their corresponding equation of linear maps under the standard interpretation holds true.  Among these properties, universality is proved in \cite{BianWang2}. Soundness can easily be checked with the standard interpretation $\llbracket \cdot \rrbracket$.  

To represent qutrit graph states in the ZX-calculus, we first show that the horizontal  nodes $H$ and $H^{\dagger}$  make sense when connected to two green nodes. 

\begin{lemma}\cite{GongWang}\label{controlz}
\begin{align}\label{controlzhold}
\beginpgfgraphicnamed{Qutrits/controlz}
\InputIfFileExists{Qutrits/controlz.tikz}{}{\input{./figures/Qutrits/controlz.tikz}}
\endpgfgraphicnamed=%
\beginpgfgraphicnamed{Qutrits/controlz2}
\InputIfFileExists{Qutrits/controlz2.tikz}{}{\input{./figures/Qutrits/controlz2.tikz}}
\endpgfgraphicnamed=: %
\beginpgfgraphicnamed{Qutrits/controlz3}
\InputIfFileExists{Qutrits/controlz3.tikz}{}{\input{./figures/Qutrits/controlz3.tikz}}
\endpgfgraphicnamed,\quad\quad\quad\quad
%\end{align*}
%\begin{align*}
%
\beginpgfgraphicnamed{Qutrits/controlzsquare}
\InputIfFileExists{Qutrits/controlzsquare.tikz}{}{\input{./figures/Qutrits/controlzsquare.tikz}}
\endpgfgraphicnamed=%
\beginpgfgraphicnamed{Qutrits/controlzsquare2}
\InputIfFileExists{Qutrits/controlzsquare2.tikz}{}{\input{./figures/Qutrits/controlzsquare2.tikz}}
\endpgfgraphicnamed=: %
\beginpgfgraphicnamed{Qutrits/controlzsquare3}
\InputIfFileExists{Qutrits/controlzsquare3.tikz}{}{\input{./figures/Qutrits/controlzsquare3.tikz}}
\endpgfgraphicnamed.
\end{align}
\end{lemma}

\begin{proof}
We will only prove the first equation, since the proof for the second one is similar. 
\begin{align*}
\beginpgfgraphicnamed{Qutrits/controlzprf}
\InputIfFileExists{Qutrits/controlzprf.tikz}{}{\input{./figures/Qutrits/controlzprf.tikz}}
\endpgfgraphicnamed.
 \end{align*}
%where we used (H2) for the first and the fifth equalities, and (P1) for the second equality. 
\end{proof}

We also list some useful properties which have been proved in \cite{GongWang}  or  \cite{BianWang2}:
\begin{equation}\label{property1}
\beginpgfgraphicnamed{Qutrits//p2s}
\InputIfFileExists{Qutrits//p2s.tikz}{}{\input{./figures/Qutrits//p2s.tikz}}
\endpgfgraphicnamed
\end{equation}
\begin{equation}\label{property2}
\beginpgfgraphicnamed{Qutrits//deri2_rotate}
\InputIfFileExists{Qutrits//deri2_rotate.tikz}{}{\input{./figures/Qutrits//deri2_rotate.tikz}}
\endpgfgraphicnamed
\end{equation}

\begin{equation}\label{property3}
\beginpgfgraphicnamed{Qutrits//controlnotslide}
\InputIfFileExists{Qutrits//controlnotslide.tikz}{}{\input{./figures/Qutrits//controlnotslide.tikz}}
\endpgfgraphicnamed
\end{equation}
\begin{equation}\label{property4}
\beginpgfgraphicnamed{Qutrits//deri1_3break}
\InputIfFileExists{Qutrits//deri1_3break.tikz}{}{\input{./figures/Qutrits//deri1_3break.tikz}}
\endpgfgraphicnamed
\end{equation}
\begin{equation}\label{property5}
\beginpgfgraphicnamed{Qutrits//quasibialgebra}
\InputIfFileExists{Qutrits//quasibialgebra.tikz}{}{\input{./figures/Qutrits//quasibialgebra.tikz}}
\endpgfgraphicnamed
\end{equation}
\begin{equation}\label{property6}
\beginpgfgraphicnamed{Qutrits//hadslidegn}
\InputIfFileExists{Qutrits//hadslidegn.tikz}{}{\input{./figures/Qutrits//hadslidegn.tikz}}
\endpgfgraphicnamed
\end{equation}
\begin{align}\label{property7}
\beginpgfgraphicnamed{Qutrits//twocontrolz3}
\InputIfFileExists{Qutrits//twocontrolz3.tikz}{}{\input{./figures/Qutrits//twocontrolz3.tikz}}
\endpgfgraphicnamed= %
\beginpgfgraphicnamed{Qutrits//controlzsquare3}
\InputIfFileExists{Qutrits//controlzsquare3.tikz}{}{\input{./figures/Qutrits//controlzsquare3.tikz}}
\endpgfgraphicnamed,\quad\quad %
\beginpgfgraphicnamed{Qutrits//twocontrolzsquare3}
\InputIfFileExists{Qutrits//twocontrolzsquare3.tikz}{}{\input{./figures/Qutrits//twocontrolzsquare3.tikz}}
\endpgfgraphicnamed= %
\beginpgfgraphicnamed{Qutrits//controlz3}
\InputIfFileExists{Qutrits//controlz3.tikz}{}{\input{./figures/Qutrits//controlz3.tikz}}
\endpgfgraphicnamed,
\quad\quad %
\beginpgfgraphicnamed{Qutrits//hhdaggern}
\InputIfFileExists{Qutrits//hhdaggern.tikz}{}{\input{./figures/Qutrits//hhdaggern.tikz}}
\endpgfgraphicnamed= %
\beginpgfgraphicnamed{Qutrits//hdaggerhn}
\InputIfFileExists{Qutrits//hdaggerhn.tikz}{}{\input{./figures/Qutrits//hdaggerhn.tikz}}
\endpgfgraphicnamed= %
\beginpgfgraphicnamed{Qutrits//parallel}
\InputIfFileExists{Qutrits//parallel.tikz}{}{\input{./figures/Qutrits//parallel.tikz}}
\endpgfgraphicnamed.
\end{align}
\begin{align}\label{property8}
\beginpgfgraphicnamed{Qutrits//zxz-xzx10}
\InputIfFileExists{Qutrits//zxz-xzx10.tikz}{}{\input{./figures/Qutrits//zxz-xzx10.tikz}}
\endpgfgraphicnamed 
\end{align} 

Note that the red-green colour swap version of these properties also holds.

\begin{remark}
In the qubit ZX-calculus, all its diagrams obey a meta-rule which has been formalized as a well-known slogan: ``only the topology matters"  \cite{CoeckeDuncan}, or more recently as ``only connectedness matters" \cite{coeckewang}. This means, 
any diagram of the qubit ZX-calculus can be transformed into a semantically equal diagram (representing the same matrix) by
moving the components around without changing their connections, thus rendering the qubit ZX-calculus unoriented. However, as one can see in (\ref{property3}), this meta-rule does not hold for all diagrams in the qutrit ZX-calculus. Therefore, one should be careful with the orientation of a qutrit ZX diagram, especially when there is a connection between green and red nodes. 
\end{remark}

\subsection{Qutrit stabilizer quantum mechanics in  the ZX-calculus}
In this subsection, we represent in ZX-calculus the qutrit stabilizer QM, which consists of  state preparations and measurements based on the computational basis $\{ \ket{0},  \ket{1},  \ket{2}\}$, as well as unitary operators belonging to the generalized Clifford group $\mathcal{C}_n$. Furthermore, we give the unique form for single qutrit Clifford operators.

 Firstly, the states and effects in computational basis can be represented as:

\begin{equation}\label{compbasis}
\ket{0} =\left\llbracket%
\beginpgfgraphicnamed{Qutrits//RGg_exd}
\InputIfFileExists{Qutrits//RGg_exd.tikz}{}{\input{./figures/Qutrits//RGg_exd.tikz}}
\endpgfgraphicnamed\right\rrbracket, \quad
\ket{1} =\left\llbracket%
\beginpgfgraphicnamed{Qutrits//rdot1}
\InputIfFileExists{Qutrits//rdot1.tikz}{}{\input{./figures/Qutrits//rdot1.tikz}}
\endpgfgraphicnamed\right\rrbracket, \quad
\ket{2} =\left\llbracket%
\beginpgfgraphicnamed{Qutrits//rdot2}
\InputIfFileExists{Qutrits//rdot2.tikz}{}{\input{./figures/Qutrits//rdot2.tikz}}
\endpgfgraphicnamed\right\rrbracket, \quad
\bra{0}=\left\llbracket%
\beginpgfgraphicnamed{Qutrits//RGg_ex}
\InputIfFileExists{Qutrits//RGg_ex.tikz}{}{\input{./figures/Qutrits//RGg_ex.tikz}}
\endpgfgraphicnamed\right\rrbracket,\quad
\bra{1} =\left\llbracket%
\beginpgfgraphicnamed{Qutrits//rcdot1}
\InputIfFileExists{Qutrits//rcdot1.tikz}{}{\input{./figures/Qutrits//rcdot1.tikz}}
\endpgfgraphicnamed\right\rrbracket, \quad
\bra{2} =\left\llbracket%
\beginpgfgraphicnamed{Qutrits//rcdot2}
\InputIfFileExists{Qutrits//rcdot2.tikz}{}{\input{./figures/Qutrits//rcdot2.tikz}}
\endpgfgraphicnamed\right\rrbracket.
\end{equation}
The generators of the generalized  Clifford group $\mathcal{C}_n$  are denoted by the following diagrams:

\begin{equation}\label{clifgenerator}
\mathcal{S}=\left\llbracket%
\beginpgfgraphicnamed{Qutrits//phasegts}
\InputIfFileExists{Qutrits//phasegts.tikz}{}{\input{./figures/Qutrits//phasegts.tikz}}
\endpgfgraphicnamed\right\rrbracket,
\qquad
H=\left\llbracket%
\beginpgfgraphicnamed{Qutrits//RGg_Hada}
\InputIfFileExists{Qutrits//RGg_Hada.tikz}{}{\input{./figures/Qutrits//RGg_Hada.tikz}}
\endpgfgraphicnamed\right\rrbracket, 
\qquad
\Lambda=\left\llbracket%
\beginpgfgraphicnamed{Qutrits//sumgt}
\InputIfFileExists{Qutrits//sumgt.tikz}{}{\input{./figures/Qutrits//sumgt.tikz}}
\endpgfgraphicnamed\right\rrbracket.
\end{equation}

\begin{lemma} \label{equivalentstabilizer}
 A pure n-qutrit quantum state $\ket{\psi}$ is  a stabilizer state if and only if there exists an n-qutrit Clifford unitary $U$  such that $\ket{\psi}=U\ket{0}^{\otimes n}$.
\end{lemma}

 \begin{proof}
 If a pure n-qutrit quantum state $\ket{\psi}$  is  a stabilizer state, then by Lemma \ref{stabilizerequivgraphstate},  there exists a Clifford operator $U^{\prime}$ such that $\ket{\psi}=U^{\prime}\ket{G}$, where 
 $\ket{G}=\mathcal{U}\ket{+}^{\otimes n}$ is a graph state, $\mathcal{U}=\prod_{\{l,m\}\in E}(C_{lm})^{\Gamma_{lm}}$, and  $C_{lm}=\Sigma_{j=0}^{2}\Sigma_{k=0}^{2}\omega^{jk}\ket{j}\bra{j}_l\otimes \ket{k}\bra{k}_m$.  It is shown in \cite{GongWang} that \[
C_{lm}=\left\llbracket%
\beginpgfgraphicnamed{Qutrits//controlz}
\InputIfFileExists{Qutrits//controlz.tikz}{}{\input{./figures/Qutrits//controlz.tikz}}
\endpgfgraphicnamed\right\rrbracket=\left\llbracket %
\beginpgfgraphicnamed{Qutrits//clm2}
\InputIfFileExists{Qutrits//clm2.tikz}{}{\input{./figures/Qutrits//clm2.tikz}}
\endpgfgraphicnamed\right\rrbracket,
\] which means $C_{lm}$ is a Clifford operator, so is $\mathcal{U}$. Let $V=\mathcal{U}H^{\otimes n}$, then 
 $V$ is also a Clifford operator, and $\ket{G}=V\ket{0}^{\otimes n}$. Therefore, $\ket{\psi}=U^{\prime}\ket{G}=U^{\prime}V\ket{0}^{\otimes n}$. Let $U=U^{\prime}V$, then $\ket{\psi}=U\ket{0}^{\otimes n}$, and $U$ is a Clifford unitary.
 
On the contrary, assume $\ket{\psi}=U\ket{0}^{\otimes n}$, where $U$ is a Clifford unitary. Let  $\ket{G}=\mathcal{U}\ket{+}^{\otimes n}$ be an n-qutrit  graph state,  
$V=\mathcal{U}H^{\otimes n}$, and $W=UV^{\dagger}$. Then $\ket{\psi}=WV\ket{0}^{\otimes n}=W\ket{G}$. Suppose $\mathfrak{S}$ is the stabilizer group of $\ket{G}$. Since 
$WSW^{\dagger}\ket{\psi}=WSW^{\dagger}W\ket{G}=WS\ket{G}=W\ket{G}=\ket{\psi},  \forall S \in \mathfrak{S}$, it follows that $W\mathfrak{S}W^{\dagger}$  is the  stabilizer group of $\ket{\psi}$, i.e., $\ket{\psi}$  is  a stabilizer state.
 \end{proof}

\begin{lemma}\label{stabaszx}
In the qutrit formalism,  any stabilizer state, Clifford unitary,  or post-selected measurement can be represented by a ZX-calculus diagram with phase angles just integer multiples of $\frac{2}{3}\pi$.
\end{lemma}
 \begin{proof}
Firstly, by Lemma \ref{equivalentstabilizer}, each stabilizer state can be  obtained by applying an n-qutrit Clifford unitary to the computational basis state $\ket{0}^{\otimes n}$. Secondly, any Clifford unitary can be generated by the gates $\mathcal{S}, H$ and 
$\Lambda$ \cite{Hostens, Shawn}, which are clearly composed of phases with angles integer multiples of $\frac{2}{3}\pi$.  Since the generators are composed of  phases with angles integer multiples of $\frac{2}{3}\pi$, so are Clifford unitaries generated by these generators. Thirdly, computational basis states and post-selected computational basis measurements are clearly with phase angles integer multiples of $\frac{2}{3}\pi$. Therefore, by  (\ref{compbasis})  and (\ref{clifgenerator}),  any stabilizer state, Clifford unitary,  or post-selected measurement can be represented by a ZX-calculus diagram with phase angles just integer multiples of $\frac{2}{3}\pi$.
 \end{proof}

\begin{lemma}\label{diagramdecom}
Each qutrit ZX-calculus diagram can be represented by a combination of the following components:
\begin{equation}\label{component}
\begin{array}{cccccccc}
\beginpgfgraphicnamed{Qutrits//RGg_ezd}
\InputIfFileExists{Qutrits//RGg_ezd.tikz}{}{\input{./figures/Qutrits//RGg_ezd.tikz}}
\endpgfgraphicnamed,& %
\beginpgfgraphicnamed{Qutrits//RGg_ez}
\InputIfFileExists{Qutrits//RGg_ez.tikz}{}{\input{./figures/Qutrits//RGg_ez.tikz}}
\endpgfgraphicnamed, & %
\beginpgfgraphicnamed{Qutrits//RGg_dz}
\InputIfFileExists{Qutrits//RGg_dz.tikz}{}{\input{./figures/Qutrits//RGg_dz.tikz}}
\endpgfgraphicnamed,
&%
\beginpgfgraphicnamed{Qutrits//RGg_dzd}
\InputIfFileExists{Qutrits//RGg_dzd.tikz}{}{\input{./figures/Qutrits//RGg_dzd.tikz}}
\endpgfgraphicnamed,
\beginpgfgraphicnamed{Qutrits//RGg_zph_ab}
\InputIfFileExists{Qutrits//RGg_zph_ab.tikz}{}{\input{./figures/Qutrits//RGg_zph_ab.tikz}}
\endpgfgraphicnamed,  &  %
\beginpgfgraphicnamed{Qutrits//RGg_xph_ab}
\InputIfFileExists{Qutrits//RGg_xph_ab.tikz}{}{\input{./figures/Qutrits//RGg_xph_ab.tikz}}
\endpgfgraphicnamed, &  %
\beginpgfgraphicnamed{Qutrits//RGg_Hada}
\InputIfFileExists{Qutrits//RGg_Hada.tikz}{}{\input{./figures/Qutrits//RGg_Hada.tikz}}
\endpgfgraphicnamed,
& %
\beginpgfgraphicnamed{Qutrits//RGg_Hadad}
\InputIfFileExists{Qutrits//RGg_Hadad.tikz}{}{\input{./figures/Qutrits//RGg_Hadad.tikz}}
\endpgfgraphicnamed,
 \end{array}
 \end{equation} where $\alpha, \beta \in [0, 2\pi)$.
\end{lemma}
The proof follows directly from the qutrit spider rule (S1) and the colour change rules (H2) and (H$2^{\prime}$).

\begin{lemma}\label{zxasstab}
Each qutrit ZX-calculus diagram with phase angles integer multiples of $\frac{2}{3}\pi$ corresponds to an element of the qutrit stabilizer QM.
\end{lemma}

\begin{proof}
By Lemma \ref{diagramdecom}, each ZX-calculus diagram can be decomposed into basic components as shown in (\ref{component}). Since composition of diagrams corresponds to composition of  corresponding matrices in QM, we only need to
consider those basic components where the phase angles are integer multiples of $\frac{2}{3}\pi$, in such a way that turns the standard interpretation of each basic component into a composition of Clifford unitaries, computational basis states, and computational basis effects. We have  tackled the basic components %
\beginpgfgraphicnamed{Qutrits//phasegts}
\InputIfFileExists{Qutrits//phasegts.tikz}{}{\input{./figures/Qutrits//phasegts.tikz}}
\endpgfgraphicnamed  and %
\beginpgfgraphicnamed{Qutrits//RGg_Hada}
\InputIfFileExists{Qutrits//RGg_Hada.tikz}{}{\input{./figures/Qutrits//RGg_Hada.tikz}}
\endpgfgraphicnamed  in (\ref{clifgenerator}), the remaining components are dealt with as follows:
\iffalse
Firstly, note that the class of ZX-calculus diagram in which all phase angles are integer multiples of $\frac{2}{3}\pi$ is closed under the rewrite rules.
Secondly, by Lemma \ref{diagramdecom}, any ZX-calculus diagram can be decomposed into four basic spiders plus phase shifts and Hadamard nodes. For each of these diagram generators, we exhibit a decomposition of the corresponding operator into the Clifford generators, computational basis states, and computational basis effects. In addition to the translations for
%
\beginpgfgraphicnamed{Qutrits//phasegts}
\InputIfFileExists{Qutrits//phasegts.tikz}{}{\input{./figures/Qutrits//phasegts.tikz}}
\endpgfgraphicnamed   and %
\beginpgfgraphicnamed{Qutrits//RGg_Hada}
\InputIfFileExists{Qutrits//RGg_Hada.tikz}{}{\input{./figures/Qutrits//RGg_Hada.tikz}}
\endpgfgraphicnamed given in (\ref{clifgenerator}),
 this gives the following decompositions:
 \fi
 %\tikzfig{Qutrits//RGg_ezd},\qquad \tikzfig{Qutrits//RGg_ez}
 
  \begin{equation}
 \left\llbracket%
\beginpgfgraphicnamed{Qutrits//RGg_ezd}
\InputIfFileExists{Qutrits//RGg_ezd.tikz}{}{\input{./figures/Qutrits//RGg_ezd.tikz}}
\endpgfgraphicnamed\right\rrbracket=
\left\llbracket%
\beginpgfgraphicnamed{Qutrits//gdotascliff}
\InputIfFileExists{Qutrits//gdotascliff.tikz}{}{\input{./figures/Qutrits//gdotascliff.tikz}}
\endpgfgraphicnamed\right\rrbracket=H\ket{0}
\end{equation}

\begin{equation}
 \left\llbracket%
\beginpgfgraphicnamed{Qutrits//RGg_ez}
\InputIfFileExists{Qutrits//RGg_ez.tikz}{}{\input{./figures/Qutrits//RGg_ez.tikz}}
\endpgfgraphicnamed\right\rrbracket=
\left\llbracket%
\beginpgfgraphicnamed{Qutrits//gcdotascliff}
\InputIfFileExists{Qutrits//gcdotascliff.tikz}{}{\input{./figures/Qutrits//gcdotascliff.tikz}}
\endpgfgraphicnamed\right\rrbracket=\bra{0}H
\end{equation}

 \begin{equation}
 \left\llbracket%
\beginpgfgraphicnamed{Qutrits//RGg_dz}
\InputIfFileExists{Qutrits//RGg_dz.tikz}{}{\input{./figures/Qutrits//RGg_dz.tikz}}
\endpgfgraphicnamed\right\rrbracket=
\left\llbracket%
\beginpgfgraphicnamed{Qutrits//gcopyascliff}
\InputIfFileExists{Qutrits//gcopyascliff.tikz}{}{\input{./figures/Qutrits//gcopyascliff.tikz}}
\endpgfgraphicnamed\right\rrbracket=\Lambda\circ(I \otimes \ket{0})
\end{equation}
 \begin{equation}
 \left\llbracket%
\beginpgfgraphicnamed{Qutrits//RGg_dzd}
\InputIfFileExists{Qutrits//RGg_dzd.tikz}{}{\input{./figures/Qutrits//RGg_dzd.tikz}}
\endpgfgraphicnamed\right\rrbracket=
\left\llbracket%
\beginpgfgraphicnamed{Qutrits//gcocopyascliff}
\InputIfFileExists{Qutrits//gcocopyascliff.tikz}{}{\input{./figures/Qutrits//gcocopyascliff.tikz}}
\endpgfgraphicnamed\right\rrbracket=(I \otimes \bra{0})\circ\Lambda(I \otimes (H\circ H))
\end{equation}
Thus we conclude the proof.
\end{proof} 
bringing together

Bringing together Lemmas \ref{stabaszx} and \ref{zxasstab}, we have 
\begin{theorem}
The ZX-calculus for pure qutrit stabilizer QM comprises exactly the diagrams with phase angles integer multiples of $\frac{2}{3}\pi$.
\end{theorem}

In the sequel, a diagram in which all phase angles are integer multiples of $\frac{2}{3}\pi$ will be called a stabilizer diagram. Now we denote $\frac{2 \pi}{3}$ and  $\frac{4 \pi}{3}$  by $1$ and $2$ (or $-1$)  respectively, and let 
 $$\mathcal{P}= \{\nststile{1}{1}, \nststile{2}{2}\}, \quad \mathcal{N}=\{\nststile{0}{1},  \nststile{1}{0}, \nststile{0}{2},  \nststile{2}{0}\}, \quad \mathcal{M}=\{\nststile{0}{0},  \nststile{1}{2}, \nststile{2}{1}\}, \quad \mathcal{Q}=\mathcal{P} \cup \mathcal{N}, \quad \mathcal{A}=\mathcal{Q} \cup \mathcal{M},$$

%First we fix the notations for some sets that will be frequently uesd. 

where the symbol $\nststile{b}{a}$ is just the denotation of the pair $(a, b)$, instead of the fraction $\frac{a}{b}$. 
Then each green or red node in a stabilizer diagram has its phase angles denoted as elements in the set  $\mathcal{A}$.  Define the addition $+$ in  $\mathcal{A}$ as $\nststile{b}{a}+\nststile{d}{c} ~:=~ \nststile{b+d (mod 3)}{a+c (mod3)}$. Then $\mathcal{A}$ is an abelian group, with $\mathcal{P}\cup \{\nststile{0}{0}\}~ \mbox{and} ~\mathcal{M}$ as subgroups. Note that the elements in  $\mathcal{P}$  resemble to the phases $\frac{\pi}{2}$ and $-\frac{\pi}{2}$ in the quibit ZX-calculus, and the non-zero elements in
 $\mathcal{M}$ are analogous to the phase $\pi$ in the quibit ZX-calculus.

With the notation of $\nststile{b}{a}$ and the addition defined above, the commutation rule (K2) has a special form for elements in  $\mathcal{P}$:

\begin{lemma}
 \begin{equation}\label{K2decom}
\beginpgfgraphicnamed{Qutrits//K2decom}
\InputIfFileExists{Qutrits//K2decom.tikz}{}{\input{./figures/Qutrits//K2decom.tikz}}
\endpgfgraphicnamed
\end{equation}
where $\nststile{m_2}{m_1}, \nststile{m_4}{m_3} \in  \mathcal{M},  \nststile{p_1}{p_1} \in \mathcal{P}.$
\end{lemma}

\begin{proof}
There are several similar cases to be verified here. We only show one case where $\nststile{m_2}{m_1}=\nststile{2}{1},  \nststile{p_1}{p_1} =\nststile{1}{1}$:
 \begin{equation*}
\beginpgfgraphicnamed{Qutrits//K2decomproof}
\InputIfFileExists{Qutrits//K2decomproof.tikz}{}{\input{./figures/Qutrits//K2decomproof.tikz}}
\endpgfgraphicnamed
\end{equation*}
\end{proof}

\begin{lemma}\label{pzxz}
Any single qutrit stabilizer diagram of form

 \begin{equation}\label{pzxzreduce}
\beginpgfgraphicnamed{Qutrits//zxz0}
\InputIfFileExists{Qutrits//zxz0.tikz}{}{\input{./figures/Qutrits//zxz0.tikz}}
\endpgfgraphicnamed
\end{equation}
 can be rewritten as
  \begin{equation}\label{pzxzreduce2}
\beginpgfgraphicnamed{Qutrits//zxz01}
\InputIfFileExists{Qutrits//zxz01.tikz}{}{\input{./figures/Qutrits//zxz01.tikz}}
\endpgfgraphicnamed \quad \mbox{or} \quad %
\beginpgfgraphicnamed{Qutrits//zxz02}
\InputIfFileExists{Qutrits//zxz02.tikz}{}{\input{./figures/Qutrits//zxz02.tikz}}
\endpgfgraphicnamed,
 \end{equation}
 where $ \nststile{p_{i2}}{p_{i1}}, \nststile{\bar{p}_{j2}}{\bar{p}_{j1}}  \in \mathcal{P}, \forall i\in\{1,2,3,4\}, \forall j\in\{1,2,3\}$. 
\end{lemma}
\begin{proof}
First we list all the possible forms of (\ref{pzxzreduce}) and their rewritten forms of (\ref{pzxzreduce2})  as follows:
\begin{align}\label{pzxzhh}
\beginpgfgraphicnamed{Qutrits//zxz-xzx1}
\InputIfFileExists{Qutrits//zxz-xzx1.tikz}{}{\input{./figures/Qutrits//zxz-xzx1.tikz}}
\endpgfgraphicnamed\quad &\quad %
\beginpgfgraphicnamed{Qutrits//zxz-xzx2}
\InputIfFileExists{Qutrits//zxz-xzx2.tikz}{}{\input{./figures/Qutrits//zxz-xzx2.tikz}}
\endpgfgraphicnamed  \quad &\quad %
\beginpgfgraphicnamed{Qutrits//zxz-xzx3}
\InputIfFileExists{Qutrits//zxz-xzx3.tikz}{}{\input{./figures/Qutrits//zxz-xzx3.tikz}}
\endpgfgraphicnamed
\quad &\quad %
\beginpgfgraphicnamed{Qutrits//zxz-xzx4}
\InputIfFileExists{Qutrits//zxz-xzx4.tikz}{}{\input{./figures/Qutrits//zxz-xzx4.tikz}}
\endpgfgraphicnamed 
\end{align}  
\begin{align}\label{pzxzhh2}
\beginpgfgraphicnamed{Qutrits//zxz-xzx5}
\InputIfFileExists{Qutrits//zxz-xzx5.tikz}{}{\input{./figures/Qutrits//zxz-xzx5.tikz}}
\endpgfgraphicnamed  \quad &\quad %
\beginpgfgraphicnamed{Qutrits//zxz-xzx6}
\InputIfFileExists{Qutrits//zxz-xzx6.tikz}{}{\input{./figures/Qutrits//zxz-xzx6.tikz}}
\endpgfgraphicnamed\quad &\quad
\beginpgfgraphicnamed{Qutrits//zxz-xzx7}
\InputIfFileExists{Qutrits//zxz-xzx7.tikz}{}{\input{./figures/Qutrits//zxz-xzx7.tikz}}
\endpgfgraphicnamed\quad &\quad %
\beginpgfgraphicnamed{Qutrits//zxz-xzx8}
\InputIfFileExists{Qutrits//zxz-xzx8.tikz}{}{\input{./figures/Qutrits//zxz-xzx8.tikz}}
\endpgfgraphicnamed  
\end{align}  

For simplicity, we just show one derivation here:
\begin{align*}
\beginpgfgraphicnamed{Qutrits//zxz-xzx9}
\InputIfFileExists{Qutrits//zxz-xzx9.tikz}{}{\input{./figures/Qutrits//zxz-xzx9.tikz}}
\endpgfgraphicnamed 
\end{align*} 
%where we used rules (Eu) and (H2$^{\prime}$) and the property (\ref{property8}). 
\end{proof}

\begin{lemma}\label{xz112}
Any single qutrit stabilizer diagram of form
\begin{equation}\label{xz111}
K=%
\beginpgfgraphicnamed{Qutrits//zxz2}
\InputIfFileExists{Qutrits//zxz2.tikz}{}{\input{./figures/Qutrits//zxz2.tikz}}
\endpgfgraphicnamed
\end{equation}

 can be rewritten as
 $$%
\beginpgfgraphicnamed{Qutrits//zxz1}
\InputIfFileExists{Qutrits//zxz1.tikz}{}{\input{./figures/Qutrits//zxz1.tikz}}
\endpgfgraphicnamed \quad \mbox{or} \quad %
\beginpgfgraphicnamed{Qutrits//zxz3}
\InputIfFileExists{Qutrits//zxz3.tikz}{}{\input{./figures/Qutrits//zxz3.tikz}}
\endpgfgraphicnamed,$$
 where $ \nststile{m_2}{m_1} \in \mathcal{M}$. 
\end{lemma}

\begin{proof}
If  $ \nststile{b_1}{a_1} \in \mathcal{M}$, then by rule $(S2)$ or $(K2)$,  this green node can be pushed down and merges with another green node thus $K$ is rewritten as  

\begin{equation}\label{xz1}
K=%
\beginpgfgraphicnamed{Qutrits//zxz4}
\InputIfFileExists{Qutrits//zxz4.tikz}{}{\input{./figures/Qutrits//zxz4.tikz}}
\endpgfgraphicnamed
\end{equation}
Otherwise,  $ \nststile{b_1}{a_1} \in \mathcal{Q}$, then $ \nststile{b_1}{a_1} =\nststile{p_{12}}{p_{11}}+\nststile{m_{12}}{m_{11}}$, where $\nststile{p_{12}}{p_{11}}\in \mathcal{P}, \nststile{m_{12}}{m_{11}}\in \mathcal{M}$ (e.g.,  $ \nststile{1}{0} =\nststile{2}{2}+\nststile{2}{1}$). Therefore, by the spider rule $(S1)$ and the commutation rule $(K2)$, the  $\mathcal{M}$ part can be separated from the  $ \nststile{b_1}{a_1}$ node and pushed down to the bottom to merge with the other green node, thus  $K$ can be rewritten as 
$$%
\beginpgfgraphicnamed{Qutrits//zxz5}
\InputIfFileExists{Qutrits//zxz5.tikz}{}{\input{./figures/Qutrits//zxz5.tikz}}
\endpgfgraphicnamed:=L
$$
Furthermore, if $\nststile{t_{12}}{t_{11}}\in \mathcal{M}$, then the red node in $L$ can be pushed up, thus $L$ is rewritten as 
\begin{equation}\label{xz2}
\beginpgfgraphicnamed{Qutrits//zxz6}
\InputIfFileExists{Qutrits//zxz6.tikz}{}{\input{./figures/Qutrits//zxz6.tikz}}
\endpgfgraphicnamed
\end{equation}
Otherwise,  $\nststile{t_{12}}{t_{11}} \in \mathcal{Q}$, then use the same trick as for $ \nststile{b_1}{a_1} \in \mathcal{Q}$, $L$ can be rewritten as 
\begin{equation*}\label{xz3}
\beginpgfgraphicnamed{Qutrits//zxz7}
\InputIfFileExists{Qutrits//zxz7.tikz}{}{\input{./figures/Qutrits//zxz7.tikz}}
\endpgfgraphicnamed:= R
\end{equation*}
where  $\nststile{p_{22}}{p_{21}}\in \mathcal{P}, \nststile{m_{22}}{m_{21}}\in \mathcal{M}$. 
We continue the similar analysis on $\nststile{s_{32}}{s_{31}}$ in $R$. If  $\nststile{s_{32}}{s_{31}}=\nststile{m_{32}}{m_{31}}\in \mathcal{M}$, then 
\begin{equation}\label{xz4}
R=%
\beginpgfgraphicnamed{Qutrits//zxz8}
\InputIfFileExists{Qutrits//zxz8.tikz}{}{\input{./figures/Qutrits//zxz8.tikz}}
\endpgfgraphicnamed
\end{equation}
where by commutation rule $(K2)$ we push the $\nststile{m_{32}}{m_{31}}$ node up and merge nodes with  same colours. 
 Otherwise, $\nststile{s_{32}}{s_{31}} \in \mathcal{Q}$, then by the same trick as for $ \nststile{b_1}{a_1} \in \mathcal{Q}$, $R$ can be rewritten as 
 
 \begin{equation}\label{xz5}
\beginpgfgraphicnamed{Qutrits//zxz9}
\InputIfFileExists{Qutrits//zxz9.tikz}{}{\input{./figures/Qutrits//zxz9.tikz}}
\endpgfgraphicnamed:= J
\end{equation}
where  $\nststile{p_{32}}{p_{31}}\in \mathcal{P}, \nststile{m_{42}}{m_{41}}\in \mathcal{M}$, and we used the fact that a geen node with phase angles in $\mathcal{M}$ must commute with a red node with phase angles also in $\mathcal{M}$.

By Lemma \ref{pzxz}, $J$ can be rewritten as 
%\begin{equation}\label{xz6}
%J=\tikzfig{Qutrits//zxz10}=\tikzfig{Qutrits//zxz1}
%\end{equation}
%or
%\begin{equation}\label{xz7}
%J= \tikzfig{Qutrits//zxz11}=\tikzfig{Qutrits//zxz3}
%\end{equation}

\begin{trivlist}\item
    \begin{minipage}{0.5\textwidth}
     \begin{equation}\label{xz6}
J=%
\beginpgfgraphicnamed{Qutrits//zxz10}
\InputIfFileExists{Qutrits//zxz10.tikz}{}{\input{./figures/Qutrits//zxz10.tikz}}
\endpgfgraphicnamed\overset{K2,S1}{=}%
\beginpgfgraphicnamed{Qutrits//zxz1}
\InputIfFileExists{Qutrits//zxz1.tikz}{}{\input{./figures/Qutrits//zxz1.tikz}}
\endpgfgraphicnamed,
\end{equation}
\end{minipage}
   \begin{minipage}{0.5\textwidth}
   \begin{equation}\label{xz7}
\mbox{or}\quad  J= %
\beginpgfgraphicnamed{Qutrits//zxz11}
\InputIfFileExists{Qutrits//zxz11.tikz}{}{\input{./figures/Qutrits//zxz11.tikz}}
\endpgfgraphicnamed\overset{\ref{property1},S1}{=}%
\beginpgfgraphicnamed{Qutrits//zxz3}
\InputIfFileExists{Qutrits//zxz3.tikz}{}{\input{./figures/Qutrits//zxz3.tikz}}
\endpgfgraphicnamed
\end{equation}
 \end{minipage}
  \end{trivlist}

\noindent where we used the commutation rule $(K2)$  to push the nodes with phase angles in $\mathcal{M}$ and the property (\ref{property1}). 

Now we finish the proof by pointing out that the single qutrit stabilizer diagrams (\ref{xz1}), (\ref{xz2}), (\ref{xz4}) are just special forms of (\ref{xz6}).
\end{proof}

\begin{corollary}\label{sinqu}
Any ZX  diagram for single qutrit Clifford operator can be rewritten in one of the forms
%\begin{equation}\label{xz8}
%\tikzfig{Qutrits//zxz1}
%\end{equation}

%\begin{equation}\label{xz9}
%\tikzfig{Qutrits//zxz12}
%\end{equation}
%where $ \nststile{m_2}{m_1} \in \mathcal{M}$. 
\begin{trivlist}\item
    \begin{minipage}{0.5\textwidth}
    \begin{equation}\label{xz8}
\beginpgfgraphicnamed{Qutrits//zxz1}
\InputIfFileExists{Qutrits//zxz1.tikz}{}{\input{./figures/Qutrits//zxz1.tikz}}
\endpgfgraphicnamed,
\end{equation}
\end{minipage}
   \begin{minipage}{0.5\textwidth}
  \begin{equation}\label{xz9}
\beginpgfgraphicnamed{Qutrits//zxz12}
\InputIfFileExists{Qutrits//zxz12.tikz}{}{\input{./figures/Qutrits//zxz12.tikz}}
\endpgfgraphicnamed
\end{equation}
 \end{minipage}
  \end{trivlist}

\end{corollary}
\begin{proof}
Because of the spider rule (S1), we only consider the case where any two adjacent nodes are of different colours. First note that single qutrit stabilizer diagram with one node or two nodes is a special form of (\ref{xz8}). If there are only three aligned nodes in the diagram,  then they are either the form of (\ref{xz8}) or (\ref{xz111}), the later can be rewritten into one of the forms (\ref{xz8}) and (\ref{xz9}) by Lemma \ref{xz112}. We can now proceed by induction: Suppose that any single qutrit stabilizer diagram $D_n$ with $n (n\geq 3)$ nodes can be rewritten into one of the forms (\ref{xz8}) and (\ref{xz9}).  For single qutrit stabilizer diagram $D_{n+1}$  with $n +1$ nodes, we consider its toppest node called $t$.  If $t$ is a red node, then by the spider rule (S1), it merges with a red node in $D_n$, thus $D_{n+1}$ is still of form  (\ref{xz8}) or (\ref{xz9}). If $t$ is a green node, after rewriting the remaining $n$ nodes  into a form of  (\ref{xz8}) or (\ref{xz9}) by the induction hypothesis, $D_{n+1}$ now has 4 interlaced green and red nodes except for Hadamard nodes. The first 3 nodes then consist a diagram of form (\ref{xz111}), applying the induction hypothesis again, $D_{n+1}$ can be reduced to a form of  (\ref{xz8}) or (\ref{xz9}), noting that we used $H^4=1$ derived from rules (H1) and (P1) in the case that 4 Hadamard notes appear in one diagram.  
\end{proof}

Now we can give a normal form for the single qutrit Clifford group $\mathcal{C}_1$. 
\begin{theorem}\label{uniformsin}
In the ZX-calculus, each element of the single qutrit Clifford group $\mathcal{C}_1$ can be uniquely represented in one of the following forms
%\begin{equation}\label{uniformsin1}
%\tikzfig{Qutrits//cliffnorformj}
%\end{equation}
%\begin{equation}\label{uniformsin2}
%\tikzfig{Qutrits//cliffnorformk}
%\end{equation}
%\begin{equation}\label{uniformsin3}
 %\tikzfig{Qutrits//cliffnorformt}
%\end{equation}
%%%%%%%%%%%%%%
\begin{trivlist}\item
    \begin{minipage}{0.295\textwidth}
      \begin{equation}
        \label{uniformsin1}
\beginpgfgraphicnamed{Qutrits/cliffnorformj}
\InputIfFileExists{Qutrits/cliffnorformj.tikz}{}{\input{./figures/Qutrits/cliffnorformj.tikz}}
\endpgfgraphicnamed,    
\end{equation}
    \end{minipage}
   \begin{minipage}{0.295\textwidth}
      \begin{equation}
        \label{uniformsin2}
\beginpgfgraphicnamed{Qutrits/cliffnorformk}
\InputIfFileExists{Qutrits/cliffnorformk.tikz}{}{\input{./figures/Qutrits/cliffnorformk.tikz}}
\endpgfgraphicnamed,    
\end{equation}
    \end{minipage}
  \begin{minipage}{0.295\textwidth}
      \begin{equation}
        \label{uniformsin3}
\beginpgfgraphicnamed{Qutrits/cliffnorformt}
\InputIfFileExists{Qutrits/cliffnorformt.tikz}{}{\input{./figures/Qutrits/cliffnorformt.tikz}}
\endpgfgraphicnamed,
\end{equation}
    \end{minipage}
  \end{trivlist}

where $\nststile{a_2}{a_1}, \nststile{a_4}{a_3}, \nststile{a_6}{a_5}, \nststile{a_8}{a_7} \in \mathcal{A}, \nststile{p_2}{p_1} \in \mathcal{P}, \nststile{q_2}{q_1} \in \mathcal{Q}, \nststile{m_2}{m_1} \in \mathcal{M}$. 
\end{theorem}

\begin{proof}
By Corollary \ref{sinqu},  any single qutrit Clifford operator is of form  (\ref{xz8}) or (\ref{xz9}). Furthermore, if   $\nststile{y_{1}}{x_{1}}\in \mathcal{M}$, then the corresponding red node  can be pushed down and merge with the other red node at the bottom, thus diagram (\ref{xz8}) becomes the form (\ref{uniformsin1}).

 If otherwise  $\nststile{y_{1}}{x_{1}}\in \mathcal{Q}$, then we can use the same trick as in the proof of Lemma\ref{xz112}: separate the $\mathcal{M}$ part and push it down to be merged. Thus diagram (\ref{xz8}) becomes the form
\begin{equation}\label{cliffnorform2}
\beginpgfgraphicnamed{Qutrits//cliffnorform22}
\InputIfFileExists{Qutrits//cliffnorform22.tikz}{}{\input{./figures/Qutrits//cliffnorform22.tikz}}
\endpgfgraphicnamed
\end{equation}
where $\nststile{p_2}{p_1} \in \mathcal{P}$. If furthermore $\nststile{s_{2}}{s_{1}}\in \mathcal{M}$, then  the corresponding green node  can be pushed up and  diagram (\ref{cliffnorform2}) becomes the form (\ref{uniformsin1}). Otherwise,  $\nststile{s_{2}}{s_{1}}\in \mathcal{Q}$,  and  diagram (\ref{cliffnorform2}) is of the form (\ref{uniformsin2}), where $\nststile{q2}{q_1} \in \mathcal{Q}$. To sum up, the  diagram (\ref{xz8}) can be rewritten into the form (\ref{uniformsin1}) or (\ref{uniformsin2}).  As a consequence, the  diagram (\ref{xz9}) can be rewritten into one of the forms
%\begin{equation}\label{uniformsin12}
%\tikzfig{Qutrits//cliffnorformj2}
%\end{equation}
%\begin{equation}\label{uniformsin22}
%\tikzfig{Qutrits//cliffnorformk2}
%\end{equation}

%%%%%%%%%%%%%%%%%%%%%%%%
\begin{trivlist}\item
    \begin{minipage}{0.5\textwidth}
 \begin{equation}\label{uniformsin12}
\beginpgfgraphicnamed{Qutrits//cliffnorformj2}
\InputIfFileExists{Qutrits//cliffnorformj2.tikz}{}{\input{./figures/Qutrits//cliffnorformj2.tikz}}
\endpgfgraphicnamed,
\end{equation}
\end{minipage}
   \begin{minipage}{0.5\textwidth}
 \begin{equation}\label{uniformsin22}
\beginpgfgraphicnamed{Qutrits//cliffnorformk2}
\InputIfFileExists{Qutrits//cliffnorformk2.tikz}{}{\input{./figures/Qutrits//cliffnorformk2.tikz}}
\endpgfgraphicnamed,
\end{equation}
 \end{minipage}
  \end{trivlist}

where $\nststile{p_2}{p_1} \in \mathcal{P}, \nststile{q_2}{q_1} \in \mathcal{Q}$.

Next we show that the diagrams  (\ref{uniformsin12}) and  (\ref{uniformsin22}) can be rewritten into the form (\ref{uniformsin1}),  (\ref{uniformsin2}) or (\ref{uniformsin3}). 
For (\ref{uniformsin12}),  if $\nststile{a_2}{a_1} \in \mathcal{M}$, it is just the form (\ref{uniformsin3}). Otherwise, $\nststile{a_2}{a_1} \in \mathcal{Q}$, thus $\nststile{a_1}{a_2} \in \mathcal{Q}$.
Let $\nststile{a_1}{a_2} =\nststile{m_2}{m_1} +\nststile{p_1}{p_1} $, where $\nststile{m_2}{m_1} \in \mathcal{M}, \nststile{p_1}{p_1} \in \mathcal{P}$.  It is clear from (\ref{pzxzhh}) and (\ref{pzxzhh2}) that 

\begin{equation}\label{singlecliffnorm1}
\beginpgfgraphicnamed{Qutrits//singlecliffnormm1}
\InputIfFileExists{Qutrits//singlecliffnormm1.tikz}{}{\input{./figures/Qutrits//singlecliffnormm1.tikz}}
\endpgfgraphicnamed,
\end{equation}
where $\nststile{p}{p},   \nststile{p^{\prime}}{p^{\prime}} \in \mathcal{P}$.
 Combining with (\ref{property1}), we have 

\begin{equation}\label{singlecliffnormm2}
\beginpgfgraphicnamed{Qutrits//singlecliffnorm2}
\InputIfFileExists{Qutrits//singlecliffnorm2.tikz}{}{\input{./figures/Qutrits//singlecliffnorm2.tikz}}
\endpgfgraphicnamed,
\end{equation}
which is just of form (\ref{uniformsin2}).

For  (\ref{uniformsin22}),  except for the top red node of phase $\nststile{p_2}{p_1}$, the remaining part is exactly the same as the case of (\ref{uniformsin12}) when  $\nststile{a_2}{a_1} \in \mathcal{Q}$,
 thus can be rewritten into the form of (\ref{uniformsin2}). Composing with the red node $\nststile{p_2}{p_1}$, we obtain a diagram of form  (\ref{uniformsin1}) or (\ref{uniformsin2}).

%Similarly, it can be shown that any diagram of form (\ref{uniformsin22}) is equal to a form of  (\ref{uniformsin1}) or (\ref{uniformsin2}).
By now we have shown that each single qutrit Clifford operator can be represented by a ZX-calculus diagram in one of the forms (\ref{uniformsin1}), (\ref{uniformsin2}) and (\ref{uniformsin3}). Next we show that these forms are unique. 
First we prove that all the forms presented in  (\ref{uniformsin1}), (\ref{uniformsin2}) and (\ref{uniformsin3}) are not equal to each other. There are several cases to be considered, 
we first compare the form of (\ref{uniformsin1}) with the form of  (\ref{uniformsin3}).
Suppose that 
\begin{align}\label{sp1}
\beginpgfgraphicnamed{Qutrits//cliffnorformj2}
\InputIfFileExists{Qutrits//cliffnorformj2.tikz}{}{\input{./figures/Qutrits//cliffnorformj2.tikz}}
\endpgfgraphicnamed=%
\beginpgfgraphicnamed{Qutrits//cliffnorformj21}
\InputIfFileExists{Qutrits//cliffnorformj21.tikz}{}{\input{./figures/Qutrits//cliffnorformj21.tikz}}
\endpgfgraphicnamed
\end{align}
Then
\begin{equation}\label{uniformsin23}
\beginpgfgraphicnamed{Qutrits//cliffnorformjt2}
\InputIfFileExists{Qutrits//cliffnorformjt2.tikz}{}{\input{./figures/Qutrits//cliffnorformjt2.tikz}}
\endpgfgraphicnamed
\end{equation}
 which can be written in a matrix form as follows:
 
\begin{equation}\label{uniformsin24}
\begin{pmatrix}
        1 & 1 & 1 \\
        1 & \omega & \bar{\omega}\\
        1 & \bar{\omega} & \omega
 \end{pmatrix}
\begin{pmatrix}
        1 & 1 & 1 \\
        1 & \omega & \bar{\omega}\\
        1 & \bar{\omega} & \omega
 \end{pmatrix}
 \begin{pmatrix}
        1 & 0 & 0 \\
        0& e^{i\alpha} & 0 \\
        0 & 0& e^{i\beta}
 \end{pmatrix}\simeq
 \begin{pmatrix}
        u & v & w \\
        w & u & v \\     
        v & w & u
 \end{pmatrix}
\end{equation}
where $\simeq$ means equal up to a non-zero scalar and the matrix on the right hand side of $\simeq$ comes from (\ref{phasematrix3}).  After simplification on the left hand side, we have
\begin{equation}\label{uniformsin25}
 \begin{pmatrix}
        1 & 0 & 0 \\
        0& 0 & e^{i\beta} \\
        0 & e^{i\alpha} & 0
 \end{pmatrix}\simeq
 \begin{pmatrix}
        u & v & w \\
        w & u & v \\     
        v & w & u
 \end{pmatrix}
\end{equation}
From the first row of the above matrices, $u\neq 0, v=w=0$, while from the third row, $w\neq 0, u=0$, a contradiction. Therefore the equality  (\ref{sp1}) is impossible, i.e., the form of (\ref{uniformsin1}) will never equal to the form of  (\ref{uniformsin3}).
 
 Then we compare the form of (\ref{uniformsin2}) with the form of  (\ref{uniformsin3}). Suppose that 
\begin{align}\label{sp2}
\beginpgfgraphicnamed{Qutrits//cliffnorformj2}
\InputIfFileExists{Qutrits//cliffnorformj2.tikz}{}{\input{./figures/Qutrits//cliffnorformj2.tikz}}
\endpgfgraphicnamed=%
\beginpgfgraphicnamed{Qutrits//cliffnorformk}
\InputIfFileExists{Qutrits//cliffnorformk.tikz}{}{\input{./figures/Qutrits//cliffnorformk.tikz}}
\endpgfgraphicnamed
\end{align}
%where the left-side diagram is just  the diagram shown in (\ref{uniformsin12}).
Then
\begin{align}\label{sp3}
\beginpgfgraphicnamed{Qutrits//cliffnorformtran}
\InputIfFileExists{Qutrits//cliffnorformtran.tikz}{}{\input{./figures/Qutrits//cliffnorformtran.tikz}}
\endpgfgraphicnamed
\end{align}
When $\nststile{p_2}{p_1} =\nststile{1}{1}$, the matrix form of  (\ref{sp3}) is as follows:
\begin{equation}\label{uniformsin26}
\begin{pmatrix}
        1 & 0 & 0 \\
        0 & 0 & 1\\
        0 & 1 & 0
 \end{pmatrix}
\begin{pmatrix}
        a & b & c \\
        c & a & b\\
        b & c & a
 \end{pmatrix}\simeq
 \begin{pmatrix}
        1 & 0 & 0 \\
        0& e^{i\alpha} & 0 \\
        0 & 0& e^{i\beta}
 \end{pmatrix}
 \begin{pmatrix}
        \omega & 1 & 1 \\
        1 &  \omega & 1 \\     
        1 & 1 &  \omega
 \end{pmatrix}
  \begin{pmatrix}
        1 & 0 & 0 \\
        0& e^{i\sigma} & 0 \\
        0 & 0& e^{i\tau}
        \end{pmatrix}
\end{equation}
where $\alpha, \beta, \sigma, \tau \in [0,  2\pi)$.
That is,
\begin{equation}\label{uniformsin27}
\begin{pmatrix}
        a & b & c \\
        c & a & b\\
        b & c & a
 \end{pmatrix}\simeq
 \begin{pmatrix}
        \omega & e^{i\sigma} & e^{i\tau} \\
        e^{i\beta} & e^{i(\beta+\sigma)} &  \omega e^{i(\beta+\tau)}\\
          e^{i\alpha} &  \omega e^{i(\alpha+\sigma)} & e^{i(\alpha+\tau)}     
 \end{pmatrix}
\end{equation}
Therefore, 
  \begin{equation*}
   \left\{\begin{array}{ccccc}
     e^{i\sigma}&= & e^{i\alpha} &= & \omega e^{i(\beta+\tau)} \\
     e^{i\tau} & = & e^{i\beta} &= & \omega e^{i(\alpha+\sigma)}\\
       \omega &= & e^{i(\alpha+\tau)} &= & e^{i(\beta+\sigma)}\\
    \end{array}\right.
  \end{equation*}
  Solving this system, we have
  $\sigma=\alpha, \tau=\beta$, 
   \begin{align}\label{uniformsin28}
   \left\{\begin{array}{l}
     \alpha=0\\
    \beta=\frac{2\pi}{3}\\
    \end{array}\right. \quad
   \left\{\begin{array}{l}
     \alpha=\frac{2\pi}{3}\\
    \beta=0\\
    \end{array}\right. \quad
     \left\{\begin{array}{l}
     \alpha=\frac{4\pi}{3}\\
    \beta=\frac{4\pi}{3}\\
    \end{array}\right. \quad
  \end{align}
%First note that there are $$ consider the matrix form of  (\ref{uniformsin12}).
Return to the equation (\ref{sp3}).
For  $ \alpha=0, ~\beta=\frac{2\pi}{3}$, we have $\nststile{q_2}{q_1} =\nststile{1}{0} ,~ \nststile{a_2}{a_1} =\nststile{2}{0} ,~ \nststile{a_4-a_5}{a_3-a_6} =\nststile{1}{0}$.
For  $ \alpha=\frac{2\pi}{3}, ~\beta=0 $, we have $\nststile{q_2}{q_1} =\nststile{0}{1} ,~ \nststile{a_2}{a_1} =\nststile{0}{2} ,~ \nststile{a_4-a_5}{a_3-a_6} =\nststile{0}{1}$.
For  $ \alpha=\frac{4\pi}{3}, ~\beta=\frac{4\pi}{3} $, we have $\nststile{q_2}{q_1} =\nststile{2}{2} ,~ \nststile{a_2}{a_1} =\nststile{1}{1} ,~ \nststile{a_4-a_5}{a_3-a_6} =\nststile{2}{2}$. To sum up, 
when $\nststile{p_2}{p_1} =\nststile{1}{1}$, it must be that  $ \nststile{a_2}{a_1} \in \{\nststile{0}{2},  \nststile{2}{0}, \nststile{1}{1}\}$ for (\ref{sp2}) to hold.    % diagram (\ref{uniformsin12}) is equal to a form of (\ref{uniformsin2}). 

%By similar analysis for $\nststile{p_2}{p_1} =\nststile{2}{2}$, 
Similarly, it can be shown that  when $\nststile{p_2}{p_1} =\nststile{2}{2}$, it must be that 
$ \nststile{a_2}{a_1} \in \{\nststile{0}{1},  \nststile{1}{0}, \nststile{2}{2}\}$ for (\ref{sp2}) to hold. 
  %diagram (\ref{uniformsin12}) is  equal to a form of (\ref{uniformsin2}). 
%Therefore, whenever  $ \nststile{a_2}{a_1} \in \mathcal{Q}$,diagram (\ref{uniformsin12}) is equal to a  form of (\ref{uniformsin2}).  We will show that
Thus for $ \nststile{a_2}{a_1} \in \mathcal{M}$,  (\ref{sp2}) does not hold,
% diagram (\ref{uniformsin12}) is never equal to  a form of  (\ref{uniformsin1}) or (\ref{uniformsin2}).
i.e., the form of (\ref{uniformsin2}) will never equal to the form of  (\ref{uniformsin3}).

%Similarly, it can be shown that any diagram of form (\ref{uniformsin22}) is equal to a form of  (\ref{uniformsin1}) or (\ref{uniformsin2}).

In the same way, we can show that the form of (\ref{uniformsin1}) is not equal to the form of  (\ref{uniformsin2}), and different elements (with different paremeters) of the same form (be it (\ref{uniformsin1}), (\ref{uniformsin2}), or (\ref{uniformsin3})) must not be equal to each other.

Now we can prove the uniqueness of these forms.
 By direct calculation, (\ref{uniformsin1}) has $9\times 9=81$ elements,  (\ref{uniformsin2}) has $2\times 6\times 9=108$ elements, and (\ref{uniformsin3}) has $3\times 9=27$ elements. So the total number elements is $81+108+27=216$, which is exactly equal to the order  $|\mathcal{C}_1|$ of the group of single qutrit Clifford operators.  Since each element of $\mathcal{C}_1$ has been shown to be representable by forms (\ref{uniformsin1}),  (\ref{uniformsin2}) or (\ref{uniformsin3}), there should be enough members belonging to these kind of forms, thus no common elements are allowed. Therefore the representation of  single qutrit Clifford operators is unique. 
\end{proof}

%\tikzfig{Qutrits//semioval2}

\section{Qutrit graph states in the ZX-calculus}

\subsection{Stabilizer state diagram and transformations of GS-LC diagrams }

The qutrit graph states have a nice representation in the ZX-calculus.

\begin{lemma}\cite{GongWang}\label{gstatezx}
A qutrit graph state $\ket{G}$, where $G = (E;V)$ is an n-vertex graph, is represented in the graphical
calculus as follows:
\begin{itemize}
 \item  for each vertex $v \in V$, a green node with one output, and
\item  for each $1$-weighted edge $\{u,v\}\in E$, an $H$ node connected to the green nodes representing vertices $u$
and $v$,
\item  for each $2$-weighted edge $\{u,v\}\in E$, an $H^{\dagger}$ node connected to the green nodes representing vertices $u$
and $v$.

\end{itemize}
\end{lemma}
A graph state $\ket{G}$ is also denoted by the diagram %
\beginpgfgraphicnamed{Qutrits/grapstg}
\InputIfFileExists{Qutrits/grapstg.tikz}{}{\input{./figures/Qutrits/grapstg.tikz}}
\endpgfgraphicnamed.

\begin{definition}\cite{Miriam2}
A diagram in the stabilizer ZX-calculus is called a GS-LC diagram if it consists of a graph state diagram with arbitrary single-qutrit Clifford operators applied to each output. These associated Clifford operators are called vertex operators.
\end{definition}
An $n$-qutrit GS-LC diagram is represented as
 \begin{equation*}
\beginpgfgraphicnamed{Qutrits/gslc}
\InputIfFileExists{Qutrits/gslc.tikz}{}{\input{./figures/Qutrits/gslc.tikz}}
\endpgfgraphicnamed
     \end{equation*}
where $G = (V,E)$ is a graph  and $U_v \in\mathcal{C}_1$ for $v\in V$.

\begin{theorem}\label{gandgst1}\cite{GongWang}
Let $G = (V,E)$ be a graph with adjacency matrix $\Gamma$ and  $G\ast_1  v $ be the graph that
results from applying to $G$ a $1$-local complementation about some $v\in V$. Then the corresponding graph states 
$\ket{G} $ and $\ket{G\ast_1  v} $ are related as follows:
\begin{equation}\label{locom1}
\beginpgfgraphicnamed{Qutrits/gandg1}
\InputIfFileExists{Qutrits/gandg1.tikz}{}{\input{./figures/Qutrits/gandg1.tikz}}
\endpgfgraphicnamed
\end{equation}
where for $1\leqslant i \leqslant n, ~i\neq v$,
 \begin{equation*}
\nststile{b_i}{a_i} =  \left\{\begin{array}{l}
    \nststile{2}{2}, ~\mbox{if}~ \Gamma_{iv}\neq 0 \vspace{0.2cm}\\
    \nststile{0}{0}, ~\mbox{if} ~\Gamma_{iv}= 0\\
    \end{array}\right. 
     \end{equation*}
\end{theorem}

\begin{corollary}\label{gandgst2}
Let $G = (V,E)$ be a graph with adjacency matrix $\Gamma$ and  $G\ast_2  v $ be the graph that
results from applying to $G$ a $2$-local complementation about some $v\in V$. Then the corresponding graph states 
$\ket{G} $ and $\ket{G\ast_2  v} $ are related as follows:
\begin{equation}\label{locom2}
\beginpgfgraphicnamed{Qutrits/gandg2}
\InputIfFileExists{Qutrits/gandg2.tikz}{}{\input{./figures/Qutrits/gandg2.tikz}}
\endpgfgraphicnamed
\end{equation}
where for $1\leqslant i \leqslant n, ~i\neq v$,
 \begin{equation*}
\nststile{b_i}{a_i} =  \left\{\begin{array}{l}
    \nststile{1}{1}, ~\mbox{if}~ \Gamma_{iv}\neq 0 \vspace{0.2cm}\\
    \nststile{0}{0}, ~\mbox{if} ~\Gamma_{iv}= 0\\
    \end{array}\right. 
     \end{equation*}
\end{corollary}
\begin{proof}
Note that $G\ast_2  v=(G\ast_1  v)\ast_1  v$.
\end{proof}

\begin{theorem}\label{gandgst3}
Let $G = (V,E)$ be a graph with adjacency matrix $\Gamma$ and let $G\circ_2  v $ be the graph that
results from applying the transformation $\circ_2  v $ about some $v\in V$. Then the corresponding graph states 
$\ket{G} $ and $\ket{G\circ_2  v} $ are related as follows:
\begin{equation}\label{locom3}
\beginpgfgraphicnamed{Qutrits//gandg3}
\InputIfFileExists{Qutrits//gandg3.tikz}{}{\input{./figures/Qutrits//gandg3.tikz}}
\endpgfgraphicnamed
\end{equation}
where the Hadamard nodes are on the output of the vertex $v$.

\end{theorem}
\begin{proof}
According to the definition of $G\circ_2  v $, we only need to consider the vertices that are connected to $v$. The effect of applying the operator $G\circ_2  v $ is to replace the $ H$ node connected to $v$  with an $H^{\dagger}(=H^{-1})$ node, and vice versa, i.e., changing from $H^{\pm 1}$ to $H^{\mp 1}$. By the rewriting rule (P1), we have %
\beginpgfgraphicnamed{Qutrits//hhdagexch}
\InputIfFileExists{Qutrits//hhdagexch.tikz}{}{\input{./figures/Qutrits//hhdagexch.tikz}}
\endpgfgraphicnamed. That means  $G\circ_2  v $ can be seen as having one more $D$ node on each edge connected to $v$ than on the corresponding edge in $G$. The equality (\ref{locom3}) then follows immediately from pushing these $D$ nodes to the output of $v$  by the 
property (\ref{property1}).
%following property \cite{GongWang}:\begin{equation}\label{locom4} \tikzfig{Qutrits//Dcopy}\end{equation}
\end{proof}

\subsection{Rewriting arbitrary stabilizer state diagram into a GS-LC diagram}
In this subsection we show that any stabilizer state diagram can be rewritten into a GS-LC diagram by rules of the ZX-calculus. We give the proof of this result following \cite{Miriam1, Miriam2}.

\begin{theorem}\label{stabtogslc1}
Any stabilizer state diagram can be rewritten into a GS-LC diagram by rules of the ZX-calculus.
\end{theorem}
\begin{proof}
By lemma \ref{diagramdecom}, any ZX-calculus diagram can be written as composed of four basic spiders together with phase shifts and Hadamard and Hadamard dagger nodes. Thus we can proceed with the proof by induction:  %
\beginpgfgraphicnamed{Qutrits//gdotsmall}
\InputIfFileExists{Qutrits//gdotsmall.tikz}{}{\input{./figures/Qutrits//gdotsmall.tikz}}
\endpgfgraphicnamed is the initial GS-LC diagram, if it can be shown that applying any basic component of the ZX-calculus diagrams to a GS-LC diagram still results in a GS-LC diagram, then any stabilizer state diagram can be rewritten into a GS-LC diagram. The following  lemmas \ref{stabtogslc2}, \ref{stabtogslc3},   \ref{stabtogslc4}, \ref{stabtogslc5} and \ref{stabtogslc6} exactly give the inductive steps, hence completes the proof.
\end{proof}

\begin{lemma}\label{stabtogslc2}
A GS-LC diagram associated with %
\beginpgfgraphicnamed{Qutrits//gdotsmall}
\InputIfFileExists{Qutrits//gdotsmall.tikz}{}{\input{./figures/Qutrits//gdotsmall.tikz}}
\endpgfgraphicnamed is still a GS-LC diagram.
\end{lemma}
\begin{proof}
It follows directly from the definition of GS-LC diagrams and lemma \ref{gstatezx}.
\end{proof}

\begin{lemma}\label{stabtogslc3}
Appying a basic single qutrit Clifford unitary to a GS-LC diagram still results in a GS-LC diagram.
\end{lemma}
\begin{proof}
This follows directly from the definition of GS-LC diagrams.
\end{proof}

\begin{lemma}\label{stabtogslc5}
If a vertex in a GS-LC diagram has no neighbours, then it must be a single-qutrit pure stabilizer state as one of the follows: 
\begin{equation}\label{singlestate}
\beginpgfgraphicnamed{Qutrits//singlestates}
\InputIfFileExists{Qutrits//singlestates.tikz}{}{\input{./figures/Qutrits//singlestates.tikz}}
\endpgfgraphicnamed
\end{equation}
\end{lemma}
\begin{proof}
A vertex with no neighbours in a GS-LC diagram must be written as
\begin{equation*}
\beginpgfgraphicnamed{Qutrits//singlestates2}
\InputIfFileExists{Qutrits//singlestates2.tikz}{}{\input{./figures/Qutrits//singlestates2.tikz}}
\endpgfgraphicnamed
\end{equation*}
where $U$ is of one of the  forms (\ref{uniformsin1}), (\ref{uniformsin2}) and (\ref{uniformsin3}).
If
 \begin{equation*}
\beginpgfgraphicnamed{Qutrits//singleu}
\InputIfFileExists{Qutrits//singleu.tikz}{}{\input{./figures/Qutrits//singleu.tikz}}
\endpgfgraphicnamed=%
\beginpgfgraphicnamed{Qutrits//cliffnorformj}
\InputIfFileExists{Qutrits//cliffnorformj.tikz}{}{\input{./figures/Qutrits//cliffnorformj.tikz}}
\endpgfgraphicnamed
\end{equation*}
then
\begin{equation}\label{singlestatejk}
\beginpgfgraphicnamed{Qutrits//singlestates2}
\InputIfFileExists{Qutrits//singlestates2.tikz}{}{\input{./figures/Qutrits//singlestates2.tikz}}
\endpgfgraphicnamed= %
\beginpgfgraphicnamed{Qutrits//singlestates3}
\InputIfFileExists{Qutrits//singlestates3.tikz}{}{\input{./figures/Qutrits//singlestates3.tikz}}
\endpgfgraphicnamed
\end{equation}
For $\nststile{a_2}{a_1}=\nststile{0}{0}$, 
\begin{equation}\label{singlestates00}
\beginpgfgraphicnamed{Qutrits//singlestates4}
\InputIfFileExists{Qutrits//singlestates4.tikz}{}{\input{./figures/Qutrits//singlestates4.tikz}}
\endpgfgraphicnamed
\end{equation}

For $\nststile{a_2}{a_1}=\nststile{2}{1}$, 
\begin{equation}\label{singlestates01}
\beginpgfgraphicnamed{Qutrits//singlestates5}
\InputIfFileExists{Qutrits//singlestates5.tikz}{}{\input{./figures/Qutrits//singlestates5.tikz}}
\endpgfgraphicnamed
\end{equation}

For $\nststile{a_2}{a_1}=\nststile{1}{2}$, 
\begin{equation}\label{singlestates02}
\beginpgfgraphicnamed{Qutrits//singlestates6}
\InputIfFileExists{Qutrits//singlestates6.tikz}{}{\input{./figures/Qutrits//singlestates6.tikz}}
\endpgfgraphicnamed
\end{equation}

For $\nststile{a_2}{a_1}=\nststile{1}{1}$, 
\begin{equation}\label{singlestates03}
\beginpgfgraphicnamed{Qutrits//singlestates7}
\InputIfFileExists{Qutrits//singlestates7.tikz}{}{\input{./figures/Qutrits//singlestates7.tikz}}
\endpgfgraphicnamed
\end{equation}
where we used the property proved in \cite{GongWang} that 
\begin{equation}\label{singlestates04}
\beginpgfgraphicnamed{Qutrits//rtog12}
\InputIfFileExists{Qutrits//rtog12.tikz}{}{\input{./figures/Qutrits//rtog12.tikz}}
\endpgfgraphicnamed
\end{equation}
Now we consider $\nststile{a_4}{a_3}$ in (\ref{singlestates03}). For $\nststile{a_4}{a_3}\in \{\nststile{0}{0}, \nststile{1}{1}, \nststile{2}{2}, \nststile{0}{2}, \nststile{2}{0}\}$, by (\ref{singlestates04}), we have
\begin{equation*}
\beginpgfgraphicnamed{Qutrits//singlestates71}
\InputIfFileExists{Qutrits//singlestates71.tikz}{}{\input{./figures/Qutrits//singlestates71.tikz}}
\endpgfgraphicnamed\in\left\{%
\beginpgfgraphicnamed{Qutrits//singlestates72}
\InputIfFileExists{Qutrits//singlestates72.tikz}{}{\input{./figures/Qutrits//singlestates72.tikz}}
\endpgfgraphicnamed, %
\beginpgfgraphicnamed{Qutrits//singlestates73}
\InputIfFileExists{Qutrits//singlestates73.tikz}{}{\input{./figures/Qutrits//singlestates73.tikz}}
\endpgfgraphicnamed, %
\beginpgfgraphicnamed{Qutrits//singlestates74}
\InputIfFileExists{Qutrits//singlestates74.tikz}{}{\input{./figures/Qutrits//singlestates74.tikz}}
\endpgfgraphicnamed, %
\beginpgfgraphicnamed{Qutrits//singlestates75}
\InputIfFileExists{Qutrits//singlestates75.tikz}{}{\input{./figures/Qutrits//singlestates75.tikz}}
\endpgfgraphicnamed, %
\beginpgfgraphicnamed{Qutrits//singlestates76}
\InputIfFileExists{Qutrits//singlestates76.tikz}{}{\input{./figures/Qutrits//singlestates76.tikz}}
\endpgfgraphicnamed\right\}.
\end{equation*}
For $\nststile{a_4}{a_3}= \nststile{1}{0}$, we have 
\begin{equation}\label{singlestates0401}
\beginpgfgraphicnamed{Qutrits//singlestates8}
\InputIfFileExists{Qutrits//singlestates8.tikz}{}{\input{./figures/Qutrits//singlestates8.tikz}}
\endpgfgraphicnamed
\end{equation}
Similarly, we have 
\begin{equation}\label{singlestates0402}
\beginpgfgraphicnamed{Qutrits//singlestates9}
\InputIfFileExists{Qutrits//singlestates9.tikz}{}{\input{./figures/Qutrits//singlestates9.tikz}}
\endpgfgraphicnamed, \quad   %
\beginpgfgraphicnamed{Qutrits//singlestates10}
\InputIfFileExists{Qutrits//singlestates10.tikz}{}{\input{./figures/Qutrits//singlestates10.tikz}}
\endpgfgraphicnamed, \quad  %
\beginpgfgraphicnamed{Qutrits//singlestates11}
\InputIfFileExists{Qutrits//singlestates11.tikz}{}{\input{./figures/Qutrits//singlestates11.tikz}}
\endpgfgraphicnamed.
\end{equation}

In the same way, for  $\nststile{a_2}{a_1}\in \{\nststile{0}{1}, \nststile{1}{0}, \nststile{2}{2}, \nststile{0}{2}, \nststile{2}{0}\}$, and $\nststile{a_4}{a_3}\in  \mathcal{A}$, we can prove that 
\begin{equation*}
\beginpgfgraphicnamed{Qutrits//singlestates71}
\InputIfFileExists{Qutrits//singlestates71.tikz}{}{\input{./figures/Qutrits//singlestates71.tikz}}
\endpgfgraphicnamed\in\left\{ %
\beginpgfgraphicnamed{Qutrits//singlestates}
\InputIfFileExists{Qutrits//singlestates.tikz}{}{\input{./figures/Qutrits//singlestates.tikz}}
\endpgfgraphicnamed\right\}.
\end{equation*}

If
 \begin{equation*}
\beginpgfgraphicnamed{Qutrits//singleu}
\InputIfFileExists{Qutrits//singleu.tikz}{}{\input{./figures/Qutrits//singleu.tikz}}
\endpgfgraphicnamed=%
\beginpgfgraphicnamed{Qutrits//cliffnorformk}
\InputIfFileExists{Qutrits//cliffnorformk.tikz}{}{\input{./figures/Qutrits//cliffnorformk.tikz}}
\endpgfgraphicnamed
\end{equation*}
then
\begin{equation*}
\beginpgfgraphicnamed{Qutrits//singlestates2}
\InputIfFileExists{Qutrits//singlestates2.tikz}{}{\input{./figures/Qutrits//singlestates2.tikz}}
\endpgfgraphicnamed= %
\beginpgfgraphicnamed{Qutrits//singlestates12}
\InputIfFileExists{Qutrits//singlestates12.tikz}{}{\input{./figures/Qutrits//singlestates12.tikz}}
\endpgfgraphicnamed
\end{equation*}
which is reduced to the case (\ref{singlestatejk}).

If
 \begin{equation*}
\beginpgfgraphicnamed{Qutrits//singleu}
\InputIfFileExists{Qutrits//singleu.tikz}{}{\input{./figures/Qutrits//singleu.tikz}}
\endpgfgraphicnamed=%
\beginpgfgraphicnamed{Qutrits//cliffnorformt}
\InputIfFileExists{Qutrits//cliffnorformt.tikz}{}{\input{./figures/Qutrits//cliffnorformt.tikz}}
\endpgfgraphicnamed
\end{equation*}
then
\begin{equation*}
\beginpgfgraphicnamed{Qutrits//singlestates2}
\InputIfFileExists{Qutrits//singlestates2.tikz}{}{\input{./figures/Qutrits//singlestates2.tikz}}
\endpgfgraphicnamed= %
\beginpgfgraphicnamed{Qutrits//singlestates13}
\InputIfFileExists{Qutrits//singlestates13.tikz}{}{\input{./figures/Qutrits//singlestates13.tikz}}
\endpgfgraphicnamed
\end{equation*}
which is also reduced to the case (\ref{singlestatejk}).

\end{proof}

\begin{definition} \cite{Miriam2}
 A vertex in a GS-LC diagram that is being acted upon by some operation is called a operand vertex.
\end{definition}

\begin{definition} \cite{Miriam2}
 A neighbour of an operand vertex in a GS-LC diagram is called a swapping partner of the operand vertex.
\end{definition}

\begin{lemma}\label{stabtogslc3to4}
 If in some GS-LC diagram an operand vertex has at least one neighbour, it is always possible to change the vertex operator on the operand vertex to the following form using $1$-local ($2$-local) complementations:
  \begin{equation}\label{chgoprt}
\beginpgfgraphicnamed{Qutrits//chgoprt2}
\InputIfFileExists{Qutrits//chgoprt2.tikz}{}{\input{./figures/Qutrits//chgoprt2.tikz}}
\endpgfgraphicnamed
\end{equation}
 where  $\nststile{m_2}{m_1}, \nststile{m_4}{m_3}\in  \mathcal{M}$.
\end{lemma}
\begin{proof}
We can pick one neighbour of the operand vertex as the swapping partner. A $1$-local complementation about the operand vertex adds %
\beginpgfgraphicnamed{Qutrits//11phasgt}
\InputIfFileExists{Qutrits//11phasgt.tikz}{}{\input{./figures/Qutrits//11phasgt.tikz}}
\endpgfgraphicnamed to its vertex
operator, while a $1$-local complementation about the swapping partner adds %
\beginpgfgraphicnamed{Qutrits//22phasgt}
\InputIfFileExists{Qutrits//22phasgt.tikz}{}{\input{./figures/Qutrits//22phasgt.tikz}}
\endpgfgraphicnamed to the vertex operator on the operand vertex. Note that local complementations about the operand vertex or its swapping partner do not remove the edge between these two vertices. Therefore, after each local complementation, the operand vertex still has a neighbour, enabling further local complementations. By Theorem \ref{uniformsin}, the vertex operator $U$ must be in one of the  form (\ref{uniformsin1}),  (\ref{uniformsin2}) or (\ref{uniformsin3}).  For
 \begin{equation}\label{chgoprt3}
\beginpgfgraphicnamed{Qutrits//singleu}
\InputIfFileExists{Qutrits//singleu.tikz}{}{\input{./figures/Qutrits//singleu.tikz}}
\endpgfgraphicnamed=%
\beginpgfgraphicnamed{Qutrits//cliffnorformj}
\InputIfFileExists{Qutrits//cliffnorformj.tikz}{}{\input{./figures/Qutrits//cliffnorformj.tikz}}
\endpgfgraphicnamed
\end{equation}
if $\nststile{a_2}{a_1}\in  \mathcal{Q}$, and $\nststile{a_4}{a_3}\in  \mathcal{Q}$, similar to the trick used in the proof of Lemma \ref{xz112}, we have  $ \nststile{a_2}{a_1} =\nststile{p_{2}}{p_{1}}+\nststile{m_{2}}{m_{1}}, \nststile{a_4}{a_3} =\nststile{p_{4}}{p_{3}}+\nststile{m_{4}}{m_{3}}$, where $\nststile{p_{2}}{p_{1}}, \nststile{p_{4}}{p_{3}}\in \mathcal{P}, \nststile{m_{2}}{m_{1}}, \nststile{m_{4}}{m_{3}}\in \mathcal{M}$.  Therefore,
 \begin{equation}
\beginpgfgraphicnamed{Qutrits//cliffnorformj}
\InputIfFileExists{Qutrits//cliffnorformj.tikz}{}{\input{./figures/Qutrits//cliffnorformj.tikz}}
\endpgfgraphicnamed=  %
\beginpgfgraphicnamed{Qutrits//cliffnorformjpmdec}
\InputIfFileExists{Qutrits//cliffnorformjpmdec.tikz}{}{\input{./figures/Qutrits//cliffnorformjpmdec.tikz}}
\endpgfgraphicnamed
\end{equation}
where $\nststile{\bar{m}_{4}}{\bar{m}_{3}}\in \mathcal{M}$.  Applying the $1$-local complementation about the operand vertex or its swapping partner, the nodes  $\nststile{p_{2}}{p_{1}}, \nststile{p_{4}}{p_{3}}$ can be removed, and the remaining part is of the form (\ref{chgoprt}). The case for
 $\nststile{a_2}{a_1}\in  \mathcal{Q}, \nststile{a_4}{a_3}\in  \mathcal{M}$ or  $\nststile{a_2}{a_1}\in  \mathcal{M}, \nststile{a_4}{a_3}\in  \mathcal{Q}$ is similar as above,  yet simpler. 
 
 For 
 \begin{equation}\label{chgoprt4}
\beginpgfgraphicnamed{Qutrits//singleu}
\InputIfFileExists{Qutrits//singleu.tikz}{}{\input{./figures/Qutrits//singleu.tikz}}
\endpgfgraphicnamed=%
\beginpgfgraphicnamed{Qutrits//cliffnorformk}
\InputIfFileExists{Qutrits//cliffnorformk.tikz}{}{\input{./figures/Qutrits//cliffnorformk.tikz}}
\endpgfgraphicnamed
\end{equation}
the node $\nststile{p_{2}}{p_{1}}$ can be removed by  the $1$-local complementation about the operand vertex or its swapping partner, then we are in the case of  (\ref{chgoprt3}).

 For 
 \begin{equation}\label{chgoprt5}
\beginpgfgraphicnamed{Qutrits//singleu}
\InputIfFileExists{Qutrits//singleu.tikz}{}{\input{./figures/Qutrits//singleu.tikz}}
\endpgfgraphicnamed=%
\beginpgfgraphicnamed{Qutrits//cliffnorformt}
\InputIfFileExists{Qutrits//cliffnorformt.tikz}{}{\input{./figures/Qutrits//cliffnorformt.tikz}}
\endpgfgraphicnamed
\end{equation}
the two Hadamard nodes can be pushed up to the top and removed by  the $1$-local complementation about the operand vertex or its swapping partner, then we are in the case of  (\ref{chgoprt3}).

The proof is similar for $2$-local complementations.
\end{proof}

\begin{lemma}\label{stabtogslc4}
Applying %
\beginpgfgraphicnamed{Qutrits//gcdotsmall}
\InputIfFileExists{Qutrits//gcdotsmall.tikz}{}{\input{./figures/Qutrits//gcdotsmall.tikz}}
\endpgfgraphicnamed to a GS-LC diagram results in a
GS-LC diagram or a zero diagram.
\end{lemma}
\begin{proof}
 There are two cases for the operand vertex when applying %
\beginpgfgraphicnamed{Qutrits//gcdotsmall}
\InputIfFileExists{Qutrits//gcdotsmall.tikz}{}{\input{./figures/Qutrits//gcdotsmall.tikz}}
\endpgfgraphicnamed to a GS-LC diagram.
 
Firstly, if the operand vertex has no neighbours, then by lemma  \ref{stabtogslc5},  it is just one of the 12 single qutrit pure stabilizer states. If the operand vertex is in state %
\beginpgfgraphicnamed{Qutrits//singlestates77}
\InputIfFileExists{Qutrits//singlestates77.tikz}{}{\input{./figures/Qutrits//singlestates77.tikz}}
\endpgfgraphicnamed or %
\beginpgfgraphicnamed{Qutrits//singlestates78}
\InputIfFileExists{Qutrits//singlestates78.tikz}{}{\input{./figures/Qutrits//singlestates78.tikz}}
\endpgfgraphicnamed, the result of applying %
\beginpgfgraphicnamed{Qutrits//gcdotsmall}
\InputIfFileExists{Qutrits//gcdotsmall.tikz}{}{\input{./figures/Qutrits//gcdotsmall.tikz}}
\endpgfgraphicnamed is the scalar zero. Otherwise the result is a non-zero global factor, which can be ignored.

Secondly, we consider the case that the operand vertex has at least one neighbour.  Note that we have the following property: %by the copy rule (B1) or the rule (K1),  together with the colour change rules (H2) and (H2$^{\prime}$), we have 
 \begin{equation}\label{singlestates05}
\beginpgfgraphicnamed{Qutrits//multycopy}
\InputIfFileExists{Qutrits//multycopy.tikz}{}{\input{./figures/Qutrits//multycopy.tikz}}
\endpgfgraphicnamed
\end{equation}
where $a_i \in \{1, -1\}, i=1, \cdots, s$, $\nststile{b}{a}\in  \mathcal{A}$, 
 $\nststile{m_2}{m_1}\in  \mathcal{M}$, 
 $\nststile{m_{i2}}{m_{i1}}\in  \{\nststile{m_2}{m_1}, \nststile{m_1}{m_2}\}, i=1, \cdots, s$.
 
Therefore, if the vertex operator of the operand vertex is
\begin{equation}\label{singlestates06}
\beginpgfgraphicnamed{Qutrits//multycopy2}
\InputIfFileExists{Qutrits//multycopy2.tikz}{}{\input{./figures/Qutrits//multycopy2.tikz}}
\endpgfgraphicnamed 
\end{equation}
then the operand vertex will be removed from the graph state when the operator %
\beginpgfgraphicnamed{Qutrits//gcdotsmall}
\InputIfFileExists{Qutrits//gcdotsmall.tikz}{}{\input{./figures/Qutrits//gcdotsmall.tikz}}
\endpgfgraphicnamed is applied.

Otherwise, we can pick one neighbour of the operand vertex as the swapping partner. By Lemma \ref{stabtogslc3to4}, it is always possible to change the vertex operator on the operand vertex to the form of (\ref{singlestates06}) using $1$-local complementations. Then we get back to the above case. This completes the proof.
\end{proof}

\begin{lemma}\label{stabtogslc5}
Applying %
\beginpgfgraphicnamed{Qutrits//gcopysmall}
\InputIfFileExists{Qutrits//gcopysmall.tikz}{}{\input{./figures/Qutrits//gcopysmall.tikz}}
\endpgfgraphicnamed  to an GS-LC diagram will still result in a
GS-LC diagram.
\end{lemma}
\begin{proof}
There are two cases for the operand vertex.

First, if the operand vertex has no neighbours, 
then by lemma  \ref{stabtogslc5},  it is just one of the 12 single qutrit pure stabilizer states which can be written as

\begin{equation*}
\beginpgfgraphicnamed{Qutrits//abandm}
\InputIfFileExists{Qutrits//abandm.tikz}{}{\input{./figures/Qutrits//abandm.tikz}}
\endpgfgraphicnamed
  \end{equation*}
where  $\nststile{b}{a}\in  \mathcal{A}$,  $\nststile{m_2}{m_1}\in  \mathcal{M}$.
Then 
\begin{equation*}
\beginpgfgraphicnamed{Qutrits//abandm2}
\InputIfFileExists{Qutrits//abandm2.tikz}{}{\input{./figures/Qutrits//abandm2.tikz}}
\endpgfgraphicnamed
  \end{equation*}
and
\begin{equation*}
\beginpgfgraphicnamed{Qutrits//abandm3}
\InputIfFileExists{Qutrits//abandm3.tikz}{}{\input{./figures/Qutrits//abandm3.tikz}}
\endpgfgraphicnamed,
  \end{equation*}
  where the right hand side of each equation is clearly a GS-LC diagram.
  
Second, if the operand vertex has at least one neighbour, we can write the green copy as
\begin{equation}\label{abandm4}
\beginpgfgraphicnamed{Qutrits//abandm4}
\InputIfFileExists{Qutrits//abandm4.tikz}{}{\input{./figures/Qutrits//abandm4.tikz}}
\endpgfgraphicnamed
  \end{equation}

Thus if there is no vertex operator on the operand vertex, then we just add a new vertex and edge to the diagram.

Otherwise, by Lemma \ref{stabtogslc3to4},  the vertex operator $U$ on the operand vertex can be changed to the form (\ref{chgoprt}) using $1$-local complementations.  Now applying %
\beginpgfgraphicnamed{Qutrits//gcopysmall}
\InputIfFileExists{Qutrits//gcopysmall.tikz}{}{\input{./figures/Qutrits//gcopysmall.tikz}}
\endpgfgraphicnamed to the  operand vertex, we have 
\begin{equation*}
\beginpgfgraphicnamed{Qutrits//abandm5}
\InputIfFileExists{Qutrits//abandm5.tikz}{}{\input{./figures/Qutrits//abandm5.tikz}}
\endpgfgraphicnamed
  \end{equation*}
which still means we just add a new vertex and edge to the graph. This completes the proof.
\end{proof}

\begin{lemma}\label{stabtogslc6}
Applying %
\beginpgfgraphicnamed{Qutrits//gcocopysmall}
\InputIfFileExists{Qutrits//gcocopysmall.tikz}{}{\input{./figures/Qutrits//gcocopysmall.tikz}}
\endpgfgraphicnamed  to an GS-LC diagram will result in a
GS-LC diagram or a zero diagram.
\end{lemma}
\begin{proof}
Now there are two operand vertices, and we have four cases to deal with. 

Firstly, if the two operand vertices are connected only to each other, then we do not need to care about all non-operand vertices. Now applying %
\beginpgfgraphicnamed{Qutrits//gcocopysmall}
\InputIfFileExists{Qutrits//gcocopysmall.tikz}{}{\input{./figures/Qutrits//gcocopysmall.tikz}}
\endpgfgraphicnamed to the  operand vertices, we have 
\begin{equation*}
\beginpgfgraphicnamed{Qutrits//abandm6}
\InputIfFileExists{Qutrits//abandm6.tikz}{}{\input{./figures/Qutrits//abandm6.tikz}}
\endpgfgraphicnamed
  \end{equation*}
where $W, V, V^{\prime}\in \mathcal{C}_1$,  $U=W\circ H^{\pm 1}\circ V^{\prime}\in \mathcal{C}_1$.% and we used property (\ref{property2}) for the second equality.  
By Theorem \ref{uniformsin},  $U$ must be in one of the  form (\ref{uniformsin1}),  (\ref{uniformsin2}) or (\ref{uniformsin3}).  For
 \begin{equation*}
\beginpgfgraphicnamed{Qutrits//singleu}
\InputIfFileExists{Qutrits//singleu.tikz}{}{\input{./figures/Qutrits//singleu.tikz}}
\endpgfgraphicnamed=%
\beginpgfgraphicnamed{Qutrits//cliffnorformj}
\InputIfFileExists{Qutrits//cliffnorformj.tikz}{}{\input{./figures/Qutrits//cliffnorformj.tikz}}
\endpgfgraphicnamed
\end{equation*}
we have 
 \begin{equation}\label{loopu1}
\beginpgfgraphicnamed{Qutrits//abandm7}
\InputIfFileExists{Qutrits//abandm7.tikz}{}{\input{./figures/Qutrits//abandm7.tikz}}
\endpgfgraphicnamed
\end{equation}
The result  is either a zero diagram or a GS-LC diagram. % where we used property (\ref{property3}) for the second equality and property (\ref{property4}) for the third equality. 

 For 
 \begin{equation*}
\beginpgfgraphicnamed{Qutrits//singleu}
\InputIfFileExists{Qutrits//singleu.tikz}{}{\input{./figures/Qutrits//singleu.tikz}}
\endpgfgraphicnamed=%
\beginpgfgraphicnamed{Qutrits//cliffnorformk}
\InputIfFileExists{Qutrits//cliffnorformk.tikz}{}{\input{./figures/Qutrits//cliffnorformk.tikz}}
\endpgfgraphicnamed
\end{equation*}
we have 
\begin{equation}\label{loopu2}
\beginpgfgraphicnamed{Qutrits//abandm8}
\InputIfFileExists{Qutrits//abandm8.tikz}{}{\input{./figures/Qutrits//abandm8.tikz}}
\endpgfgraphicnamed
\end{equation}
The result  is  a GS-LC diagram, where we used property (\ref{property2}) for the second equality, property (\ref{property5}) for the fourth equality, and (\ref{singlestates04}),  (\ref{singlestates0401}) and (\ref{singlestates0402}) for the fifth equality.

 For 
 \begin{equation*}
\beginpgfgraphicnamed{Qutrits//singleu}
\InputIfFileExists{Qutrits//singleu.tikz}{}{\input{./figures/Qutrits//singleu.tikz}}
\endpgfgraphicnamed=%
\beginpgfgraphicnamed{Qutrits//cliffnorformt}
\InputIfFileExists{Qutrits//cliffnorformt.tikz}{}{\input{./figures/Qutrits//cliffnorformt.tikz}}
\endpgfgraphicnamed
\end{equation*}
we have 
\begin{equation}\label{loopu3}
\beginpgfgraphicnamed{Qutrits//abandm9}
\InputIfFileExists{Qutrits//abandm9.tikz}{}{\input{./figures/Qutrits//abandm9.tikz}}
\endpgfgraphicnamed
\end{equation}

The result is  a GS-LC diagram. This complete the first case.
%, where we used rules (P1) and (H2), and  properties (\ref{property1}) and (\ref{property6}).

Secondly, if one operand vertex has no neighbours,  then it must be one of the 12 single qutrit states given in (\ref{singlestate}).
For  $\nststile{a_2}{a_1}\in  \mathcal{A}$ and 
 $\nststile{m_2}{m_1}\in  \mathcal{M}$, 

\begin{equation*}
\beginpgfgraphicnamed{Qutrits//abandm10}
\InputIfFileExists{Qutrits//abandm10.tikz}{}{\input{./figures/Qutrits//abandm10.tikz}}
\endpgfgraphicnamed
\end{equation*}
 \begin{equation*}
\beginpgfgraphicnamed{Qutrits//abandm11}
\InputIfFileExists{Qutrits//abandm11.tikz}{}{\input{./figures/Qutrits//abandm11.tikz}}
\endpgfgraphicnamed
\end{equation*}
Thus we are back into the application situation of lemma \ref{stabtogslc3} and lemma \ref{stabtogslc4}, which means the resulting diagram is still a GS-LC diagram.

Thirdly, if both operand vertices have non-operand neighbours,  denoted by $a$ and $b$ respectively, then by applying %
\beginpgfgraphicnamed{Qutrits//gcocopysmall}
\InputIfFileExists{Qutrits//gcocopysmall.tikz}{}{\input{./figures/Qutrits//gcocopysmall.tikz}}
\endpgfgraphicnamed to $a$ and $b$, we have 

\begin{equation*}
\beginpgfgraphicnamed{Qutrits//bhvnonb1}
\InputIfFileExists{Qutrits//bhvnonb1.tikz}{}{\input{./figures/Qutrits//bhvnonb1.tikz}}
\endpgfgraphicnamed
\end{equation*}
or
\begin{equation*}
\beginpgfgraphicnamed{Qutrits//bhvnonb2}
\InputIfFileExists{Qutrits//bhvnonb2.tikz}{}{\input{./figures/Qutrits//bhvnonb2.tikz}}
\endpgfgraphicnamed
\end{equation*}
where $U, V\in \mathcal{C}_1$. From now on, we will always denote by %
\beginpgfgraphicnamed{Qutrits//hnode}
\InputIfFileExists{Qutrits//hnode.tikz}{}{\input{./figures/Qutrits//hnode.tikz}}
\endpgfgraphicnamed the node %
\beginpgfgraphicnamed{Qutrits//RGg_Hada}
\InputIfFileExists{Qutrits//RGg_Hada.tikz}{}{\input{./figures/Qutrits//RGg_Hada.tikz}}
\endpgfgraphicnamed
 or %
\beginpgfgraphicnamed{Qutrits//RGg_Hadad}
\InputIfFileExists{Qutrits//RGg_Hadad.tikz}{}{\input{./figures/Qutrits//RGg_Hadad.tikz}}
\endpgfgraphicnamed when it is unnecessary to state explicitly.

By lemma \ref{stabtogslc3to4},  applying local complementations to $a$ and its swapping partner, the vertex operator $U$ can be changed to  the form (\ref{chgoprt}). So we have 

\begin{equation*}
\beginpgfgraphicnamed{Qutrits//bhvnonb3}
\InputIfFileExists{Qutrits//bhvnonb3.tikz}{}{\input{./figures/Qutrits//bhvnonb3.tikz}}
\endpgfgraphicnamed
\end{equation*}
or 
\begin{equation*}
\beginpgfgraphicnamed{Qutrits//bhvnonb4}
\InputIfFileExists{Qutrits//bhvnonb4.tikz}{}{\input{./figures/Qutrits//bhvnonb4.tikz}}
\endpgfgraphicnamed
\end{equation*}
where $V_1\in \mathcal{C}_1$. 
Note that 
\begin{equation*}
\beginpgfgraphicnamed{Qutrits//mslide}
\InputIfFileExists{Qutrits//mslide.tikz}{}{\input{./figures/Qutrits//mslide.tikz}}
\endpgfgraphicnamed
\end{equation*}
Therefore we have

\begin{equation*}
\beginpgfgraphicnamed{Qutrits//bhvnonb5}
\InputIfFileExists{Qutrits//bhvnonb5.tikz}{}{\input{./figures/Qutrits//bhvnonb5.tikz}}
\endpgfgraphicnamed
\end{equation*}
or 
\begin{equation}\label{bhvnonb66}
\beginpgfgraphicnamed{Qutrits//bhvnonb6}
\InputIfFileExists{Qutrits//bhvnonb6.tikz}{}{\input{./figures/Qutrits//bhvnonb6.tikz}}
\endpgfgraphicnamed
\end{equation}
where $W, V_2\in \mathcal{C}_1$. 
In the same way as we just did for $a$, we can repeat the procedure for b by choosing its own swapping partner. Note that we will get extra phases of the form %
\beginpgfgraphicnamed{Qutrits//22phasgt}
\InputIfFileExists{Qutrits//22phasgt.tikz}{}{\input{./figures/Qutrits//22phasgt.tikz}}
\endpgfgraphicnamed  or  %
\beginpgfgraphicnamed{Qutrits//green11gt}
\InputIfFileExists{Qutrits//green11gt.tikz}{}{\input{./figures/Qutrits//green11gt.tikz}}
\endpgfgraphicnamed added to $a'$s vertex operator thus merging into the operator $W$ to be $W_1$, in the case that $a$ is connected to b or to $b'$s swapping partner. Now we have
  
\begin{equation*}
\beginpgfgraphicnamed{Qutrits//bhvnonb7}
\InputIfFileExists{Qutrits//bhvnonb7.tikz}{}{\input{./figures/Qutrits//bhvnonb7.tikz}}
\endpgfgraphicnamed
\end{equation*}
or 
\begin{equation*}
\beginpgfgraphicnamed{Qutrits//bhvnonb8}
\InputIfFileExists{Qutrits//bhvnonb8.tikz}{}{\input{./figures/Qutrits//bhvnonb8.tikz}}
\endpgfgraphicnamed
\end{equation*}
where $W_1, W_2, W_3 \in \mathcal{C}_1$,  $\nststile{m_4}{m_3},  \nststile{m_{40}}{m_{30}}, \nststile{m_{41}}{m_{31}}, \cdots \nststile{m_{4s}}{m_{3s}}\in  \mathcal{M}$. Note that these green nodes  $\nststile{m_{41}}{m_{31}}, \cdots \nststile{m_{4s}}{m_{3s}}$ can be merged with the vertex operators of $b'$s neighbours by the spider rule.

Furthermore, we can merge the nodes $a$, $b$ and $c$ to get a GS-LC diagram. In fact, there are three cases.  

1) Neither $a$ and $b$ are connected by a Hadamard node nor they have common non-operand neighbours. In this case $a$, $b$ and $c$ can be directly merged, thus we have 
\begin{equation*}
\beginpgfgraphicnamed{Qutrits//bhvnonb9}
\InputIfFileExists{Qutrits//bhvnonb9.tikz}{}{\input{./figures/Qutrits//bhvnonb9.tikz}}
\endpgfgraphicnamed
\end{equation*}
which is a GS-LC diagram.

2)  $a$ and $b$ are not connected but they have common non-operand neighbours.  For any common non-operand neighbour $d$ we have 
\begin{equation*}
\beginpgfgraphicnamed{Qutrits//bhvnonb10}
\InputIfFileExists{Qutrits//bhvnonb10.tikz}{}{\input{./figures/Qutrits//bhvnonb10.tikz}}
\endpgfgraphicnamed \mbox{if}~ h_1 h_2=1  ~ \mbox{or} ~%
\beginpgfgraphicnamed{Qutrits//bhvnonb11}
\InputIfFileExists{Qutrits//bhvnonb11.tikz}{}{\input{./figures/Qutrits//bhvnonb11.tikz}}
\endpgfgraphicnamed ~\mbox{othwise},
\end{equation*}
where in either sub-case we get  a GS-LC diagram.% and we used property (\ref{property7}).

3)  $a$ and $b$ are connected. Then we have
\begin{equation*}
\beginpgfgraphicnamed{Qutrits//bhvnonb12}
\InputIfFileExists{Qutrits//bhvnonb12.tikz}{}{\input{./figures/Qutrits//bhvnonb12.tikz}}
\endpgfgraphicnamed
\end{equation*}
where $W_4\in \mathcal{C}_1$,   $\nststile{p}{p} =\nststile{1}{1} \mbox{or} \nststile{2}{2}$, and we used the rule (EU),  property   (\ref{property3}) and property   (\ref{property4}). Clearly the result is a GS-LC diagram.

Finally, if one operand vertex is connected only to the other, but the latter has a non-operand neighbour, then we have

\begin{equation*}
\beginpgfgraphicnamed{Qutrits//bhvnonb13}
\InputIfFileExists{Qutrits//bhvnonb13.tikz}{}{\input{./figures/Qutrits//bhvnonb13.tikz}}
\endpgfgraphicnamed
\end{equation*}
where $V, U\in \mathcal{C}_1$.
We can use the same strategy as we have applied for (\ref{bhvnonb66}) in the third case to cancel out the vertex operator $U$ of the second operand vertex $b$. Therefore we have
\begin{equation*}
\beginpgfgraphicnamed{Qutrits//bhvnonb14}
\InputIfFileExists{Qutrits//bhvnonb14.tikz}{}{\input{./figures/Qutrits//bhvnonb14.tikz}}
\endpgfgraphicnamed
\end{equation*}
where $V_1, T\in \mathcal{C}_1$.
If in this process the operand vertex $a$ gains one or more non-operand neighbours, then we can proceed as above. Otherwise,
\begin{equation*}
\beginpgfgraphicnamed{Qutrits//bhvnonb15}
\InputIfFileExists{Qutrits//bhvnonb15.tikz}{}{\input{./figures/Qutrits//bhvnonb15.tikz}}
\endpgfgraphicnamed
\end{equation*}
where $W=V_1\circ h\in \mathcal{C}_1$.
By equalities (\ref{loopu1}), (\ref{loopu2}) and (\ref{loopu3}), and lemma \ref{stabtogslc5}, 
$K= %
\beginpgfgraphicnamed{Qutrits//bhvnonb16}
\InputIfFileExists{Qutrits//bhvnonb16.tikz}{}{\input{./figures/Qutrits//bhvnonb16.tikz}}
\endpgfgraphicnamed$  is either a zero diagram or a stabilizer state of the form (\ref{singlestate}). If $K$ is a stabilizer state represented by a green node in (\ref{singlestate}), then the green phase operator is added to the operator $T$  by the spider rule (S1), thus we get a   GS-LC diagram. Otherwise, by the copy rule (B1) and the rule (K1),  $K$ can be copied, and it is easy to see that we still get a   GS-LC diagram. 
This completes the proof. 
\end{proof}

\subsection{Reduced GS-LC diagrams}
Further to GS-LC diagrams, we can define a more reduced form as in \cite{Miriam2}.

\begin{definition}\label{rgslcdef}

A stabilizer state diagram is called to be in a reduced GS-LC (or rGS-LC) form if it is a GS-LC diagram satisfying the following conditions:
\begin{itemize}
\item All vertex operators belong to the set
\begin{equation}\label{rgslc1}
  \mathcal{R}=\left\{\quad %
\beginpgfgraphicnamed{Qutrits/singlephases}
\InputIfFileExists{Qutrits/singlephases.tikz}{}{\input{./figures/Qutrits/singlephases.tikz}}
\endpgfgraphicnamed\quad\right\}.
  \end{equation}

\item Two adjacent vertices must not both have vertex operators including red nodes.
\end{itemize}
\end{definition}

\begin{theorem}\label{gslctorgslc}
In the qutrit ZX-calculus, each qutrit stabilizer state diagram can be rewritten into some rGS-LC diagram.
  \end{theorem}

\begin{proof}
By theorem \ref{stabtogslc1}, any stabilizer state diagram can be rewrite into a GS-LC diagram. If a vertex in the GS-LC diagram has no neighbours, by Lemma \ref{stabtogslc5}, it can be brought into one of the forms in (\ref{singlestate}), where the nine green nodes can be clearly seen as having vertex operators belong to the set $\mathcal{R}$. For the other three red nodes, note that
\begin{equation*}
\beginpgfgraphicnamed{Qutrits//vxrgslc1}
\InputIfFileExists{Qutrits//vxrgslc1.tikz}{}{\input{./figures/Qutrits//vxrgslc1.tikz}}
\endpgfgraphicnamed
\end{equation*}
\begin{equation*}
\beginpgfgraphicnamed{Qutrits//vxrgslc2}
\InputIfFileExists{Qutrits//vxrgslc2.tikz}{}{\input{./figures/Qutrits//vxrgslc2.tikz}}
\endpgfgraphicnamed
\end{equation*}
\begin{equation*}
\beginpgfgraphicnamed{Qutrits//vxrgslc3}
\InputIfFileExists{Qutrits//vxrgslc3.tikz}{}{\input{./figures/Qutrits//vxrgslc3.tikz}}
\endpgfgraphicnamed
\end{equation*}
%where we used equalities (\ref{singlestates0401}) and (\ref{singlestates0402}).
Therefore,  the vertex operators on these three red nodes can be  brought into  elements of  $\mathcal{R}$. 

From now on, we can assume that each vertex in the GS-LC diagram has at least one neighbour. Our strategy is to prove the theorem in three steps: First we prove that each vertex operator in the GS-LC diagram can be brought into the following form:
\begin{equation}\label{rprimeform}
  \mathcal{R^{\prime}}=\left\{\quad %
\beginpgfgraphicnamed{Qutrits//rprime}
\InputIfFileExists{Qutrits//rprime.tikz}{}{\input{./figures/Qutrits//rprime.tikz}}
\endpgfgraphicnamed\quad\right\}.
  \end{equation}
where $ \nststile{q_{2}}{q_{1}} \in \mathcal{Q}, \nststile{p_{2}}{p_{1}}  \in \mathcal{P}$. Then we show that any GS-LC diagram with vertex operators all belonging to $\mathcal{R^{\prime}}$ can be further brought into a form where any two adjacent vertices must not both have vertex operators including red nodes while the vertex operators still resides in $\mathcal{R^{\prime}}$.  Finally we prove that the GS-LC diagram obtained in the second step can be rewritten into a rGS-LC diagram.

For the first step of the strategy, we pick  an arbitrary node $v$ that has at least one neighbour. By theorem \ref{uniformsin}, the vertex operator $U$ of the vertex $v$  has the form  (\ref{uniformsin1}),  (\ref{uniformsin2}) or (\ref{uniformsin3}). We deal with these three forms separately.

 Firstly consider that $U$ has the form  (\ref{uniformsin1}):
 \begin{equation}\label{u2nodes0}
\beginpgfgraphicnamed{Qutrits//singleu}
\InputIfFileExists{Qutrits//singleu.tikz}{}{\input{./figures/Qutrits//singleu.tikz}}
\endpgfgraphicnamed=%
\beginpgfgraphicnamed{Qutrits//cliffnorformj}
\InputIfFileExists{Qutrits//cliffnorformj.tikz}{}{\input{./figures/Qutrits//cliffnorformj.tikz}}
\endpgfgraphicnamed.
\end{equation}
We analyse all possible cases of the phase $ \nststile{a_{4}}{a_{3}}$.  

If  $ \nststile{a_{4}}{a_{3}}\in \mathcal{M} $, then either $\nststile{a_{4}}{a_{3}}=\nststile{0}{0}$, where clearly $U\in   \mathcal{R^{\prime}}$, or
$ \nststile{a_{4}}{a_{3}}\in \{ \nststile{2}{1}, \nststile{1}{2}\}$. In the latter case, by the commutation rule (K2) , the copy rule (K1) and the colour change rules (H2) and (H2$^{\prime}$), the red node  $ \nststile{a_{4}}{a_{3}}$ can be pushed up through the node $ \nststile{a_{2}}{a_{1}}$, copied  by $v$ into connected edges with its neighbours, then colour-changed from  red to green by Hadamard nodes and finally merged into the vertex operators of neighbours of $v$. Therefore the  vertex operator $U$ now is just a green node, thus belongs to the set  $\mathcal{R^{\prime}}$. 

If $ \nststile{a_{4}}{a_{3}}\in \mathcal{N}=\{\nststile{0}{1},  \nststile{1}{0}, \nststile{0}{2},  \nststile{2}{0}\}$, then $ \nststile{a_{4}}{a_{3}}=\nststile{p_{2}}{p_{1}} + \nststile{m_{2}}{m_{1}}$, where $\nststile{p_{2}}{p_{1}}  \in \mathcal{P},   \nststile{m_{2}}{m_{1}} \in \{ \nststile{2}{1}, \nststile{1}{2}\}$. Thus by the commutation rule (K2),
 \begin{equation*}
\beginpgfgraphicnamed{Qutrits//singleu}
\InputIfFileExists{Qutrits//singleu.tikz}{}{\input{./figures/Qutrits//singleu.tikz}}
\endpgfgraphicnamed=%
\beginpgfgraphicnamed{Qutrits//cliffnorformju3}
\InputIfFileExists{Qutrits//cliffnorformju3.tikz}{}{\input{./figures/Qutrits//cliffnorformju3.tikz}}
\endpgfgraphicnamed.
\end{equation*}

In a similar way as above, the node  $\nststile{m_{2}}{m_{1}}$  can be merged into the vertex operators of neighbours of $v$. Now $U$ has the form
\begin{equation}\label{u2nodes}
\beginpgfgraphicnamed{Qutrits//cliffnorformju32}
\InputIfFileExists{Qutrits//cliffnorformju32.tikz}{}{\input{./figures/Qutrits//cliffnorformju32.tikz}}
\endpgfgraphicnamed.
\end{equation}

We need to consider all the possible cases of  $\nststile{\bar{a}_2}{\bar{a}_1}$  now. If $\nststile{\bar{a}_2}{\bar{a}_1} = \nststile{0}{0} $, then applying $1$-local complementations to $v$, the red node   $\nststile{p_{2}}{p_{1}}$ can be removed, so $U$ is changed to the identity, which belongs to 
set  $\mathcal{R^{\prime}}$.  If $\nststile{\bar{a}_2}{\bar{a}_1} \in \{ \nststile{2}{1}, \nststile{1}{2}\}$,  then
 \begin{equation*}
\beginpgfgraphicnamed{Qutrits//singleu}
\InputIfFileExists{Qutrits//singleu.tikz}{}{\input{./figures/Qutrits//singleu.tikz}}
\endpgfgraphicnamed=%
\beginpgfgraphicnamed{Qutrits//cliffnorformju33}
\InputIfFileExists{Qutrits//cliffnorformju33.tikz}{}{\input{./figures/Qutrits//cliffnorformju33.tikz}}
\endpgfgraphicnamed
\end{equation*}
where $\nststile{n_{2}}{n_{1}}  \in \mathcal{N},  \nststile{\bar{m}_2}{\bar{m}_1} \in \{ \nststile{2}{1}, \nststile{1}{2}\},  \nststile{\bar{p}_2}{\bar{p}_1} \in  \mathcal{P}$. As  we did above, the red node $ \nststile{\bar{m}_2}{\bar{m}_1}$  merged into the vertex operators of neighbours of $v$, and the red node $ \nststile{\bar{p}_2}{\bar{p}_1}$ can be removed by applying $1$-local complementations to $v$. So $U$ is changed to be just the green node $\nststile{\bar{a}_2}{\bar{a}_1}$, which clearly belongs to  the
set  $\mathcal{R^{\prime}}$. To sum up, for $\nststile{\bar{a}_2}{\bar{a}_1}  \in \mathcal{M}$ in the case of (\ref{u2nodes}), the vertex operator $U$ can be changed to have no red nodes. For the remaining case where 
 $\nststile{\bar{a}_2}{\bar{a}_1} \in \mathcal{Q}$, it is clear that the vertex operator $U$ in form (\ref{u2nodes}) now  belongs  to the
set  $\mathcal{R^{\prime}}$. 

Therefore, for $ \nststile{a_{4}}{a_{3}}\in \mathcal{N}$ in the form of  (\ref{u2nodes0}), the vertex operator $U$ can be brought into a form of  $\mathcal{R^{\prime}}$.

Now the remaining case of $ \nststile{a_{4}}{a_{3}}$ in the form of  (\ref{u2nodes0})  is that  $ \nststile{a_{4}}{a_{3}}  \in  \mathcal{P}$, this case is exactly the same as (\ref{u2nodes}), which has already been proved. Therefore, for 
 \begin{equation*}
\beginpgfgraphicnamed{Qutrits//singleu}
\InputIfFileExists{Qutrits//singleu.tikz}{}{\input{./figures/Qutrits//singleu.tikz}}
\endpgfgraphicnamed=%
\beginpgfgraphicnamed{Qutrits//cliffnorformj}
\InputIfFileExists{Qutrits//cliffnorformj.tikz}{}{\input{./figures/Qutrits//cliffnorformj.tikz}}
\endpgfgraphicnamed
\end{equation*}
 $U$ can always be brought into the form of  $\mathcal{R^{\prime}}$.
 
  Secondly we consider that $U$ has the form  (\ref{uniformsin2}):
  \begin{equation*}
\beginpgfgraphicnamed{Qutrits//singleu}
\InputIfFileExists{Qutrits//singleu.tikz}{}{\input{./figures/Qutrits//singleu.tikz}}
\endpgfgraphicnamed=%
\beginpgfgraphicnamed{Qutrits//cliffnorformk}
\InputIfFileExists{Qutrits//cliffnorformk.tikz}{}{\input{./figures/Qutrits//cliffnorformk.tikz}}
\endpgfgraphicnamed.
\end{equation*}
We first apply $1$-local complementations to $v$, then the red node $ \nststile{p_2}{p_1}$  is removed, thus we are back to the case (\ref{u2nodes0}), which has already been proved.

Thirdly we consider that $U$ has the form  (\ref{uniformsin3}):
  \begin{equation*}
\beginpgfgraphicnamed{Qutrits//singleu}
\InputIfFileExists{Qutrits//singleu.tikz}{}{\input{./figures/Qutrits//singleu.tikz}}
\endpgfgraphicnamed=%
\beginpgfgraphicnamed{Qutrits//cliffnorformt}
\InputIfFileExists{Qutrits//cliffnorformt.tikz}{}{\input{./figures/Qutrits//cliffnorformt.tikz}}
\endpgfgraphicnamed.
\end{equation*}
Here we can push the two Hadamard nodes up by property (\ref{property1}), then apply the graphical transformation $\circ_2  v $,  the two Hadamard nodes can be removed by theorem \ref{gandgst3} and property  (\ref{property1}). Therefore we get back to the case (\ref{u2nodes0}), which has already been proved.

So far we are done with the chosen node $v$ in the first step of the proof strategy. To proceed to deal with other nodes than $v$ in the GS-LC diagram, we need to consider the impact on $v$'s neighbours whose vertex operators belong to the set $\mathcal{R^{\prime}}$ during the above operatings on $v$. First note that in the above process of proof, there is not any red node added to any neighbour of $v$, only a green node $ \nststile{p_2}{p_1}$ or  $ \nststile{m_2}{m_1}$ is possibly added to   $v$'s neighbours, where $\nststile{p_2}{p_1} \in \mathcal{P},   \nststile{m_2}{m_1}\in \{ \nststile{2}{1}, \nststile{1}{2}\}$. If  it is the green node $ \nststile{m_2}{m_1}$ that is added to $v$'s neighbours, then their vertex operators still belong to the set $\mathcal{R^{\prime}}$.  If the green node $ \nststile{p_2}{p_1}$  is added to $v$'s neighbours, then the resulted vertex operator that beyond the set $\mathcal{R^{\prime}}$ must be the sub-case of (\ref{u2nodes}) where  $\nststile{\bar{a}_2}{\bar{a}_1}  \in \mathcal{M}$, which we have shown that the red node can be removed. Therefore, each time when we have brought a vertex operator in to an element of  $\mathcal{R^{\prime}}$, we need to check whether the vertex operators of  neighbours of the operand vertex are still belong to $\mathcal{R^{\prime}}$. If not, then the red node will be removed. Since each vertex operator needs to be considered at most twice, the process will eventually terminate. Therefore all vertex operators must belong to the set $\mathcal{R^{\prime}}$ after finite steps.

By now we have finished the first step of the proof strategy: we have proved that any stabilizer state diagram is equal to some GS-LC diagram such that  all vertex operators belong to the set $\mathcal{R^{\prime}}$.

For the second step of the strategy, we are going to prove that each GS-LC diagram with vertex operators all belonging to $\mathcal{R^{\prime}}$ can be further brought into a form where any two adjacent vertices must not both have vertex operators including red nodes,  with all vertex operators being still in $\mathcal{R^{\prime}}$. 

Suppose there are two adjacent qutrits $a$ and $b$ which have red
nodes in their vertex operators, i.e. there is a subdiagram of the form
 \begin{equation}\label{2redside}
\beginpgfgraphicnamed{Qutrits//2rednodes2}
\InputIfFileExists{Qutrits//2rednodes2.tikz}{}{\input{./figures/Qutrits//2rednodes2.tikz}}
\endpgfgraphicnamed
\end{equation}
where  $\nststile{p_{2}}{p_{1}},  \nststile{p_{4}}{p_{3}} \in \mathcal{P},  \nststile{q_{2}}{q_{1}},  \nststile{q_{4}}{q_{3}} \in \mathcal{Q}$.
We  operate on the vertex $b$ to remove the red node from its vertex operator.
Firstly, 
 \begin{equation}\label{2redside2}
\beginpgfgraphicnamed{Qutrits//2rednodes3}
\InputIfFileExists{Qutrits//2rednodes3.tikz}{}{\input{./figures/Qutrits//2rednodes3.tikz}}
\endpgfgraphicnamed
\end{equation}
where $\nststile{m_2}{m_1}\in \mathcal{M}$ and  we used local complementations about 
$a$ to remove the green node $\nststile{p_6}{p_5} $ on the output of $b$ for the second equality. Below we analyse all possible values of $\nststile{m_2}{m_1}$ in (\ref{2redside2}).

 If $\nststile{m_2}{m_1}=\nststile{0}{0}$, then we  can apply  
local complementations about 
$b$ to remove the red node $\nststile{p_4}{p_3} $, and we have 
 \begin{equation}\label{2redside3}
\beginpgfgraphicnamed{Qutrits//2rednodes4}
\InputIfFileExists{Qutrits//2rednodes4.tikz}{}{\input{./figures/Qutrits//2rednodes4.tikz}}
\endpgfgraphicnamed
\end{equation}
Now we proceed on simplifying the vertex operator of $a$:
 \begin{equation}\label{2redside4}
\beginpgfgraphicnamed{Qutrits//2rednodes5}
\InputIfFileExists{Qutrits//2rednodes5.tikz}{}{\input{./figures/Qutrits//2rednodes5.tikz}}
\endpgfgraphicnamed
\end{equation}
where  $\nststile{p_i}{p_j}\in \mathcal{P},  \nststile{m_4}{m_3}, \nststile{m_6}{m_5}\in \{ \nststile{2}{1}, \nststile{1}{2}\} ~ \mbox{or}   \nststile{m_4}{m_3}=\nststile{m_6}{m_5}=\nststile{0}{0}$. By lemma \ref{pzxz}, the first three nodes on the top of the output of $a$ can be rewritten in one of two forms. If we use the rewritten form with two Hadamard nodes, then we have 
 \begin{equation}\label{2redside5}
\beginpgfgraphicnamed{Qutrits//2rednodes6}
\InputIfFileExists{Qutrits//2rednodes6.tikz}{}{\input{./figures/Qutrits//2rednodes6.tikz}}
\endpgfgraphicnamed
\end{equation}
where we apply  the graphical transformation $\circ_2  a $ and local complementations about $a$ for the second equality, perform local complementations about $a$ for the third equality, and make a copy of red node $\nststile{m_6}{m_5}$ through $a$ for the fourth equality. Obviously we get a desired form at the end of (\ref{2redside5}). Otherwise we use the rewritten form without Hadamard node,  then
 \begin{equation}\label{2redside6}
\beginpgfgraphicnamed{Qutrits//2rednodes7}
\InputIfFileExists{Qutrits//2rednodes7.tikz}{}{\input{./figures/Qutrits//2rednodes7.tikz}}
\endpgfgraphicnamed
\end{equation}
where $\nststile{\bar{p}_6}{\bar{p}_5}  \in \mathcal{P},  \nststile{p_8}{p_7}+\nststile{p_2}{p_1} = \nststile{p_{10}}{p_{9}}\in  \{ \nststile{0}{0}, \nststile{1}{1}, \nststile{2}{2}\}$, and  we apply  local complementations about $a$ for the second equality. Here we need to consider all the possible values of $ \nststile{p_{10}}{p_{9}}$ in (\ref{2redside6}).  If $ \nststile{p_{10}}{p_{9}}=\nststile{0}{0}$,  then
 
 \begin{equation}\label{2redside7}
\beginpgfgraphicnamed{Qutrits//2rednodes8}
\InputIfFileExists{Qutrits//2rednodes8.tikz}{}{\input{./figures/Qutrits//2rednodes8.tikz}}
\endpgfgraphicnamed
\end{equation}
where $\nststile{\bar{p}_6}{\bar{p}_5}  +\nststile{m_4}{m_3} = \nststile{q_6}{q_5}\in  \mathcal{Q}, \nststile{q_8}{q_7}\in  \mathcal{Q},  \nststile{\bar{m}_6}{\bar{m}_5}  \in \mathcal{M}$, and we make a copy of red node $\nststile{m_6}{m_5}$ through $a$ for the third equality. Otherwise $ \nststile{p_{10}}{p_{9}} \in \mathcal{P}$ in (\ref{2redside6}),  then
 \begin{equation}\label{2redside8}
\beginpgfgraphicnamed{Qutrits//2rednodes9}
\InputIfFileExists{Qutrits//2rednodes9.tikz}{}{\input{./figures/Qutrits//2rednodes9.tikz}}
\endpgfgraphicnamed
\end{equation}
where $\nststile{q_{10}}{q_{9}}\in  \mathcal{Q}, \nststile{q_{10}}{q_{9}}+\nststile{m_4}{m_3}=\nststile{q_{12}}{q_{11}}\in  \mathcal{Q}, \nststile{q_{14}}{q_{13}}\in  \mathcal{Q},  \nststile{\bar{m}_6}{\bar{m}_5}, \nststile{m_8}{m_7},  \nststile{\bar{m}_8}{\bar{m}_7}  \in \mathcal{M}$, and we make copies of red nodes $\nststile{m_6}{m_5}$ and $\nststile{m_8}{m_7}$  through $a$ for the third equality and the sixth equality respectively. 

By now,  for  $\nststile{m_2}{m_1}=\nststile{0}{0}$, we have proved that the resulting diagram in (\ref{2redside2}) can be brought into a form that at least one qutrit of $a$ and $b$ has no red
nodes in its vertex operator while both $a$ and $b$ have vertex operators belonging to  $\mathcal{R^{\prime}}$.

The remaining case is that $\nststile{m_2}{m_1}\in \{ \nststile{2}{1}, \nststile{1}{2}\} $  in (\ref{2redside2}). In this case, 

\begin{equation}\label{2redside9}
\beginpgfgraphicnamed{Qutrits//2rednodes10}
\InputIfFileExists{Qutrits//2rednodes10.tikz}{}{\input{./figures/Qutrits//2rednodes10.tikz}}
\endpgfgraphicnamed
\end{equation}
where $\nststile{m_4}{m_3},  \nststile{m_6}{m_5}, \nststile{m_8}{m_7}, \nststile{m_{10}}{m_{9}} \in \{ \nststile{2}{1}, \nststile{1}{2}\},   \nststile{m_8}{m_7}+\nststile{m_{10}}{m_{9}} = \nststile{m_{12}}{m_{11}} \in  \mathcal{M}$, $\nststile{q_{4}}{q_{3}}\in  \mathcal{Q}$.

In the resulting diagram at the end of (\ref{2redside9}), the first four nodes on the top of the output of $a$ are of the same type as the corresponding part of (\ref{2redside3}). So we can use the obtained results of (\ref{2redside3}),  push the red node $ \nststile{m_{12}}{m_{11}} $  through green nodes and make it  copied and added to  vertex operators of $a$'s neighbours in green colour by the commutation rule (K2) and colour-change rules (H2) and (H2$^{\prime}$). In this way, for $\nststile{m_2}{m_1}\in \{ \nststile{2}{1}, \nststile{1}{2}\} $  in (\ref{2redside2}), the resulting diagram  can be brought into a form that at least one qutrit of $a$ and $b$ has no red
nodes in its vertex operator, while both $a$ and $b$ have vertex operators belonging to  $\mathcal{R^{\prime}}$. 

Therefore, the diagram (\ref{2redside}) can always be brought into a form such that at least one qutrit of $a$ and $b$ has no red
nodes in its vertex operator, while both $a$ and $b$ have vertex operators belonging to  $\mathcal{R^{\prime}}$. This finishes the second step of the proof strategy.

The third step of the proof strategy is to show that the vertex operators not only belong to $\mathcal{R^{\prime}}$, but further resides in $\mathcal{R}$. To do this, it suffices to prove that the red node in the following diagram can be removed:

\begin{equation}\label{2redside10}
\beginpgfgraphicnamed{Qutrits//2rednodes11}
\InputIfFileExists{Qutrits//2rednodes11.tikz}{}{\input{./figures/Qutrits//2rednodes11.tikz}}
\endpgfgraphicnamed
\end{equation}
where the vertex operator of $a$ belongs to the set
\begin{equation}\label{maset}
  \mathcal{T}=\left\{\quad %
\beginpgfgraphicnamed{Qutrits//alterphases}
\InputIfFileExists{Qutrits//alterphases.tikz}{}{\input{./figures/Qutrits//alterphases.tikz}}
\endpgfgraphicnamed\quad\right\}.
  \end{equation}
For the sake of conciseness, we just show one case, the others are similar:
\begin{equation}\label{2redside11}
\beginpgfgraphicnamed{Qutrits//2rednodes12}
\InputIfFileExists{Qutrits//2rednodes12.tikz}{}{\input{./figures/Qutrits//2rednodes12.tikz}}
\endpgfgraphicnamed
\end{equation}
where we  apply  $1$-local complementations about $a$ for the first equality, use equality  (\ref{pzxzhh}) for the second equality, and apply  the graphical transformation $\circ_2  a $  for the third equality.

Each time when we have brought a diagram of form (\ref{2redside}) into a diagram that satisfies the conditions of rGS-LC diagram for the two connected nodes, we need to check whether the vertex operators of  neighbours of the operand vertex are still belong to $\mathcal{R}$. If not, then the red node will be removed. Since the number of vertex operator beyond $\mathcal{R}$ in such a diagram is always decreasing,  all vertex operators must belong to the set $\mathcal{R}$ after finite steps. Therefore, any GS-LC diagram is equal to some rGS-LC diagram within the ZX-calculus.

  \end{proof}

\subsection{Transformations of rGS-LC diagrams}
 In this subsection we show how to  transform one rGS-LC diagrams into another rGS-LC diagrams. Note that we call the 
graphical transformation $\circ_2  v $ a doubling-neighbour-edge transformation.
\begin{lemma}\label{equivrgslc}
Suppose there is a rGS-LC diagram which has a pair of neighbouring qutrits $a$ and $b$  as follows:\begin{equation}\label{equivrgslcs1}
\beginpgfgraphicnamed{Qutrits/equivrgslct1}
\InputIfFileExists{Qutrits/equivrgslct1.tikz}{}{\input{./figures/Qutrits/equivrgslct1.tikz}}
\endpgfgraphicnamed
\end{equation}
where  $\nststile{a_2}{a_1}\in \mathcal{A}, \nststile{p_1}{p_1}\in \mathcal{P}, \nststile{m_2}{m_1}\in \mathcal{M}, \nststile{q_2}{q_1} = \nststile{p_1}{p_1} +\nststile{m_2}{m_1}$, $h$ stands for either an $H$ node or an $H^{\dagger}$ node.   Then a  rGS-LC diagram with the  following pair can be obtained by  performing firstly a $p_1$-local complementation about $b$, followed by a
$(-p_1)$-local complementation about $a$, and possibly  a further  
$(-p_1)$-local complementation about $b$, in addition with some doubling-neighbour-edge operations $\circ_2  a(b) $ and copying operations on red nodes with phase angles in $ \mathcal{M}$:
\begin{equation}\label{equivrgslcs2}
\beginpgfgraphicnamed{Qutrits/equivrgslct2}
\InputIfFileExists{Qutrits/equivrgslct2.tikz}{}{\input{./figures/Qutrits/equivrgslct2.tikz}}
\endpgfgraphicnamed
\end{equation}
where $\nststile{a^{\prime}_2}{a^{\prime}_1}\in \mathcal{A}, \nststile{p^{\prime}_1}{p^{\prime}_1}\in \mathcal{P}, \nststile{m^{\prime}_2}{m^{\prime}_1}\in \mathcal{M}$.
\end{lemma}

\begin{proof}
We first show the process of transformation from (\ref{equivrgslcs1}) to (\ref{equivrgslcs2}), then check that all the involved operations, including  the two local complementations, doubling-neighbour-edge operations and copying operations,  will transform all vertex operators to allowed ones.

Firstly consider the case that  $\nststile{m_2}{m_1}=\nststile{0}{0}, \nststile{a_2}{a_1}=\nststile{m_4}{m_3} \in \mathcal{M}$ in  (\ref{equivrgslcs1}). We have
\begin{equation}\label{equivrgslcs33}
\beginpgfgraphicnamed{Qutrits//equivrgslct33}
\InputIfFileExists{Qutrits//equivrgslct33.tikz}{}{\input{./figures/Qutrits//equivrgslct33.tikz}}
\endpgfgraphicnamed
\end{equation}
where  $\nststile{m_j}{m_i} \in \mathcal{M}, $ we applied a $p_1$-local complementation about $b$ for the first equality, then a $(-p_1)$-local complementation about $a$ for the second equality, and copied the red node $\nststile{m_6}{m_5}$  in the last equality.  The resulting diagram has the two properties in the definition of  rGS-LC diagram.

Secondly consider the case that  $\nststile{m_2}{m_1}  \in \mathcal{M}, \nststile{a_2}{a_1}=\nststile{m_4}{m_3} \in \mathcal{M}$ in  (\ref{equivrgslcs1}). We have

\begin{equation}\label{equivrgslcs4}
\beginpgfgraphicnamed{Qutrits//equivrgslct4}
\InputIfFileExists{Qutrits//equivrgslct4.tikz}{}{\input{./figures/Qutrits//equivrgslct4.tikz}}
\endpgfgraphicnamed
\end{equation}
where  $\nststile{m_j}{m_i} \in \mathcal{M}, \nststile{m_{14}}{m_{13}} = \nststile{m_{8}}{m_{7}} +\nststile{m_{2}}{m_{1}},    \nststile{m_{12}}{m_{11}} = \nststile{m^{\prime}_{6}}{m^{\prime}_{5}} +\nststile{m_{4}}{m_{3}}$, we copied the red node $\nststile{m_6}{m_5}$ in the third equality. Now we can proceed base on the result of (\ref{equivrgslcs33}). That is
\begin{equation}\label{equivrgslcs5}
\beginpgfgraphicnamed{Qutrits//equivrgslct5}
\InputIfFileExists{Qutrits//equivrgslct5.tikz}{}{\input{./figures/Qutrits//equivrgslct5.tikz}}
\endpgfgraphicnamed
\end{equation}
where  $\nststile{m_j}{m_i} \in \mathcal{M},  \nststile{m_{20}}{m_{19}} = \nststile{m_{14}}{m_{13}} +\nststile{m_{16}}{m_{15}}, \nststile{m_{24}}{m_{23}} = \nststile{m_{18}}{m_{17}} +\nststile{m_{22}}{m_{21}}$,  we copied the red node $\nststile{m_{10}}{m_9}$  in the third equality. The resulting diagram has the two properties in the definition of  rGS-LC diagram.

\iffalse
Thirdly consider the case that  $\nststile{m_2}{m_1} =\nststile{0}{0}, \nststile{a_2}{a_1}=\nststile{p_2}{p_2} \in \mathcal{P}$ in  (\ref{equivrgslcs1}). Note that either $p_1=p_2$ or $p_1=-p_2$ since $p_1, p_2 \in \mathbb{Z}_3 $. If $p_1=p_2$,  we have

\begin{equation}\label{equivrgslcs6}
  %
\beginpgfgraphicnamed{Qutrits//equivrgslct6}
\InputIfFileExists{Qutrits//equivrgslct6.tikz}{}{\input{./figures/Qutrits//equivrgslct6.tikz}}
\endpgfgraphicnamed
\end{equation}
where  we applied a $p_1$-local complementation about $b$ for the first equality and the last equality,  a $(-p_1)$-local complementation about $a$ for the second equality, and used the property (\ref{property8}) for the third equality.  The resulted diagram has the two properties in the definition of  rGS-LC diagram.

If $p_1=-p_2$, then $p_2=-p_1=2p_1 (mod 3)$, and we have
\begin{equation}\label{equivrgslcs7}
  %
\beginpgfgraphicnamed{Qutrits//equivrgslct7}
\InputIfFileExists{Qutrits//equivrgslct7.tikz}{}{\input{./figures/Qutrits//equivrgslct7.tikz}}
\endpgfgraphicnamed
\end{equation}
where $h_1=H~\mbox{or}~H^{-1}$,  we applied a $p_1$-local complementation about $b$ for the first equality,  a $(-p_1)$-local complementation about $a$ for the second equality, and doubling-neighbour-edge operation $\circ_2  b$ for the last equality.  The resulted diagram has the two properties in the definition of  rGS-LC diagram.
\fi

Finally consider the case that  $\nststile{m_2}{m_1}  \in \mathcal{M}, \nststile{a_2}{a_1}=\nststile{q_2}{q_1} =\nststile{p_2}{p_2} +\nststile{m_4}{m_3}, \in \mathcal{M}$ in  (\ref{equivrgslcs1}), where $\nststile{p_2}{p_2} \in \mathcal{P},  \nststile{m_4}{m_3} \in \mathcal{M}$. By (\ref{equivrgslcs4}) and (\ref{equivrgslcs5}), we have

\begin{equation}\label{equivrgslcs8}
\beginpgfgraphicnamed{Qutrits//equivrgslct8}
\InputIfFileExists{Qutrits//equivrgslct8.tikz}{}{\input{./figures/Qutrits//equivrgslct8.tikz}}
\endpgfgraphicnamed
\end{equation}
where $\nststile{m_j}{m_i} \in \mathcal{M}$.  Note that either $p_1=p_2$ or $p_1=-p_2$ since $p_1p_2\neq 0, p_1, p_2 \in \mathbb{Z}_3 $. If $p_1=p_2$,  we have

\begin{equation}\label{equivrgslcs9}
\beginpgfgraphicnamed{Qutrits//equivrgslct9}
\InputIfFileExists{Qutrits//equivrgslct9.tikz}{}{\input{./figures/Qutrits//equivrgslct9.tikz}}
\endpgfgraphicnamed
\end{equation}
where  $\nststile{m_j}{m_i} \in \mathcal{M}$, we used the property (\ref{property8}) for the first equality,   copied the red node $\nststile{m_{10}}{m_9}$  in the third equality,  and applied a $(-p_1)$-local complementation about $b$ for the last equality.  The resulting diagram has the two properties in the definition of  rGS-LC diagram.

If $p_1=-p_2$, then $p_2=-p_1=2p_1 (mod 3)$, and we have
\begin{equation}\label{equivrgslcs10}
\beginpgfgraphicnamed{Qutrits//equivrgslct10}
\InputIfFileExists{Qutrits//equivrgslct10.tikz}{}{\input{./figures/Qutrits//equivrgslct10.tikz}}
\endpgfgraphicnamed
\end{equation}
where $h_1=H~\mbox{or}~H^{-1}$,  and we used doubling-neighbour-edge operation $\circ_2  b$ for the third equality.  The resulting diagram has the two properties in the definition of  rGS-LC diagram.

Next we need to check that all the operations applied above will transform all vertex operators to allowed ones. First note that we can neglect all vertices which are not connected to $a$ or $b$, since their vertex operators won't be changed under the transformation. 

Besides, the neighbouring vertices of the operand vertex will gain some green phase operators through copying red nodes with phase angles in $ \mathcal{M}$. Since the vertex operator of a  neighbouring vertex  either is merely a green phase or contains a green phase in the group $ \mathcal{M}$, this copying operation  preserves property 1 of rGS-LC diagrams.

As the doubling-neighbour-edge operation does not change any connectivity, but only add a $D$ node to the vertex operator of the operand vertex whenever it is necessary, it helps for   preserving the first property of rGS-LC diagrams.

Furthermore we consider the effect of local complementations. First, each vertex operator consisting of only green phases on a vertex other than $a$ and $b$ will continue to be a green phase under local complementations, since local complementations merely give green phases to such kind of vertex. Second, we consider vertices adjacent to $a$ or $b$ with vertex operator containing a red node. By the definition of  rGS-LC diagram, such vertices must not be adjacent to $a$.  Thus we can assume that $w$ is a vertex whose vertex operator contains a red node and $\{w, b\}$ is an edge of the rGS-LC diagram before the transformation by local complementation. Then the $p_1$-local complementation about $b$ adds a phase  %
\beginpgfgraphicnamed{Qutrits//_p1-p1}
\InputIfFileExists{Qutrits//_p1-p1.tikz}{}{\input{./figures/Qutrits//_p1-p1.tikz}}
\endpgfgraphicnamed to the vertex operator of $w$ and produces an edge between $w$ and $a$ with weight $\Gamma^{\prime}_{wa}=0+p_1\Gamma_{wb}\Gamma_{ab}=p_1\Gamma_{wb}\Gamma_{ab}\neq 0 (mod 3)$. And the $(-p_1)$-local complementation about $a$  adds %
\beginpgfgraphicnamed{Qutrits//p1p1}
\InputIfFileExists{Qutrits//p1p1.tikz}{}{\input{./figures/Qutrits//p1p1.tikz}}
\endpgfgraphicnamed  to $w$ with the weight 
$\Gamma^{\prime\prime}_{wb}=\Gamma^{\prime}_{wb}+(-p_1)\Gamma^{\prime}_{wa}\Gamma^{\prime}_{ab}=\Gamma_{wb}-(p_1)^2\Gamma_{wb}\Gamma_{ab}^2=\Gamma_{wb}-\Gamma_{wb} = 0 (mod 3)$, where $\Gamma^{\prime}_{wb}=\Gamma_{wb}, ~\Gamma^{\prime}_{ab}=\Gamma_{ab}$, thus removes the edge between $w$ and $b$. So if there further follows a  $(-p_1)$-local complementation about $b$, nothing will be added to  the vertex operator of $w$.  Therefore the vertex operator of $w$ still resides in the set $ \mathcal{R}$, which means the transformation preserves the first property of the rGS-LC diagram.

Finally we consider two vertices $v,w$ being adjacent to $b$ which simultaneously have red nodes in their vertex operators. Since $v,w$ lie in a rGS-LC diagram, there is no edge between $v$ and $w$. After the $p_1$-local  complementation about $b$, there will be an edge between $v$ and $w$ and two edges $\{a, v\}$  and $\{a, w\}$ with weight $\Gamma^{\prime}_{vw}=p_1\Gamma_{wb}\Gamma_{vb},~\Gamma^{\prime}_{aw}=p_1\Gamma_{wb}\Gamma_{ab}, ~\Gamma^{\prime}_{av}=p_1\Gamma_{vb}\Gamma_{ab}$ respectively.  Also the $(-p_1)$-local complementation about $a$ results in $\Gamma^{\prime\prime}_{vw}=\Gamma^{\prime}_{vw}+(-p_1)\Gamma^{\prime}_{wa}\Gamma^{\prime}_{va}=p_1\Gamma_{wb}\Gamma_{vb}-p_1(p_1)^2\Gamma_{wb}\Gamma_{vb}\Gamma_{ab}^2= 0 (mod 3)$, which means the edge $\{v, w\}$ is removed, thus  $v$ and $w$ are still non-adjacent. By the above calculation, the edges  $\{v, b\}$ and  $\{w, b\}$ are also removed,  so if there further follows a  $(-p_1)$-local complementation about $b$, $v$ and $w$ will be non-adjacent. Clearly, edges connecting vertices which are not adjacent to $a$ or $b$ remain unchanged. Therefore the the second property of the rGS-LC diagram is also retained after transformation. To sum up, the resulting diagram is still a rGS-LC diagram.
 \end{proof}

\section{Completeness}

\subsection{Comparing rGS-LC diagrams}
In this subsection, we show that a pair of rGS-LC diagrams can be transformed into such a form that they are equal under the standard interpretation if and only if they  are identical. 

 \begin{definition}\cite{Miriam2}
A pair of rGS-LC diagrams on the same number of qutrit is called simplified if there are no pairs of qutrits $a, b$,  such that $a$ has a red node in its vertex operator in the first diagram but not in the second, $b$ has a red node in the second diagram but not in the first, and $a$ and $b$ are adjacent in at least one of the diagrams.
 \end{definition}

\begin{lemma}\label{rgspsym}
Each pair of rGS-LC diagrams on the same number of qutrits can be made into a simplified form.
\end{lemma}
\begin{proof}
The proof is same as that of the qubit case presented in  \cite{Miriam1, Miriam2}.
 \end{proof}

\begin{lemma}\label{unpaired}
Any component  diagrams of a simplified pair of rGS-LC with  an unpaired red node are unequal, where this red node resides as a vertex operator only  in one of the pair of diagrams. 
\end{lemma}

\begin{proof}
We denote by $a$ the vertex which has the red node as a vertex operator, by $D_1$ the diagram where $a$ resides,  and by $D_2$ the other diagram of the simplified pair.  Then we give the proof by considering three cases.

Firstly, if $a$ has no neighbours in either $D_1$ or $D_2$, then $a$ along with its vertex operators must be a single qutrit state. Thus $D_1$ and $D_2$ are equal only if the two states of $a$ in each diagram are the same.
But in fact,
\begin{equation}\label{gnntrd}
\beginpgfgraphicnamed{Qutrits//gnntred1}
\InputIfFileExists{Qutrits//gnntred1.tikz}{}{\input{./figures/Qutrits//gnntred1.tikz}}
\endpgfgraphicnamed,
\end{equation}
where  $\nststile{a_2}{a_1} \in \mathcal{A}, p \in \{ 1, 2\}, m \in \{0, 1, -1\}$, therefore $D_1$ and $D_2$ are unequal.

Secondly, we consider the case that $a$ is separated in just one of $D_1$ and $D_2$. By theorem \ref{graphstateeuivalence},  two graph states are equivalent if and only if there exists a sequence of local complementations $\ast$ and doulbing-neighbour-edge operations $\circ$ acting on one of them to obtain the other. But a local complementation or a doulbing-neighbour-edge operation will never change the connectivity of 
a separated single qutrit state, thus $D_1$ and $D_2$ cannot be equal.

Thirdly, we consider the case that $a$ has neighbours in both $D_1$ and $D_2$. Denote by $N_1$ the set of all vertices that are adjacent to $a$ in $D_1$, and by $N_2$ the set of all vertices that are adjacent to $a$ in $D_2$. Since $D_1$ is a rGS-LC diagram and $a$ has a red vertex operator, the vertex operators of any vertex in $N_1$ must be green phases. Also the vertex operators of any vertex in $N_2$ must be green phases, otherwise that vertex and $a$ would be a pair which is not allowed in the simplified pair of rGS-LC composed of  $D_1$ and $D_2$.

Let $n = |N_1|, m = |N_1 \cap N_2|$. Without loss of generality, we can assume that the first $m$ elements are those exactly from $N_1 \cap N_2$. 
Now the diagram $D_1$ has the following form:
\begin{equation}\label{d1form}
\beginpgfgraphicnamed{Qutrits//d1fm}
\InputIfFileExists{Qutrits//d1fm.tikz}{}{\input{./figures/Qutrits//d1fm.tikz}}
\endpgfgraphicnamed
\end{equation}
where $\nststile{b_i}{a_i} \in \mathcal{A}, h_i=H ~\mbox{or}~ H^{-1},   i \in \{ 1, \cdots,  n\}, \nststile{m_2}{m_1} \in \mathcal{M}, \nststile{p}{p} \in \mathcal{P} $. Let
\begin{equation*}
\beginpgfgraphicnamed{Qutrits//hfm}
\InputIfFileExists{Qutrits//hfm.tikz}{}{\input{./figures/Qutrits//hfm.tikz}}
\endpgfgraphicnamed
\end{equation*}

\noindent then  $h_i=h^{a_i},  a_i \in \{ 1,  -1\},  i \in \{ 1, \cdots,  n\}$.

We now define an operation depending on $D_1$ as follows:
\begin{equation}\label{udef}
U_{D_1}=\left(\bigotimes_{v_i\in N_1} C_{X, v_i\rightarrow a}\right)\circ \left( R_{Z,a}\bigotimes_{v_i\in N_1} D^{\frac{1-a_i}{2}}\right)
\end{equation}
where $D=H^2=(H^{-1})^2$,  $R_{Z,a}$ is the operation %
\beginpgfgraphicnamed{Qutrits//ppgt}
\InputIfFileExists{Qutrits//ppgt.tikz}{}{\input{./figures/Qutrits//ppgt.tikz}}
\endpgfgraphicnamed acting on $a$, and $C_{X, v_i\rightarrow a}$ is a generalized controlled-NOT gate with control $v_i$ and target $a$. The operation $U_{D_1}$ is well-defined as all the $C_{X, v_i\rightarrow a}$ commute with each other. Since each component is invertible, $U_{D_1}$ must be invertible as well, therefore $U_{D_1}\circ D_1 = U_{D_1}  \circ D_2 \Leftrightarrow D_1 = D_2$, which means if  $U_{D_1}\circ D_1 \neq U_{D_1}  \circ D_2$, then $D_1 \neq D_2$.

 Below we will show that the qutrit $a$ is in state  %
\beginpgfgraphicnamed{Qutrits//m1m2st}
\InputIfFileExists{Qutrits//m1m2st.tikz}{}{\input{./figures/Qutrits//m1m2st.tikz}}
\endpgfgraphicnamed in $U_{D_1}\circ D_1$,  while being either entangled with other qutrits or in a state unequal to %
\beginpgfgraphicnamed{Qutrits//m1m2st}
\InputIfFileExists{Qutrits//m1m2st.tikz}{}{\input{./figures/Qutrits//m1m2st.tikz}}
\endpgfgraphicnamed in $U_{D_1}\circ D_2$,
  where $\nststile{\bar{m}_2}{\bar{m}_1} \in \mathcal{M}$. By the proof for the first two cases, this would entail that  $U_{D_1}\circ D_1 \neq U_{D_1}  \circ D_2$, thus $D_1 \neq D_2$.

%\begin{itemize}
 %\item the qutrit $a$ is in state  \tikzfig{Qutrits//m1m2st} in $U_{D_1}\circ D_1$, where $\nststile{\bar{m}_2}{\bar{m}_1} \in \mathcal{M}$;
%\item  the qutrit $a$ in $U_{D_1}\circ D_2$ is either entangled with other qutrits, or in a state unequal to \tikzfig{Qutrits//m1m2st}.
%\end{itemize}

First, for $U_{D_1}\circ D_1$, we have
   \begin{equation}\label{d1uop}
\beginpgfgraphicnamed{Qutrits//d1uopt}
\InputIfFileExists{Qutrits//d1uopt.tikz}{}{\input{./figures/Qutrits//d1uopt.tikz}}
\endpgfgraphicnamed
\end{equation}
where for the first equality, we applied the doubling-neighbour-edge operation to $v_i$  whenever $a_i=-1$, and used the  colour change rules (H2), (H2$^{\prime}$) for the second and the last equalities,   the property (\ref{property7}) for the third equality.

Second, we consider  $U_{D_1}\circ D_2$. There are three cases for vertices about their adjacency to $a$ as well as application situation of controlled-NOT gates: 1) vertices are adjacent to $a$ but have no controlled-NOT gates applied to them; 2) vertices are not adjacent to $a$ but have controlled-NOT gates applied to them; 3) vertices are adjacent to $a$ and also have controlled-NOT gates applied to them. Here we ignore edges not connected to $a$ and edges without vertices in $N_1$. Then we have

  \begin{equation}\label{d1uop2}
\beginpgfgraphicnamed{Qutrits//d1uopt2}
\InputIfFileExists{Qutrits//d1uopt2.tikz}{}{\input{./figures/Qutrits//d1uopt2.tikz}}
\endpgfgraphicnamed
\end{equation}
where the phase $\nststile{c}{b} \in \mathcal{A}$ comes from the fusion of the vertex operator on $a$ and the $R_Z$ part of  $U_{D_1}$. 

To proceed, we investigate on different cases according to the value of $\nststile{c}{b}$.

If $\nststile{c}{b}=\nststile{p_1}{p_1 }+ \nststile{m_4}{m_3}$, where $\nststile{p_1}{p_1 }\in \mathcal{P}, \nststile{m_4}{m_3}\in \mathcal{M}$, we apply the doubling-neighbour-edge operation to $v_i$  for $a_i=-1, 1\leqslant i \leqslant m$. Then we have

  \begin{equation}\label{d1uop3}
\beginpgfgraphicnamed{Qutrits//d1uopt3}
\InputIfFileExists{Qutrits//d1uopt3.tikz}{}{\input{./figures/Qutrits//d1uopt3.tikz}}
\endpgfgraphicnamed
\end{equation}
Furthermore, we apply $p_1$-local complementation about $a$, then the vertex operator of $a$ is

  \begin{equation}\label{d1uop4}
\beginpgfgraphicnamed{Qutrits//d1uopt4}
\InputIfFileExists{Qutrits//d1uopt4.tikz}{}{\input{./figures/Qutrits//d1uopt4.tikz}}
\endpgfgraphicnamed
\end{equation}
where $\nststile{m_6}{m_5}, \nststile{m_8}{m_7}  \in \mathcal{M}$. The red node $\nststile{m_6}{m_5}$ can be copied and changed in colour to fuse into vertex operators of $a$'s neighbours. So the diagram (\ref{d1uop3}) is equal to

\begin{equation}\label{d1uop5}
\beginpgfgraphicnamed{Qutrits//d1uopt5}
\InputIfFileExists{Qutrits//d1uopt5.tikz}{}{\input{./figures/Qutrits//d1uopt5.tikz}}
\endpgfgraphicnamed
\end{equation}
If $N_1=N_2$, by property (\ref{property7}), $a$ is either adjacent to some neighbours or in the state    %
\beginpgfgraphicnamed{Qutrits//q1q2st}
\InputIfFileExists{Qutrits//q1q2st.tikz}{}{\input{./figures/Qutrits//q1q2st.tikz}}
\endpgfgraphicnamed where $\nststile{q_2}{q_1}  \in \mathcal{Q}$, which is clearly unequal to %
\beginpgfgraphicnamed{Qutrits//m1m2st}
\InputIfFileExists{Qutrits//m1m2st.tikz}{}{\input{./figures/Qutrits//m1m2st.tikz}}
\endpgfgraphicnamed. Otherwise, $a$ will always be connected to some neighbours after the application of $U_{D_1}$.

If $\nststile{c}{b}=\nststile{m_4}{m_3}\in \mathcal{M}$, there are two sub-cases. First, if $N_2-N_1\neq \emptyset$, then there exists $v \in N_2$ such that $v \notin N_1$. We apply a local
complementation to this $v$, which will add a green phase %
\beginpgfgraphicnamed{Qutrits//p1p1}
\InputIfFileExists{Qutrits//p1p1.tikz}{}{\input{./figures/Qutrits//p1p1.tikz}}
\endpgfgraphicnamed to the vertex operator on $a$. The edges connected with $a$ are changed as well, while there remains at least one vertex adjacent to $a$. 
Therefore we come back to the case $\nststile{c}{b}=\nststile{p_1}{p_1 }+ \nststile{m_4}{m_3}$. Second, if $N_2-N_1= \emptyset$, since $N_2\neq \emptyset$, it must be that $N_2 \subseteq N_1$, and thus $m =| N_2|>0$.
 Not considering the edges unconnected to $a$,  now the diagram is shown as follows:
\begin{equation}\label{d1uop6}
\beginpgfgraphicnamed{Qutrits//d1uopt6}
\InputIfFileExists{Qutrits//d1uopt6.tikz}{}{\input{./figures/Qutrits//d1uopt6.tikz}}
\endpgfgraphicnamed
\end{equation}
where for the first equality, we applied the doubling-neighbour-edge operation to $v_i$  whenever $a_i=-1$. Note that if $a_i=-1$ while $v_i$ has no neighbours, then we just use the property that   %
\beginpgfgraphicnamed{Qutrits//dcopyg}
\InputIfFileExists{Qutrits//dcopyg.tikz}{}{\input{./figures/Qutrits//dcopyg.tikz}}
\endpgfgraphicnamed.

As we have mentioned exactly before dealing with $U_{D_1}\circ D_1$, to show that $D_1\neq D_2$ it suffices to show that the qutrit $a$ is either entangled with other qutrits or in a state unequal to %
\beginpgfgraphicnamed{Qutrits//m1m2st}
\InputIfFileExists{Qutrits//m1m2st.tikz}{}{\input{./figures/Qutrits//m1m2st.tikz}}
\endpgfgraphicnamed in $U_{D_1}\circ D_2$. 
For this purpose, we only need to pay attention to vertices $a, v$ and the qutrit version of the controlled-Z gates between them.
 Also the last $H^{\dagger}$ gate and the green phase $\nststile{m_4}{m_3}$ on $a$ are irrelevant here, so will be ignored. Therefore we have
\begin{equation}\label{d1uop7}
\beginpgfgraphicnamed{Qutrits//d1uopt7}
\InputIfFileExists{Qutrits//d1uopt7.tikz}{}{\input{./figures/Qutrits//d1uopt7.tikz}}
\endpgfgraphicnamed
\end{equation}
where $\nststile{m}{m}=\nststile{1}{1} \mbox{or} \nststile{0}{0}$,  we used the Euler decomposition rule (EU) and applied a $2$-local complementation to $v$ for the third equality, a $1$-local complementation on $a$ for the fourth equality,  and the following property for the last equality which can be easily proved by rule (EU) and the property \ref{singlestates04} :
\begin{equation}\label{d1uopt8}
\beginpgfgraphicnamed{Qutrits//d1uopt8}
\InputIfFileExists{Qutrits//d1uopt8.tikz}{}{\input{./figures/Qutrits//d1uopt8.tikz}}
\endpgfgraphicnamed
\end{equation}
Now it is obvious that $a$ is entangled with $v$. Thus for the second sub-case of the case  $\nststile{c}{b}=\nststile{m_4}{m_3}\in \mathcal{M}$, we also proved $D_1\neq D_2$.
 
So far we have shown in all cases that $D_1\neq D_2$, hence finished the proof that any two component diagrams of a simplified pair of rGS-LC with an unpaired red node must be unequal.

 \end{proof}
 
%\fi

%two diagrams are unequal (equal) means the quantum states they represented are not equal(equal). (it doesn't mean that  the two diagrams are equal cannot be derived from the rewriting rules. )
%So here we don't consider diagrams in a category, where diagrams are morphisms, thus are equal modulo rewrite rules.
\begin{theorem}\label{identical}
The two component diagrams of any simplified pair of rGS-LC  are equal under the standard interpretation if and only if they are identical. 
\end{theorem}

 \begin{proof} The necessity is obvious. We only prove the sufficiency.
 Denote the two component diagrams by $D_1$ and $D_2$ respectively. Suppose $D_1 = D_2$ in the sense that they represent the same quantum state. By lemma  \ref{unpaired},  $D_1\neq D_2$ if there exits an unpaired red node. So we can assume that there are no unpaired red nodes in the simplified pair of rGS-LC under consideration.
 Let the graph underlying $D_1$ be $G_1 = (V, E_1 $), and the graph underlying $D_2$ be $G_2 = (V, E_2)$. Without loss of generality, assume that $V = \{1,2, \cdots, n\}$. 
Then $D_1$ and $D_2$ can be depicted as follows:
\begin{equation}\label{dgmidn}
\beginpgfgraphicnamed{Qutrits//dgmidnty}
\InputIfFileExists{Qutrits//dgmidnty.tikz}{}{\input{./figures/Qutrits//dgmidnty.tikz}}
\endpgfgraphicnamed
\end{equation}

where $\nststile{d_v}{c_v},  \nststile{t_v}{s_v} \in \{ \nststile{0}{0}, \nststile{1}{1}, \nststile{2}{2}\}$, 
 \begin{equation*}
\nststile{b_v}{a_v} \in  \left\{\begin{array}{l}
    \mathcal{A}, ~\mbox{if}~ \nststile{d_v}{c_v}= \nststile{0}{0}\vspace{0.2cm}\\
     \{ \nststile{1}{1}, \nststile{2}{0}, \nststile{0}{2}\}, ~\mbox{if}~ \nststile{d_v}{c_v}= \nststile{1}{1}\vspace{0.2cm}\\
     \{ \nststile{2}{2}, \nststile{1}{0}, \nststile{0}{1}\}, ~\mbox{if}~ \nststile{d_v}{c_v}= \nststile{2}{2}\\
    \end{array}\right. \quad\quad
\nststile{l_v}{k_v}  \in  \left\{\begin{array}{l}
   \mathcal{A}, ~\mbox{if}~ \nststile{t_v}{s_v}= \nststile{0}{0}\vspace{0.2cm}\\
     \{ \nststile{1}{1}, \nststile{2}{0}, \nststile{0}{2}\}, ~\mbox{if}~ \nststile{t_v}{s_v}= \nststile{1}{1}\vspace{0.2cm}\\
     \{ \nststile{2}{2}, \nststile{1}{0}, \nststile{0}{1}\}, ~\mbox{if}~ \nststile{t_v}{s_v}= \nststile{2}{2}\\    \end{array}\right. 
     \end{equation*}
$1\leqslant v \leqslant n$.

Since  all red nodes are paired up, it follows that $\nststile{d_v}{c_v}= \nststile{0}{0} \Leftrightarrow \nststile{t_v}{s_v}= \nststile{0}{0}$. If  $\nststile{d_v}{c_v}\neq \nststile{0}{0},  \nststile{t_v}{s_v}\neq \nststile{0}{0}$, but $\nststile{d_v}{c_v}\neq  \nststile{t_v}{s_v}$, w.l.o.g., assume $\nststile{d_v}{c_v}= \nststile{1}{1} ,\nststile{t_v}{s_v}= \nststile{2}{2}$.  Let 

 \begin{equation*}
U_v=%
\beginpgfgraphicnamed{Qutrits//dgmidnty2}
\InputIfFileExists{Qutrits//dgmidnty2.tikz}{}{\input{./figures/Qutrits//dgmidnty2.tikz}}
\endpgfgraphicnamed
\end{equation*}
Since $U_v$ is unitary, $U_{v}\circ D_1 = U_{v}  \circ D_2 \Leftrightarrow D_1 = D_2$. Now

 \begin{equation*}
U_v\circ D_1=%
\beginpgfgraphicnamed{Qutrits//uvd1}
\InputIfFileExists{Qutrits//uvd1.tikz}{}{\input{./figures/Qutrits//uvd1.tikz}}
\endpgfgraphicnamed \quad\quad \quad U_v\circ D_2=%
\beginpgfgraphicnamed{Qutrits//uvd2}
\InputIfFileExists{Qutrits//uvd2.tikz}{}{\input{./figures/Qutrits//uvd2.tikz}}
\endpgfgraphicnamed 
\end{equation*}
Clearly $U_v\circ D_1$ and $U_v\circ D_2$ are unpaired, so $U_v\circ D_1\neq U_v\circ D_2$, thus 
$ D_1\neq D_2$. Therefore,  $D_1 = D_2$ implies  $ \nststile{d_v}{c_v} = \nststile{t_v}{s_v},  \forall v\in \{1,2, \cdots, n\}$. Thus $D_2$ can be represented as 
 \begin{equation*}
\beginpgfgraphicnamed{Qutrits//d2new}
\InputIfFileExists{Qutrits//d2new.tikz}{}{\input{./figures/Qutrits//d2new.tikz}}
\endpgfgraphicnamed
\end{equation*}

%Let $V^{\prime}=\{v\in V | \nststile{d_v}{c_v}\neq \nststile{0}{0}\}$ and define the operators:
Define two operators
 \begin{equation}
U=\bigotimes_{v\in V} R_{X, v}^{-1}  \quad\quad \mbox{and}  \quad W=\bigotimes_{\{u, w \}\in E_1} C_{Z, uw}^{-1},   
\end{equation}
where $R_{X, v}=$ %
\beginpgfgraphicnamed{Qutrits//cvdvps}
\InputIfFileExists{Qutrits//cvdvps.tikz}{}{\input{./figures/Qutrits//cvdvps.tikz}}
\endpgfgraphicnamed,  $ C_{Z, uw}$ is the edge %
\beginpgfgraphicnamed{Qutrits//ctozuw}
\InputIfFileExists{Qutrits//ctozuw.tikz}{}{\input{./figures/Qutrits//ctozuw.tikz}}
\endpgfgraphicnamed connecting $u$ and $w$ in $G_1$ with $h=H$ or $H^{-1}$.
Clearly, both $U$ and $W$ are invertible, Thus  $(W \circ U)\circ D_1 = (W \circ U) \circ D_2 \Leftrightarrow D_1 = D_2$. 
After applying the operator $U$ to $ D_1$ and $ D_2$, all the red nodes are cancelled out, only green nodes left as vertex operators.  Based on this operation, further composition with $W$ will result in $(W \circ U)\circ D_1= %
\beginpgfgraphicnamed{Qutrits//wud1}
\InputIfFileExists{Qutrits//wud1.tikz}{}{\input{./figures/Qutrits//wud1.tikz}}
\endpgfgraphicnamed$.
Since $D_1 = D_2$, it must be that $(W \circ U)\circ D_2= %
\beginpgfgraphicnamed{Qutrits//wud1}
\InputIfFileExists{Qutrits//wud1.tikz}{}{\input{./figures/Qutrits//wud1.tikz}}
\endpgfgraphicnamed$. If there is an edge in $E_1$ but not belong to $E_2$, or an edge in $E_2$ but not belong to $E_1$, then $(W \circ U)\circ D_2$ must be a entangled state. 
So  $(E_1-E_2)\cup (E_2-E_1)=\emptyset$, i.e.,  $E_1=E_2$.  Furthermore, the weight of each edge  in $G_1$ should be the same as that of the corresponding edge in $G_2$, otherwise  $(W \circ U)\circ D_2$ is impossible to be a product of single qutrit states. Therefore we have $G_1=G_2$. It follows immediately that $(W \circ U)\circ D_2= %
\beginpgfgraphicnamed{Qutrits//wud12}
\InputIfFileExists{Qutrits//wud12.tikz}{}{\input{./figures/Qutrits//wud12.tikz}}
\endpgfgraphicnamed$. Again by  $D_1 = D_2$, we have $\nststile{b_v}{a_v} = \nststile{l_v}{k_v},  \forall v\in V$. Thus $D_1$ and $D_2$ are identical. This completes the proof.

 \end{proof}
 
 \subsection{Completeness for qutrit stabilizer quantum mechanics}
To achieve the proof of completeness for qutrit stabilizer QM, we will proceed in two main steps.   Firstly, we show the completeness for stabilizer states. 

\begin{theorem}\label{complst}
The ZX-calculus is complete for pure qutrit stabilizer states.
\end{theorem}

\begin{proof}
We only need to show that any two ZX-calculus diagrams that represent the same qutrit stabilizer state can be rewritten from one to the other by the ZX rules. 
Suppose $D_1$ and $D_2$ are such two diagrams. By theorem \ref{gslctorgslc},  $D_1$ and $D_2$ can be rewritten into rGS-LC diagrams $D^{\prime}_1$ and $D^{\prime}_2$ respectively. Clearly $D^{\prime}_1$ and $D^{\prime}_2$ are a pair of 
 rGS-LC diagrams on the same number of qutrits. Then by lemma \ref{rgspsym}, they can be transformed to  a simplified pair of 
 rGS-LC diagrams $D^{\prime\prime}_1$ and $D^{\prime\prime}_2$ in the ZX-calculus while still representing the same quantum state.  Now it follows from theorem \ref{identical} that $D^{\prime\prime}_1$ and $D^{\prime\prime}_2$ are identical diagrams, we denote this diagram by $D^{\prime\prime}$. As described above, we have show the steps of rewriting  $D_1$ and $D_2$  into $D^{\prime\prime}$ respectively. If we invert the rewriting processes from $D_2$  to $D^{\prime\prime}$ and compose with the rewriting processes from $D_1$  to $D^{\prime\prime}$, then we get the rewriting processes from $D_1$ to $D_2$. This completes the proof.
 \end{proof}
 
Secondly, we use the following map-state duality to relate quantum states and linear operators:

\begin{equation}\label{mapstate}
\beginpgfgraphicnamed{Qutrits/mapstatedual}
\InputIfFileExists{Qutrits/mapstatedual.tikz}{}{\input{./figures/Qutrits/mapstatedual.tikz}}
\endpgfgraphicnamed
\end{equation}

\noindent Then by theorem \ref{complst}  and the map-state duality (\ref{mapstate}), we have the main result: 
\begin{theorem}\label{complall}
The ZX-calculus is complete for qutrit stabilizer quantum mechanics.
\end{theorem}

A natural question arises at the end of this chapter: is there a general proof of completeness of the ZX-calculus for arbitrary dimensional (qudit) stabilizer QM?  This is the problem we would like to address next, but we should also mention some challenges we may face. With exception of the increase of the order of local Clifford groups, the main difficulty comes from the fact that that it is not known whether any stabilizer state is equivalent to a graph state under local Clifford group for the dimension $d$ having multiple prime factors \cite{Hostens}. As far as we know, stabilizer states are equivalent to graph states under local Clifford group for the dimension  $d$ which has only single prime factors  \cite{ShiangGr}.  Furthermore, the sufficient and necessary condition for two qudit graph states to be  equivalent under local Clifford group in terms of operations on graphs is unknown in the case that $d$ is non-prime. Finally, the generalised Euler decomposition rule for the generalised Hadamard gate can not be trivially derived, and other uncommon rules might be needed for general $d$.

%% file: Qutrits/RGg_Hadad.tikz
\begin{tikzpicture}
	\begin{pgfonlayer}{nodelayer}
		\node [style={H box}] (0) at (0, -0) {$H^\dagger$};
		\node [style=none] (1) at (0, -0.5) {};
		\node [style=none] (2) at (0, 0.5) {};
	\end{pgfonlayer}
	\begin{pgfonlayer}{edgelayer}
		\draw (2.center) to (0);
		\draw (1.center) to (0);
	\end{pgfonlayer}
\end{tikzpicture}

%% file: Qutrits/RGg_Hada.tikz
\begin{tikzpicture}
	\begin{pgfonlayer}{nodelayer}
		\node [style={H box}] (0) at (0, -0) {$H$};
		\node [style=none] (1) at (0, 0.5) {};
		\node [style=none] (2) at (0, -0.5) {};
	\end{pgfonlayer}
	\begin{pgfonlayer}{edgelayer}
		\draw (1.center) to (0);
		\draw (2.center) to (0);
	\end{pgfonlayer}
\end{tikzpicture}

%% file: Qutrits/RGg_zph_ab.tikz
\begin{tikzpicture}
	\begin{pgfonlayer}{nodelayer}
		\node [style=gsn] (0) at (0, -0) {$\alpha$\nodepart{lower}$\beta$};
		\node [style=none] (1) at (0, 0.5) {};
		\node [style=none] (2) at (0, -0.5) {};
	\end{pgfonlayer}
	\begin{pgfonlayer}{edgelayer}
		\draw (1.center) to (0);
		\draw (2.center) to (0);
	\end{pgfonlayer}
\end{tikzpicture}

%% file: Qutrits/RGg_xph_ab.tikz
\begin{tikzpicture}
	\begin{pgfonlayer}{nodelayer}
		\node [style=rsn] (0) at (0, -0) {$\alpha$\nodepart{lower}$\beta$};
		\node [style=none] (1) at (0, 0.5) {};
		\node [style=none] (2) at (0, -0.5) {};
	\end{pgfonlayer}
	\begin{pgfonlayer}{edgelayer}
		\draw (1.center) to (0);
		\draw (2.center) to (0);
	\end{pgfonlayer}
\end{tikzpicture}

%% file: Qutrits/parallel.tikz
\begin{tikzpicture}
	\begin{pgfonlayer}{nodelayer}
		\node [style=none] (0) at (-0.25, 0.75) {};
		\node [style=none] (1) at (-0.25, -0.75) {};
		\node [style=none] (2) at (0.25, 0.75) {};
		\node [style=none] (3) at (0.25, -0.75) {};
	\end{pgfonlayer}
	\begin{pgfonlayer}{edgelayer}
		\draw (0.center) to (1.center);
		\draw (2.center) to (3.center);
	\end{pgfonlayer}
\end{tikzpicture}

%% file: Qutrits/RGg_exd.tikz
\begin{tikzpicture}
	\begin{pgfonlayer}{nodelayer}
		\node [style=none] (0) at (0, -0.25) {};
		\node [style=rn] (1) at (0, 0.25) {};
	\end{pgfonlayer}
	\begin{pgfonlayer}{edgelayer}
		\draw (0.center) to (1);
	\end{pgfonlayer}
\end{tikzpicture}

%% file: Qutrits/rdot1.tikz
\begin{tikzpicture}
	\begin{pgfonlayer}{nodelayer}
		\node [style=rsn] (0) at (0, 0.25) {$2$\nodepart{lower}$1$};
		\node [style=none] (1) at (0, -0.25) {};
	\end{pgfonlayer}
	\begin{pgfonlayer}{edgelayer}
		\draw (1.center) to (0);
	\end{pgfonlayer}
\end{tikzpicture}

%% file: Qutrits/rdot2.tikz
\begin{tikzpicture}
	\begin{pgfonlayer}{nodelayer}
		\node [style=rsn] (0) at (0, 0.25) {$1$\nodepart{lower}$2$};
		\node [style=none] (1) at (0, -0.25) {};
	\end{pgfonlayer}
	\begin{pgfonlayer}{edgelayer}
		\draw (1.center) to (0);
	\end{pgfonlayer}
\end{tikzpicture}

%% file: Qutrits/RGg_ex.tikz
\begin{tikzpicture}
	\begin{pgfonlayer}{nodelayer}
		\node [style=none] (0) at (0, 0.25) {};
		\node [style=rn] (1) at (0, -0.25) {};
	\end{pgfonlayer}
	\begin{pgfonlayer}{edgelayer}
		\draw (0.center) to (1);
	\end{pgfonlayer}
\end{tikzpicture}

%% file: Qutrits/rcdot1.tikz
\begin{tikzpicture}
	\begin{pgfonlayer}{nodelayer}
		\node [style=none] (0) at (0, 0.25) {};
		\node [style=rsn] (1) at (0, -0.25) {$1$\nodepart{lower}$2$};
	\end{pgfonlayer}
	\begin{pgfonlayer}{edgelayer}
		\draw (0.center) to (1);
	\end{pgfonlayer}
\end{tikzpicture}

%% file: Qutrits/rcdot2.tikz
\begin{tikzpicture}
	\begin{pgfonlayer}{nodelayer}
		\node [style=none] (0) at (0, 0.25) {};
		\node [style=rsn] (1) at (0, -0.25) {$2$\nodepart{lower}$1$};
	\end{pgfonlayer}
	\begin{pgfonlayer}{edgelayer}
		\draw (0.center) to (1);
	\end{pgfonlayer}
\end{tikzpicture}

%% file: Qutrits/phasegts.tikz
\begin{tikzpicture}
	\begin{pgfonlayer}{nodelayer}
		\node [style=none] (0) at (0, 0.5) {};
		\node [style=none] (1) at (0, -0.5) {};
		\node [style=gsn] (2) at (0, 0) {$0$\nodepart{lower}$1$};
	\end{pgfonlayer}
	\begin{pgfonlayer}{edgelayer}
		\draw (0.center) to (2);
		\draw (1.center) to (2);
	\end{pgfonlayer}
\end{tikzpicture}

%% file: Qutrits/RGg_ezd.tikz
\begin{tikzpicture}
	\begin{pgfonlayer}{nodelayer}
		\node [style=none] (0) at (0, -0.5) {};
		\node [style=gn] (1) at (0, 0) {};
	\end{pgfonlayer}
	\begin{pgfonlayer}{edgelayer}
		\draw (0.center) to (1);
	\end{pgfonlayer}
\end{tikzpicture}

%% file: Qutrits/RGg_ez.tikz
\begin{tikzpicture}
	\begin{pgfonlayer}{nodelayer}
		\node [style=none] (0) at (0, 0.25) {};
		\node [style=gn] (1) at (0, -0.25) {};
	\end{pgfonlayer}
	\begin{pgfonlayer}{edgelayer}
		\draw (0.center) to (1);
	\end{pgfonlayer}
\end{tikzpicture}

%% file: Qutrits/RGg_dz.tikz
\begin{tikzpicture}
	\begin{pgfonlayer}{nodelayer}
		\node [style=none] (0) at (0, 0.5) {};
		\node [style=none] (1) at (0.25, -0.5) {};
		\node [style=none] (2) at (-0.25, -0.5) {};
		\node [style=gn] (3) at (0, -0) {};
	\end{pgfonlayer}
	\begin{pgfonlayer}{edgelayer}
		\draw (0.center) to (3);
		\draw (3) to (2.center);
		\draw (3) to (1.center);
	\end{pgfonlayer}
\end{tikzpicture}

%% file: Qutrits/RGg_dzd.tikz
\begin{tikzpicture}
	\begin{pgfonlayer}{nodelayer}
		\node [style=gn] (0) at (0, -0) {};
		\node [style=none] (1) at (0, -0.5) {};
		\node [style=none] (2) at (0.25, 0.5) {};
		\node [style=none] (3) at (-0.25, 0.5) {};
	\end{pgfonlayer}
	\begin{pgfonlayer}{edgelayer}
		\draw (0) to (1.center);
		\draw (0) to (2.center);
		\draw (0) to (3.center);
	\end{pgfonlayer}
\end{tikzpicture}

%% file: Qutrits/gdotascliff.tikz
\begin{tikzpicture}
	\begin{pgfonlayer}{nodelayer}
		\node [style=H box] (0) at (0, 0) {$H$};
		\node [style=rn] (1) at (0, 0.5) {};
		\node [style=none] (2) at (0, -0.5) {};
	\end{pgfonlayer}
	\begin{pgfonlayer}{edgelayer}
		\draw (2.center) to (1);
	\end{pgfonlayer}
\end{tikzpicture}

%% file: Qutrits/gcdotascliff.tikz
\begin{tikzpicture}
	\begin{pgfonlayer}{nodelayer}
		\node [style=H box] (0) at (0, 0) {$H$};
		\node [style=rn] (1) at (0, -0.5) {};
		\node [style=none] (2) at (0, 0.5) {};
	\end{pgfonlayer}
	\begin{pgfonlayer}{edgelayer}
		\draw (2.center) to (1);
	\end{pgfonlayer}
\end{tikzpicture}

%% file: Qutrits/gdotsmall.tikz
\begin{tikzpicture}
	\begin{pgfonlayer}{nodelayer}
		\node [style=none] (0) at (0, -0.25) {};
		\node [style=gn] (1) at (0, 0) {};
	\end{pgfonlayer}
	\begin{pgfonlayer}{edgelayer}
		\draw (0.center) to (1);
	\end{pgfonlayer}
\end{tikzpicture}

%% file: Qutrits/singlestates2.tikz
\begin{tikzpicture}
	\begin{pgfonlayer}{nodelayer}
		\node [style=none] (0) at (0, -0.5) {};
		\node [style=gn] (1) at (0, 0.5) {};
		\node [style=square box] (2) at (0, 0) {$U$};
	\end{pgfonlayer}
	\begin{pgfonlayer}{edgelayer}
		\draw (0.center) to (1);
	\end{pgfonlayer}
\end{tikzpicture}

%% file: Qutrits/singleu.tikz
\begin{tikzpicture}
	\begin{pgfonlayer}{nodelayer}
		\node [style=square box] (0) at (0, 0) {$U$};
		\node [style=none] (1) at (0, -0.5) {};
		\node [style=none] (2) at (0, 0.5) {};
	\end{pgfonlayer}
	\begin{pgfonlayer}{edgelayer}
		\draw (2.center) to (1.center);
	\end{pgfonlayer}
\end{tikzpicture}

%% file: Qutrits/singlestates72.tikz
\begin{tikzpicture}
	\begin{pgfonlayer}{nodelayer}
		\node [style=rn] (0) at (0, 0.25) {};
		\node [style=none] (1) at (0, -0.25) {};
	\end{pgfonlayer}
	\begin{pgfonlayer}{edgelayer}
		\draw (1.center) to (0);
	\end{pgfonlayer}
\end{tikzpicture}

%% file: Qutrits/singlestates73.tikz
\begin{tikzpicture}
	\begin{pgfonlayer}{nodelayer}
		\node [style=none] (0) at (0, -0.25) {};
		\node [style=gsn] (1) at (0, 0.25) {$1$\nodepart{lower}$1$};
	\end{pgfonlayer}
	\begin{pgfonlayer}{edgelayer}
		\draw (0.center) to (1);
	\end{pgfonlayer}
\end{tikzpicture}

%% file: Qutrits/singlestates74.tikz
\begin{tikzpicture}
	\begin{pgfonlayer}{nodelayer}
		\node [style=none] (0) at (0, -0.25) {};
		\node [style=gsn] (1) at (0, 0.25) {$2$\nodepart{lower}$2$};
	\end{pgfonlayer}
	\begin{pgfonlayer}{edgelayer}
		\draw (0.center) to (1);
	\end{pgfonlayer}
\end{tikzpicture}

%% file: Qutrits/singlestates75.tikz
\begin{tikzpicture}
	\begin{pgfonlayer}{nodelayer}
		\node [style=none] (0) at (0, -0.25) {};
		\node [style=rsn] (1) at (0, 0.25) {$1$\nodepart{lower}$2$};
	\end{pgfonlayer}
	\begin{pgfonlayer}{edgelayer}
		\draw (0.center) to (1);
	\end{pgfonlayer}
\end{tikzpicture}

%% file: Qutrits/singlestates76.tikz
\begin{tikzpicture}
	\begin{pgfonlayer}{nodelayer}
		\node [style=rsn] (0) at (0, 0.25) {$2$\nodepart{lower}$1$};
		\node [style=none] (1) at (0, -0.25) {};
	\end{pgfonlayer}
	\begin{pgfonlayer}{edgelayer}
		\draw (1.center) to (0);
	\end{pgfonlayer}
\end{tikzpicture}

%% file: Qutrits/chgoprt2.tikz
\begin{tikzpicture}
	\begin{pgfonlayer}{nodelayer}
		\node [style=rsn] (0) at (0, 0.75) {$m_3$\nodepart{lower}$m_4$};
		\node [style=gsn] (1) at (0, 0) {$m_{1}$\nodepart{lower}$m_{2}$};
		\node [style=none] (2) at (0, 1.25) {};
		\node [style=none] (3) at (0, -0.5) {};
	\end{pgfonlayer}
	\begin{pgfonlayer}{edgelayer}
		\draw (2.center) to (3.center);
	\end{pgfonlayer}
\end{tikzpicture}

%% file: Qutrits/11phasgt.tikz
\begin{tikzpicture}
	\begin{pgfonlayer}{nodelayer}
		\node [style=none] (0) at (0, 0.35) {};
		\node [style=none] (1) at (0, -0.35) {};
		\node [style=rsn] (2) at (0, 0) {$1$\nodepart{lower}$1$};
	\end{pgfonlayer}
	\begin{pgfonlayer}{edgelayer}
		\draw (0.center) to (1.center);
	\end{pgfonlayer}
\end{tikzpicture}

%% file: Qutrits/22phasgt.tikz
\begin{tikzpicture}
	\begin{pgfonlayer}{nodelayer}
		\node [style=none] (0) at (0, 0.35) {};
		\node [style=gsn] (1) at (0, 0) {$2$\nodepart{lower}$2$};
		\node [style=none] (2) at (0, -0.35) {};
	\end{pgfonlayer}
	\begin{pgfonlayer}{edgelayer}
		\draw (0.center) to (2.center);
	\end{pgfonlayer}
\end{tikzpicture}

%% file: Qutrits/gcdotsmall.tikz
\begin{tikzpicture}
	\begin{pgfonlayer}{nodelayer}
		\node [style=none] (0) at (0, 0.25) {};
		\node [style=gn] (1) at (0, 0) {};
	\end{pgfonlayer}
	\begin{pgfonlayer}{edgelayer}
		\draw (0.center) to (1);
	\end{pgfonlayer}
\end{tikzpicture}

%% file: Qutrits/singlestates77.tikz
\begin{tikzpicture}
	\begin{pgfonlayer}{nodelayer}
		\node [style=gsn] (0) at (0, 0.25) {$1$\nodepart{lower}$2$};
		\node [style=none] (1) at (0, -0.25) {};
	\end{pgfonlayer}
	\begin{pgfonlayer}{edgelayer}
		\draw (1.center) to (0);
	\end{pgfonlayer}
\end{tikzpicture}

%% file: Qutrits/singlestates78.tikz
\begin{tikzpicture}
	\begin{pgfonlayer}{nodelayer}
		\node [style=none] (0) at (0, -0.25) {};
		\node [style=gsn] (1) at (0, 0.25) {$2$\nodepart{lower}$1$};
	\end{pgfonlayer}
	\begin{pgfonlayer}{edgelayer}
		\draw (0.center) to (1);
	\end{pgfonlayer}
\end{tikzpicture}

%% file: Qutrits/gcopysmall.tikz
\begin{tikzpicture}
	\begin{pgfonlayer}{nodelayer}
		\node [style=none] (0) at (0.25, -0.35) {};
		\node [style=gn] (1) at (0, 0) {};
		\node [style=none] (2) at (-0.25, -0.35) {};
		\node [style=none] (3) at (0, 0.35) {};
	\end{pgfonlayer}
	\begin{pgfonlayer}{edgelayer}
		\draw (2.center) to (1);
		\draw (0.center) to (1);
		\draw (3.center) to (1);
	\end{pgfonlayer}
\end{tikzpicture}

%% file: Qutrits/gcocopysmall.tikz
\begin{tikzpicture}
	\begin{pgfonlayer}{nodelayer}
		\node [style=none] (0) at (1, 0.35) {};
		\node [style=gn] (1) at (0.75, 0) {};
		\node [style=none] (2) at (0.5, 0.35) {};
		\node [style=none] (3) at (0.75, -0.35) {};
	\end{pgfonlayer}
	\begin{pgfonlayer}{edgelayer}
		\draw (2.center) to (1);
		\draw (0.center) to (1);
		\draw (3.center) to (1);
	\end{pgfonlayer}
\end{tikzpicture}

%% file: Qutrits/hnode.tikz
\begin{tikzpicture}
	\begin{pgfonlayer}{nodelayer}
		\node [style=none] (0) at (0, -0.25) {};
		\node [style=H box] (1) at (0, 0) {$h$};
		\node [style=none] (2) at (0, 0.25) {};
	\end{pgfonlayer}
	\begin{pgfonlayer}{edgelayer}
		\draw (2.center) to (0.center);
	\end{pgfonlayer}
\end{tikzpicture}

%% file: Qutrits/green11gt.tikz
\begin{tikzpicture}
	\begin{pgfonlayer}{nodelayer}
		\node [style=gsn] (0) at (0.5, 0) {$1$\nodepart{lower}$1$};
		\node [style=none] (1) at (0.5, 0.35) {};
		\node [style=none] (2) at (0.5, -0.35) {};
	\end{pgfonlayer}
	\begin{pgfonlayer}{edgelayer}
		\draw (1.center) to (2.center);
	\end{pgfonlayer}
\end{tikzpicture}

%% file: Qutrits/cliffnorformju32.tikz
\begin{tikzpicture}
	\begin{pgfonlayer}{nodelayer}
		\node [style=none] (0) at (0, 1) {};
		\node [style=none] (1) at (0, -0.75) {};
		\node [style=rsn] (2) at (0, -0.25) {$p_1$\nodepart{lower}$p_2$};
		\node [style=gsn] (3) at (0, 0.5) {$\bar{a}_1$\nodepart{lower}$\bar{a}_2$};
	\end{pgfonlayer}
	\begin{pgfonlayer}{edgelayer}
		\draw (0.center) to (1.center);
	\end{pgfonlayer}
\end{tikzpicture}

%% file: Qutrits/_p1-p1.tikz
\begin{tikzpicture}
	\begin{pgfonlayer}{nodelayer}
		\node [style=none] (0) at (0, -0.5) {};
		\node [style=gsn] (1) at (0, 0) {$-p_1$\nodepart{lower}$-p_1$};
		\node [style=none] (2) at (0, 0.5) {};
	\end{pgfonlayer}
	\begin{pgfonlayer}{edgelayer}
		\draw (2.center) to (0.center);
	\end{pgfonlayer}
\end{tikzpicture}

%% file: Qutrits/p1p1.tikz
\begin{tikzpicture}
	\begin{pgfonlayer}{nodelayer}
		\node [style=gsn] (0) at (0, 0) {$p_1$\nodepart{lower}$p_1$};
		\node [style=none] (1) at (0, -0.5) {};
		\node [style=none] (2) at (0, 0.5) {};
	\end{pgfonlayer}
	\begin{pgfonlayer}{edgelayer}
		\draw (2.center) to (1.center);
	\end{pgfonlayer}
\end{tikzpicture}

%% file: Qutrits/ppgt.tikz
\begin{tikzpicture}
	\begin{pgfonlayer}{nodelayer}
		\node [style=none] (0) at (0, 0.5) {};
		\node [style=none] (1) at (0, -0.5) {};
		\node [style=gsn] (2) at (0, 0) {$p$\nodepart{lower}$p$};
	\end{pgfonlayer}
	\begin{pgfonlayer}{edgelayer}
		\draw (0.center) to (1.center);
	\end{pgfonlayer}
\end{tikzpicture}

%% file: Qutrits/m1m2st.tikz
\begin{tikzpicture}
	\begin{pgfonlayer}{nodelayer}
		\node [style=none] (0) at (0, 0.25) {};
		\node [style=none] (1) at (0, -0.25) {};
		\node [style=rsn] (2) at (0, 0.25) {$\bar{m}_1$\nodepart{lower}$\bar{m}_2$};
	\end{pgfonlayer}
	\begin{pgfonlayer}{edgelayer}
		\draw (0.center) to (1.center);
	\end{pgfonlayer}
\end{tikzpicture}

%% file: Qutrits/q1q2st.tikz
\begin{tikzpicture}
	\begin{pgfonlayer}{nodelayer}
		\node [style=none] (0) at (0, 0.25) {};
		\node [style=none] (1) at (0, -0.25) {};
		\node [style=rsn] (2) at (0, 0.25) {$q_1$\nodepart{lower}$q_2$};
	\end{pgfonlayer}
	\begin{pgfonlayer}{edgelayer}
		\draw (0.center) to (1.center);
	\end{pgfonlayer}
\end{tikzpicture}

%% file: Qutrits/cvdvps.tikz
\begin{tikzpicture}
	\begin{pgfonlayer}{nodelayer}
		\node [style=none] (0) at (0, -0.35) {};
		\node [style=none] (1) at (0, 0.35) {};
		\node [style=rsn] (2) at (0, 0) {$c_v$\nodepart{lower}$d_v$};
	\end{pgfonlayer}
	\begin{pgfonlayer}{edgelayer}
		\draw (1.center) to (0.center);
	\end{pgfonlayer}
\end{tikzpicture}

%% file: chapter6.tex
\chapter{Some applications of the ZX-calculus}\label{chapply}

%\section{Applications of the ZX-calculus in quantum information}
%Quatum bit commitment protocol

%\section{Applications of the ZX-calculus in quantum computation}
%Equivalence checking for Toffoli based quantum circuits, classification of 3-entanglement qubit states.
In the previous chapters, we mainly explored the theoretical aspects of the ZX-calculus: giving complete axiomatisations for the entire qubit quantum mechanics and other fragments of the QM. Now we turn to the application side of the ZX-calculus. Since its invention, the ZX-calculus has been applied to quantum foundations, such as non-locality \cite{CDKW,CDKW2}, and quantum computing \cite{Duncanpx,  Horsman, ckzh, domniel}. It has also been incorporated into the Quantomatic  software for automatic reasoning \cite{Quanto}. However, none of these applications use any of the rules involved with the new generators--triangle and $\lambda$ box.   

In this chapter, we show some applications of the ZX-calculus with rules involving the new generators. These include four parts: proof of the generalise supplementarity,  equivalent forms of 3-entanglement qubit states,  representation of Toffoli  gate, and equivalence-checking for the UMA gate.

\section{Proof of the generalised supplementarity}
%Next we illustrate the feature of the  $\lambda$ box. 
In Chapter \ref{chfull}, the  $\lambda$ box is restricted to be parameterised by a non-negative real number.  However, we can define green and red spiders parameterised by arbitrary complex numbers in terms of generators of the $ZX_{ full}$-calculus. In fact, for 
any complex number $a$, let $a=\lambda e^{\alpha}$, where $\lambda  \geq 0, \alpha \in [0,~2\pi)$. Let  %
\beginpgfgraphicnamed{diagrams//lambdanooutput}
\InputIfFileExists{diagrams//lambdanooutput.tikz}{}{\input{./figures/diagrams//lambdanooutput.tikz}}
\endpgfgraphicnamed. Then by rules (S1), (L1) and (L5) of the $ZX_{ full}$-calculus, the following definition is well-defined:
 \begin{equation}\label{generalgreenspiderdef}
\beginpgfgraphicnamed{diagrams//generalgreenspiderdefini}
\InputIfFileExists{diagrams//generalgreenspiderdefini.tikz}{}{\input{./figures/diagrams//generalgreenspiderdefini.tikz}}
\endpgfgraphicnamed
\end{equation}
%While in  \cite{coeckewang}, it has been generalised to a general green phase of form \tikzfig{diagrams//greenbxa} with arbitrary complex number as a parameter.
Thus by the spider rule (S1) and the rules (L1), (L4) and (L5),  we have the generalised  green spider rule:

 \begin{equation}\label{generalgrspiderrule}
\beginpgfgraphicnamed{diagrams//generalgreenspiderfuse}
\InputIfFileExists{diagrams//generalgreenspiderfuse.tikz}{}{\input{./figures/diagrams//generalgreenspiderfuse.tikz}}
\endpgfgraphicnamed
\end{equation}
where $a, b$ are arbitrary complex numbers.

By the colour change rule (H),  the generalised red spider can be defined as follows:

\begin{equation}\label{generalredspiderdef}
\beginpgfgraphicnamed{diagrams//generalredspiderdefini}
\InputIfFileExists{diagrams//generalredspiderdefini.tikz}{}{\input{./figures/diagrams//generalredspiderdefini.tikz}}
\endpgfgraphicnamed
\end{equation}

Therefore, we have the generalised  red spider rule:

 \begin{equation}\label{generalredspiderrule}
\beginpgfgraphicnamed{diagrams//generalredspiderfuse}
\InputIfFileExists{diagrams//generalredspiderfuse.tikz}{}{\input{./figures/diagrams//generalredspiderfuse.tikz}}
\endpgfgraphicnamed
\end{equation}
where $a, b$ are arbitrary complex numbers.

 %Similarly, we have the general red phase \tikzfig{diagrams//redbxb}  \cite{coeckewang}. 
 %Below we give the spider form of general phase which are interpreted in Hilbert spaces:
  The generalised  green and red spiders have the following interpretation in Hilbert spaces: 
 \begin{equation}\label{gphaseinter}
 \begin{array}{c}
\left\llbracket %
\beginpgfgraphicnamed{diagrams//generalgreenspider}
\InputIfFileExists{diagrams//generalgreenspider.tikz}{}{\input{./figures/diagrams//generalgreenspider.tikz}}
\endpgfgraphicnamed \right\rrbracket=\ket{0}^{\otimes m}\bra{0}^{\otimes n}+a\ket{1}^{\otimes m}\bra{1}^{\otimes n}, \\
\left\llbracket %
\beginpgfgraphicnamed{diagrams//generalredspider}
\InputIfFileExists{diagrams//generalredspider.tikz}{}{\input{./figures/diagrams//generalredspider.tikz}}
\endpgfgraphicnamed \right\rrbracket=\ket{+}^{\otimes m}\bra{+}^{\otimes n}+a\ket{-}^{\otimes m}\bra{-}^{\otimes n},
\end{array}
\end{equation}
where $a$ is an arbitrary complex number. 
%The generalised spider rules and colour change rule are depicted in the following:
 %\begin{equation}\label{generalspider}
 %\begin{array}{c}
 %\tikzfig{diagrams//generalgreenspiderfuse}, \quad   \tikzfig{diagrams//generalredspiderfuse},\\
 %\tikzfig{diagrams//generalcolorchange}.
%\end{array}
%\end{equation}
%where $a, b$ are arbitrary complex numbers.

Now we consider the generalised supplementarity-- also called cyclotomic supplementarity, with supplementarity as a special case--which is interpreted as merging $n$ subdiagrams if the $n$ phase angles divide the circle uniformly \cite{jpvw}.  The diagrammatic form of the generalised supplementarity is as follows:
\begin{equation}\label{generalpsgt0}
\beginpgfgraphicnamed{diagrams//generalsupp}
\InputIfFileExists{diagrams//generalsupp.tikz}{}{\input{./figures/diagrams//generalsupp.tikz}}
\endpgfgraphicnamed
\end{equation}
where there are $n$ parallel wires in the diagram at the right-hand side.

Next we show that the generalised supplementarity can be seen as a special form of the generalised spider rule as shown in (\ref{generalredspiderrule}). For simplicity, we ignore scalars in the rest of this section.

%First note that by comparing the normal form translated from the ZW-calculus \cite{amar}, we  have 
First, it can be directly calculated that 
\begin{equation*}
\left\llbracket %
\beginpgfgraphicnamed{diagrams//generalpsgtelt}
\InputIfFileExists{diagrams//generalpsgtelt.tikz}{}{\input{./figures/diagrams//generalpsgtelt.tikz}}
\endpgfgraphicnamed \right\rrbracket= \left\llbracket %
\beginpgfgraphicnamed{diagrams//generalpsgtert}
\InputIfFileExists{diagrams//generalpsgtert.tikz}{}{\input{./figures/diagrams//generalpsgtert.tikz}}
\endpgfgraphicnamed \right\rrbracket
\end{equation*}
where $a\in \mathbb{C}, a\neq -1$. 

Then by the completeness of the $ZX_{ full}$-calculus, we have
\begin{equation}\label{generalpsgt}
\beginpgfgraphicnamed{diagrams//generalpsgte}
\InputIfFileExists{diagrams//generalpsgte.tikz}{}{\input{./figures/diagrams//generalpsgte.tikz}}
\endpgfgraphicnamed
\end{equation}
%where $a\in \mathbb{C}, a\neq 1$. 

In particular,
\begin{equation}\label{generalpsgt2}
\beginpgfgraphicnamed{diagrams//generalpsgte2}
\InputIfFileExists{diagrams//generalpsgte2.tikz}{}{\input{./figures/diagrams//generalpsgte2.tikz}}
\endpgfgraphicnamed
\end{equation}
 where $\alpha \in [0, 2\pi), \alpha\neq \pi$. For $\alpha= \pi$, we can use the $\pi$ copy rule directly. 
  
  Then
  \begin{equation}\label{generalpsgt3}
\beginpgfgraphicnamed{diagrams//generalpsgte3}
\InputIfFileExists{diagrams//generalpsgte3.tikz}{}{\input{./figures/diagrams//generalpsgte3.tikz}}
\endpgfgraphicnamed
\end{equation}
where we used the following formula given in  \cite{jpvw}:
  \begin{equation}\label{fracidentyty}
\prod_{j=0}^{n-1}(1+e^{i(\alpha+\frac{j}{n}2\pi)})=1+e^{i(n\alpha+(n-1)\pi)}
\end{equation}

  Note that if $n$ is odd, then
  \begin{equation}\label{generalpsgt4}
\beginpgfgraphicnamed{diagrams//generalpsgte4}
\InputIfFileExists{diagrams//generalpsgte4.tikz}{}{\input{./figures/diagrams//generalpsgte4.tikz}}
\endpgfgraphicnamed
\end{equation}

If $n$ is even, then
 \begin{equation}\label{generalpsgt5}
\beginpgfgraphicnamed{diagrams//generalpsgte5}
\InputIfFileExists{diagrams//generalpsgte5.tikz}{}{\input{./figures/diagrams//generalpsgte5.tikz}}
\endpgfgraphicnamed
\end{equation}

It is not hard to see that if we compute the parity of $n$ in the right diagram of (\ref{generalpsgt0}) not considering the scalars, then by Hopf law we get the same result as shown in (\ref{generalpsgt4}) and (\ref{generalpsgt5}).

\section{ 3-entanglement qubit states}
It is well known that there are two SLOCC (Stochastic Local Operations and Classical Communication) equivalent classes of 3-entanglement qubit states  \cite{Dur} with representative members $\ket{GHZ}=\ket{000}+\ket{111}, \ket{W}=\ket{100}+\ket{010}+\ket{001}$. In fact, we have
\begin{theorem} \cite{Dur} 
Two n-partite states $\ket{\psi}$  and $\ket{\phi}$ are SLOCC equivalent if and only if there exist invertible linear maps $L_1, L_2, \ldots, L_n$  such that
 \begin{equation}\label{SLOCCequivalent}
\ket{\psi}=(L_1\otimes L_2\otimes \cdots \otimes L_n)\ket{\phi}
\end{equation}
\end{theorem} 
In \cite{CoeckeEdwards}, Coecke and Edwards have represented the W state with phases in the ZX-calculus as follows:
 \begin{equation}\label{wstatephases}
\beginpgfgraphicnamed{diagrams//wstatewithphases}
\InputIfFileExists{diagrams//wstatewithphases.tikz}{}{\input{./figures/diagrams//wstatewithphases.tikz}}
\endpgfgraphicnamed
\end{equation}

While another representation of the W state by a triangle was essentially given in \cite{Emmanuel},  and explicitly given in  \cite{ngwang}:
\begin{equation}\label{wstatetriangle}
\beginpgfgraphicnamed{diagrams//wstatewithtrianle}
\InputIfFileExists{diagrams//wstatewithtrianle.tikz}{}{\input{./figures/diagrams//wstatewithtrianle.tikz}}
\endpgfgraphicnamed
\end{equation}

%In this section, we give equivalent forms for these kinds of states represented by the new generators.
Here we build a bridge between these two representations of the W state by  $\lambda$ box.  

\begin{lemma}
For the $W$ state, up to non-zero scalars, we have

\beginpgfgraphicnamed{diagrams//w2forms}
\InputIfFileExists{diagrams//w2forms.tikz}{}{\input{./figures/diagrams//w2forms.tikz}}
\endpgfgraphicnamed
\end{lemma}

\begin{proof}

\beginpgfgraphicnamed{diagrams//w2formsderive}
\InputIfFileExists{diagrams//w2formsderive.tikz}{}{\input{./figures/diagrams//w2formsderive.tikz}}
\endpgfgraphicnamed
 
% where we used the decomposition of the triangle.
\end{proof}

Furthermore, Coecke and Edwards give explicit conditions for  a GHZ-class state:
\begin{lemma}\cite{CoeckeEdwards}
The following diagram
 \begin{equation}\label{ghzgeneral}
\beginpgfgraphicnamed{diagrams//ghzgeneralorigin}
\InputIfFileExists{diagrams//ghzgeneralorigin.tikz}{}{\input{./figures/diagrams//ghzgeneralorigin.tikz}}
\endpgfgraphicnamed
\end{equation}
 represents a  GHZ-class state if $\alpha, \beta, \gamma$ violate one of the following equalities:
 \[
  \begin{array}{c}
  \alpha+\beta+\gamma=\pi \\
   \alpha+\beta-\gamma=\pi \\
   \alpha-\beta+\gamma=\pi \\
   \alpha-\beta-\gamma=\pi \\
\end{array}
 \]
\end{lemma}

However, if we use the generalised spiders defined in the previous section, %generalise the $\lambda$ box to a general phase parameterised by a complex number \cite{coeckewang}, 
then we can obtain an exact solution for transformation of  GHZ-class state into simpler form. Note that we will use standard interpretations for deriving diagrammatical identities in the following proofs because of the  completeness of the $ZX_{ full}$-calculus.
\begin{lemma}\label{ghznorm}
For GHZ state, we have
 \begin{equation}\label{ghznormalform}
\beginpgfgraphicnamed{diagrams//ghznormal}
\InputIfFileExists{diagrams//ghznormal.tikz}{}{\input{./figures/diagrams//ghznormal.tikz}}
\endpgfgraphicnamed
\end{equation}
where $\lambda_i$ and $x_i$ are complex numbers, 
$(1+\lambda_1\lambda_2\lambda_3)(\lambda_1+\lambda_2\lambda_3)(\lambda_2+\lambda_1\lambda_3)(\lambda_3+\lambda_1\lambda_2)\neq 0$,
\[
x_1=\pm\sqrt{\frac{(\lambda_2+\lambda_1\lambda_3)(\lambda_1+\lambda_2\lambda_3)}{(1+\lambda_1\lambda_2\lambda_3)(\lambda_3+\lambda_1\lambda_2)}},
 x_2= \pm\sqrt{\frac{(\lambda_1+\lambda_2\lambda_3)(\lambda_3+\lambda_1\lambda_2)}{(1+\lambda_1\lambda_2\lambda_3)(\lambda_2+\lambda_1\lambda_3)}},
 \]
 \[
x_3=\pm\sqrt{\frac{(\lambda_2+\lambda_1\lambda_3)(\lambda_3+\lambda_1\lambda_2)}{(1+\lambda_1\lambda_2\lambda_3)(\lambda_1+\lambda_2\lambda_3)}}.
\]
%%%%%%%%%%%%%%%%%%%%%%%%%%%%%%%%%%%%%%%%%%%%%%%%%%%%%%%%%%%%%%%%%%%%%%%%%%%%
%%%%%%%%%%%%%%%%%%%%%%%%%%%%%%%%%%%%%%%%%%%%%%%%%%%%%%%%%%%%%%%%%%%%%%%%%%%%
\iffalse
\[
%
\beginpgfgraphicnamed{diagrams//lambdai}
\InputIfFileExists{diagrams//lambdai.tikz}{}{\input{./figures/diagrams//lambdai.tikz}}
\endpgfgraphicnamed=\begin{pmatrix}
        1 & 0 \\
        0 & \lambda_i
         \end{pmatrix},
 %
\beginpgfgraphicnamed{diagrams//xi}
\InputIfFileExists{diagrams//xi.tikz}{}{\input{./figures/diagrams//xi.tikz}}
\endpgfgraphicnamed=\begin{pmatrix}
        1 & 0 \\
        0 &x_i
 \end{pmatrix}.
 \]
 \fi
\end{lemma}
\begin{proof}
The matrix of the left-hand side of (\ref{ghznormalform}) is
\[
\begin{pmatrix}
        1+\lambda_1\lambda_2\lambda_3 & 0 \\
        0 &\lambda_2+\lambda_1\lambda_3\\
        0 & \lambda_1+\lambda_2\lambda_3 \\
        \lambda_3+\lambda_1\lambda_2 & 0
 \end{pmatrix}.
\]
The matrix of the right-hand side of (\ref{ghznormalform}) is
\[
\begin{pmatrix}
        1 & 0 \\
        0 &x_1x_3\\
        0 &x_1x_2 \\
    x_2x_3 & 0
 \end{pmatrix}.
\]
These two matrices are equal up to a scalar, thus
\[
x_1x_3=\frac{\lambda_2+\lambda_1\lambda_3}{1+\lambda_1\lambda_2\lambda_3},
x_1x_2=\frac{\lambda_1+\lambda_2\lambda_3}{1+\lambda_1\lambda_2\lambda_3},
x_2x_3=\frac{\lambda_3+\lambda_1\lambda_2}{1+\lambda_1\lambda_2\lambda_3}.
\]
Then
\[
(x_1x_2x_3)^2=\frac{(\lambda_2+\lambda_1\lambda_3)(\lambda_1+\lambda_2\lambda_3)(\lambda_3+\lambda_1\lambda_2)}{(1+\lambda_1\lambda_2\lambda_3)^3},
\]
i.e.,
\[
x_1x_2x_3=\pm\sqrt{\frac{(\lambda_2+\lambda_1\lambda_3)(\lambda_1+\lambda_2\lambda_3)(\lambda_3+\lambda_1\lambda_2)}{(1+\lambda_1\lambda_2\lambda_3)^3}}.
\]
Therefore,
\[
x_1=\frac{x_1x_2x_3}{x_2x_3}=\pm\sqrt{\frac{(\lambda_2+\lambda_1\lambda_3)(\lambda_1+\lambda_2\lambda_3)}{(1+\lambda_1\lambda_2\lambda_3)(\lambda_3+\lambda_1\lambda_2)}},
 x_2=\frac{x_1x_2x_3}{x_1x_3}= \pm\sqrt{\frac{(\lambda_1+\lambda_2\lambda_3)(\lambda_3+\lambda_1\lambda_2)}{(1+\lambda_1\lambda_2\lambda_3)(\lambda_2+\lambda_1\lambda_3)}},
 \]
 \[
x_3=\frac{x_1x_2x_3}{x_1x_2}=\pm\sqrt{\frac{(\lambda_2+\lambda_1\lambda_3)(\lambda_3+\lambda_1\lambda_2)}{(1+\lambda_1\lambda_2\lambda_3)(\lambda_1+\lambda_2\lambda_3)}}.
\]

\end{proof}

As an example, we have

\begin{lemma}\label{trianglebroken}
\begin{equation}\label{trianglebroke}
\beginpgfgraphicnamed{diagrams//pbct6new}
\InputIfFileExists{diagrams//pbct6new.tikz}{}{\input{./figures/diagrams//pbct6new.tikz}}
\endpgfgraphicnamed
\end{equation}
where 
\[
A=\begin{pmatrix}
        1 & 0 \\
        0 & \pm \sqrt{3}i
 \end{pmatrix},
 B=\begin{pmatrix}
        1 & 0 \\
        0 & \mp  \sqrt{\frac{1}{3}}(\sqrt{2}+i)
 \end{pmatrix},
 C=\begin{pmatrix}
        1 & 0 \\
        0 & \pm  \sqrt{\frac{1}{3}}(1+\sqrt{2}i)
 \end{pmatrix}.
 \]
\end{lemma}
\begin{proof}
The matrix of the left-hand side of (\ref{trianglebroke}) is
\[
\begin{pmatrix}
        1 & 0 \\
        0 &-\sqrt{2}+i\\
        0 &1-\sqrt{2}i \\
    -i & 0
 \end{pmatrix}.
\]
Then the result follows from Lemma  \ref{ghznorm}.
\end{proof}

Finally, we have two more equivalent transformations of GHZ states from a loop form to a non-loop form like the bialgebra rule (B2), which can be verified by matrix calculations:
\begin{equation}\label{triangleghz}
\beginpgfgraphicnamed{diagrams//triangleghzel}
\InputIfFileExists{diagrams//triangleghzel.tikz}{}{\input{./figures/diagrams//triangleghzel.tikz}}
\endpgfgraphicnamed
\end{equation}

\begin{equation}\label{triangleghz}
\beginpgfgraphicnamed{diagrams//andgtsimple2forms}
\InputIfFileExists{diagrams//andgtsimple2forms.tikz}{}{\input{./figures/diagrams//andgtsimple2forms.tikz}}
\endpgfgraphicnamed
\end{equation}

It is interesting to point out that the first diagram in (\ref{triangleghz}) has been used by Jeandel, Perdrix, and Vilmart to construct the Toffoli gate \cite{Emmanuel}, while the third diagram in (\ref{triangleghz}) was used in (\ref{toffoligate}) in this chapter and in \cite{ngwang2} to construct the same gate.
The equalities of (\ref{triangleghz}) were obtained before the upload of the paper  \cite{ngwang2},  we only choose the third diagram to construct the Toffoli gate because we think it makes the Toffoli gate simpler--there is no loop in the construction-- and is easy to generalise to a multiple control Toffoli gate of form (\ref{toffoligatemultiple}).

\section{Representation of Toffoli  gate}\label{toffolirepresent}
The Toffoli gate is  known as the ``controlled-controlled-not" gate  \cite{Nielsen}. The standard circuit form of Toffoli gate is given in  \cite{Vivek} as follows:
\begin{center}
{\includegraphics[scale=0.65]{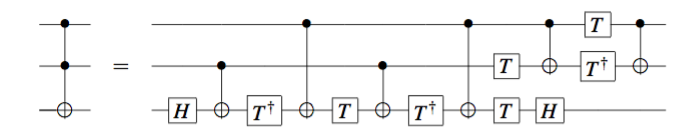}}
  \end{center}

We express the circuit form of Toffoli gate in ZX-calculus as follows: 
\begin{equation}\label{toffoli1eq}
\beginpgfgraphicnamed{diagrams//toffoli1}
\InputIfFileExists{diagrams//toffoli1.tikz}{}{\input{./figures/diagrams//toffoli1.tikz}}
\endpgfgraphicnamed
\end{equation}

In contrast to circuit form, the Toffoli gate can be expressed by the new generator-- triangle--in a nice way (without phases). Here we follow the way of section 12.1.3 in  \cite{Coeckebk} to construct the Toffoli gate.
First, we need to specify the AND gate in ZX-calculus. The AND gate was given by a GHZ/W-pair in \cite{Herrmann}  as follows:
\begin{center}
{\includegraphics[scale=0.65]{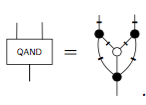}}
  \end{center}
  The right-side diagram above is a GHZ/W-calculus diagram \cite{Coeckealeksmen}.
  If we translate this diagram into ZX-calculus, then we get
\begin{equation}\label{andgate0}
\beginpgfgraphicnamed{diagrams//andgt}
\InputIfFileExists{diagrams//andgt.tikz}{}{\input{./figures/diagrams//andgt.tikz}}
\endpgfgraphicnamed
%\tikzfig{diagrams//andgtsimple}
%\tikzfig{diagrams//qand}=\tikzfig{diagrams//andgtsimplest}
\end{equation}
By further calculation, we have
\begin{equation}\label{andgate}
%\tikzfig{diagrams//andgt}
%\tikzfig{diagrams//andgtsimple}
%
\beginpgfgraphicnamed{diagrams//qand}
\InputIfFileExists{diagrams//qand.tikz}{}{\input{./figures/diagrams//qand.tikz}}
\endpgfgraphicnamed=%
\beginpgfgraphicnamed{diagrams//andgtsimplest}
\InputIfFileExists{diagrams//andgtsimplest.tikz}{}{\input{./figures/diagrams//andgtsimplest.tikz}}
\endpgfgraphicnamed
\end{equation}
where the triangle with a $-1$ on the top-left corner is the inverse of the normal triangle.

We check the correctness of the AND gate as follows:
\begin{equation}\label{andgtverify0}
%\tikzfig{diagrams//andgtvery0}
%
\beginpgfgraphicnamed{diagrams//andgtvery02}
\InputIfFileExists{diagrams//andgtvery02.tikz}{}{\input{./figures/diagrams//andgtvery02.tikz}}
\endpgfgraphicnamed
\end{equation}

\iffalse
\begin{equation}\label{andgtverify1}
%
\beginpgfgraphicnamed{diagrams//andgtvery1}
\InputIfFileExists{diagrams//andgtvery1.tikz}{}{\input{./figures/diagrams//andgtvery1.tikz}}
\endpgfgraphicnamed
\end{equation}

The simplified  AND gate can be verified as follows:
\begin{equation}\label{andgtsimpleverify1}
%
\beginpgfgraphicnamed{diagrams//andgtsimpleverify}
\InputIfFileExists{diagrams//andgtsimpleverify.tikz}{}{\input{./figures/diagrams//andgtsimpleverify.tikz}}
\endpgfgraphicnamed
\end{equation}

However, AND gate can be simplified further in two forms in the following, we will see which form is better.
\begin{equation}\label{andgtsimple2form}
%
\beginpgfgraphicnamed{diagrams//andgtsimple2forms}
\InputIfFileExists{diagrams//andgtsimple2forms.tikz}{}{\input{./figures/diagrams//andgtsimple2forms.tikz}}
\endpgfgraphicnamed
\end{equation}
\fi

Note that the matrix form of the diagram %
\beginpgfgraphicnamed{diagrams//andcheck}
\InputIfFileExists{diagrams//andcheck.tikz}{}{\input{./figures/diagrams//andcheck.tikz}}
\endpgfgraphicnamed in (\ref{andgtverify0})   is $\sqrt{2}\ket{0}(\bra{0}+\bra{1})$, which means the two rows of diagrammatical equalities of (\ref{andgtverify0}) have the same scalars under the standard basis. Therefore we prove the correctness of  (\ref{andgate}) for  the AND gate.

It turns out that the $``/"$ box ( Exercise 12.10 in \cite{Coeckebk}) has the same matrix form as the new generator triangle (up to a scalar), thus the  AND gate there is in a form similar to  (\ref{andgate}).

Then the Toffoli gate can be constructed as
%\begin{equation}\label{toffoligate}
%\tikzfig{diagrams//toffoligt}
%\tikzfig{diagrams//toffoligtsimplest}
%\end{equation}
\begin{equation}\label{toffoligate}
%\tikzfig{diagrams//toffoligt}
%
\beginpgfgraphicnamed{diagrams//toffoligtsimplest}
\InputIfFileExists{diagrams//toffoligtsimplest.tikz}{}{\input{./figures/diagrams//toffoligtsimplest.tikz}}
\endpgfgraphicnamed
\end{equation}

Now we prove the correctness of the Toffoli gate. The key point to remember here is that the Toffoli gate is a ``controlled-CNOT" gate  \cite{Coeckebk}.
\begin{equation}\label{toffoligatecrt}
%\tikzfig{diagrams//toffoligtctns00}
%
\beginpgfgraphicnamed{diagrams//toffoligtctns002}
\InputIfFileExists{diagrams//toffoligtctns002.tikz}{}{\input{./figures/diagrams//toffoligtctns002.tikz}}
\endpgfgraphicnamed
\end{equation}
\begin{equation}\label{toffoligatecrt2}
%\tikzfig{diagrams//toffoligtctns01}
%
\beginpgfgraphicnamed{diagrams//toffoligtctns012}
\InputIfFileExists{diagrams//toffoligtctns012.tikz}{}{\input{./figures/diagrams//toffoligtctns012.tikz}}
\endpgfgraphicnamed
\end{equation}
The matrix form of %
\beginpgfgraphicnamed{diagrams//cnotgt}
\InputIfFileExists{diagrams//cnotgt.tikz}{}{\input{./figures/diagrams//cnotgt.tikz}}
\endpgfgraphicnamed is $\frac{1}{\sqrt{2}}(\ket{00}\bra{00}+\ket{01}\bra{01}+\ket{11}\bra{10}+\ket{10}\bra{11})$, which means the final diagrams of (\ref{toffoligatecrt}) and  (\ref{toffoligatecrt2}) have the same scalars under the standard basis. Therefore we prove the correctness of  (\ref{toffoligate}) for  the Toffoli gate.

It is easy to see that the AND gate can be generalised as follows:
\begin{equation}\label{andgategeneral}
\beginpgfgraphicnamed{diagrams//andgategeneralise}
\InputIfFileExists{diagrams//andgategeneralise.tikz}{}{\input{./figures/diagrams//andgategeneralise.tikz}}
\endpgfgraphicnamed
\end{equation}

Therefore the multiple control Toffoli gate (up to a scalar) can be represented as:
\begin{equation}\label{toffoligatemultiple}
\beginpgfgraphicnamed{diagrams//toffoligtmultiple}
\InputIfFileExists{diagrams//toffoligtmultiple.tikz}{}{\input{./figures/diagrams//toffoligtmultiple.tikz}}
\endpgfgraphicnamed
\end{equation}

We end this section with pointing out that it is not easy to find an efficient way to prove in the ZX-calculus the equivalence of the two forms of Toffoli gate (\ref{toffoli1eq}) and (\ref{toffoligate}),  although
this can, in principle, be done by rewriting them into normal forms induced from the ZW-calculus.
% this can be done by rewriting them into normal forms in principle. 

\section{Equivalence-checking for  the UMA gate}
In this section, we consider applying the ZX-calculus to equivalence-checking  \cite{Markov} for two special quantum circuits. 
 These two circuits are  two forms of the so-called ``UnMajority and Add''  (UMA) quantum gate  given in \cite{Cuccaro} as follows:
 %compute the same function on qubits 
%We will show this application by the example of a quantum gate called ``UnMajority and Add''  (UMA) gate.
%There are two forms of the UMA gate given in \cite{Cuccaro} as follows:

{\includegraphics[scale=0.65]{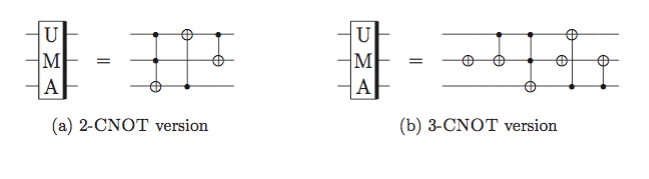}}

We check the equivalence of the two kind of UMA gate diagrammatically in terms of the  representation of Toffoli  gate (\ref{toffoligate}). The translation from  the UMA gate into ZX-calculus is direct: Toffoli  gate in UMA translated to diagram (\ref{toffoligate}), CNOT gate in UMA to CNOT in ZX, and NOT gate in UMA to the red $\pi$ gate in the ZX-calculus. 
\begin{proposition}
The two versions of the UMA gate can be proved to be equal in the ZX-calculus:
\begin{equation}\label{umagt2form}
\beginpgfgraphicnamed{diagrams//umagt2forms}
\InputIfFileExists{diagrams//umagt2forms.tikz}{}{\input{./figures/diagrams//umagt2forms.tikz}}
\endpgfgraphicnamed
\end{equation}
%where the diagrams are read from top to bottom.
\end{proposition}

\begin{proof}
\begin{equation}\label{umagt2formprf}
\beginpgfgraphicnamed{diagrams//2umaequivproof}
\InputIfFileExists{diagrams//2umaequivproof.tikz}{}{\input{./figures/diagrams//2umaequivproof.tikz}}
\endpgfgraphicnamed
\end{equation}

where for the second equality we used lemma \ref{umapartlm}, and for the fourth equality we used the bialgebra rule.
\end{proof}

\begin{lemma}\label{umapartlm}
\begin{equation}\label{umapart}
\beginpgfgraphicnamed{diagrams//umaparttrans}
\InputIfFileExists{diagrams//umaparttrans.tikz}{}{\input{./figures/diagrams//umaparttrans.tikz}}
\endpgfgraphicnamed
\end{equation}
\end{lemma}

\begin{proof}
\begin{equation}\label{umagatedr}
\beginpgfgraphicnamed{diagrams//UMAgatederive}
\InputIfFileExists{diagrams//UMAgatederive.tikz}{}{\input{./figures/diagrams//UMAgatederive.tikz}}
\endpgfgraphicnamed
\end{equation}
where for the third equality we used the rule (TR11),
for the last equality we made the following transformation by commutativity of the red and green spiders as well as the naturality of diagrams (free move without changing any connections):
\begin{equation}\label{umagatedr2}
\beginpgfgraphicnamed{diagrams//UMAgatederive2}
\InputIfFileExists{diagrams//UMAgatederive2.tikz}{}{\input{./figures/diagrams//UMAgatederive2.tikz}}
\endpgfgraphicnamed
\end{equation}

\end{proof}

%% file: diagrams/generalpsgtelt.tikz
\begin{tikzpicture}
	\begin{pgfonlayer}{nodelayer}
		\node [style=rn] (0) at (0, 0) {};
		\node [style=gbox] (1) at (-0.5, 0.5) {$a$};
		\node [style=none] (2) at (0, 0.75) {};
		\node [style=none] (3) at (0, -0.75) {};
	\end{pgfonlayer}
	\begin{pgfonlayer}{edgelayer}
		\draw (1) to (0);
		\draw (2.center) to (0);
		\draw (0) to (3.center);
	\end{pgfonlayer}
\end{tikzpicture}

%% file: diagrams/generalpsgtert.tikz
\begin{tikzpicture}
	\begin{pgfonlayer}{nodelayer}
		\node [style=rbox] (0) at (0, 0) {$\frac{1-a}{1+a}$};
		\node [style=none] (1) at (0, -0.75) {};
		\node [style=none] (2) at (0, 0.75) {};
	\end{pgfonlayer}
	\begin{pgfonlayer}{edgelayer}
		\draw (2.center) to (0);
		\draw (0) to (1.center);
	\end{pgfonlayer}
\end{tikzpicture}

%% file: diagrams/andcheck.tikz
\begin{tikzpicture}
	\begin{pgfonlayer}{nodelayer}
		\node [style=gn] (0) at (0, 0.25) {};
		\node [style=none] (1) at (0, -0.5) {};
		\node [style=rn] (2) at (0, -0.25) {};
		\node [style=none] (3) at (0, 0.5) {};
	\end{pgfonlayer}
	\begin{pgfonlayer}{edgelayer}
		\draw (2) to (1.center);
		\draw (0) to (3.center);
	\end{pgfonlayer}
\end{tikzpicture}

%% file: conclusions.tex
\chapter{Conclusions and further work}
In this thesis, we first show that the ZX-calculus is complete for the overall pure qubit QM, with the aid of completeness of ZW-calculus for the whole qubit QM.  Thus qubit quantum computing can be done purely diagrammatically in principle. 

Based on this universal completeness, we directly  obtain a complete axiomatisation of the ZX-calculus for the Clifford+T quantum mechanics
by restricting the ring of complex numbers to its subring corresponding to the Clifford+T fragment resting on the completeness theorem of the ZW-calculus for arbitrary commutative ring.

Furthermore, we prove the completeness of the ZX-calculus (with just 9 rules) for 2-qubit Clifford+T circuits by verifying the complete set of 17 circuit relations in diagrammatic rewriting.  This is an important step towards efficient simplification of general n-qubit  Clifford+T circuits.

In addition to completeness results within the qubit related formalism, we extend the completeness of the ZX-calculus for qubit stabilizer quantum mechanics  to the qutrit stabilizer system. This generalisation is far from trivial as one can see that, for example, the local Clifford group for qubits has only 24 elements, while the local Clifford group for qutrits has 216 elements.

Finally, we we show by some examples the application of the ZX-calculus to the proof of generalised supplementarity, the representation of entanglement classification and Toffoli gate, as well as equivalence-checking for the UMA gate. 

The results of this chapter has been published in the paper \cite{coeckewang},  with the coauthor Bob Coecke. 

\section*{Further work}

An obvious next step is to generalise the completeness of the ZX-calculus for qubit to that for qudit  with arbitrary dimension $d$. One possible way to do this is to establish a completeness result for the qudit version of the ZW-calculus. Though there is a ZW-calculus universal for qudit quantum mechanics  \cite{amar}, it is not sure whether the completeness result for qudit can be finally obtained. To be honest, we have tried very hard along this direction, but have not succeeded in the end. Another way to have the completeness of the qudit ZX-calculus is to  prove it directly without any resort to the completeness of the ZW-calculus. There is already a proof of completeness of the qubit ZX-calculus independent of the completeness of  the ZW-calculus \cite{JPV3n}, however we have no idea on how to generalise this result from qubit to qudit. Furthermore, we would like to mention that a normal form of qubit ZX diagrams can be obtained if we resort to elementary transformations represented in ZX-calculus and the map-state duality, thus completeness could follow up. But no elegant ZX rewriting rules have been found in this way to date, so continuous work is still required, including generalisation to qudit case.  

The standard interpretation of the  ZX-calculus maps diagrams to matrices in a  category with objects being tensor powers of a fixed finite dimensional Hilbert space. But in quantum computing,  we often deal with the category  $\mathbf{FdHilb}$, whose objects could be  Hilbert space of arbitrary finite dimension. So it is natural to construct a  ZX-calculus that can deal with quantum computing for various dimensions in the same framework. In fact, we could have such a ZX-calculus by indicating each dimension of the type of a system and introducing the isomorphism between composite  space and its component spaces. Once this mixed version of ZX-calculus is  obtained, then one could fill in all the boxes that one usually encounters in categorical quantum mechanics with theses kind of ZX diagrams. However, the completeness problem of this mixed  version of ZX-calculus keeps open. Even more, one could try to generalise the ZX-calculus to the case of infinite dimension.

Since now we have a set of complete ZX rules for Clifford+T quantum mechanics, it would be of great interest to develop efficient strategies in the ZX-calculus for optimising Clifford+T quantum circuits. One simple idea for this direction is to firstly take the ZX-calculus as a learning machine to learn existing excellent strategies and algorithms by encoding everything in the language of the ZX-calculus, and then try to improve the results obtained.  

As we demonstrated in Section \ref{toffolirepresent},  the Toffoli gate has a simple representation in the ZX-calculus. Then it would be interesting to apply this version of Toffoli gate to the filed of Toffoli based quantum circuits, for example, quantum boolean circuits \cite{Iwamaqbc}.  Meanwhile, Toffoli and Hadamard are universal for quantum computation \cite{Shiyao}, so one could employ the ZX version of Toffoli gate to explore on Toffoli+H quantum circuits instead of Clifford+T circuits.

The completeness of ZX-calculus for 2-qubit Clifford+T circuits depends on the  proof of complete relations given in \cite{ptbian}.  It is unknown whether one can derive a direct proof  from  the universal completeness of the ZX-calculus. Furthermore, can we obtain the completeness of ZX-calculus for 3-qubit Clifford+T circuits,  or even arbitrary n-qubit Clifford+T circuits?

As for the stabilizer formalism, it  is natural to ask whether there is a general proof of completeness of the ZX-calculus for arbitrary dimensional (qudit) stabilizer quantum mechanics. As in the qubit case, we can't expect to use a triangle as a generator to have a complete axiomatisation of the qudit stabilizer quantum mechanics, so the techniques from qudit graph states may still be needed. 
%Except for the increase of the order of local Clifford groups, the main difficulty  of this problem comes from the fact that that it is not known whether any stabilizer state is equivalent to a graph state under local Clifford group for the dimension $d$ having multiple prime factors \cite{Hostens}. As far as we know, it is true in the case that $d$ has only single prime factors  \cite{ShiangGr}.  Furthermore, the sufficient and necessary condition for two qudit graph states to be  equivalent under local Clifford group in terms of operations on graphs is unknown in the case that $d$ is non-prime. Finally, the generalised Euler decomposition rule for the generalised Hadamard gate can not be trivially derived, and other uncommon rules might be needed for general $d$.

Another question suggested by Bob Coecke is to embed the qubit ZX-calculus into  qutrit ZX-calculus, which is not obvious since a representation of some special non-stabilizer  phase is involved.

Lastly, it is also interesting to incorporate the rules of the universally complete ZX-calculus in the automated graph rewriting system Quantomatic  \cite{Quanto}.

%% file: thesismain.bbl
\begin{thebibliography}{10}

\bibitem{cqmwiki}
\url{http://cqm.wikidot.com/zx-completeness}.

\bibitem{Coeckesamson}
Samson Abramsky and Bob Coecke.
\newblock A categorical semantics of quantum protocols.
\newblock In {\em Proceedings of the 19th Annual IEEE Symposium on Logic in
  Computer Science}, LICS '04, pages 415--425, Washington, DC, USA, 2004. IEEE
  Computer Society.
\newblock \doi {10.1109/LICS.2004.1}.

\bibitem{Miriam1}
Miriam Backens.
\newblock The {ZX}-calculus is complete for stabilizer quantum mechanics.
\newblock {\em New Journal of Physics}, 16(9):093021, September 2014.
\newblock \doi {10.1088/1367-2630/16/9/093021}.

\bibitem{Miriam1ct}
Miriam Backens.
\newblock The {ZX}-calculus is complete for the single-qubit {C}lifford+{T}
  group.
\newblock {\em Electronic Proceedings in Theoretical Computer Science},
  172:293--303, 2014.
\newblock \doi {10.4204/eptcs.172.21}.

\bibitem{Miriam2}
Miriam Backens.
\newblock {\em Completeness and the ZX-calculus}.
\newblock PhD thesis, February 2016.
\newblock \href{http://arxiv.org/abs/1602.08954}{arXiv:1602.08954}.

\bibitem{bpw}
Miriam Backens, Simon Perdrix, and Quanlong Wang.
\newblock A {Simplified} {Stabilizer} {ZX}-calculus.
\newblock {\em EPTCS}, 236:1--20, January 2017.
\newblock \doi{10.4204/EPTCS.236.1}.

\bibitem{BaezCR}
John~C. Baez, Brandon Coya, and Franciscus Rebro.
\newblock Props in network theory.
\newblock \href{http://arxiv.org/abs/1707.08321}{arXiv:1707.08321}, 2017.

\bibitem{Mohsen}
Mohsen Bahramgiri and Salman Beigi.
\newblock Graph states under the action of local clifford group in non-binary
  case.
\newblock
  \href{https://arxiv.org/abs/quant-ph/0610267}{arXiv:quant-ph/0610267}.

\bibitem{BianWang1}
Xiaoning Bian and Quanlong Wang.
\newblock Graphical calculus for qutrit systems.
\newblock {\em IEICE Transactions on Fundamentals of Electronics,
  Communications and Computer Sciences}, E98.A(1):391--399, 2015.
\newblock \doi{10.1587/transfun.E98.A.391}.

\bibitem{BONCHI2017144}
Filippo Bonchi, Pawe{\l} Soboci{\'n}ski, and Fabio Zanasi.
\newblock Interacting {Hopf} algebras.
\newblock {\em Journal of Pure and Applied Algebra}, 221(1):144 -- 184, 2017.
\newblock \doi {https://doi.org/10.1016/j.jpaa.2016.06.002}.

\bibitem{borceux_1994}
Francis Borceux.
\newblock {\em Handbook of Categorical Algebra}, volume~2 of {\em Encyclopedia
  of Mathematics and its Applications}.
\newblock Cambridge University Press, 1994.
\newblock \doi {10.1017/CBO9780511525865}.

\bibitem{Boykin}
P.~O. Boykin, T.~Mor, M.~Pulver, V.~Roychowdhury, and F.~Vatan.
\newblock On universal and fault-tolerant quantum computing: a novel basis and
  a new constructive proof of universality for shor's basis.
\newblock In {\em 40th Annual Symposium on Foundations of Computer Science
  (Cat. No.99CB37039)}, pages 486--494, 1999.
\newblock \doi {10.1109/SFFCS.1999.814621}.

\bibitem{ckzh}
Nicholas Chancellor, Aleks Kissinger, Joschka Roffe, Stefan Zohren, and Dominic
  Horsman.
\newblock Coherent parity check construction for quantum error correction.
\newblock \href{https://arxiv.org/abs/1611.08012}{arXiv:1611.08012}.

\bibitem{coeckeduncan2008}
Bob Coecke and Ross Duncan.
\newblock Interacting quantum observables.
\newblock In {\em Automata, Languages and Programming}, volume 5126, pages
  298--310. Springer Berlin Heidelberg, 2008.
\newblock \url {https://doi.org/10.1007/978-3-540-70583-3_25}.

\bibitem{CoeckeDuncan}
Bob Coecke and Ross Duncan.
\newblock Interacting quantum observables: categorical algebra and
  diagrammatics.
\newblock {\em New Journal of Physics}, 13(4):043016, 2011.
\newblock \doi {10.1088/1367-2630/13/4/043016}.

\bibitem{CDKW}
Bob Coecke, Ross Duncan, Aleks Kissinger, and Quanlong Wang.
\newblock Strong complementarity and non-locality in categorical quantum
  mechanics.
\newblock In {\em Proceedings of the 2012 27th Annual IEEE/ACM Symposium on
  Logic in Computer Science}, LICS '12, pages 245--254. IEEE Computer Society,
  2012.
\newblock \doi{10.1109/LICS.2012.35}.

\bibitem{CDKW2}
Bob Coecke, Ross Duncan, Aleks Kissinger, and Quanlong Wang.
\newblock {\em Generalised Compositional Theories and Diagrammatic Reasoning},
  pages 309--366.
\newblock Springer Netherlands, 2016.
\newblock \doi{10.1007/978-94-017-7303-4-10}.

\bibitem{CoeckeEdwards}
Bob Coecke and Bill Edwards.
\newblock Three qubit entanglement within graphical z/x-calculus.
\newblock 52:22--33, 01 2010.
\newblock \doi { 10.4204/EPTCS.52.3}.

\bibitem{Coeckealeksmen}
Bob Coecke and Aleks Kissinger.
\newblock The compositional structure of multipartite quantum entanglement.
\newblock In Samson Abramsky, Cyril Gavoille, Claude Kirchner, Friedhelm Meyer
  auf~der Heide, and Paul~G. Spirakis, editors, {\em Automata, Languages and
  Programming}, pages 297--308, Berlin, Heidelberg, 2010. Springer Berlin
  Heidelberg.
\newblock \doi {10.1007/978-3-642-14162-1_25}.

\bibitem{Coeckebk}
Bob Coecke and Aleks Kissinger.
\newblock {\em Picturing quantum processes}.
\newblock Cambridge University Press, 2017.

\bibitem{coeckewang}
Bob Coecke and Quanlong Wang.
\newblock {ZX}-rules for 2-qubit {Clifford}+{T} quantum circuits.
\newblock In {\em Proceedings of the 10th International Conference, Reversible
  Computation 2018}, LNCS, pages 144--161, 2018.
\newblock \doi {10.1007/978-3-319-99498-7_10}.

\bibitem{Cuccaro}
Steven~A. Cuccaro, Thomas~G. Draper, Samuel~A. Kutin, and David~Petrie Moulton.
\newblock A new quantum ripple-carry addition circuit.
\newblock
  \href{https://arxiv.org/abs/quant-ph/0410184}{arXiv:quant-ph/0410184}, June
  2004.

\bibitem{Shawn}
Shawn~X. Cui and Zhenghan Wang.
\newblock Universal quantum computation with metaplectic anyons.
\newblock {\em Journal of Mathematical Physics}, 56(3):032202, 2015.
\newblock \doi {10.1063/1.4914941}.

\bibitem{domniel}
Niel de~Beaudrap and Dominic Horsman.
\newblock The zx calculus is a language for surface code lattice surgery.
\newblock 06 2017.
\newblock \href{http://arxiv.org/abs/1704.08670}{arXiv:1704.08670}.

\bibitem{DuncanLucas}
Ross Duncan and Maxime Lucas.
\newblock Verifying the steane code with quantomatic.
\newblock In {\em Proceedings of the 10th International Workshop on Quantum
  Physics and Logic, {QPL} 2013, Castelldefels (Barcelona), Spain, July 17-19,
  2013.}, pages 33--49, 2013.
\newblock \doi {10.4204/EPTCS.171.4}.

\bibitem{duncan_graph_2009}
Ross Duncan and Simon Perdrix.
\newblock Graph states and the necessity of {Euler} decomposition.
\newblock {\em Mathematical Theory and Computational Practice}, 5635:167--177,
  2009.
\newblock \doi { 10.1007/978-3-642-03073-4\_18}.

\bibitem{Duncanpx}
Ross Duncan and Simon Perdrix.
\newblock Rewriting measurement-based quantum computations with generalised
  flow.
\newblock In {\em Automata, Languages and Programming}, volume 6199, pages
  285--296. Springer Berlin Heidelberg, Berlin, Heidelberg, 2010.
\newblock \doi{10.1007/978-3-642-14162-1\_24}.

\bibitem{duncanperdrix}
Ross Duncan and Simon Perdrix.
\newblock Pivoting makes the {ZX}-calculus complete for real stabilizers.
\newblock {\em Electronic Proceedings in Theoretical Computer Science},
  171:50--62, December 2014.
\newblock \doi {10.4204/EPTCS.171.5}.

\bibitem{Dur}
W.~Dur, G.~Vidal, and J.~I. Cirac.
\newblock Three qubits can be entangled in two inequivalent ways.
\newblock {\em Phys. Rev.}, A62:062314, 2000.
\newblock \doi {10.1103/PhysRevA.62.062314}.

\bibitem{GilesSelinger}
Brett Giles and Peter Selinger.
\newblock Exact synthesis of multiqubit {Clifford}+{T} circuits.
\newblock {\em Phys. Rev. A}, 87:032332, Mar 2013.
\newblock \doi {10.1103/PhysRevA.87.032332}.

\bibitem{GongWang}
Xiaoyan Gong and Quanlong Wang.
\newblock Equivalence of local complementation and {Euler} decomposition in the
  qutrit {ZX}-calculus.
\newblock \href{http://arxiv.org/abs/1704.05955}{arXiv:1704.05955}.

\bibitem{amar1}
Amar Hadzihasanovic.
\newblock A diagrammatic axiomatisation for qubit entanglement.
\newblock In {\em 2015 30th Annual ACM/IEEE Symposium on Logic in Computer
  Science}, pages 573--584, July 2015.
\newblock \doi {10.1109/LICS.2015.59}.

\bibitem{amar}
Amar Hadzihasanovic.
\newblock {\em The algebra of entanglement and the geometry of composition}.
\newblock PhD thesis, University of Oxford, 2017.
\newblock \href{http://arxiv.org/abs/1709.08086}{arXiv:1709.08086}.

\bibitem{amarngwanglics}
Amar Hadzihasanovic, Kang~Feng Ng, and Quanlong Wang.
\newblock Two complete axiomatisations of pure-state qubit quantum computing.
\newblock In {\em Proceedings of the 33rd Annual ACM/IEEE Symposium on Logic in
  Computer Science}, LICS '18, pages 502--511. ACM, 2018.
\newblock \doi {10.1145/3209108.3209128}.

\bibitem{Herrmann}
Michael Herrmann.
\newblock Models of multipartite entanglement.
\newblock {\em Master thesis, University of Oxford}, 2010.
\newblock \url {http://www.cs.ox.ac.uk/people/bob.coecke/Michael.pdf}.

\bibitem{Horsman}
Clare Horsman.
\newblock Quantum picturalism for topological cluster-state computing.
\newblock {\em New Journal of Physics}, 13(9):095011, 2011.
\newblock \doi{10.1088/1367-2630/13/9/095011}.

\bibitem{Hostens}
Erik Hostens, Jeroen Dehaene, and Bart De~Moor.
\newblock Stabilizer states and clifford operations for systems of arbitrary
  dimensions and modular arithmetic.
\newblock {\em Phys. Rev. A}, 71:042315, Apr 2005.
\newblock \doi{10.1103/PhysRevA.71.042315}.

\bibitem{Mark}
Mark Howard, Eoin Brennan, and Jiri Vala.
\newblock Quantum contextuality with stabilizer states.
\newblock {\em Entropy}, 15(6):2340--2362, 2013.
\newblock \doi {10.3390/e15062340}.

\bibitem{Dan}
Dan Hu, Weidong Tang, Meisheng Zhao, Qing Chen, Sixia Yu, and C.~H. Oh.
\newblock Graphical nonbinary quantum error-correcting codes.
\newblock {\em Phys. Rev. A}, 78:012306, Jul 2008.
\newblock \doi {10.1103/PhysRevA.78.012306}.

\bibitem{CQC}
Cambridge Quantum~Computing Inc.
\newblock
  \url{https://cambridgequantum.com/2017/10/20/collaboration-with-university-of-oxford-computer-science-department/}.

\bibitem{Iwamaqbc}
K.~Iwama, Y.~Kambayashi, and S.~Yamashita.
\newblock Transformation rules for designing cnot-based quantum circuits.
\newblock In {\em Proceedings 2002 Design Automation Conference (IEEE Cat.
  No.02CH37324)}, pages 419--424, June 2002.
\newblock \doi{10.1109/DAC.2002.1012662}.

\bibitem{Emmanuel}
Emmanuel Jeandel, Simon Perdrix, and Renaud Vilmart.
\newblock A complete axiomatisation of the {ZX}-calculus for {Clifford+T}
  quantum mechanics.
\newblock \href{https://arxiv.org/abs/1705.11151}{arXiv:1705.11151v1}.

\bibitem{JPV2}
Emmanuel Jeandel, Simon Perdrix, and Renaud Vilmart.
\newblock Diagrammatic reasoning beyond {Clifford+T} quantum mechanics.
\newblock \href{https://arxiv.org/abs/1801.10142}{arXiv:1801.10142}.

\bibitem{JPV3n}
Emmanuel Jeandel, Simon Perdrix, and Renaud Vilmart.
\newblock A generic normal form for {ZX}-diagrams and application to the
  rational angle completeness.
\newblock \href{https://arxiv.org/abs/1805.05296}{arXiv:1805.05296}.

\bibitem{jpvw}
Emmanuel Jeandel, Simon Perdrix, Renaud Vilmart, and Quanlong Wang.
\newblock {ZX}-calculus: Cyclotomic supplementarity and incompleteness for
  {Clifford+T} quantum mechanics.
\newblock In {\em 42nd International Symposium on Mathematical Foundations of
  Computer Science, {MFCS} 2017, August 21-25, 2017 - Aalborg, Denmark}, pages
  11:1--11:13, 2017.
\newblock \doi {10.4230/LIPIcs.MFCS.2017.11}.

\bibitem{Joyal}
Andre Joyal and Ross Street.
\newblock The geometry of tensor calculus, {I}.
\newblock {\em Advances in Mathematics}, 88:55--112, 1991.
\newblock \doi {10.1016/0001-8708(91)90003-P}.

\bibitem{Quanto}
Aleks Kissinger, Alex Merry, Ben Frot, Bob Coecke, David Quick, Lucas Dixon,
  Matvey Soloviev, Ross Duncan, and Vladimir Zamdzhiev.
\newblock Quantomatic.
\newblock \url{https://quantomatic.github.io/}.

\bibitem{ShiangGr}
Shiang~Yong Looi and Robert~B. Griffiths.
\newblock Tripartite entanglement in qudit stabilizer states and application in
  quantum error correction.
\newblock {\em Phys. Rev. A}, 84:052306, Nov 2011.
\newblock \doi{10.1103/PhysRevA.84.052306}.

\bibitem{Griffiths}
Shiang~Yong Looi, Li~Yu, Vlad Gheorghiu, and Robert~B. Griffiths.
\newblock Quantum-error-correcting codes using qudit graph states.
\newblock {\em Phys. Rev. A}, 78:042303, Oct 2008.
\newblock \doi{10.1103/PhysRevA.78.042303}.

\bibitem{MacLane}
Saunders Mac~Lane.
\newblock {\em Categories for the Working Mathematician}, volume~5.
\newblock Springer New York, 1978.
\newblock \url {http://dx.doi.org/10.1007/978-1-4757-4721-8}.

\bibitem{maclane1965}
Saunders MacLane.
\newblock Categorical algebra.
\newblock {\em Bulletin of the American Mathematical Society}, 71(1):40--106,
  01 1965.
\newblock \url{https://projecteuclid.org:443/euclid.bams/1183526392}.

\bibitem{Malik}
Mehul Malik, Manuel Erhard, Marcus Huber, Mario Krenn, Robert Fickler, and
  Anton Zeilinger.
\newblock Multi-photon entanglement in high dimensions.
\newblock In {\em Frontiers in Optics 2016}, page FW3F.1. Optical Society of
  America, 2016.
\newblock \doi{10.1364/FIO.2016.FW3F.1}.

\bibitem{mmp}
Anne Marin, Damian Markham, and Simon Perdrix.
\newblock {Access Structure in Graphs in High Dimension and Application to
  Secret Sharing}.
\newblock In Simone Severini and Fernando Brandao, editors, {\em {TQC 2013 -
  8th Conference on the Theory of Quantum Computation, Communication and
  Cryptography}}, volume~22 of {\em Leibniz International Proceedings in
  Informatics (LIPIcs)}, pages 308--324, Guelph, Canada, May 2013. {Schloss
  Dagstuhl}.
\newblock 18 pages, 2 figures.

\bibitem{Muthukrishnan}
Ashok Muthukrishnan and C.~R. Stroud.
\newblock Multivalued logic gates for quantum computation.
\newblock {\em Phys. Rev. A}, 62:052309, Oct 2000.
\newblock \doi{10.1103/PhysRevA.62.052309}.

\bibitem{Nagarjuna}
N$\bar{a}$g$\bar{a}$rjuna.
\newblock {\em The Fundamental Wisdom of the Middle Way}.
\newblock Oxford University Press, 1995.
\newblock Translated by Garfield, Jay L.

\bibitem{ngwang}
Kang~Feng Ng and Quanlong Wang.
\newblock A universal completion of the {ZX}-calculus.
\newblock 2017.
\newblock arXiv:1706.09877.

\bibitem{ngwang2}
Kang~Feng Ng and Quanlong Wang.
\newblock Completeness of the {ZX}-calculus for pure qubit clifford+t quantum
  mechanics.
\newblock 2018.
\newblock \href{http://arxiv.org/abs/1801.07993}{arXiv:1801.07993}.

\bibitem{Nielsen}
Michael~A. Nielsen and Isaac~L. Chuang.
\newblock {\em {Quantum Computation and Quantum Information}}.
\newblock Cambridge University Press, Cambridge, 2010.
\newblock \doi{ 10.1017/CBO9780511976667}.

\bibitem{PerdrixWang}
Simon Perdrix and Quanlong Wang.
\newblock {Supplementarity is Necessary for Quantum Diagram Reasoning}.
\newblock In {\em 41st International Symposium on Mathematical Foundations of
  Computer Science (MFCS 2016)}, volume~58 of {\em LIPIcs}, pages 76:1--76:14,
  2016.
\newblock \doi {10.4230/LIPIcs.MFCS.2016.76}.

\bibitem{Ranchin}
Andre Ranchin.
\newblock {Depicting qudit quantum mechanics and mutually unbiased qudit
  theories}.
\newblock In {\em Electronic Proceedings in Theoretical Computer Science},
  volume 172 of {\em EPTCS}, pages 68--91, 2014.
\newblock \doi {10.4204/EPTCS.172.6}.

\bibitem{Vladimir}
Christian Schr\"{o}der~de Witt and Vladimir Zamdzhiev.
\newblock The {ZX}-calculus is incomplete for quantum mechanics.
\newblock {\em EPTCS}, 172:285--292, December 2014.
\newblock \doi {10.4204/EPTCS.172.20}.

\bibitem{selinger_selfdual_2010}
Peter Selinger.
\newblock Autonomous categories in which {$A\cong A^*$}: ({Extended}
  {Abstract}).
\newblock {\em Proceedings of the 7th International Workshop on Quantum Physics
  and Logic (QPL 2010)}, pages 151--160, 2010.

\bibitem{SELINGER2011113}
Peter Selinger.
\newblock Finite dimensional hilbert spaces are complete for dagger compact
  closed categories (extended abstract).
\newblock {\em Electronic Notes in Theoretical Computer Science}, 270(1):113 --
  119, 2011.
\newblock \url{https://doi.org/10.1016/j.entcs.2011.01.010}.

\bibitem{ptbian}
Peter Selinger and Xiaoning Bian.
\newblock Relations for {Clifford}+{T} operators on two qubits.
\newblock {\em Quantum Programming and Circuits Workshop}, June 2015.
\newblock \url{https://www.mathstat.dal.ca/~xbian/talks/slides_cliffordt2.pdf}.

\bibitem{Vivek}
Vivek~V. Shende and Igor~L. Markov.
\newblock On the cnot-cost of {TOFFOLI} gates.
\newblock {\em Quantum Information {\&} Computation}, 9(5):461--486, 2009.
\newblock \url {http://www.rintonpress.com/xxqic9/qic-9-56/0461-0486.pdf}.

\bibitem{Shiyao}
Yaoyun Shi.
\newblock Both toffoli and controlled-not need little help to do universal
  quantum computing.
\newblock {\em Quantum Info. Comput.}, 3(1):84--92, January 2003.

\bibitem{Smith}
A.~Smith, B.~E. Anderson, H.~Sosa-Martinez, C.~A. Riofr\'{\i}o, Ivan~H.
  Deutsch, and Poul~S. Jessen.
\newblock Quantum control in the cs $6{S}_{1/2}$ ground manifold using
  radio-frequency and microwave magnetic fields.
\newblock {\em Phys. Rev. Lett.}, 111:170502, Oct 2013.
\newblock \doi{10.1103/PhysRevLett.111.170502}.

\bibitem{Celso}
Celso Vieira.
\newblock Which is more fundamental: processes or things?
\newblock {\em Aeon}, 2017.
\newblock
  \url{https://aeon.co/ideas/which-is-more-fundamental-processes-or-things}.

\bibitem{qlwang}
Quanlong Wang.
\newblock Qutrit {ZX}-calculus is {complete} for {Stabilizer} {Quantum}
  {Mechanics}.
\newblock {\em Electronic Proceedings in Theoretical Computer Science},
  266:58--70, 2018.
\newblock \doi { 10.4204/EPTCS.266.3}.

\bibitem{BianWang2}
Quanlong Wang and Xiaoning Bian.
\newblock {Qutrit Dichromatic Calculus and Its Universality}.
\newblock {\em EPTCS}, June 2014.
\newblock \doi{10.4204/EPTCS.172.7}.

\bibitem{Markov}
Shigeru Yamashita and Igor~L. Markov.
\newblock Fast equivalence - checking for quantum circuits.
\newblock {\em Quantum Information {\&} Computation}, 10(9{\&}10):721--734,
  2010.
\newblock \url {http://www.rintonpress.com/xxqic10/qic-10-910/0721-0734.pdf}.

\end{thebibliography}
